\documentstyle[preprint,aps,epsfig,floats]{revtex}
\draft

\newcommand{\RG}{renormalization group}
\newcommand{\SCS}{standard convolution scheme}
\newcommand{\MCS}{modified convolution scheme}
\newcommand{\OPE}{operator product expansion}
\newcommand{\DA}{distribution amplitude}
\newcommand{\IMF}{infinite momentum frame}
\newcommand{\MSR}{moment sum rules}
\newcommand{\kv}{{\bf k}_{\perp}}

\newcommand{\bv}{{\bf b}_{\perp}}

\tightenlines
\begin{document}
\date{\today}
\preprint{RUB-TPII-07/99}
\title{THE PHYSICS OF EXCLUSIVE REACTIONS IN QCD: THEORY AND
       PHENOMENOLOGY\\}
\author{N.\ G.\ STEFANIS~\thanks{E-mail:
        stefanis@tp2.ruhr-uni-bochum.de}
       }
\address{Institut f\"ur Theoretische Physik II      \\
         Ruhr-Universit\"at Bochum                  \\
         D-44780 Bochum, Germany                    \\
        }
\maketitle


\begin{abstract}
The modern formulation of exclusive reactions within Quantum
Chromodynamics is reviewed, the emphasis being placed on
the pivotal ideas and methods pertaining to perturbative and
non-perturbative topics. Specific problems, related to scale
locality, infrared safety, gluonic radiative corrections
(Sudakov effects), and the role of hadronic size effects
(intrinsic transverse momentum), are studied. These issues are more
precisely analyzed in terms of the essential mechanisms of momentum
transfer to a hadron while remaining intact. Different factorization
schemes are considered and the conceptual lacunas are pointed out.
The quite technical subject of renormalization-group evolution is
given a detailed account. By combining analytical and numerical
algorithms, the one-gluon exchange nucleon evolution equation is
diagonalized and next-to-leading eigenfunctions are calculated in
terms of Appell polynomials. The corresponding anomalous dimensions
of trilinear quark operators are found to form a degenerate
system whose envelope shows logarithmic large-order behavior.
Selected applications of this framework are presented, focusing on
the helicity-conserving elastic form factors of the pion and the
nucleon. The theoretical constraints imposed by QCD sum rules on the
moments of nucleon distribution amplitudes are used to determine a
whole spectrum of optional solutions. They organize themselves along
an ``orbit'' characterized by a striking scaling relation between the
form-factor ratio $R=|G_{\rm M}^{\rm n}|/G_{\rm M}^{\rm p}$ and the
projection coefficient $B_{4}$ on to the corresponding eigensolution.
The main reasons for the failure of the present theoretical
predictions to match the experimental data are discussed and
workable explanations are sketched.
\end{abstract}

\newpage   

\tableofcontents

\newpage   

\input amssym.def
\input amssym.tex

\section{PREFACE}
\label{sec:pre}
Quantum Chromodynamics (QCD) is that part of the Standard Model which
is supposed to describe the strong interactions at the microscopic
level.
However, despite the numerous phenomenological successes of this
theory in regimes where perturbation theory applies, several
conceptual and technical challenges concerning the large-distance
domain still remain.

Understanding the questions and fixing the problems is possibly not
enough to yield convincing quantitative answers towards an analytic
description of confinement and the formation of hadronic bound
states.
Especially the theoretical analysis of exclusive processes, in which
{\it intact} hadrons appear in the initial and final states,
involves the detailed calculation of hadronic wave functions --
incalculable within a perturbative framework.
Nevertheless, it is tempting to analyze the current status of
knowledge by exposing physical scenarios pertinent to understanding
some basic features in the transition from the perturbative to the
non-perturbative phase.
Probably the greatest technical barrier here is the mutation of light
current quarks -- the degrees of freedom of high-energy QCD -- to
massive constituent quarks, operative after gluon bosonization in
low-energy effective theories.

By now the pure perturbative QCD option to large-momentum transfer
exclusive processes has been proved elusive. The troubles may be a
reflexion of the poorly known complex substructure of hadrons
(intrinsically linked to the non-perturbative QCD vacuum); or the
faults may lie elsewhere, e.g., in the limited knowledge of
higher-order perturbative corrections and the possible lack of
convergence of the perturbative expansion.
A common key element of these difficulties is infrared (IR)
sensitivity due to incomplete factorization of short from
large-distance effects.
As a consequence, either there are intrusions in forbidden
kinematical regimes -- invalidating the initial factorization
assumptions -- or in order to preserve IR finiteness, severe
IR-cutoff prescriptions have to be employed -- resulting in a
depletion of the perturbative contributions.

This report focuses on these problems from a user's perspective,
identifying the principal drawbacks, and trying to set benchmarks for
future developments.
We concentrate on two major subjects: (i) hadron distribution
amplitudes and elastic form factors, and (ii) implementation of the
renormalization group within appropriate factorization schemes.
To set the stage, we discuss and review problems relating to the
momentum (mass) scales involved in form factor calculations: scale
locality, IR safety, gluonic radiative corrections (Sudakov-type form
factors), and the role of hadronic size effects (intrinsic transverse
momenta).
In addition, we give a systematic non-perturbative analysis for
determining hadron distribution amplitudes in conjunction with
QCD sum rules.
These issues are related to distinct mechanisms by which a large
space-like momentum is transferred to an intact hadron and we use
detailed calculations to investigate how these mechanisms influence
the predictions for
$F_{\pi }(Q^{2})$, $G_{\rm M}^{\rm p}(Q^{2})$,
$G_{\rm M}^{\rm n}(Q^{2})$, and
$G_{\rm M}^{*}(Q^{2})$ relative to existing data.
Other phenomenological applications in this context are also
discussed.

\section{INTRODUCTION}
\label{sec:intro}
Lack of precise knowledge about the large-scale inter-quark forces and
the mathematical intractability of the nonlinear structure of the QCD
dynamics preclude an {\it ab initio} analytical calculation of
quantities like the hadron wave functions.
Of the various approaches now being studied, those employing
quark/gluon condensates, the order parameters of non-perturbative QCD,
are among the most successful -- albeit not without their loose ends.
In particular, nonlocal condensates \cite{Shu82} which ascribe a
nonzero average virtuality to vacuum quarks and gluons (a technical
review with earlier references is given in \cite{Gro95}) may enable
one to do more realistic calculations of hadron wave functions, but
such techniques are still in an embryonic phase of development.
Thus far the non-locality of quark/gluon condensates appears only in
model form, though it was found by lattice calculations \cite{CGM84}
that it decreases exponentially with the coordinates.

Frameworks which are capable of simulating the non-perturbative regime
of QCD, either in the continuum (e.g., via QCD sum rules \cite{SVZ79}),
or numerically on the lattice \cite{Wil74}, have been widely used
during the last decade.
Although modern lattice calculations in the strong-coupling regime
are very promising, there are conceptual and technical limitations:
e.g., the properties of a single proton cannot be determined properly
at present.
Especially the computation of higher-order moments (corresponding to
higher derivatives acting on local operators) of light-cone distribution
amplitudes appears very difficult at present \cite{Dan92}.
On the other hand, since hadron wave functions enter only {\it
integrated} quantities, like form factors, their extraction directly
from the data is still in its infancy, and much more experimental input
is needed to make significant progress here.

In the past few years, there have been several theoretical attempts
to emulate low-energy QCD by using effective chiral ($\chi$) actions,
e.g., the Nambu-Jona-Lasinio model (for a recent review, see, e.g.,
\cite{Chr96} and references cited therein).
Such approaches implement chiral symmetry breaking and are, in
principle, deducible from the instanton model of the QCD
vacuum \cite{DP86}, albeit a rigorous derivation is still
rudimentary, and the range of their validity is limited to momentum
values below an ultraviolet (UV) cutoff scale of about 1~GeV.

\begin{figure}
\begin{picture}(0,200)
  \put(79,0){\psboxscaled{500}{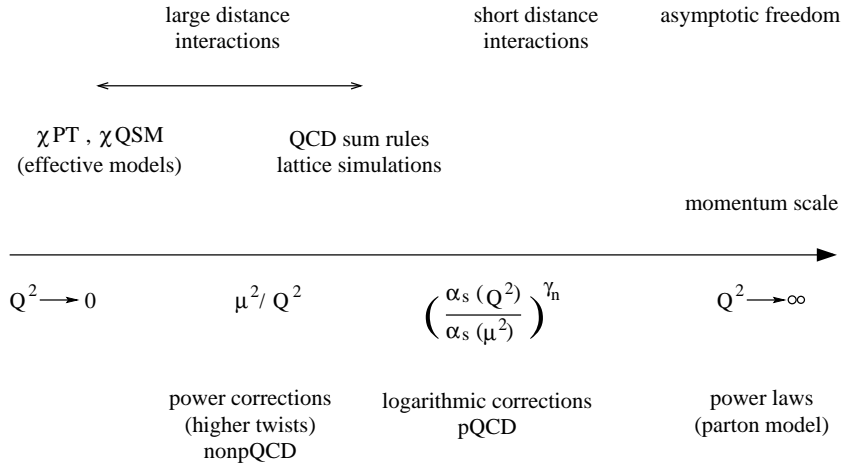}}
\end{picture}
\caption[fig:momflow]
        {\tenrm Perception of strong-interaction dynamics with momentum
         flow.
\label{fig:flow}}
\end{figure}

The perception of strong-interaction dynamics at different resolution
(energy) scales is illustrated in Fig.~\ref{fig:flow}.

%
\input psbox.tex
%

Exclusive processes in QCD are therefore particularly interesting
because they provide the possibility of analyzing hadron wave
functions in terms of their quark and gluon degrees of freedom.
In order to take advantage of perturbative QCD in exclusive
processes,\footnote{``Exclusive'' means that the momenta of all
individual hadrons participating in the process are measured.}
a short-distance amplitude has to be isolated and proved to be IR
finite.
This provides a useful handle on strong-interaction amplitudes and
one is in the position to make quantitative computations in a
systematic fashion.
Formally, this corresponds to the procedure of multiplicative
separation (factorization) of regimes, ubiquitous in many areas of
theoretical physics.
It is the property of cross sections (in inclusive processes) or
amplitudes (in exclusive processes) that high-momenta (``hard'') and
low-momenta (``soft'') regions can be disentangled in such a way that
the factorized parts depend only on the dynamics specific for the
corresponding scale.

Proofs of factorization theorems \cite{APV78,CSS84}, both for
inclusive cross sections and exclusive amplitudes, have reached a
certain level of sophistication, but the subject is still open (for
a comprehensive review, we refer to \cite{CSS89}).
For instance, the generalization of factorization theorems beyond
leading twist faces severe problems due to incomplete Bloch-Nordsieck
(BN) cancellation \cite{BN37} of IR divergences \cite{DFT80,LGHS81}.
Along similar lines of argument, uncanceled IR divergences near the
boundary of phase space (the so-called endpoint region) have to be
re-summed by improving {\RG} techniques to render the perturbative
calculation sound.
Completing this procedure, the second step is to interphase the
factorized parts to renormalization.

According to these ideas, the asymptotic, i.e., large $Q^{2}$-behavior
of electroweak form factors can be calculated with perturbative QCD by
simply assuming free valence quarks entering and exiting the
hard-scattering region, without any recourse to confinement.
It is clear that as the external momentum becomes smaller, the
resolution scale decreases, quantum modes with corresponding
wavelengths emerge and the pure perturbative picture of quasi-free
quarks bound together by single-gluon exchange breaks down.
At this point, fluctuations of the background fields of the
non-trivial QCD vacuum interfere and a self-consistent computation
must take them into account.
Where this transition of the pure perturbative phase to the
non-perturbative one takes place is a point of ongoing
controversy \cite{ILS84,Rad84,Rad90,Kro95}.

A useful framework for describing exclusive processes along these
premises is the convolution scheme of Brodsky and
Lepage \cite{LB79mes,LB79bar,LB80}.
[A somewhat different approach to hard processes was developed
independently by Efremov and Radyushkin \cite{Rad84,ER78,ER80a}.]
Within this scheme, the reaction amplitude becomes the product (the
convolution) of two or more factors, each depending only on the
dynamics specific for that particular momentum (or distance scale),
and the evolution of the factorized parts is {\RG} controlled.
More important, the hard part of the process becomes amenable to
the methodology of QCD perturbation theory.
To enforce IR finiteness, subtractions of IR singularities are imposed.
The guiding idea is that the time scale involved in the hard part of
the amplitude and that one for the formation of the intact hadron in
the final state(s) are disparate, so that the corresponding dynamics
are uncorrelated (impulse approximation).
As a consequence of the sudden character of the hard interactions,
the perturbed quantum mechanical state remains unchanged and time
evolution is governed by the perturbed Hamiltonian.

In leading order of the coupling constant, large-momentum transfer
quark-gluon subprocesses can be adequately described by one-gluon
exchange kernels, the justification being provided by
asymptotic freedom \cite{GW73}.
We will call this approach in the following ``the standard convolution
scheme'' of exclusive processes.
Detailed applications of this type of QCD analysis to hadronic form
factors and decay amplitudes are given in later sections of this
survey.

\begin{figure}
\begin{picture}(0,180)
  \put(70,-3){\psboxscaled{500}{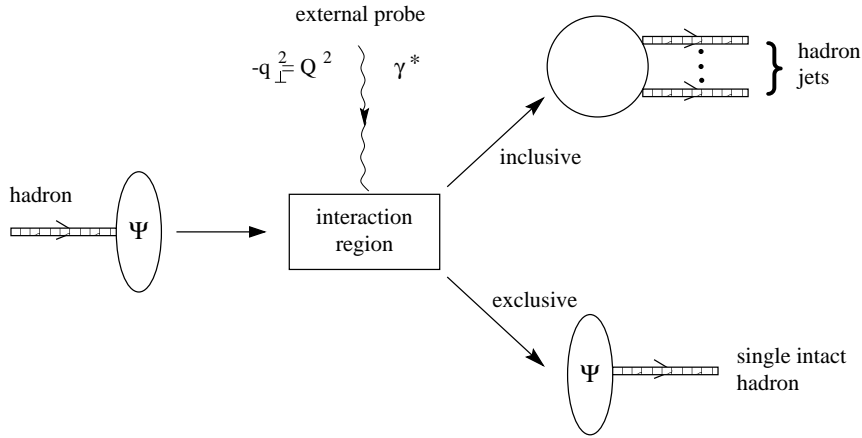}}
\end{picture}
\caption[fig:exclincl]
        {\tenrm Inclusive versus exclusive reactions in QCD.
\label{fig:exinc}}
\end{figure}

Having extracted a (process-dependent) hard-scattering amplitude,
the remaining soft contributions responsible for the bound-state
dynamics are encapsulated in universal (but {\it factorization-scheme
dependent} \cite{Ste95}) hadron {\DA}s.
In axial gauges (e.g., the light-cone gauge, $A^{+}=0$), the {\DA} is
the probability amplitude for the hadron to consist of valence quarks
with (longitudinal) momentum fractions
\hbox{$0\le x_{i}\le 1$}, $\sum_{i}x_{i}=1$
(in an infinite momentum frame)
moving collinearly up to the factorization scale.
Restricting the flow of soft transverse momenta into the {\DA}s amounts
to subtracting their high-momentum tales, and is tantamount to avoid
double counting and ensure that all vertices and propagators entering
the microscopic processes of a Feynman graph are entirely governed by
perturbative QCD.\footnote{A more technical discussion of these points
is postponed to a following section.}
The practical utility of this picture, besides being physically
appealing, resides in the fact that presciently dissecting the
reaction amplitude in the form of a convolution of soft and hard
parts, its evolution behavior, i.e., the variation of the parts with
momentum, is completely controlled by the large scale of the process.
In fact, if the momentum transfer $Q^{2}$ is very large, the UV
regularization scale of the {\RG} can be safely traded for $Q^{2}$ to
give rise to {\RG} evolution.
As the momentum transfer imparted to the hadron by the external
electromagnetic probe increases, inclusive channels are gradually
triggered, and the probability that the hadron rebounds without going
apart diminishes.
This behavior is illustrated in Fig.~\ref{fig:exinc}.

%
%

A theoretical tool to calculate hadron wave functions is provided
by the operator product expansion \cite{Wil69}.
In the {\OPE} language, the hard-scattering amplitude (small-distance
interaction) corresponds to the Wilson coefficient function, whereas
the initial- and final-state hadron wave functions (large-distance
interactions) derive from the matrix elements of local operators
with appropriate quantum numbers for each channel.
A complicating factor in the {\OPE} analysis is that the identification
of a short-distance part in the exclusive amplitude is not {\it a
priori} obvious and has to be determined process by process.

At leading-twist level, the soft parts of the factorized exclusive
amplitude are represented by the valence-state wave functions on the
light cone and describe the distribution in longitudinal momentum
fractions, being averaged over transverse momenta up to values
of the factorization scale.
Contributions from higher Fock states with additional $q\bar q$-pairs
and gluons (higher twists) are suppressed by powers of the momentum
transfer.
Such contributions correspond in the {\OPE} to composite quark-gluon
operators with progressively higher dimensions, the inclusion of
which introduces new order parameters that have to be estimated.

Applications of QCD sum rules to exclusive processes started already
in the early eighties by computing the meson and nucleon form
factors \cite{IS82,NR82}, as well as the first moment
(i.e., the $f_{\pi}$ decay constant) of the pion {\DA}.
Following a different strategy, V. L. Chernyak, A. R. Zhitnitsky, and
I.R. Zhitnitsky (CZ) attempted to reconstruct model {\DA}s from their
few first moments for the pion \cite{CZ82}, the nucleon \cite{CZ84a},
and other hadrons \cite{CZ84b}.
The values of the moments were restricted in their approach by
constraints extracted from QCD sum rules, and model {\DA}s as
polynomials in the longitudinal momenta of the valence quarks were
derived by means of moment inversion.
Fixing the second and fourth moment of the pion {\DA}, CZ proposed a
``double-humped'' model which brings the value of the pion form
factor close to the data for a value of the strong coupling
constant $\alpha _{\rm s}$ as low as 0.3.
This was considered a remarkable success because the asymptotic
prediction yields the value
$Q^{2}F_{\pi}(Q^{2})\approx 0.13$
which falls short a factor of three compared to the
experimental data \cite{Beb76} (though the accuracy of the present
data is rather poor to be conclusive).
A similar analysis for the nucleon on the basis of the first- and
second-order moments, led the same authors to propose a nucleon {\DA}
which shows considerable asymmetry in the distribution of the
longitudinal momentum of the valence quarks.

Soon an alternative nucleon {\DA} was suggested by Gari and Stefanis
(GS) \cite{GS86}, constructed with the aim to yield
helicity-conserving nucleon form factors which account for the
possibility that the electron-neutron differential cross section is
dominated by $G_{\rm E}^{n}$, while $G_{\rm M}^{\rm n}$ is
asymptotically small or equivalently that there is a sizeable neutron
Pauli form factor overwhelming the Dirac one at all $Q^{2}$ values.
The GS model gives very good agreement with the latest high-$Q^{2}$
SLAC data \cite{SLAC86} on $G_{\rm M}^{\rm p}$ (or $F_{1}^{\rm p}$) and
makes realistic predictions for the corresponding neutron form factor
in the high-momentum region \cite{Ste89}.
In contrast, as effected in \cite{Ste89}, the CZ model overestimates
both form factors almost by a factor of two in the region of
10-20~GeV${}^{2}$, if realistic values of $\Lambda _{\rm QCD}$ around
200~MeV are used.\footnote{In the original CZ analysis, the value
$\Lambda _{{\rm QCD}}=100$~MeV was used. Such a low value is now
de facto excluded by experiment.}
But, on the theoretical side, a heavy price is paid: some moments of
the GS model {\DA} cannot match the requirements set by the CZ moment
sum rules in the allowed saturation range \cite{Ste89}.
Moreover, as it was shown later by Chernyak and coworkers \cite{COZ89b},
this model leads to a prediction for the ${}^{3}S_{1} \to p{\bar p}$
decay width of charmonium which is several orders of magnitude smaller
than the experimental value.\footnote{Provided one uses again the
favored value of the strong coupling constant $\alpha _{\rm s}=0.3$.}

In 1987 the moment sum rules were re-evaluated by King and Sachrajda
(KS) \cite{KS87}, spotting the gaps in the CZ analysis and shifting
the range of the {\MSR}, albeit the gross features of the method were
confirmed as was the basic shape of the nucleon {\DA}.

A couple of years later, Chernyak, Ogloblin, and I. R. Zhitnitsky
(COZ) \cite{COZ89a} refined their previous {\MSR} for the second-order
moments and extended their method to third-order moments.
Their new {\MSR} comprise 18 terms with restricted margins of
uncertainty relative to the previous CZ analysis and, in general,
comply with the results of the KS computation, but contradict those
obtained on the lattice for the lowest two moments \cite{RSS87}.
The same authors have also proposed a new model {\DA} for the nucleon
-- still restricted to polynomials of second degree in the longitudinal
momentum -- which satisfies all, but 6 of the new {\MSR}, whereas the
CZ amplitude and the GS one violate, respectively, 13 and 14 of them.
The KS amplitude provides almost the same quality as that of COZ,
with only 7 broken {\MSR}.
In the same year, Sch\"afer \cite{Sch89} presented a variety of model
{\DA}s for the nucleon, which incorporate polynomials of degree three
in the longitudinal momentum, and found that such contributions play
a token role, if properly incorporated.

The essence of these investigations is that {\DA}s extracted from
{\MSR} are much broader than the asymptotic solution (to the evolution
equation) and have a rich structure that is reflected in a quite
asymmetric balance in the distribution of longitudinal momentum
fractions among the valence quarks, accentuated by nodes.
The latter may be understood in the following way:
A valence quark embedded in the non-perturbative QCD vacuum rebounds
differently from one in ``free space''.
Hence, quark/gluon condensates act somewhat like a dispersive medium
in which different components of a wave propagate with different
speeds and tend to change phase with respect to each other
\cite{Ste89}.
Regions of different phases in the nucleon {\DA} may be interpreted as
evidence for binding effects inside the nucleon, amounting perhaps to
diquark formation.
A full-fledged discussion of these issues will be given in the
course of this work.

\begin{figure}
\begin{picture}(0,220)
  \put(136,25){\psboxscaled{600}{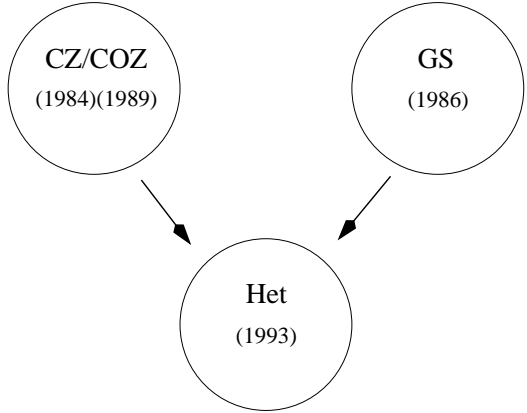}}
\end{picture}
\caption[fig:hetcon]
        {\tenrm Heterotic conception of the nucleon distribution
         amplitude. The heterotic model amalgamates typical
         characteristics of COZ-like and GS-like amplitudes.
\label{fig:hetconcept}}
\end{figure}

The next major step in the determination of nucleon {\DA}s was done in
the early nineties by Stefanis and Bergmann with the invention of the
{\it heterotic conception} \cite{CORFU92,SB93nuc}
(Fig.~\ref{fig:hetconcept}).\footnote{``Heterosis'' in Greek means
increased vigor due to cross-breeding.}
Previously, the CZ model (or its descendants: KS and COZ models) on
one hand and the GS model on the other hand were treated in the
literature as competing alternatives -- mutually excluding each other.
In addition, either way, it was not possible to reconcile the
theoretical {\MSR} constraints with the experimental data because none
of these models is able to give, simultaneously, a quantitatively
satisfactory agreement with the form-factor and charmonium-decay data.

%
%

Perhaps understandably, in view of such distinctive models and
predictions, the conventional view has been one of fragmentation.
But the new idea, underlying the nucleon heterotic model, makes it
possible to amalgamate the best features of COZ-type and GS-type
{\DA}s into a single mold, thus lifting the disparity between theory
and experiment.
As it will become more transparent in the following sections, the
heterotic {\DA} is a ``hybrid'' -- sort of -- and seems to have a foot
in each of the previous models.
This duality is also reflected in its profile which, though
distinctive in overall shape from both the COZ and the GS {\DA}s, bears
geometrical characteristics typical for both models.

Similar ideas of heterosis were then applied by the same authors
\cite{SB93del} to the {\DA} of the $\Delta ^{+}(1232)$ isobar, treating
the {\MSR} of Farrar et al. (FZOZ) \cite{Far88} and those
by Carlson and Poor (CP) \cite{CP88} in combination.
Again a heterotic {\DA} was determined in between the CP and the FZOZ
model {\DA}s.
Using the heterotic {\DA}s for the nucleon and the $\Delta$, the
transition form factor $G_{\rm M}^{*}$ was calculated
\cite{SB93nuc,SB93del} within the {\SCS} and remarkable agreement with
the available data was found.
Even more, a recent reanalysis by Stuart et al. \cite{Stu93,Stu96},
of the inclusive $e-p$ data in the $\Delta (1232)$ region by the SLAC
experiment NE11, combined with low $Q^{2}$ data, high $Q^{2}$ data
from the SLAC experiment E133, and missing mass squared data,
finds results for the transition form factor systematically higher
then the previous data analysis by Stoler \cite{Sto91} and confirms
within the errors the ``heterotic'' predictions \cite{ELBA93}.

These developments gave rise to a more general picture (Stefanis and
Bergmann) \cite{ELBA93,BANSKA93,BS93,BS94}, which deals with {\it
global} features of nucleon {\DA}s.
The crucial question is:
If it is possible to synthesize {\it one} hybrid-type {\DA}, namely the
heterotic one, is it then possible to find also other {\DA}s with
elements belonging to different classes of solutions to the {\MSR}?
That the {\MSR} are not stringent enough to fix the shape of the
nucleon {\DA} uniquely was already pointed out by Gari and Stefanis
\cite{GS87}, and in more mathematical detail by Stefanis \cite{Ste89}.
A similar analysis for the pion {\DA} was given by Mikhailov and
Radyushkin \cite{MR90}.

To determine the possible variation of {\DA}s allowed by {\MSR}, one
has first to identify appropriate order parameters in order to classify
the obtained solutions.
It turns out that the crucial parameters are: (1) the expansion
coefficient $B_{4}$ in the eigenfunctions decomposition of the
nucleon {\DA}, (2) the form-factor ratio
$R\equiv |G_{\rm M}^{\rm n}|/G_{\rm M}^{\rm p}$,
and (3) a $\chi ^{2}$ criterion in the {\MSR} analysis that accounts
for the higher stability of the lower {\MSR} relative to the higher
ones \cite{Ste89}.
In addition, one can define a ``hybridity'' angle \cite{BS94} which
quantifies the information about the mixing of geometrical
characteristics of {\DA}s associated with different sorts of
(asymmetric) longitudinal momentum balance.
The upshot of this treatment is a pattern of unity in diversity.
One is able to track the ``metamorphosis'' of the heterotic {\DA}
across an orbit in the plane spanned by $B_{4}$ and $R$ as it
transforms into the COZ one.
Furthermore, the orbit turns out to be finite, starting at small
$R$-values, associated with the heterotic {\DA}, and terminating at a
COZ-like {\DA} which is a fixed point of the transformation.
Interpolating solutions with comparable $\chi ^{2}$-values are also
determined, showing various degrees of ``hybridity''.
In effect, the emerging orbit provides at a glance a global, and
perhaps more perspicacious, perspective on the information
encapsulated in available {\MSR} \cite{COZ89a,KS87} for the structure
of nucleon {\DA}s.

Regardless of the degree of conviction of particular model {\DA}s,
problems are lurking in the endpoint region of phase space, where
the typical gluon virtualities are much smaller than the external
high momentum.
Indeed, Isgur and Llewellyn-Smith \cite{ILS84}, and also Radyushkin
\cite{Rad84,Rad90}, have pointed out that the contributions from this
region in fact dominate the pion and nucleon form factors, rendering a
perturbative treatment at accessible momentum transfers questionable.
Their main conclusion is that the observed power-law behavior of the
hadron form factors is not due to the perturbative contribution via
hard-gluon exchange, but rather a reflection of the finite size of the
hadrons.
The perturbative contribution takes over, they argue, at
momentum-transfer values too high to be measured in the foreseen
future.
Similar thoughts were more recently expressed by Bolz and
Kroll \cite{Kro95,BK96} on a phenomenological basis.
These authors claim that the broad flat maximum of the (soft) overlap
contributions to the pion and nucleon form factors in the intermediate
momentum-transfer region ``mimics'' the $Q^{2}$-dependence of the
short-distance term.
Again the conclusion is that the bulk of the existing data can be
explained by the soft contribution alone without resort to the
small perturbative contribution.
Such statements could be construed as indicating that perturbative
QCD is irrelevant for this sort of exclusive processes at laboratory
$Q^{2}$.

A serious criticism in this context concerns whether asymmetric
{\DA}s, as those discussed above, are reliable.
It was noted by Stefanis \cite{Ste89} that a {\it finite} set of
low-order moments cannot provide a unique solution, and that the
ensuing variation is inherent in the technique of moments inversion
that is unable to render the diversity of possible solutions small.
It is exactly because of this reason that a ``hierarchical''
$\chi ^{2}$ criterion, which weighs the {\MSR} according to their
order, was used in the ``heterotic'' approach, proposed in
\cite{SB93nuc,SB93del}.
Imposing such a {\it global} criterion, the sensitivity to
disregarded higher-order terms is {\it de  facto} averted \cite{BS93}.

Another objection, intimately tied to the QCD sum-rule method, was
raised by Radyushkin and expounded in several articles (see, e.g.,
\cite{Rad90}).
The point is that {\it local} condensates assign zero virtuality to
vacuum quarks corresponding, equivalently, to infinite correlation
lengths of vacuum field fluctuations and, since the higher-order
moments are sensitive to the size of the non-locality (or inverse
quark virtuality), this effect should be taken into account.
This is perhaps a turning point in the conceptual evolution of QCD
sum rules, but crucial technical challenges remain, e.g., the {\OPE}
has to be generalized to nonlocal operators.
Nevertheless, following this rationale, Mikhailov and Radyushkin
\cite{MR86}, and Bakulev and Radyushkin \cite{BR91} (see also
\cite{Mik93}) were able to obtain pion wave functions, modeling the
vacuum quark virtuality distribution by a Gaussian around an average
virtuality ranging from 0.4~GeV${}^{2}$ \cite{BI82} to
1.2~GeV${}^{2}$ \cite{Shu89}.
A more recent analysis by Dorokhov, Esaibegyan, and Mikhailov
\cite{DEM97} attempts to calculate quark and gluon virtualities within
the model of the liquid instanton vacuum \cite{DP86,Shu89}.
The wave functions derived this way are broader than the asymptotic
solution of perturbative QCD, but, unlike the CZ amplitude, have no
dip in the central region.
It goes without saying that the corresponding second and fourth order
moments are close to their asymptotic values.
These investigations seem to confirm again the dominance of the soft
contribution to the pion form factor at intermediate values of momentum
transfer.

Unfortunately, the extension of this method to the nucleon case is
not straightforward.
New nonlocal quark-gluon condensates of higher dimension emerge, which
have to be estimated, but cannot be reduced to condensates of lower
dimension because they explicitly violate the ``factorization
hypothesis'' \cite{SVZ79}.

Putting our comments so far together, we draw two important conclusions:
(1) the endpoint region of exclusive processes is IR sensitive and has
to be considered very carefully; (2) the {\MSR} have defined a much more
profitable goal to work towards than an elaborate formalism for
calculating reliable {\DA}s.
But however incompletely we may comprehend the details of how the
complex structure of the non-perturbative QCD vacuum works, we all
understand that until more advanced techniques are really available,
the {\MSR} and the herewith derived model {\DA}s are revealing.
In succeeding sections, we focus on the ramifications and
implications of issue (1) considering assessment (2) as a sound
working hypothesis.

While the non-perturbative non-locality of condensates may eliminate
a source of errors in the determination of wave-function moments,
another type of non-locality, namely in transverse configuration
space, may help out in restoring the validity of the perturbative
treatment in the endpoint region by permitting the divergences due
to soft gluons to be {\it screened} by means of (re-summed) Sudakov
effects.
We do not propose to review these extensive investigations in great
detail here, but we do sketch the key ideas in connection with
form-factor calculations and complete this review in
Sec.~\ref{sec:mod} with some concrete examples.

In Abelian theories, like Quantum ElectroDynamics (QED), the emission
of soft real ({\it bremsstrahlung}) quanta prevents the vanishing of
the cross section by giving rise to BN cancellation between (soft)
real and virtual photons \cite{BN37}.
This cancellation does not occur within perturbation theory because
these two different sorts of soft quanta correspond to different
powers of the coupling constant, rent by the iteration procedure.
Hence, one needs closed-form all-order expressions to accomplish
cancellation.
In the IR regime of non-Abelian theories (like QCD), where the ratios
of all physical scales relative to the IR-cutoff scale are large,
cross sections of non-forward processes involving {\it isolated}
colored particles, whether or not an infinite number of soft gauge
bosons (gluons) are included, vanish in the IR limit.
In contrast, processes involving only neutral (composite) particles
have non-vanishing cross sections in this limit.
It transpires that colored amplitudes at large momentum transfers
are suppressed by damping exponential (Sudakov) factors -- one for
each non-color-singlet, near mass-shell particle \cite{CT76}.

The Sudakov form factor is the probability for no emission of soft
photons (gluons) with increasing momentum transfer in hard
photon-electron scattering within QED (QCD).
Sudakov found \cite{Sud56} that the one-loop result exponentiates
to give a double logarithm of the form
$
 \exp \left[
            - (\alpha /c \, \pi )
              \ln ^{2} \left( Q^{2}/\mu ^{2} \right)
      \right]
$,
$\mu$ being the small scale of the system; e.g., the invariant
squared mass of the electron \cite{Sud56} ($c=2$), or an auxiliary
photon mass in the case of on-shell external fermions ($c=4$)
\cite{Jac68,FS71,AP71}.
The dominance of the leading double-logarithmic term in the
high-momentum limit was confirmed by Mueller \cite{Mue79}, and
Collins \cite{Col80} by re-summing non-leading single logarithms to
all orders of the fine-coupling constant $\alpha$.
The upshot is that, asymptotically, the Sudakov form factor drops to
zero faster than any power of $Q^{2}$.

The generalization of this type of calculations to QCD was initiated
by Cornwall and Tiktopoulos \cite{CT76}.
They computed the non-Abelian vertex function up to the three-loop
order and conjectured exponentiation of the leading double-logarithmic
result.
The formal proof of exponentiation was provided by Belokurov and
Ussyukina \cite{BU80}, and by Dahmen and Steiner \cite{DS80}.
The inclusion of single-logarithmic terms was supplied by
Sen \cite{Sen81}, along the lines of Collins' previous QED
analysis \cite{Col80}.
An elaborate and systematic re-summation approach for leading and
non-leading logarithms was conducted by Collins and Soper \cite{CS81}
(see also \cite{CSS85,Ster96}), using eikonalization and {\RG} type
equations in the axial gauge, with recourse to the Grammer-Yennie
method \cite{GY73}.
They found that the leading contributions stem from integration
regions where either all four-momentum components of gluons are
small (``soft region'') or from gluons collinear to external
(massless) quark lines.\footnote{Whether a gluon belongs to the
soft or the collinear set depends on the gauge.}
When such loop momenta regions overlap they give rise to double
logarithms of the generic form
$
 \exp \left\{
             - c \ln \left( \frac{Q^{2}}{\mu ^{2}} \right)
                 \ln \left[
                 \frac{\ln \left( Q^{2}/\Lambda _{{\rm QCD}}^{2}
                           \right)}
                      {\ln \left( \mu ^{2}/\Lambda _{{\rm QCD}}^{2}
                           \right)}
                     \right]
      \right\}
$,
in which the constant $c$ depends on the color algebra via the {\RG}
effective coupling
$\beta _{0}=\left( 11-2n_{\rm f}/3 \right)/4$,
where $n_{\rm f}$ is the number of quark flavors.
In our considerations to follow $n_{\rm f}$ is taken to be equal to 3.

\begin{figure}
\begin{picture}(0,330)
  \put(78,25){\psboxscaled{600}{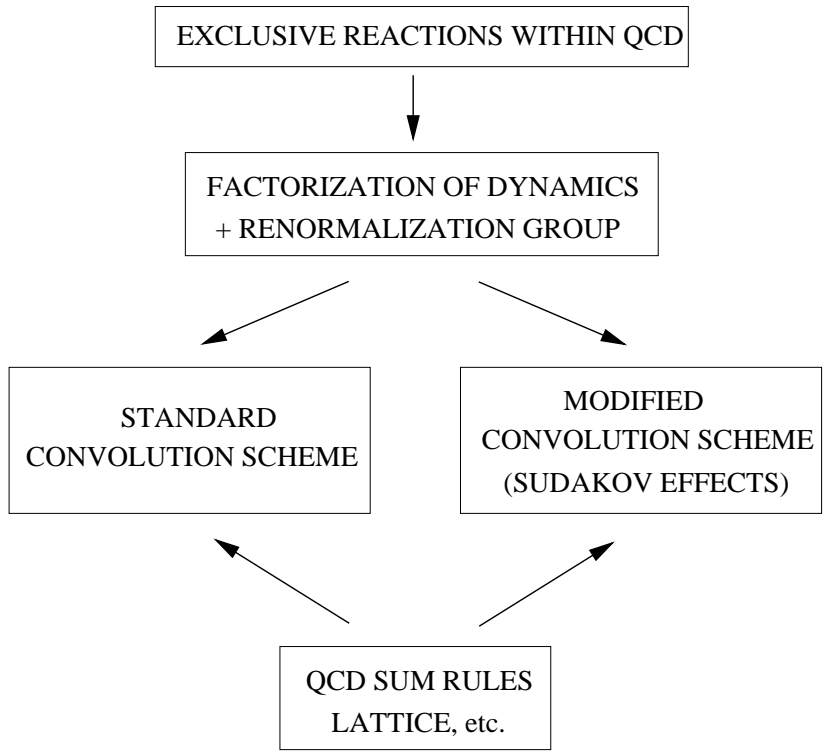}}
\end{picture}
\caption[fig:chartschemes]
        {\tenrm Flow chart illustrating two possible ways of realizing
         factorization of exclusive reactions in convolution
         form within QCD.
\label{fig:chart}}
\end{figure}

One crucial clue of Sudakov effects in exclusive processes came
into view, largely as the result of work by Botts and Sterman
\cite{BS89}.
They realized that the IR cutoff in the Sudakov function can be
traded for the inter-quark separation in transverse configuration
space (after Fourier transformation), thus providing an {\it in situ}
regularization of IR divergences without introducing external IR
regulators.
Conventionally, transverse momenta in the hard-scattering amplitude
are dispensed with on the assumption that they are negligible
compared to the large scale $Q^{2}$.
This is true -- no doubt -- as long as all momentum flows within
a Feynman graph are large.
However, in the endpoint region, as already mentioned, some
momentum can formally come close to zero, becoming smaller than the
neglected transverse momentum.
Retaining the flow of transverse momenta into the hard-scattering
region, is actually tantamount to modifying factorization because
the momentum region at the interface between the true confinement
regime -- parameterized in the wave functions -- and the true hard
regime -- expressed through the short-distance part -- is explicitly
taken into account in the convolution of the reaction amplitude.
In the axial gauge, all Sudakov effects can be encapsulated
in wave-function-like factors, one for each incoming and out-coming
quark line, that link the soft wave functions with the hard
scattering region.
This amounts to a {\it finite} wave function renormalization
(Stefanis \cite{Ste95}).
The modified convolution (factorization) scheme is illustrated
in Fig.~\ref{fig:chart}.

%
%

In this vein, Li and Sterman \cite{LS92} were able to show that
in the pion case there is strong Sudakov suppression for large
inter-quark separation as $Q^{2}$ increases.
Whence, contributions from the endpoint region, where perturbation
theory fails, become less and less important.
Indeed, the bulk of the pion form factor accumulates in regions
where the coupling constant is less than about 0.7 and the transverse
distances are moderate.
This is a remarkable result, for no external regulators are needed
to saturate the strong coupling at small momenta, the IR protection
being intrinsically provided by the transverse inter-quark separation
alone.

As regards three-quark states, the situation is more complicated
because several transverse scales are involved and therefore IR
protection is not automatically accomplished.
It was shown by Kroll and Stefanis, and their respective
collaborators \cite{BJKBS95pro}, that the simple extension of the
pion analysis to the nucleon case by Li \cite{Li93} fails.
The main reason is that there exist kinematical regimes where none
of the Sudakov factors provides suppression for arbitrary choices
of the inter-quark distances.
One has to correlate the different transverse scales and adopt a
{\it common} IR cutoff scale that is taken to be the maximum
transverse separation (``MAX'' prescription).
The underlying physical idea is that due to the color neutrality
of a hadron, its quark distribution cannot be resolved by gluons
with a wavelength much larger than an average inter-quark separation
scale.
Thus, gluons with wavelengths large compared to the (transverse)
hadron size probe the hadron as a whole, i.e., in a color-singlet
state and decouple.
As a result, quarks in such configurations act {\it coherently} and
therefore (soft) gluon radiation is dynamically inhibited.

The particular advantage of this approach \cite{BJKBS95pro,Ste95} is
that (1) it suffices to protect the amplitudes from becoming singular
for all possible kinematical configurations, and (2) it yields a
perturbative contribution to the nucleon form factors which {\it
saturates}, i.e., which is rather insensitive to distances of order
$1/\Lambda _{{\rm QCD}}$.
Other IR regularization choices, for instance that adopted by Li, may
lead to un-compensated singularities (see for more details, in
Sect.~\ref{sec:mod}).
Results for both the proton \cite{BJKBS95pro} as well as the neutron
form factors \cite{BJKBS95neu} were obtained, taking also into
account the intrinsic transverse momenta in the nucleon wave function
in the realm of the pion analysis of Jakob and Kroll \cite{JK93}.
The dark side of these theoretical improvements is that the
calculated nucleon form factors fall short by at least a factor of
two relative to the experimental data.
The possible reasons for this depletion will be discussed in
subsequent sections.

The remaining part of this report is divided into two main parts.
The first part deals with hadron {\DA}s (Sect. \ref{sec:con}) and
form factors (Sect. \ref{sec:elff}) within the {\SCS}.
In the second part (Sect. \ref{sec:mod}) we treat similar topics
within the {\MCS}.
We close this review with a further in-depth discussion of the
whole picture, attempting at providing a stimulating outlook for
the future.
Some important technical issues and useful formulae are collected in
the appendixes.

\section{STANDARD CONVOLUTION SCHEME}
\label{sec:con}
The theoretical tools for the description of exclusive processes
are the hard-scattering amplitude -- which describes the
process-dependent quark-gluon interactions within perturbative QCD --
and the probability amplitude for finding the lowest-twist Fock
state ({\it alias} the {\DA} for the valence state) in a hadron.
The total exclusive amplitude is then represented by the convolution
of these three factors \cite{LB80,ER80a}, assuming factorization of
large-momentum flow regions from those of soft transverse momenta,
necessary to form bound states.
In terms of the {\OPE}, this corresponds to truncation at leading twist
($t$=dimension--spin) level, namely $t=2$ (meson case) and $t=3$
(nucleon case).
Higher-twist components, corresponding to a higher number of partons
(quark-pairs and gluons), are suppressed by powers of the large scale,
i.e., the momentum transfer $Q^{2}$.
To clarify the physical meaning of these statements, we now turn to
factorization.

\subsection{FACTORIZATION}
\label{subsec:fact}
Factorization theorems are of central importance in quantum field
theory.
The basic idea is that one can separate high-momentum from low-momentum
dependence in a multiplicative way.
For example, proving that UV divergences occurring in Feynman graphs can
be absorbed into multiplicative renormalization factors (infinite
constants) is instrumental in establishing renormalizability.
The technical challenge is to prove factorization of a particular
QCD process to {\it all orders} in the coupling constant going
beyond leading logarithms \cite{CSS89}.
These difficulties derive from the fact that in QCD a new type of
IR-divergence is encountered, the {\it collinear} divergence, and
that in higher orders the self-coupling of gluons becomes important
in the exponentiation of IR divergences.

The realization of factorization when applying to elastic form
factors can be written in the form of a convolution of the
hard-scattering amplitude (dubbed $T_{\rm H}$) and two soft wave
functions corresponding to the incoming and out-coming
hadron (termed $\Phi$).
Generically (i.e., absorbing all integrations over internal variables
into $\otimes$), one has
\begin{equation}
  F(Q^{2})
=
  \Phi ^{\rm out}(m/\mu )
  \otimes
  T_{\rm H}(\mu /Q )
  \otimes
  \Phi ^{\rm in}(m/\mu ) \; ,
\label{eq:F(Q^2)}
\end{equation}
where $m$ sets the typical (small) virtuality in the soft parts and
$Q$ is the (external) large scale, characteristic of the hard
(parton) subprocesses.

The matching scale $\mu$ at which factorization has been performed
is arbitrary and, assuming that $\mu \gg m$, it can be safely
identified with the renormalization scale -- unavoidable in any
perturbative calculation -- by virtue of the {\RG} equations.
In this way, $F$ can be rewritten as a function only of the
coupling constant operative at that same scale, the latter being
identified with the large external scale.

As long as {\it scale locality} is preserved, i.e., as long as the
variation of the effective coupling constant with $\mu$ is governed
by the {\it same} momentum scale, and the limit $m \to 0$ is finite,
Eq.~(\ref{eq:F(Q^2)}) is valid because {\it intrusions} from the
hard into the soft regime are prohibited.
Potential IR divergences in $T_{\rm H}$ are removed by subtractions on
account of the properties of the wave function parts.
This means that $T_{\rm H}$ in leading order is by definition
insensitive to long-distance interactions, i.e., it is {\it IR safe}.
All IR-sensitivity resides in the hadron distribution amplitudes
$\Phi ^{{\rm in}({\rm out})}$ which are peaked around small transverse
momenta of the order $m^{-1}$ with their large-momentum tails removed
by cutting off from above the integration over transverse momenta.
Their dependence on the cutoff scale is mild and is governed by
{\RG}-evolution.
Both the subtraction procedure of the large momentum tails in the
soft parts and the cancellation of IR divergences in the hard part
are not unique.
They actually define the factorization procedure adopted, thus
imposing an {\it implicit} factorization-scheme dependence on the
hadron {\DA}s \cite{Stenew}.
Of course the asymptotic behavior of $F$ should not be affected and
still be determined by the asymptotic {\DA} evolved with the leading
(i.e., lowest-order) anomalous dimension, associated with vertex and
quark self-energy corrections.

\subsection{SHORT DISTANCE PART
\label{subsec:short}}
Invoking factorization, the leading order expression for the
helicity-conserving hadron form factor can be cast in the form
\begin{equation}
  F(Q^{2})
=
  \int_{0}^{1}[dx]\int_{0}^{1}[dy] \,
  \Phi ^{*}(y_{i},\tilde{Q}_{y}) \,
  T_{\rm H}(x_{i}, y_{i}, Q) \,
  \Phi (x_{i}, \tilde{Q}_{x}) \; ,
\label{eq:G(Q^2)}
\end{equation}
where
\begin{equation}
 [dx] = \delta \, \left(
                       1 - \sum_{i=1}^{n} x_{i}
                 \right)
        \prod_{j=1}^{n} \, dx_{j}
\end{equation}
\label{eq:xmeasure)}
($n=2$ for mesons and $n=3$ for baryons), and
\hbox{
$
 \tilde{Q}_{x} = {\rm min}(x_{i}Q)
$
     }
or
\hbox{
$
 \tilde{Q}_{x} = {\rm min}\left((1-x_{i})Q\right)
$
     }
with analogous definitions for the $y$ variable.
The form factor is the probability for the hadron to absorb large
transverse momentum while remaining intact.
The bound-state (i.e., confining) dynamics is encoded in $\Phi$,
while in the hard-scattering amplitude $T_{\rm H}$ the hadron is
simulated by on-mass-shell (but off-light-cone energy) valence quarks
with negligible mass and transverse momentum.
To leading order $\alpha _{\rm s}$, $T_{\rm H}$ is the sum of all Born
diagrams contributing to the particular process.
The transition from the initial to the final state is supposed to go
via hard gluon re-scattering which involves a factor
$
 (\alpha _{\rm s}(Q^{2})/Q^{2})
$,
($Q^{2}\equiv q_{\perp}^{2} = -q^{2}$)
after one quark was derailed from the initial to the final direction.
To leading order, the contributions of $q\bar q$-irreducible diagrams
to the pion form factor amount to the hard-scattering amplitude
\begin{equation}
  T(x,y,Q)
=
  \frac{16\pi C_{\rm F} \alpha _{\rm s}(Q^{2})}{{\bar x}{\bar y}Q^{2}}
\; ,
\label{eq:pihard}
\end{equation}
where the abridged notation ${\bar x}=1-x, {\bar y}=1-y$ has been used,
$C_{\rm F}=(N_{\rm c}^{2}-1)/2N_{\rm c}=4/3$
is the Casimir operator of the fundamental representation of
$SU(3)_{\rm c}$, and
$
 \alpha _{\rm s}(Q^{2})
=
 \left(4\pi /\beta _{0}\ln (Q^{2}/\Lambda _{{\rm QCD}}^{2})\right)
$
is the running coupling constant in the one-loop approximation.

\begin{figure}
\begin{picture}(0,200)
  \put(130,15){\psboxscaled{400}{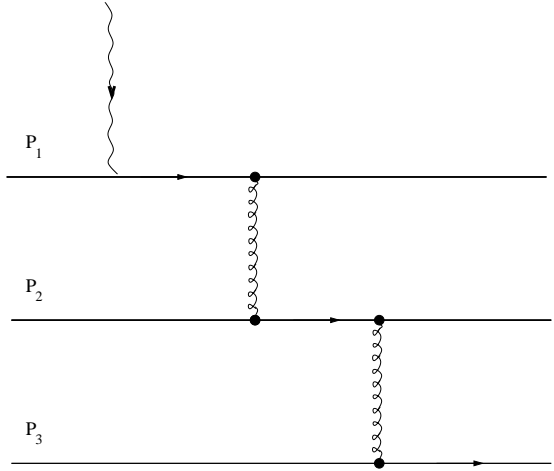}}
\end{picture}
\caption[fig:feynman]
        {\tenrm Example of a Feynman graph contributing to the nucleon
         form factor at tree level.
\label{fig:nucfey}}
\end{figure}

In the nucleon case, there is a total of 14 Born diagrams with
different topologies out of which only 8 give non-zero
contributions.
A complete list of these diagrams is compiled in \cite{CZ84b}; the
Feynman rules for light-cone perturbation theory are given
in \cite{LB80}.
A typical diagram contributing to the nucleon form factor is
depicted in Fig.~\ref{fig:nucfey}.
%
%
%
%
The quark-propagator denominators along the first and the second quark
lines are, respectively:
$-Q^{2}{\bar x}_{1}$ and $-Q^{2}x_{3}{\bar y}_{1}$; those of the
gluon propagators which channel the external momentum flow from the
struck quark to the spectator quarks are:
$-Q^{2}{\bar x}_{1}{\bar y}_{1}$ and $-Q^{2}x_{3}y_{3}$,
where all quark masses and $k_{\perp}$-momenta have been set equal
to zero (collinear approximation).
Recall that in the light-cone frame, parton ``i'' in a hadron has
four-momentum
\hbox{
$
 p_{i} = x_{i}P+k_{i} = (p^{+},p^{-},p_{\perp})
$
     },
where
$x_{i} = \frac{p_{i}^{+}}{P^{+}}$
and
$k_{i} = (0,k^{-},{\bf k}_{\perp})$.
Momentum conservation implies
\begin{equation}
  \sum_{i=1}^{n}x_{i}=1 \; ; \;\;\;\;\;
  \sum_{i=1}^{n}{\bf k}_{\perp i}=0 \; .
\label{eq:momcon}
\end{equation}
Quarks are on-mass shell, i.e.,
$p_{i}^{2}=m_{i}^{2}$ but off the light-cone energy:
\begin{equation}
  p_{i}^{-}
=
  \frac{\left(
        {\bf p}_{\perp i} + {\bf k}_{\perp i}
        \right)^{2}
  + m_{i}^{2}}{2p_{i}^{+}} \; .
\label{eq:offen}
\end{equation}
In a frame where $P=(P^{+}=1,0,0_{\perp})$ and for massless quarks,
this relation simplifies to
\begin{equation}
  p_{i}^{-}
=
  \frac{{\bf k}_{\perp i}^{2}}{2x_{i}} \; .
\label{eq:offenrest}
\end{equation}
Note that the sum over all $p_{i}^{-}$ is not equal $P^{-}$.
The difference is a boost-invariant measure of how far off energy shell
a Fock state is \cite{LB80,BL89}.
This off-shellness is large in the kinematic endpoint region, i.e., when
${\bf k}_{\perp i}^{2}$ or $x_{i}$ is small and, as a consequence, the
hadron wave function should vanish in these limits.
Hence, formally, all wave functions should satisfy the boundary
conditions
\begin{eqnarray}
& &
  {\bf k}_{\perp i}^{2}\,
  \psi _{n}(x_{i},{\bf k}_{\perp i},\lambda _{i})
\rightarrow 0
\quad\quad\quad {\rm as}
\quad\quad\quad
  {\bf k}_{\perp i}^{2}
\rightarrow \infty
\nonumber \\ & \\
\label{eq:boundcond}
& &
  \psi _{n}(x_{i},{\bf k}_{\perp i},\lambda _{i})
\rightarrow 0
\quad\quad\quad {\rm as}
\quad\quad\quad
  x_{i}
\rightarrow 0
\nonumber
\end{eqnarray}
if the free-particle Hamiltonian is to have a finite expectation
value ($\lambda _{i}$ denotes here parton's ``i'' helicity).
None of these conditions is generally satisfied in the absence of
UV (${\bf k}_{\perp} \to \infty$) and IR ($x \to 0$)
regulators (for a further discussion of these subtle issues, we refer
to \cite{BL89}).

\subsection{LARGE DISTANCE PART: PERTURBATIVE ASPECTS}
\label{subsec:largepert}

\subsubsection{Renormalization}
\label{subsubsec:renorm}
The function $\Phi (x_{i},\tilde Q)$ is the probability distribution
amplitude for finding the hadron to consist of valence quarks each
carrying a fraction
$
0 \le x_{i} \le 1
$
of the hadron's longitudinal momentum
$P^{+}$ ($P^{\pm}=(P^{0}\pm P^{3})/\sqrt{2}$)
at transverse separations (relative to the hadron's $P^{+}$ momentum)
not smaller than ${\tilde{Q}}^{-1}$.
We work in the infinite momentum frame, where
$\vec{P}=(0,0,P^{3})$ and $P^{3}=|\vec{P}|\to \infty$.
In this frame the $p^{+}$ component is large and conserved (the same
for $\vec{p}$), and the $p^{-}$ component is small and not conserved
(the same for $E$).
Furthermore, boosts along the $+z$-direction (``+3''-direction)
leave the transverse momenta unchanged.
In leading order, $\tilde{Q}_{x(y)}$ can be replaced by $Q$, which
is then the only large scale involved.
The small scale of the system, called $m$ in Eq.~(\ref{eq:F(Q^2)}),
does not enter the game explicitly (provided it is much smaller than
$Q$); it nevertheless plays a crucial role in the determination of
$\Phi$ and is the non-perturbative counterpart of the perturbative
scale $\Lambda _{{\rm QCD}}$.
Physically, the existence of such a scale indicates that gluons with
wavelengths larger than $m^{-1}$ do not ``see'' individual quarks and
decouple.
One should eventually be able to relate $m$ with typical
non-perturbative scales, like the average quark virtuality in the
vacuum \cite{BI82}, mentioned in the Introduction, or the size of
instantons \cite{DP86,Shu89}.

Dispensing with the $k_{\perp}$-dependence in $T_{\rm H}$, allows to
integrate out this variable in the wave function parts up to the
factorization scale $\mu ^{2}$ and replace the (renormalized) wave
function by the distribution amplitude
\begin{equation}
  \Phi (x_{i},\mu ^{2})
\equiv
  \left(
        \ln \frac{\mu ^{2}}{\Lambda _{{\rm QCD}}^{2}}
  \right)^{- c\,\gamma _{\rm F}/\beta _{0}}
  \int_{}^{} \prod_{i=1}^{n} [d^{2}{\bf k}_{\perp i}]\,
  \psi (x_{i},{\bf k}_{\perp i})\,
  \Theta \left(\mu ^{2} - |{\bf k}_{\perp i}|^{2}\right) \; ,
\label{eq:phidef}
\end{equation}
where
\begin{equation}
  [d^{2}{\bf k}_{\perp i}]
\equiv
  16 \pi ^{3} \delta ^{(2)}\left(\sum_{i}^{n}{\bf k}_{\perp i}\right)
     \prod_{j}^{n}\frac{d^{2}{\bf k}_{\perp j}}{16 \pi ^{3}}
\label{eq:kperpmeas} \; ,
\end{equation}
and $\gamma _{\rm F}$ is the anomalous dimension associated with quark
self-energy in the light-cone gauge:
\begin{equation}
  \gamma _{\rm F}
=
  C_{\rm F}\left(
             1 + 4 \int_{0}^{1}dx\, \frac{x}{1-x}
       \right) \; .
\label{eq:gammaf}
\end{equation}

The logarithm in front of the integral in Eq.~(\ref{eq:phidef})
stems from UV divergences due to gluon radiative corrections in the
hard-scattering amplitude.
Hence, all vertices and propagators have to be renormalized.
However, $Z_{3}^{\rm g}$, the gluon renormalization constant, and
$Z_{1}^{\rm qg}$, the quark-gluon renormalization constant, can be
absorbed (removed) by replacing $\alpha _{\rm s}$ by its renormalized
value
$
 \alpha _{\rm s}^{{\rm ren}}
=
 (Z_{2}^{\rm q}/Z_{1}^{\rm qg})^{2}\, Z_{3}^{\rm g}\, \alpha _{\rm s}
$
so that the only renormalization constant to be determined is
$Z_{2}^{\rm q}$ which renormalizes the quark propagator (or, by taking
its square root, each incoming and outgoing quark line).
But $Z_{2}^{\rm q}$ can be computed from the gluon radiative
corrections to the quark-photon vertex by virtue of the QED Ward
identity
$Z_{1}^{{\rm q}\gamma}=Z_{2}^{\rm q}$.
It is evident from our factorization prescription that only momenta
$\tilde Q \lesssim Q$
are included in the wave function, meaning in turn that the virtual
photon's momentum probing it is also limited to $\tilde Q$.
Hence, in computing the UV-divergent part of the photon-quark vertex,
the photon momentum can be ignored, keeping only gluon loop momenta
greater than $\tilde Q$.
Then $Z_{2}^{\rm q} \to Z_{2}^{\rm q}(\tilde Q)$ and as $\tilde Q$
increases, $Z_{2}^{\rm q}(\tilde Q)$ (i.e., the probability to find a
bare quark) decreases.
The asymptotic result is
\begin{equation}
  Z_{2}^{\rm q}
=
  \lim_{\mu ^{2} \to \infty} \ln
  \left(
        \frac{\mu ^{2}}{\Lambda _{{\rm QCD}}^{2}}
  \right)^{- c \gamma _{\rm F}/\beta _{0}} \; ,
\label{eq:quarkrencon}
\end{equation}
where we have used the fact that in leading order
$
 k_{\perp} \ll \tilde{Q}_{x}
=
 {\rm min}\{x_{i}Q\},
$
(or
$
 k_{\perp} \ll \tilde{Q}_{x}
=
 {\rm min}\{{\bar x}_{i}Q\}
$)
so that each individual ``tilded'' wave-function scales can be traded
for a common factorization scale $\mu ^{2}$.
This renormalization procedure \cite{BL89} yields $c=1$ for mesons
(two quark lines) and $c=3/2$ for baryons (three quark lines).

The gauge-invariant distribution amplitude
$\Phi ^{(\rm H)}(x_{i},\mu ^{2})$
is intrinsically non-perturbative and -- provided {\it the same
factorization scheme is used to cast exclusive amplitudes for
different processes in convolution form} \cite{Stenew} -- universal.
The large-momentum behavior of these functions can be analyzed either
using {\OPE} techniques or, equivalently, by evolution equations
analogous to DGLAP equations \cite{Dok77,GL72,AP77} in deep-inelastic
scattering.
Following the second approach, one takes derivatives with respect
to $Q^{2}$ of Eq.~(\ref{eq:phidef}) to arrive at evolution equations
of the generic form \cite{LB80}
\begin{equation}
  \frac{\partial \,\Phi ^{(\rm H)}(x_{i},Q^{2})}{\partial \ln Q^{2}}
=
  \int_{0}^{1} [dy]\, V\left(x_{i},y_{i},\alpha _{\rm s}(Q^{2})\right)
  \Phi ^{(\rm H)}(x_{i},Q^{2})
\label{eq:evolequ}
\end{equation}
with distinct kernels
$V\left(x_{i},y_{i},\alpha _{\rm s}(Q^{2})\right)$
for each process at hand, which, to leading order in $\alpha _{\rm s}$,
are computable from the single-gluon-exchange kernel.

To solve the evolution equation, $\Phi ^{(\rm H)}$ for hadron H has to
be expressed as an orthogonal expansion in terms of appropriate
functions which constitute an eigenfunction basis of the particular
gluon-exchange kernel, i.e.,
\begin{equation}
  \Phi ^{(\rm H)}(x_{i},Q^{2})
=
  \Phi _{\rm as}^{(\rm H)}(x_{i})
  \sum_{n=0}^{\infty} B_{n}^{(\rm H)}(\mu ^{2})
  \tilde{\Phi} _{n}^{(\rm H)}(x_{i})
  \exp
      \left\{
             \int_{\mu ^{2}}^{Q^{2}}
             \frac{d\bar{\mu}^{2}}{\bar{\mu}^{2}}
             \gamma _{\rm F}(g(\bar{\mu} ^{2}))
      \right\} \; ,
\label{eq:Phieigen}
\end{equation}
where $\Phi _{\rm as}^{(\rm H)}$ is the {\RG} asymptotic {\DA} (see
below) being proportional to the weight $w(x_{i})$ of the particular
orthogonal basis, and $\tilde{\Phi} _{n}^{(\rm H)}$ denotes the
corresponding eigenfunctions.
The coefficients $B_{n}^{(\rm H)}$ of this expansion are associated
with matrix elements of composite lowest-twist operators with definite
anomalous dimensions (after diagonalization of the evolution kernel)
taken between the vacuum and the external hadron.
They represent the non-perturbative input (integration constants of
the {\RG} equation) in Eq.~(\ref{eq:Phieigen}) and have to be
determined at some initial scale of evolution $\mu ^{2}$ by
non-perturbative techniques.
The exponential factor in Eq.~(\ref{eq:Phieigen}) takes care of
momentum evolution according to the {\RG} and is governed by the quark
anomalous dimension
\begin{equation}
  \gamma _{\rm F}
=
  \frac{\mu}{Z_{2}^{\rm q}}\,
  \frac{\partial Z_{2}^{\rm q}}{\partial \alpha _{\rm s}} \,
  \frac{\partial \alpha _{\rm s}}{\partial \mu}
\label{eq:defandim}
\end{equation}
which in the axial gauge is \cite{Sop80}
\begin{equation}
    \gamma _{q}
=
  - \frac{\alpha _{\rm s}}{\pi} + O(\alpha _{\\rm s}^{2}) \; .
\label{eq:andimaxg}
\end{equation}

\subsubsection{Meson evolution equation}
\label{subsubsec:mesevoleq}
The advantage of using an eigenfunctions decomposition is that the
evolution equation can be solved by diagonalization.
Consider, for example, the meson evolution equation.
Using the evolution ``time'' parameter \cite{GL72,LB80}
\begin{equation}
  \xi
\equiv
  \frac{\beta _{0}}{4\pi}
  \int_{\mu ^{2}}^{Q^{2}} \frac{dk_{\perp}^{2}}{k_{\perp}^{2}} \,
                   \alpha _{\rm s}(k_{\perp}^{2})
=
  \ln \frac{\alpha _{\rm s}(\mu ^{2})}{\alpha _{\rm s}(Q^{2})}
=
  \ln \frac{\ln Q^{2}/\Lambda _{{\rm QCD}}^{2}}
           {\ln \mu ^{2}/\Lambda _{{\rm QCD}}^{2}},
\label{eq:xi}
\end{equation}
and the relation
\begin{equation}
  \frac{\partial}{\partial Q^{2}}
=
  \frac{\partial\xi}{\partial Q^{2}} \, \frac{\partial}{\partial\xi}
=
  \frac{\beta _{0}}{4\pi} \,
  \frac{\alpha _{\rm s}(Q^{2})}{Q^{2}} \,
  \frac{\partial}{\partial \xi} \; ,
\label{eq:qxi}
\end{equation}
the meson evolution equation reads
\begin{equation}
  x_{1}x_{2} \,
  \left[
          \frac{\partial}{\partial \xi}\, \tilde{\Phi} (x_{i},Q^{2})
        + \frac{C_{\rm F}}{\beta _{0}}\, \tilde{\Phi} (x_{i},Q^{2})
  \right]
=
  \frac{C_{\rm F}}{\beta _{0}}\,
  \int_{0}^{1}[dy]\, V(x_{i},y_{i})\, \tilde{\Phi} (x_{i},Q^{2}) \; ,
\label{eq:evolmes}
\end{equation}
where
$
 \Phi (x_{i},Q^{2})
=
 x_{1}x_{2}\tilde{\Phi} (x_{i},Q^{2})
$.
Expanding in terms of eigenfunctions
$\Phi _{n}$, or $x_{1}x_{2}\tilde{\Phi} _{n}$, one has
\begin{equation}
    \frac{\partial}{\partial \xi} \, \Phi_{n}(x_{i}, Q^{2})
=
  - \gamma _{n} \, \Phi_{n}(x_{i}, Q^{2})
\label{eq:diagpi}
\end{equation}
from which one readily obtains
\begin{eqnarray}
  \Phi_{n}(x_{i},Q^{2})
& = &
  \Phi_{n}(x_{i}, \mu ^{2}) \, {\rm e}^{-\gamma _{n}\, \xi}
\nonumber \\
& = &
  \Phi_{n}(x_{i}, \mu ^{2}) \,
  \exp \left[
           -\gamma _{n}\,
            \ln\frac{\alpha _{\rm s}(\mu ^{2})}{\alpha _{\rm s}(Q^{2})}
       \right]
\nonumber \\
& \simeq &
  x_{1}x_{2}\tilde{\Phi}_{n}(x_{i}, \mu ^{2})\,
  \left(
        \ln \frac{Q^{2}}{\Lambda _{{\rm QCD}}^{2}}
  \right)^{-\gamma _{n}} \; .
\label{eq:Eigensolpi}
\end{eqnarray}
Then the equation to be solved becomes
\begin{equation}
  x_{1}x_{2}
  \left(
        \frac{C_{\rm F}}{\beta _{0}} - \gamma _{n}
  \right)
  \tilde{\Phi}_{n}(x_{i})
=
  \frac{C_{\rm F}}{\beta _{0}}
  \int_{0}^{1} [dy]\, V(x_{i}, y_{i})\, \tilde{\Phi}_{n}(y_{i}) \; ,
\label{eq:diagpievoleq}
\end{equation}
where $V(x_{i}, y_{i})$ is both real and symmetric, and
$\{\tilde{\Phi}_{n}(x_{i})\}$ form a set of orthogonal functions
with respect to the weight $w(x_{i})=x_{1}x_{2}$:
\begin{equation}
  \int_{0}^{1}[dx]\, w(x_{i})\, \Phi _{n}(x_{i})\, \Phi _{m}(x_{i})
=
  K_{n}\, \delta _{nm} \; .
\label{eq:orthopion}
\end{equation}
Within this basis of eigenfunctions, the pion {\DA} has a convergent
expansion for all $Q^{2}$, {\it viz.}:
\begin{equation}
  \Phi ^{(\pi)}(x_{i},Q^{2})
=
  w(x_{i}) \, \sum_{n=0}^{\infty}B_{n}^{(\pi)}
  \left(
        \ln \frac{Q^{2}}{\Lambda _{{\rm QCD}}^{2}}
  \right)^{-\gamma _{n}}
  \tilde{\Phi}_{n}(x_{i})
\label{eq:piexp}
\end{equation}
with expansion coefficients
\begin{equation}
  B_{n}^{(\pi)} \,
  \left(
        \ln \frac{Q^{2}}{\Lambda _{{\rm QCD}}^{2}}
  \right)^{-\gamma _{n}}
=
  K_{n}^{-1} \, \int_{0}^{1}[dx] \, w(x_{i})
  \tilde{\Phi}_{n}^{(\pi)}(x_{i})
  \tilde{\Phi}_{n}^{(\pi)}(x_{i}, Q^{2}) \; .
\label{eq:piexpcoef}
\end{equation}

To determine the anomalous dimensions $\gamma _{n}$ and the
corresponding eigenfunctions, it is convenient to introduce the
relative coordinate $-1 \leq \zeta =x_{1} -x_{2} \leq 1$ and express
the kernel $V$ in terms of a matrix $U_{jn}$ via a monomial basis
$|\zeta ^{n} \rangle$,
\begin{equation}
  V(\zeta )|\zeta ^{n}\rangle
=
  w(\zeta ) \, \sum_{j=0}^{n} |\zeta ^{j} \rangle \, U_{jn} \; ,
\label{eq:kernmatr}
\end{equation}
where $w(\zeta )=x_{1}x_{2}=\frac{1}{4}(1-\zeta ^{2})$.
The matrix $U_{jn}$ turns out to be triangular \cite{LB80}:
\begin{equation}
  U_{jn}
=
  0
\quad \quad \quad \quad \quad
 j > n
\label{eq:U_j>n}
\end{equation}

\begin{equation}
  U_{nn}
=
    \frac{2\delta _{h_{1}{\bar h}_{2}}}{(n+1)(n+2)}\,
  - 4\, \sum_{k=2}^{n+1} \frac{1}{k}
\quad \quad \quad \quad \quad
 j = n
\label{eq:U_j=n}
\end{equation}

\begin{equation}
  U_{jn}
=
  \frac{1 + (-1)^{n-j}}{2}
\left[
        \frac{2\delta _{h_{1}{\bar h}_{2}}}{(n+1)(n+2)}\,
      + \frac{4(j+1)}{(n-j)(n+1)}
\right]
\quad \quad \quad \quad \quad
  j < n \; .
\label{eq:U_j<n}
\end{equation}

\begin{figure}
\begin{picture}(0,370)
  \put(74,-10){\psboxscaled{460}{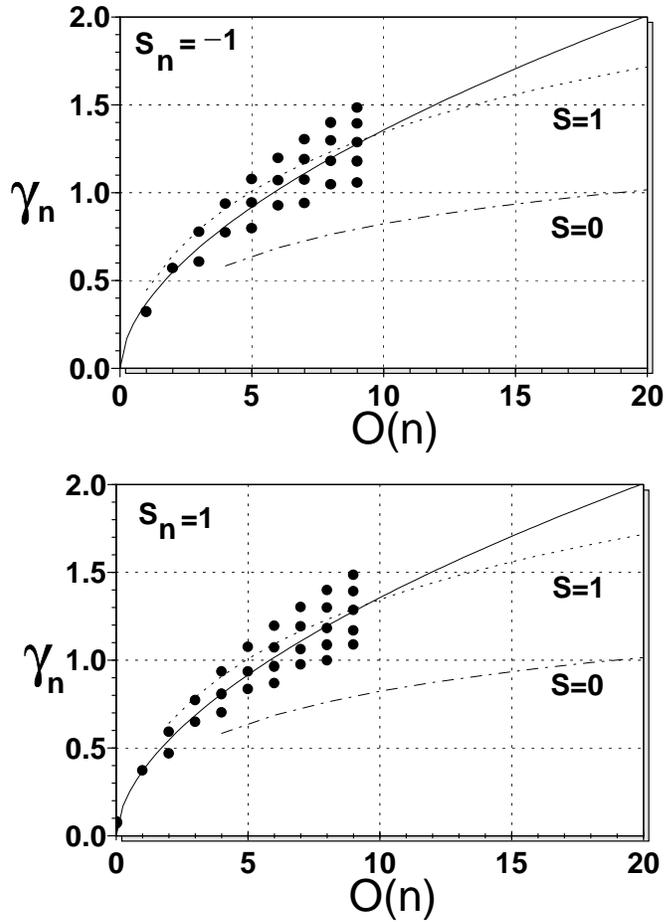}}
\end{picture}
\caption[fig:anomdimord9]
        {\tenrm Anomalous dimensions of three-quark operators up to
         polynomial order $M=9$ in comparison with those of two-quark
         (meson-like) operators. Curves are shown for antisymmetric
         ($S_{n}=-1$) and symmetric ($S_{n}=1$) eigenfunctions of the
         diagonalized nucleon evolution equation. The solid line is an
         empirical fit which is compatible with power-low behavior.
\label{fig:anomdim9}}
\end{figure}

Hence, the eigenvalues are just $U_{nn}$ and the eigenfunctions are
polynomials of finite order.
The only polynomials in the interval $[-1, 1]$ orthogonal with respect
to
$w(\zeta )=\frac{1}{4}(1-\zeta ^{2})$
are the Gegenbauer polynomials\footnote{The Gegenbauer polynomials
correspond to the conformally invariant {\OPE} for two
spin-$\frac{1}{2}$ operators \cite{ER80a,ER80b,BDFL85,BFL86}.}
$C_{n}^{3/2}(\zeta )$ (see, e.g., \cite{Erd53}), normalized by
\begin{equation}
  \int_{-1}^{1} d\zeta\, w(\zeta )\,
  C_{n}^{3/2}(\zeta ) \, C_{m}^{3/2}(\zeta )
=
  K_{n}\, \delta _{nm} \; ,
\label{eq:gegenpol}
\end{equation}
where
\begin{equation}
  K_{n}
=
  \frac{(2+n)(1+n)}{2(2n+3)} \; .
\label{eq:normgegen}
\end{equation}
Then, on account of
\begin{equation}
  \frac{C_{\rm F}}{\beta _{0}} - \gamma _{n}
=
  \frac{C_{\rm F}}{\beta _{0}}\, U_{nn} \; ,
\end{equation}
the associated anomalous dimensions become
\begin{equation}
  \gamma _{n}
=
  \frac{C_{\rm F}}{\beta _{0}}
\left[
      1 + 4 \sum_{k=2}^{n+1} \frac{1}{k}
        -   \frac{2\delta _{h_{1}{\bar h}_{2}}}{(n+1)(n+2)}
\right]
\geq 0 \quad\quad (n\;\; {\rm even}) \; ,
\label{eq:gamma_pi}
\end{equation}
where, for the pion, $\delta _{h_{1}{\bar h}_{2}}=1$ (see
Fig.~\ref{fig:anomdim9}).
The corresponding eigenfunctions are
($
  \tilde{\Phi}_{n}^{(\pi )}(x_{i})
=
  C_{n}^{3/2}(x_{i}) = C_{n}^{3/2}(x_{1}-x_{2})
$) so that
\begin{equation}
  \Phi _{n}^{(\pi )}(x_{i}, Q^{2})
=
  x_{1}x_{2}\, C_{n}^{3/2}(x_{1} - x_{2})\,
  \left(
        \ln \frac{Q^{2}}{\Lambda _{{\rm QCD}}^{2}}
  \right)^{-\gamma _{n}}
\label{eq:pieigengegen}
\end{equation}
with coefficients going like (cf. Eq.~(\ref{eq:piexpcoef}))
\begin{eqnarray}
  B_{n}^{(\pi )}
       \left(
             \ln \frac{Q^{2}}{\Lambda _{{\rm QCD}}^{2}}
       \right)^{-\gamma _{n}}
& = &
  \frac{2(2n+3)}{(2+n)(1+n)}\int_{-1}^{1}d\zeta\,
  C_{n}^{3/2}(\zeta )\; \Phi ^{(\pi )}\!(\zeta ,Q^{2}) ,
\nonumber \\
& = &
  \frac{{\sqrt 2}(2n+3)}{(2+n)(1+n)}\,
  \frac{1}{{\sqrt N_{\rm c}}}\,
  \langle 0|\bar{\psi}(0)\, \gamma ^{+}\gamma _{5}\,
            C_{n}^{3/2}\left( i \tensor{D}^{+} \right)\,
            \psi (0)|\pi
  \rangle _{(Q^{2})} \; ,
\label{eq:expcoefpi}
\end{eqnarray}
where in the last step we switched from the relative variable $\zeta$
back to $x$, and the matrix element of the local operator is
evaluated with UV cutoff $Q^{2}$.
Here $\tensor{D}_{\mu} = \roarrow{D} - \loarrow{D}$ with
$\roarrow{D}=\partial _{\mu} - i g \sum_{i=1}^{8}t^{i}A_{\mu}^{i}$
and
$\loarrow{D}=\partial _{\mu} + i g \sum_{i=1}^{8}t^{i}A_{\mu}^{i}$,
and
$D^{+}=\partial^{+}$ in the light-cone gauge.
From this equation one sees that the expansion coefficients
$B_{n}^{(\pi )}$ are matrix elements of local operators and
decrease like $1/n^{2}$, provided
$
 \Phi ^{(\pi )}(x_{i}, Q^{2})\leq K x_{i}^{\epsilon}
$
as $x_{i} \to 0$ for some $\epsilon > 0$ \cite{LB80}.
Thus to leading order the pion {\DA} becomes ($P_{\pi}^{+}=1$)
\begin{eqnarray}
  \Phi ^{(\pi)}(x_{i},Q^{2})
=
& \phantom{} & \!\!\!\!\!
  x(1-x)
  \sum_{n=0}^{\infty} \frac{{\sqrt 2}(2n+3)}{(2+n)(1+n)}
  \frac{1}{{\sqrt N_{\rm c}}}
  C_{n}^{3/2}(2x-1)
\nonumber \\
& \times &
  \langle 0|\bar{\psi}(0)\, \gamma ^{+}\gamma _{5}
            C_{n}^{3/2}\left( i
  \protect\stackrel{\leftrightarrow \protect\atop
  {\protect\textstyle D}}{{\textstyle{}}}^{+} \right)\,
            \psi (0)|\pi
  \rangle _{(Q^{2})} \; .
\label{eq:piDAlo}
\end{eqnarray}
For asymptotic values of $Q^{2}$ only the leading logarithm in
Eq.~(\ref{eq:expcoefpi}) with the least anomalous dimension
$\gamma _{0}=0$ survives ($C_{0}^{3/2}=1$), so that
\begin{equation}
  \Phi ^{(\pi)}(x_{i},Q^{2})
\to
  \frac{3}{{\sqrt N_{\rm c}}} \, f_{\pi} \, x(1-x) \, ,
\label{eq:piDAasQ}
\end{equation}
where $f_{\pi}=93$~MeV is the pion weak decay constant.
From this, one infers the normalization condition (sum rule)
\begin{equation}
  \int_{0}^{1} [dx]\, \Phi ^{(\pi)}(x_{i},Q^{2})
=
  \frac{f_{\pi}}{2{\sqrt N_{\rm c}}} \; ,
\label{eq:normpidim}
\end{equation}
which, given the shape of $\Phi ^{(\pi)}(x_{i},Q^{2})$, normalizes
it for every value of $Q^{2}$.

%
%

The above treatment can be readily used to provide information on the
asymptotic behavior of the pion form factor as well.
The electromagnetic pion form factor in the space-like region is
defined by
\begin{equation}
  \langle \pi ^{\pm}(P^{\prime})|
  J_{\mu}^{{\rm em}}(0)|
  \pi ^{\pm}(P) \rangle
=
  (P^{\prime} + P)_{\mu}\, F_{\pi}(Q^{2}) \; ,
\label{eq:piffspacelike}
\end{equation}
where
$
 J_{\mu}^{{\rm em}}
=
 \sum_{\rm f}^{} e_{\rm f} {\bar q}_{\rm f} \gamma _{\mu} q_{\rm f}
$
is the electromagnetic current, and the index f denotes different
flavors.
To relate the pion form factor to the Fock state wave functions,
discussed above, the $\mu = +$ component of this equation is taken.

With the spectrum of eigenfunctions and corresponding anomalous
dimensions in hand, we can write Eq.~(\ref{eq:G(Q^2)}) in the form
($n$ even because of $C$-parity)
\begin{equation}
  F_{\pi}(Q^{2})
=
  \frac{16\pi C_{\rm F}\alpha _{\rm s}(Q^{2})}{Q^{2}}\, e_{1}
  \left|
        \sum_{n}^{\infty} B_{n}^{(\pi)} \,
        \left( \ln \frac{Q^{2}}{\Lambda _{{\rm QCD}}^{2}}
        \right)^{-\gamma _{n}}
        \int_{0}^{1} dx_{1} \, x_{1}\, C_{n}^{3/2}(x_{1} - x_{2})
  \right|^{2}
\; + \; (1 \leftrightarrow 2) \; .
\label{eq:asypiff.s1}
\end{equation}
Recalling that for $\pi ^{+}$ the charges of the struck quarks sum to
one, and that for even Gegenbauer polynomials,
$
 \int_{0}^{1} dx_{1} \, x_{1}\, C_{n}^{3/2}(x_{1} - x_{2})
=
 \frac{1}{2}
 \int_{0}^{1} dx_{1} \,
 (x_{1} + x_{2}) \, C_{n}^{3/2}(x_{1} - x_{2})
=
 \frac{1}{2}
$,
the pion form factor becomes ($n$ even)
\begin{equation}
  F_{\pi}(Q^{2})
=
  \frac{4\pi C_{\rm F}\alpha _{\rm s}(Q^{2})}{Q^{2}}\,
  \left|
        \sum_{n}^{\infty} B_{n}^{(\pi)} \,
        \left( \ln \frac{Q^{2}}{\Lambda _{{\rm QCD}}^{2}}
        \right)^{-\gamma _{n}}
  \right|^{2} \; .
\label{eq:asypiff.s2}
\end{equation}
In the limit of asymptotic values of momentum transfer, only the
zeroth order term $C_{0}^{3/2}=1$ contributes with anomalous dimension
$\gamma _{0}=0$, so that
\begin{equation}
  \lim_{Q^{2} \to \infty}
  F_{\pi}(Q^{2})
=
  \frac{4\pi C_{\rm F}\alpha _{\rm s}(Q^{2})}{Q^{2}}\,
  [B_{0}^{(\pi)}]^{2} \; .
\label{eq:asypiff.s3}
\end{equation}
This result can be couched in the final form \cite{FJ79}
\begin{equation}
  \lim_{Q^{2} \to \infty}
  F_{\pi}(Q^{2})
=
  \frac{16\pi f_{\pi}^{2}\alpha _{\rm s}(Q^{2})}{Q^{2}}
\label{eq:asypiff.s4}
\end{equation}
by virtue of the decay process
$\pi^{+} \longrightarrow l^{+}\nu _{l}$,
($l=\mu ^{+}, e^{+}$)
which fixes
$B_{0}^{(\pi)}=\frac{3f_{\pi}}{\sqrt N_{\rm c}}$
in terms of the pion decay constant $f_{\pi}$, independent of the
momentum variance.
Unfortunately, the asymptotic prediction is not supported by the
existing data \cite{Beb76}.
Indeed, evaluating the above expression, say, at
$Q^{2}=6$~GeV${}^{2}$
for $\Lambda _{{\rm QCD}}=0.2$~GeV, one finds
$Q^{2}F_{\pi}(Q^{2}) = 0.12$ which is about a factor of three below
the experimental value (and well below the error bars).
The conclusion is that additional contributions of Gegenbauer terms in
the expansion of the pion {\DA} have to be included, or that one has
to take into account higher twist contributions (power corrections).
In order to increase the value of the pion form factor at moderate
$Q^{2}$-values, a broader pion {\DA}, relative to
$x_{1}x_{2}=\frac{1}{4}(1-x^{2})$, is needed with at least
$B_{2}^{(\pi)}/B_{0}^{(\pi)}>0$.
This goal can be achieved using QCD sum rules \cite{CZ82,MR86,BM95}.
The literature on this subject is vast and the selection of
references \cite{Zhi94,DM87,BF89,BH94,AMN95,JKR96,Dor96}
is by no means complete.
We apologize to those authors whose works have not been included.

\subsubsection{Nucleon evolution equation}
\label{subsubsec:nucevoleq}
Let us now turn to the nucleon case.
The evolution equation is
\begin{equation}
  x_{1}x_{2}x_{3} \,
  \left[
          \frac{\partial}{\partial \xi} {\tilde\Phi}(x_{i},Q^{2})
        + \frac{3}{2}\frac{C_{\rm F}}{\beta _{0}}\,
          \tilde{\Phi}(x_{i},Q^{2})
  \right]
=
  \frac{C_{\rm F}}{\beta _{0}}\,
  \int_{0}^{1}[dy]\, V(x_{i},y_{i})\, \tilde{\Phi}(y_{i},Q^{2}) \; ,
\label{eq:evolnuc}
\end{equation}
where
$\Phi = x_{1}x_{2}x_{3} \, \tilde{\Phi}$,
$C_{\rm F}$ is defined below Eq.~(\ref{eq:pihard}), and
\begin{equation}
  V(x_{i},y_{i})
=
  2 x_{1}x_{2}x_{3} \,
  \sum_{i \not= j}^{}\Theta (y_{i} - x_{i}) \, \delta (y_{k} - x_{k}) \,
  \frac{y_{j}}{x_{j}}
  \left[
          \frac{\delta _{h_{i}{\bar h}_{j}}}{x_{i} + x_{j}}
        + \frac{\Delta}{y_{i} - x_{i}}
  \right] \; .
\label{eq:nuckern}
\end{equation}
Note that $V(x_{i},y_{i})=V(y_{i},x_{i})$ is the sum over
single-gluon interactions between quark pairs $\{i,j\}$, and the
subtraction prescription
$
 \Delta \tilde{\Phi}(y_{i},Q^{2})
\equiv
 \tilde{\Phi}(y_{i},Q^{2}) - \tilde{\Phi}(x_{i},Q^{2})
$
ensures IR finiteness at $x_{i}=y_{i}$, i.e., $V(x_{i},y_{i})$ is not
a function but a distribution.
For antiparallel spins $\delta _{h_{i}{\bar h}_{j}}=1$ and for
parallel spins it equals 0.
Note also that if no gluon is exchanged, each $x_{k}$ in the initial
and final wave function is the same because no longitudinal momentum
is introduced by $q^{\mu}$ ($q^{+}=0$).

We propose to solve the evolution equation by employing factorization
of the dependence on longitudinal momentum from that on the external
(large) momentum scale $Q^{2}$ (cf. Eq.~(\ref{eq:Phieigen})).
The latter is {\RG} controlled according to
\begin{equation}
    \frac{\partial}{\partial \xi}\,
    \tilde{\Phi}_{n}(x_{i}, Q^{2})
=
  - \gamma _{n}\, \tilde{\Phi}_{n}(x_{i}, Q^{2})
\label{eq:eigenvaleqnuc}
\end{equation}
with solutions
\begin{equation}
  \tilde{\Phi}_{n}(x_{i}, Q^{2})
\simeq
  \tilde{\Phi}_{n}(x_{i})\,
  \left(
        \ln \frac{Q^{2}}{\Lambda _{{\rm QCD}}^{2}}
  \right)^{-\gamma _{n}} \; .
\label{eq:eigenfuncnuc}
\end{equation}
This allows us to write the full nucleon {\DA} in the form
\begin{equation}
  \Phi (x_{i}, Q^{2})
\sim
  x_{1}x_{2}x_{3} \,
  \sum_{n=0}^{\infty} B_{n} \,
  \tilde{\Phi}_{n}(x_{i})\,
  \left(
        \ln \frac{Q^{2}}{\Lambda _{{\rm QCD}}^{2}}
  \right)^{-\gamma _{n}} \; ,
\label{eq:solnuc}
\end{equation}
where $\tilde{\Phi}_{n}(x_{i})$ are appropriate but not tabulated
polynomials, and the expansion coefficients $B_{n}$ encode the
non-perturbative input of the bound-states dynamics at the
factorization (renormalization) scale.
Their determination will concern us in the next section.

From the factorized form of $\tilde{\Phi}_{n}(x_{i}, Q^{2})$ in
Eq.~(\ref{eq:eigenfuncnuc}), it follows that the evolution equation
for the $x$-dependence reduces to the characteristic equation
\begin{equation}
  x_{1}x_{2}x_{3} \,
  \left[
          \frac{3}{2}\frac{C_{\rm F}}{\beta _{0}}\,
        - \gamma _{n}
  \right] \,
         \tilde{\Phi}(x_{i})
=
  \frac{C_{\rm B}}{\beta _{0}}\,
  \int_{0}^{1} [dy] \,
  \frac{V(x_{i},y_{i})}{w(x_{i})} \,
         \tilde{\Phi}(y_{i}) \, ,
\label{eq:evolxpart}
\end{equation}
where
$w(x_{i})=x_{1}x_{2}x_{3}=x_{1}(1-x_{1}-x_{3})x_{3}$
is the weight function of the orthogonal basis and
$C_{\rm B}=(N_{\rm c}+1)/2N_{\rm c}=2/3$ the Casimir operator of the
adjoint representation of $SU(3)_{\rm c}$.
To proceed, it is convenient to conceive of the kernel
$V(x_{i},y_{i})$ as being an operator expanded over the polynomial
basis \cite{LB80}
$
 |x_{1}^{k}\, x_{3}^{l} \rangle
\equiv
 |k\,l\rangle
$,
(recall that because of momentum conservation, only two out of three
$x_{i}$ variables are linearly independent) i.e., to write
\begin{equation}
  \hat{V}
\equiv
  \int_{0}^{1}[dy] \, V(x_{i},y_{i})
\label{eq:voperx1x3}
\end{equation}
and convert Eq.~(\ref{eq:evolxpart}) into the algebraic equation
\begin{equation}
  \left[
            \frac{3}{2}\frac{C_{\rm F}}{\beta _{0}}\,
        - 2 \frac{C_{B}}{\beta _{0}}\,
            \frac{\hat{V}}{2w(x_{i})}
  \right]
  \tilde{\Phi}_{n}(x_{i})
=
  \gamma _{n}\, \tilde{\Phi}_{n}(x_{i}) \; .
\label{eq:algnucevoleq}
\end{equation}
In this way, the action of the operator $\hat{V}$ can be completely
determined by a matrix, namely:
\begin{equation}
  \frac{\hat{V}|k \, l>}{2 w(x_{i})}
=
  \frac{1}{2} \sum_{i,j}^{i+j\leq M} |i\, j \rangle U_{ij,kl} \; .
\label{eq:kernux1x3}
\end{equation}
The corresponding eigenvalues are then determined by the roots
$\eta _{n}$ of the characteristic polynomial that diagonalizes the
matrix $U$:
\begin{equation}
   \hat{V} \, \tilde{\Phi}_{n}(x_{i})
=
  - \eta _{n}\, w(x_{i})\, \tilde{\Phi}_{n}(x_{i}) \; ,
\end{equation}
\label{eq:rootsofVhat}
so that the anomalous dimensions for order $M$ are given by
\begin{equation}
 \gamma _{n}(M)
=
 \frac{1}{\beta_{0}}
 \left(
       \frac{3}{2}C_{\rm F} + 2\eta _{n}(M)C_{\rm B}
 \right) \; ,
\label{eq:gamma_N}
\end{equation}
where the orthogonalization prescription
\begin{equation}
  \int_{0}^{1}[dx]\, w\left(x_{i}\right)\,
  \tilde{\Phi}_{m}(x_{i}) \tilde{\Phi}_{n}(x_{i})
=
  \frac{1}{N_{m}}\, \delta _{mn}
\label{eq:orthonuc}
\end{equation}
has been employed with $N_{m}$ being appropriate normalization constants
(see, Table \ref{tab:eigen}).

The explicit form of the matrix $U$ was derived by Lepage and
Brodsky \cite{LB80} and is reproduced in Appendix \ref{ap:nucevolkern}.
Within the basis $|k\, l\rangle$, the matrix $U$ can be diagonalized
to provide eigenfunctions, which are polynomials of degree
$M=k+l=0,1,2,3 \ldots$, with $M+1$ eigenfunctions for each $M$.
This was done in \cite{LB80} by diagonalizing the $(M+1)\times(M+1)$
matrix $U_{ij,kl}$ with $i+j=k+l=M$ and results up to $M=2$ were
obtained.
In our approach, reported in several meetings \cite{ELBA93,KYF93,COMO94}
and worked out in \cite{Ste94,Ber94}, we make use of another method that
is based on symmetrized Appell polynomials \cite{Erd53}.

Appell polynomials are special hypergeometric functions (see Appendix
{\ref{ap:nucevolkern}) of the form
\begin{equation}
 {\cal F}_{mn}^{(M)}(5,2,2;x_{1},x_{3})
\equiv
 {\cal F}_{mn}(x_{1},x_{3}) \; ,
\label{eq:appellpol}
\end{equation}
which constitute an orthogonal polynomial set on the triangle
$T = T(x_{1},x_{3})$ with $x_{1}>0$, $x_{3}>0$, $x_{1} + x_{3}<1$.
They provide a suitable basis for solving the eigenvalue equations
for the nucleon because within this basis $\hat V$ is block diagonal
for different polynomial orders.
Moreover, introducing a ``symmetrized'' basis of such polynomials
according to \cite{Ste94}
\begin{eqnarray}
 \tilde{{\cal F}}_{mn}(x_{1},x_{3})
& = &
 \frac{1}{2}[{\cal F}_{mn}(x_{1},x_{3})
\pm
 {\cal F}_{nm}(x_{1},x_{3})]
\nonumber \\
& = &
  \sum_{k,l=0}^{k+l\leq m+n} Z_{kl}^{mn} |k\, l\rangle
\label{eq:symApp}
\end{eqnarray}
(where $+$ refers to $m\geq n$ and $-$ to $m<n$), $\hat V$ commutes
with the permutation operator $P_{13}=[321]$ and thus becomes
block diagonal within each sector of (definite) permutation-symmetry
class of eigenfunctions for fixed order $M$.
As a result, the kernel $\hat V$ can be analytically diagonalized up
to order seven.
This is related to the fact that the characteristic polynomial of
matrices with rank four can be solved analytically.
Beyond that order, its roots have to be determined numerically.
We present here results up to order $M=4$.
Still higher-order eigenfunctions (up to $M=9$) were obtained by
Bergmann \cite{Ber94}.
The differential equation defining the symmetrized Appell polynomials
is given in Appendix~\ref{ap:nucevolkern}.

%
\begin{table}
\squeezetable
\caption{Orthogonal eigenfunctions
         $\tilde{\Phi}_n(x_{i})=\sum_{kl}^{}\, a_{kl}^{n}\,
         x_{1}^{k}x_{3}^{l}$ of the nucleon evolution equation up to
         polynomial order $M=4$ in terms of the coefficient matrix
         $a_{kl}^{n}$ ($a_{kl}^{n} = S_{n}\, a_{lk}^{n}$ -- no
         summation over $n$ implied) and the corresponding anomalous
         dimensions $\gamma _{n}$ defined in the text.
         The numerical results for $n\geq 12$ have been obtained with
         a much higher numerical accuracy than shown in this table.
\label{tab:eigen}}
\begin{tabular}{rc|rcccc}
$ n  $&$ M $&$ S_n $&$\gamma_n$&$\eta_n$&$ N_n$&$a_{00}^{n}$\\
\hline
 $ 0$&$ 0$&$1$&${2\over {27}}$&$-1$&$120$&$1$\\
 $ 1$&$ 1$&$-1$&${{26}\over {81}}$&${2\over 3}$&$1260$&$0$\\
 $ 2$&$ 1$&$1$&${{10}\over {27}}$&$1$&$420$&$-2$\\
 $ 3$&$ 2$&$1$&${{38}\over {81}}$&${5\over 3}$&$756$&$2$\\
 $ 4$&$ 2$&$-1$&${{46}\over {81}}$&${7\over 3}$&$34020$&$0$\\
 $ 5$&$ 2$&$1$&${{16}\over {27}}$&${5\over 2}$&$1944$&$2$\\
 $ 6$&$ 3$&$1$&${{115 - {\sqrt{97}}}\over {162}}$&${{-\left( -79 + {\sqrt{97}} \right) }\over {24}}$&${{4620\,\left( 485 + 11\,{\sqrt{97}} \right) }\over {97}}$&$1$\\
 $ 7$&$ 3$&$1$&${{115 + {\sqrt{97}}}\over {162}}$&${{79 + {\sqrt{97}}}\over {24}}$&${{4620\,\left( 485 - 11\,{\sqrt{97}} \right) }\over {97}}$&$1$\\
 $ 8$&$ 3$&$-1$&${{559 - {\sqrt{4801}}}\over {810}}$&${{-\left( -379 + {\sqrt{4801}} \right) }\over {120}}$&${{27720\,\left( 33607 - 247\,{\sqrt{4801}} \right)}\over {4801}}$&$0$\\
 $ 9$&$ 3$&$-1$&${{559 + {\sqrt{4801}}}\over {810}}$&${{379 + {\sqrt{4801}}}\over {120}}$&${{27720\,\left( 33607 + 247\,{\sqrt{4801}} \right) }\over {4801}}$&$0$\\
 $ 10$&$4$&$-1$&${{346 - {\sqrt{1081}}}\over {405}}$&${{-\left( -256 + {\sqrt{1081}} \right) }\over {60}}$&${{196560\,\left( 7567 - 13\,{\sqrt{1081}} \right) }\over {1081}}$&$0$\\
 $ 11$&$4$&$-1$&${{346 + {\sqrt{1081}}}\over {405}}$&${{256 + {\sqrt{1081}}}\over {60}}$&${{196560\,\left( 7567 + 13\,{\sqrt{1081}} \right)}\over{1081}}$&$0$\\
 $ 12$&$4$&$1$&$0.70204$&$3.23876$&$1$&$153.37061$ \\
 $ 13$&$4$&$1$&$0.80651$&$3.94397$&$1$&$332.500864$ \\
 $ 14$&$4$&$1$&$0.93589$&$4.81727$&$1$&$-137.11538$
\end{tabular}
\vspace{-7 pt}
\begin{tabular}{r|cccccccc}
$ n  $&$a_{10}^n$&$a_{20}^n$&$a_{11}^n$&$a_{30}^n$&
       $a_{21}^n$&$a_{40}^n$&$a_{31}^n$&$a_{22}^n$ \\
\hline
 $  0$&$0$&$0$&$0$&$0$&$0$&$0$&$0$&$0$\\
 $  1$&$1$&$0$&$0$&$0$&$0$&$0$&$0$&$0$\\
 $  2$&$3$&$0$&$0$&$0$&$0$&$0$&$0$&$0$\\
 $  3$&$-7$&$8$&$4$&$0$&$0$&$0$&$0$&$0$\\
 $  4$&$1$&$-{4\over 3}$&$0$&$0$&$0$&$0$&$0$&$0$\\
 $  5$&$-7$&${{14}\over 3}$&$14$&$0$&$0$&$0$&$0$&$0$\\
 $  6$&$-6$&${{41 + {\sqrt{97}}}\over 4}$&${{3\,\left( 31 - {\sqrt{97}} \right) }\over 4}$&${{-5\,\left( 17 + {\sqrt{97}} \right) }\over {16}}$&${{-5\,\left( 31 - {\sqrt{97}} \right) }\over 8}$&$0$&$0$&$0$\\
 $  7$&$-6$&${{41 - {\sqrt{97}}}\over 4}$&${{3\,\left( 31 + {\sqrt{97}} \right) }\over 4}$&${{-5\,\left( 17 - {\sqrt{97}} \right) }\over {16}}$&${{-5\,\left( 31 + {\sqrt{97}} \right) }\over 8}$&$0$&$0$&$0$\\
 $  8$&$1$&$-3$&$0$&${{601 + {\sqrt{4801}}}\over {264}}$&${{59 - {\sqrt{4801}}}\over {44}}$&$0$&$0$&$0$\\
 $  9$&$1$&$-3$&$0$&${{601 - {\sqrt{4801}}}\over {264}}$&${{59 + {\sqrt{4801}}}\over {44}}$&$0$&$0$&$0$\\
 $ 10$&$1$&$-5$&$0$&${{379 + {\sqrt{1081}}}\over {48}}$&${{61 - {\sqrt{1081}}}\over 8}$&${{-\left( 159 + {\sqrt{1081}} \right) }\over {40}}$&${{-\left( 61 - {\sqrt{1081}} \right) }\over 8}$&$0$\\
 $ 11$&$1$&$-5$&$0$&${{379 - {\sqrt{1081}}}\over {48}}$&${{61 + {\sqrt{1081}}}\over 8}$&${{-\left( 159 - {\sqrt{1081}} \right) }\over {40}}$&${{-\left( 61 + {\sqrt{1081}} \right) }\over 8}$&$0$\\

$12$&$-1380.33552$&$5232.86956$&$5006.42414$&$-8063.85349$&$-9178.44426$&$4345.63139$&$4926.80699$&$8503.27454$\\

$13$&$-2992.50778$&$9240.51876$&$17166.06044$&$-11695.76593$&$-31471.11081$&$5068.49438$&$19489.65169$&$23962.91822$\\

$14$&$1234.03849$&$-1843.05428$&$-12981.41464$&$-587.61051$&$23799.26017$&$1382.85660$&$-10302.90296$&$-26992.71442$\\
\end{tabular}
\end{table}
%

To appreciate the usefulness of this type of approach, some historical
remarks ought to be made before we proceed.
The solution of the eigenvalue equations for the nucleon beyond
leading order is a long-standing problem.
As already mentioned, results up to order two were obtained by Lepage
and Brodsky \cite{LB80}.
The anomalous dimensions they computed were subsequently confirmed
by Peskin \cite{Pes79} who considered composite three-quark operators
containing derivatives and having baryon quantum numbers.
Such operators interpolate between the nucleon (or the $\Delta$
resonance) and the vacuum at leading twist-three.
Their anomalous dimension was extracted from the divergent parts of
their matrix elements by diagonalizing the renormalization matrix
of these operators (see in this context also \cite{Kre79}).
Then Tesima \cite{Tes82} analyzed the light-cone behavior of the
Bethe-Salpeter wave functions of three-quark bound states with the
use of the {\OPE} in terms of a conformally invariant operator basis.
He presented results for twist-three operators up to order two which
in general coincide with Peskin's results and also with those obtained
by Lepage and Brodsky.
In addition, he obtained results for higher orders, but published
only the eigenvalues and eigenfunctions corresponding to order three.
However, his higher-order results are not supported by our
calculation.
Moreover, it was shown by Ohrndorf \cite{Ohr82} that collinear
conformal covariance alone does not suffice to fix trilinear quark
operators with derivatives uniquely.

We shall now give a brief description of our approach and present
the main results.
All nucleon eigenfunctions can be represented as linear combinations
of (symmetrized) Appell polynomials of the same order $M$:
\begin{equation}
  \tilde{\Phi}_{k}(x_{i})
=
  \sum_{m,n=0}^{m+n=M}\,c_{mn}^{k}{\cal F}_{mn}(5,2,2;x_{1},x_{3})
\label{eq:Appellphi}
\end{equation}
since Appell polynomials of the same order are not orthogonal
to each other.
For instance, for $M=1$, one finds
\begin{eqnarray}
  \tilde{\Phi}_{1}(x_{i})
& = &
      {\cal F}_{01}^{(1)}(5,2,2;x_{1},x_{ 3})
    -
      {\cal F}_{10}^{(1)}(5,2,2;x_{1},x_{ 3})
\nonumber \\
& = &
  x_{1} - x_{3}
\label{eq:phi2}
\end{eqnarray}
and
\begin{eqnarray}
  \tilde{\Phi}_{2}(x_{i})
& = &
       {\cal F}_{10}^{(1)}(5,2,2;x_{1},x_{ 3})
     +
       {\cal F}_{01}^{(1)}(5,2,2;x_{1},x_{ 3})
\nonumber \\
& = &
    - 2 + 3(x_{1} + x_{3}) \; ,
\label{eq:phi3}
\end{eqnarray}
where the notations and conventions of \cite{Ste89} are
adopted.\footnote{In particular, the coefficient $B_{2}$ has the
opposite sign relative to \cite{LB80}.}
Before diagonalizing, it is convenient to rearrange
$\tilde{{\cal F}}_{mn}$, which belongs to a definite symmetry class
$S_{n}=\pm 1$ within order $M$, in the form of an (arbitrary) vector:
\begin{equation}
  \tilde{{\cal F}}_{mn}(x_{1},x_{3})
\longmapsto
  \tilde{{\cal F}}_{q}(x_{1},x_{3}) \; .
\label{eq:symAppvec}
\end{equation}
Then Hilbert-Schmidt orthogonalization yields a basis
\begin{equation}
  |\tilde{{\cal F}}_{q}^{\prime} \rangle
=
  \sum_{k,l}^{}Z_{kl}^{q}|k \, l \rangle
\label{eq:orthobasis}
\end{equation}
with
\begin{equation}
  \int_{0}^{1} [dx]\, w(x_{i})\, \tilde{{\cal F}}_{q}^{\prime}
  \tilde{{\cal F}}_{q^{\prime}}^{\prime}
\propto
  \delta _{qq^{\prime}} \; ,
\label{eq:orthorel}
\end{equation}
so that
\begin{equation}
  \frac{\hat{V}|\tilde{{\cal F}}_{q}^{\prime} \rangle}{2w(x_{i})}
=
  \frac{1}{2}
  \sum_{i,j,k,l}^{}Z_{kl}^{q}\, U_{ij,kl} |i\, j \rangle \; .
\label{eq:hatVinF_q}
\end{equation}
Note that the construction of polynomials depending on two variables
via the Hilbert-Schmidt method has no unique solution, but depends on
the order in which the orthogonalization is performed.
Since beyond order $M=3$ neither the eigenvalues nor the
normalization factors are rational numbers, one has to find which
representation is more convenient for calculations.

The last step in determining the eigenfunctions and eigenvalues of
$\hat{V}$ is to define the matrix
\begin{equation}
  {\cal M}_{q^{\prime}q}
=
  \int_{0}^{1}[dx]\,w\left(x_{1},(1-x_{1}-x_{3}),x_{3}\right)
  \tilde{{\cal F}}_{q^{\prime}}^{\prime}(x_{1},x_{3})
  \hat{V}(x_{1},x_{3})\tilde{{\cal F}}_{q}^{\prime}(x_{1},x_{3})
\label{eq:eigenmatrix}
\end{equation}
and calculate the roots of the characteristic polynomial
\begin{equation}
  {\cal P}(\eta )
=
  \det \left[{\cal M}_{q^{\prime}q} - \eta I_{q^{\prime}q}\right] \; .
\label{eq:P}
\end{equation}
Consequently, in terms of the eigenvectors
$\bbox{m}_{q} = \left(m_{q}^{1}, \ldots m_{q}^{q^{\prime}}\right)$
of ${\cal M}_{q^{\prime}q}$,
the eigenfunctions of the evolution equation are given by
\begin{eqnarray}
  \tilde{\Phi}_{q}(x_{1},x_{3})
& \propto &
  \sum_{q^{\prime}}^{}m_{q}^{q^{\prime}}
  \tilde{{\cal F}}_{q^{\prime}}^{\prime}(x_{1},x_{3})
\nonumber \\
& = &
  \sum_{k,l}^{} a_{kl}^{q}\, |k\, l \rangle \; .
\label{eq:eigenfuns}
\end{eqnarray}
For every order $M$, there are $M+1$ eigenfunctions of the same order
with an excess of symmetric terms by one for even orders.
The total number of eigenfunctions up to order $M$ is
$n_{\rm max}(M)=\frac{1}{2}(M+1)(M+2)$ and the corresponding $(M+1)$
eigenvalues are obtained by diagonalizing the $(M+1)\times (M+1)$
matrix.
A compendium of the results up to order $M=4$, meaning a total of
15 eigenfunctions and associated anomalous dimensions, is given in
Table~\ref{tab:eigen}.
The precision of orthogonality is at least $10^{-8}$.
To get acquaintance with the use of Table~\ref{tab:eigen}, we write
out explicitly one of the eigenfunctions contributing to order $M=3$:
\begin{eqnarray}
  \tilde{\Phi}_{9}
& = &
    a_{00}^{9} + a_{10}^{9} \left(x_{1} -x_{3}\right)
  + a_{11}^{9} x_{1}x_{3}
  + a_{20}^{9} \left(x_{1}^{2} - x_{3}^{2}\right)
  + a_{21}^{9} x_{1}x_{3} \left(x_{1} -x_{3}\right)
  + a_{30}^{9} \left(x_{1}^{3} - x_{3}^{3}\right)
\nonumber \\
& = &
    \left(x_{1} - x_{3}\right) - 3\left(x_{1}^{2} - x_{3}^{2}\right)
  + \frac{59 + {\sqrt{4801}}}{44}
    x_{1} x_{3} \left(x_{1} - x_{3}\right)
  + \frac{601 - {\sqrt{4801}}}{264}
    \left(x_{1}^{3} - x_{3}^{3}\right) \; .
\label{eq:Phi9}
\end{eqnarray}

It turns out that the eigenfunctions $\{\tilde{\Phi}_{k}\}$
of the nucleon evolution equation satisfy a commutative algebra
subject to the triangular condition
$
 |{\cal O}(k) - {\cal O}(l)| \leq {\cal O}(m) \leq {\cal O}(k) +
{\cal O}(l):
$
\begin{equation}
  \tilde{\Phi}_{k}(x_{i}) \tilde{\Phi}_{l}(x_{i})
=
  \sum_{m=0}^{\infty}\,F_{kl}^{m}\tilde{\Phi}_{m}(x_{i})
\label{eq:algebra}
\end{equation}
with structure coefficients $F_{kl}^{m}$ given by
\begin{equation}
  F_{kl}^{m}
=
  N_{m} \int_{0}^{1}[dx]\, x_{1}x_{3}(1-x_{1}-x_{3})
  \tilde{\Phi}_{m}(x_{i})
  \tilde{\Phi}_{k}(x_{i})
  \tilde{\Phi}_{l}(x_{i}) \; ,
\label{eq:structurecoefficients}
\end{equation}
${\cal O}(k)$ being defined by
\begin{equation}
  {\cal O}(k) = \left\{
\begin{array}{ll}
       0  & \quad\quad k = 0 \cr
       1  & \quad\quad 1 \leq k \leq 2 \cr
       2  & \quad\quad 3 \leq k \leq 5 \cr
       3  & \quad\quad 6 \leq k \leq 9 \cr
       4  & \quad\quad k = 10,11       \cr
  \vdots  & \quad\quad \ldots
\end{array} \right.
\label{eq:counter}
\end{equation}
The structure coefficients are symmetric, i.e.,
$
 F_{kl}^{m} = F_{lk}^{m}.
$
Furthermore, $F_{kk}^{0}=\frac{N_{0}}{N_{k}}$.
The utility of this algebra derives from the fact that once the
structure coefficients have been computed, they can be used to express
any function $f(x_{1},x_{3})$ in terms of the nucleon eigenfunctions.
For example, one can calculate the Husini function
\begin{equation}
  h
\sim
  \int_{0}^{1}[dx]\,
  \left|
        \tilde{\Phi}_{m}(x_{i})\, \Phi ^{({\rm N})}(x_{i})
  \right|^{2}
\label{eq:husini}
\end{equation}
which gives the probability for finding $\Phi ^{(\rm N)}$ in a
particular solution $\tilde{\Phi}_{m}$.
The values of $F_{lk}^{m}$ up to ${\cal O}(k)=11$ are tabulated
in \cite{Ber94}.

The obtained results for the exponents (anomalous dimensions)
governing the scaling behavior of the nucleon {\DA}
(cf. Eqs.~(\ref{eq:eigenfuncnuc}), (\ref{eq:solnuc})) are shown
in a series of four figures associated with successively increasing
order $M$.\footnote{This part was done in collaboration with Michael
Bergmann.}
As outlined above, the eigenfunctions correspond to trilinear quark
operators with definite anomalous dimension.
Such operators can be renormalized in a multiplicative way.
But as the order $M$ increases, so increases also the number of
derivatives in these operators entailing a strong amount of mixing
owing to the fact that they carry the same quantum numbers.
As a consequence, the anomalous dimensions are degenerate within
fixed symmetry classes and hence a ``multiplet'' structure emerges.
Because all $\gamma _{n}$ are positive fractional numbers increasing
with $n$ (i.e., order $M$), higher terms in the eigenfunctions
decomposition are gradually suppressed.
Fig.~\ref{fig:anomdim9} displays the anomalous dimensions
$\gamma _{n}$ up to order $M=9$, distinguishing between symmetric and
antisymmetric eigenfunctions.
The trend line of this pattern seems to follow the empirical law
(solid line)
$\gamma _{n} = 0.37{\cal O}(n)^{0.565}$.
The illustration of the spectrum up to order 20 is displayed in
Fig. \ref{fig:anomdim20}.

%
\begin{figure}
\begin{picture}(0,330)
  \put(82,-110){\psboxscaled{600}{fig7.ps}}
\end{picture}
\caption[fig:anomdimen20]
        {\tenrm Pattern of anomalous dimensions of three-quark operators
         up to $M=20$. Values associated with symmetric eigenfunctions
         are denoted by open circles. Those corresponding to
         antisymmetric eigenfunctions are marked by black dots.
\label{fig:anomdim20}}
\end{figure}
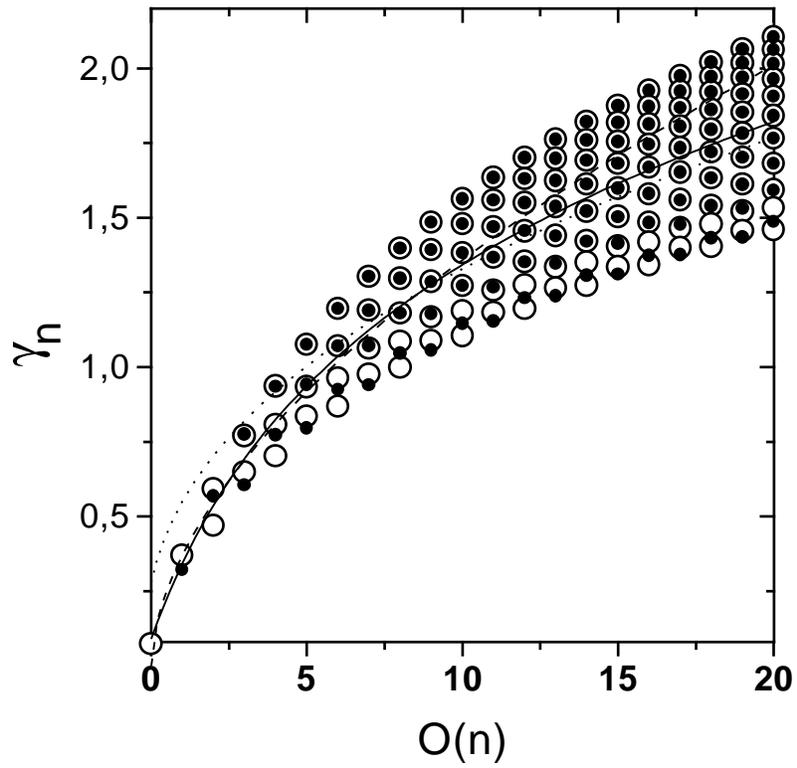
%

%
\begin{figure}
\begin{picture}(0,240)
  \put(35,-270){\psboxscaled{700}{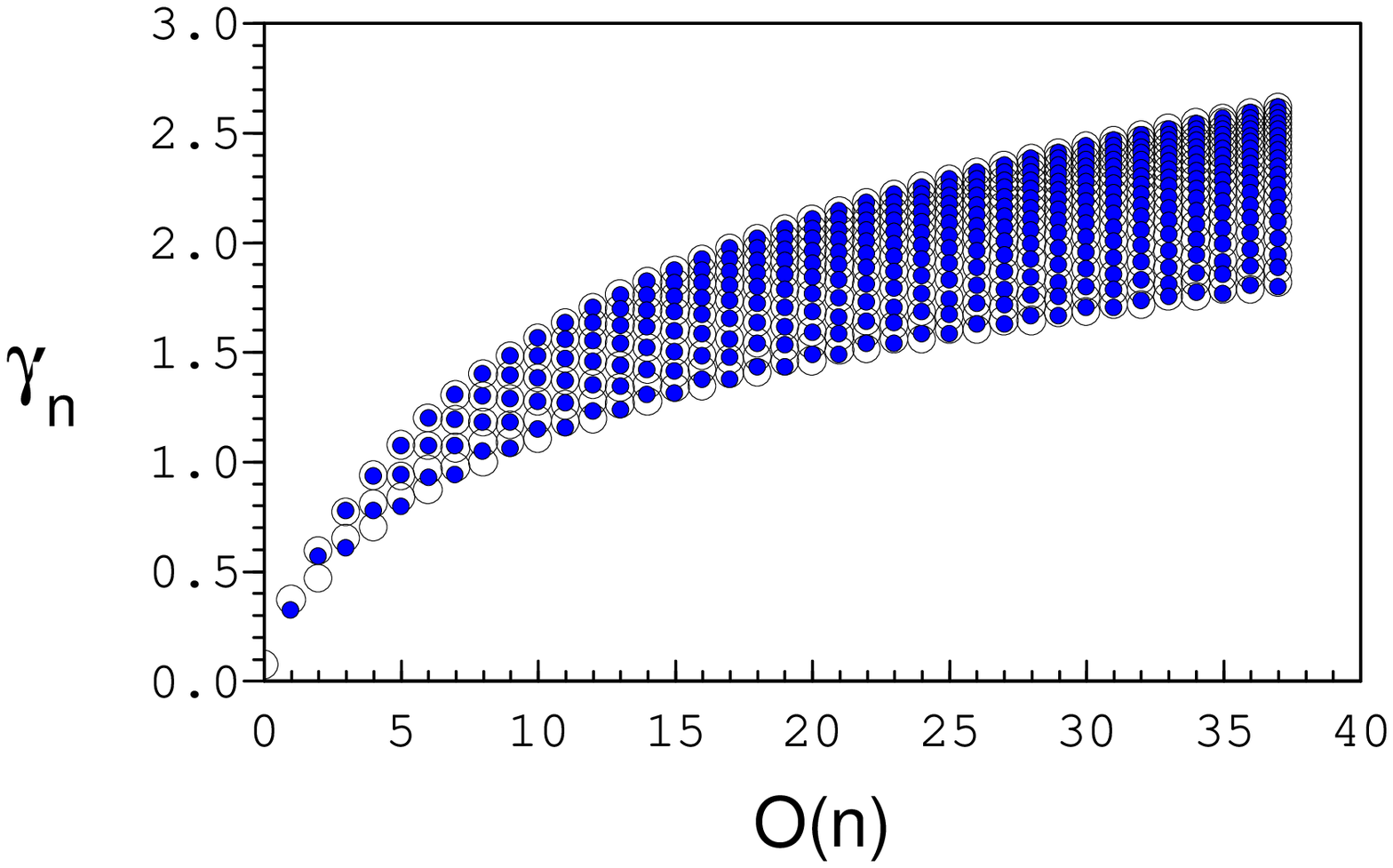}}
\end{picture}
\caption[fig:anomdimen35]
        {\tenrm Pattern of anomalous dimensions of three-quark operators
         up to $M=37$. Values associated with symmetric eigenfunctions
         are denoted by open circles. Those associated with
         antisymmetric eigenfunctions are marked by black dots. The
         solid line is an empirical fit already compatible with
         logarithmic behavior.
\label{fig:anomdim37}}
\end{figure}
%

The next two figures, Fig.~\ref{fig:anomdim37} and
Fig.~\ref{fig:anomdim150}, address the large-order behavior of the
anomalous dimensions.
Also here, both symmetry classes under the permutation $P_{13}$ are
shown in unison: open circles for values belonging to $S_{n}=1$ and
black dots for those belonging to $S_{n}=-1$.
As the order of eigenfunctions increases, a different picture for the
large-order behavior of the spectrum of anomalous dimensions
develops, namely, one of logarithmic increase.
Indeed, in the limit of very high polynomial orders, symmetric
eigenfunctions tend to degenerate with their antisymmetric partners
as one observes from Figs.~\ref{fig:anomdim20} and
\ref{fig:anomdim37}, where, with increasing order, the dots tend to
enter the centers of the circles.
At the same time, the eigenvalues $\gamma_{n}$ increase logarithmically.
Hence, to order $20$, the spectrum of anomalous dimensions as a whole
is better described by a logarithmic fit (solid line) of the form
$\gamma _{n}=1.25[\log _{10}\left({\cal O}(n) + 1.37\right)]^{1.32}$.
The dotted curve shows the power-law fit
$\gamma _{n} =0.52 {\cal O}(n)^{0.417}$
which is no more supported by the higher-order values.
A more accurate prediction for the asymptotic behavior of
$\gamma _{n}$ is provided by Fig.~\ref{fig:anomdim150} where the
upper envelope of the spectrum is well-described by
$\gamma _{n} = [\log _{10}(2.13{\cal O}(n) +1.4)]^{1.48}$.
These findings, presented here for the first time, confirm
expectations based on exponentiation assumptions of leading-loop
contributions \cite{CT76,DM80,Kor89,ABH91}.
Physically, the logarithmic rise is due to enhanced emission of soft
gluons, reflecting the fact that the probability for finding bare
quarks decreases.
However, in order to establish this trend, still higher orders have to
be computed, at least up to $M=10^{3}$ (three points on a logarithmic
plot).

%
\begin{figure}
%
\centering
\epsfig{figure=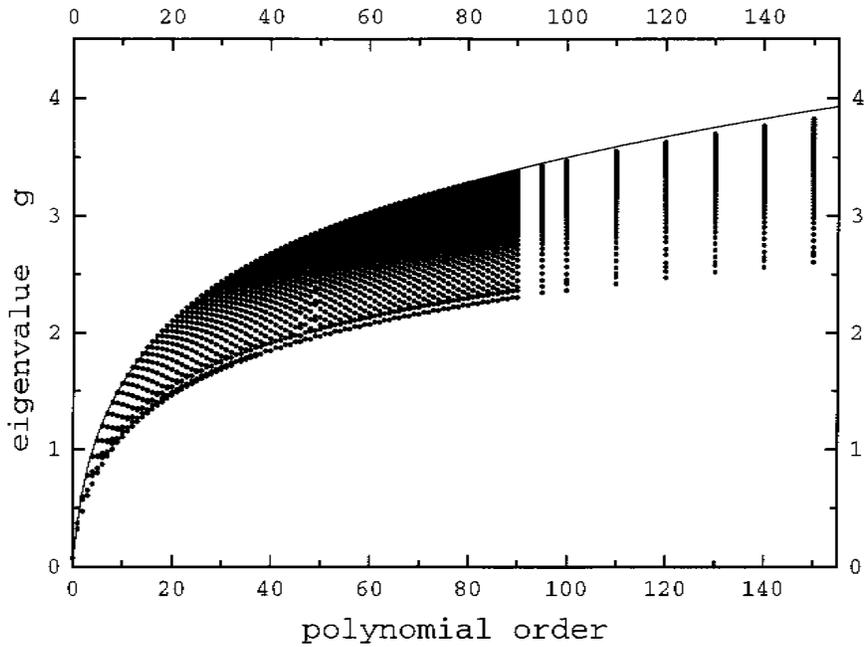,width=12cm,angle=90,silent=}
\vspace{0.4cm}
\caption[fig:anomdimen150]
        {\tenrm Pattern of anomalous dimensions of three-quark operators
         up to $M=150$. For technical reasons, antisymmetric
         eigenfunctions are included only up to order $M=37$. This has
         no substantial influence on the large-order structure. The
         obstruction at order $M=45$ is caused by a change in the
         numerical precision. The solid line is a logarithmic fit to the
         upper envelope of the spectrum up to order $M=90$.
\label{fig:anomdim150}}
\end{figure}
%

\subsection{LARGE DISTANCE PART: NON-PERTURBATIVE ASPECTS}
\label{subsec:largenonpert}
Given the eigenfunctions and associated anomalous dimensions of
the nucleon evolution equation, the only unknown quantities
entering the description of the nucleon {\DA} are the expansion
coefficients $B_{n}$, i.e., the set of matrix elements of tri-local
operators interpolating between the one-nucleon state and the vacuum.
The QCD sum rules method is presently the best approach for
extracting these quantities, albeit there are conceptual limitations.
We do not discuss the pion here since it has been studied
extensively elsewhere \cite{Rad90,CZ84b,Zhi94,Kro94}.

Specifically, we consider the following gauge-invariant proton wave
function
\begin{equation}
  \langle 0| \left[C(z_{1},z_{3}|A)\, u_{\alpha}(z_{1})\right]^{a}
             \left[C(z_{2},z_{3}|A)\, u_{\beta} (z_{2})\right]^{b}
             d_{\gamma}^{c}(z_{3})
             |P \rangle \frac{1}{\sqrt{ N_{\rm c}!}} \, \epsilon ^{abc},
\label{eq:gaugeinvnuvwf}
\end{equation}
where $a,b,c$ and $\alpha ,\beta , \gamma$ are color and spinor
labels, respectively,
$A_{\mu}(x)=\sum_{i=1}^{8}A_{\mu}(x)t^{i}$ are the Lie-algebra valued
gluon fields ($t^{i}$ being the generators of $SU(3)_{\rm c}$).
The operator
\begin{equation}
  C(z_{i},z_{j}|A)
=
  {\cal P}\exp \left[
                     i g \int_{z_{i}}^{z_{j}} d\omega _{\mu}\,
                     A_{\mu}(\omega )
               \right]
\label{eq:connector}
\end{equation}
denotes a path-dependent phase factor (``connector'' \cite{Ste84})
in which the expansion of the exponential along a contour $C$
joining the points $z_{i}$ and $z_{j}$ is controlled by the
path-ordering prescription ${\cal P}$.
For a short-distance expansion, the connector can be evaluated along
straight lines with lengths
$
 \propto |z_{i}-z_{j}|^{1/2}
 \approx |z_{\perp i}-z_{\perp j}|^{1/2}
\to 0
$
as $Q^{2} \to \infty$.
Working in the light cone frame, one can use the axial gauge
$A^{+}=0$ by virtue of which the connector at short distances can
be replaced by unity since all inter-quark separations along the light
cone are short.
A covariant gauge in which the connector can also be neglected is
provided by using the gauge parameter $a=-3$ \cite{Ste84,CD81}.
By virtue of the light-cone gauge, the contour factors reduce to
unity and the path-dependent matrix element simplifies to a
tri-local quantity of leading twist-three which can be written in
terms of three Lorentz invariant functions of positive parity
$V$ (vector $\leftrightarrow \gamma _{\mu}C$),
$A$ (axial vector $\leftrightarrow \gamma _{\mu}\gamma _{5}C$), and
$T$ (tensor $\leftrightarrow \sigma _{\mu\nu}C$) \cite{HKM75}:
\begin{eqnarray}
  \langle 0| u_{\alpha}^{a}(z_{1}) u_{\beta}^{b}(z_{2})
             d_{\gamma}^{c}(z_{3})
  | P,+ \rangle
  \frac{\epsilon ^{abc}}{\sqrt{ N_{\rm c}!}}
& = &
  \frac{1}{4\sqrt{ N_{\rm c}!}}f_{\rm N}
  [
     ({\not\! p} C)_{\alpha\beta}
     (\gamma _{5}N)_{\gamma}
     V(z_{i} \cdot p) +
     ({\not\! p}\gamma _{5} C)_{\alpha\beta}
     N_{\gamma} A(z_{i} \cdot p)
\nonumber \\
& &
   - (\sigma _{\mu\nu}p_{\nu}C)_{\alpha\beta}
     (\gamma _{\mu}\gamma _{5}N)_{\gamma}
     T(z_{i} \cdot p)
  ] \; .
\label{eq:trilocal}
\end{eqnarray}
Here $\vert P, + \rangle$ is the proton state with momentum $P$ and
positive helicity, $N$ denotes the proton spinor, and
$C=i\gamma _{0}\gamma _{2}$ is the charge-conjugation matrix.
The ``proton decay constant'' $f_{\rm N}$ is a dimensionful quantity
determining the value of the nucleon {\DA} at the
origin \cite{CZ84a,CZ84b}.

To perform the actual calculations of the nucleon {\DA}s, it is
convenient to define the functions $V$, $A$, and $T$ in the space
of longitudinal momentum fractions:
\begin{equation}
  F(x_{1}, x_{2}, x_{3})
=
  \prod_{i=1}^{3}\, \frac{d(z_{i}\cdot P)}{2\pi}\,
  \exp
      \left[
            i \sum_{j=1}^{3} x_{j} (z_{j} \cdot P)
      \right]
  F(z_{i}\cdot P) \; .
\label{eq:DAmomsp}
\end{equation}
Note that these functions depend implicitly on the factorization
scale which is supposed to serve as the starting point of evolution.
Furthermore, {\RG} controlled evolution fixes the asymptotic limits of
these functions to be
$
 \lim_{\mu ^{2}\to\infty} V
=
 \lim_{\mu ^{2}\to\infty} T
=
 \phi _{\rm as}
=
 120 x_{1}x_{2}x_{3}
$,
and
$
 \lim_{\mu ^{2}\to\infty} A
=
0
$,
the latter because of the Pauli principle.
The next step is to obtain solutions for these functions at finite
evolution scales by determining the coefficients $B_{n}$ and then to
represent them as a series expansion over a finite number of their
eigenfunctions.

Before we proceed, let us first discuss the symmetry properties of
these functions in the {\IMF}.
Because of the identity of the two $u$-quarks in the nucleon, the
functions $V$ and $T$ transform under the scalar symmetric
representation and the function $A$ under the scalar antisymmetric
representation of the permutation group $S_{3}$.
Then one has (in obvious shorthand notation \cite{AKC81}:
$P_{123}\equiv [123]$, etc.)
\vbox{
\begin{eqnarray}
  [213] V(1,2,3)
& = &
  V(1,2,3)
\label{eq:VATsym}
\cr
  [213] T(1,2,3)
& = &
  T(1,2,3)
\cr
  [213] A(1,2,3)
& = &
  -A(1,2,3) \; .
\end{eqnarray}
     }
On the other hand, the requirement that the total isospin of the
three-quark bound state should equal $1/2$ yields one more relation:
\begin{equation}
  2T(1,2,3)
=
  V(1,3,2) - A(1,3,2) + V(2,3,1) - A(2,3,1) \; .
\label{eq:relforT}
\end{equation}
Thus the system of the functions $V$, $A$, and $T$ is redundant and
the whole information they contain can be expressed by a single
function, termed $\Phi _{N}$ \cite{AKC81,CZ84a}:
\begin{equation}
  \Phi _{N} (x_{i})
\equiv
  V(x_{i}) - A(x_{i}) \; .
\label{eq:defPhiN}
\end{equation}
This function is mixed symmetric, i.e., it transforms under the
2-dimensional (matrix) representation of $S_{3}$.
Knowing $\Phi _{N}$ it is sufficient to determine all other nucleon
{\DA}s.
For the time being, there is actually no method of computing hadron
{\DA}s as a {\it whole}.\footnote{A promising approach, applied so far
only to the pion, was recently presented in \cite{BM95}.}
Hence, we must content ourselves with descriptions in terms of
truncated eigenfunctions series demanding that they (1) comply with
theoretical constraints, set for example by QCD sum rules or lattice
calculations, (2) are {\RG} controlled, i.e., satisfy the nucleon
evolution equation at every order of truncation, and (3) eventually
match the available experimental data.

In the following exposition the nucleon {\DA}s will be determined via
their moments
\begin{equation}
  F^{(n_{1}n_{2}n_{3})}
=
  \int_{0}^{1}[dx]
  x_{1}^{n_{1}}x_{2}^{n_{2}}x_{3}^{n_{3}}
  F(x_{1},x_{2},x_{3}) \; ,
\label{eq:moments}
\end{equation}
where $F(x_{i})$ stands for one of the amplitudes $V$, $T$, $A$, or
linear combinations of them.
These moments are related to the covariant derivatives of the
invariant functions with respect to the light-cone positions
$z_{i} \cdot p$.
The latter are the Fourier conjugate variables to the longitudinal
momenta $x_{i}$.
Hence we have
\begin{eqnarray}
  \prod_{i=1}^{3}
  \left(
        iz\cdot \frac{\partial}{\partial z_{i}}
  \right)^{n_{i}}
        \hat{F}(z_{i}\cdot p)\big\vert _{z_{i}=0}
& = &
  \prod_{i=1}^{3}
  \left(
        iz\cdot \frac{\partial}{\partial z_{i}}
  \right)^{n_{i}}
  \int_{0}^{1} [dx] \exp
  \left[
        -i \sum_{j=1}^{3} x_{j} (z_{j} \cdot P)
  \right]
  \hat{F}(x_{i})
\nonumber \\
& = &
  (z \cdot P)^{n_{1}+n_{2}+n_{3}}
  \int_{0}^{1} [dx] x_{1}^{n_{1}}x_{2}^{n_{2}}x_{3}^{n_{3}}\,
  \hat{F}(x_{i})
\nonumber \\
& = &
  (z \cdot P)^{n_{1}+n_{2}+n_{3}} \, \hat{F}^{(n_{1}n_{2}n_{3})} \; ,
\label{eq:derinvfun}
\end{eqnarray}
where $z$ is an arbitrary auxiliary vector ($z^{2}=0$) and the caret,
$\hat{}\;\, ,$ serves to distinguish operators from scalar functions.
Taking matrix elements between the proton state and the vacuum, we get
\begin{equation}
  \langle 0 | \hat{F}_{\gamma}^{(n_{1}n_{2}n_{3})}(0) | P \rangle
=
  f_{\rm N}(z\cdot P)^{\,{n_{1} + n_{2}+n_{3}+1}}N_{\gamma}\,
  F^{\,(n_{1}n_{2}n_{3})} \; ,
\label{eq:matelofF}
\end{equation}
where the moments on the rhs are defined in Eq.~(\ref{eq:moments}).
More precisely, we consider matrix elements (moments) of the following
operators:
\begin{equation}
  \hat{V}_{\gamma}^{(n_{1}n_{2}n_{3})}(0)
\equiv
  (iz \cdot D)^{n_{1}} u(0)
  [C\, \gamma _{\mu}\, z_{\mu}]
  (iz \cdot D)^{n_{2}} u(0)\,
  (iz \cdot D)^{n_{3}}
  \gamma _{5} d_{\gamma}(0) \; ,
\label{eq:Vder}
\end{equation}

\begin{equation}
  \hat{A}_{\gamma}^{(n_{1}n_{2}n_{3})}(0)
\equiv
  (iz \cdot D)^{n_{1}} u(0)
  [C\, \gamma _{5}\, \gamma _{\mu}\, z_{\mu}]
  (iz \cdot D)^{n_{2}} u(0)\,
  (iz \cdot D)^{n_{3}} d_{\gamma}(0) \; ,
\label{eq:Ader}
\end{equation}

\begin{equation}
  \hat{T}_{\gamma}^{(n_{1}n_{2}n_{3})}(0)
\equiv
  (iz \cdot D)^{n_{1}} u(0)
  [C\, (-\sigma _{\mu\nu})\, z_{\nu}]
  (iz \cdot D)^{n_{2}} u(0)\,
  (iz \cdot D)^{n_{3}}
  \gamma _{5}\, \gamma _{\mu}\, d_{\gamma}(0) \; .
\label{eq:Tder}
\end{equation}
The determination of the moments of these operators derives from
correlators of the generic form
\begin{eqnarray}
  I^{\,(n_{1}n_{2}n_{3},m)}(q,z)
& = &
  i\int_{}^{}
  d^{4}x \, {\rm e}^{iq\cdot x}
  \langle 0 | T\bigl (F_{\gamma}^{\,(n_{1}n_{2}n_{3})}(0)
  \hat J_{\gamma\prime}^{\,(m)}(x)\bigr ) | 0 \rangle
  (z\cdot \gamma )_{\gamma \gamma\prime}
\nonumber \\
& = &
  (z\cdot q)^{\,{n_{1}+n_{2}+n_{3}+m+3}}
  I^{\,(n_{1}n_{2}n_{3},m)}(q^{2}) \; ,
\label{eq:correlator}
\end{eqnarray}
where $z$ is again a light-like auxiliary vector and the factor
$(z\cdot \gamma )_{\gamma\gamma '}$ serves to project out the
leading-twist structure of the correlator.
The computation of the Wilson coefficients on the quark side of the
correlator amounts to the perturbative evaluation of diagrams
involving local quark/gluon condensates \cite{CZ84a}.
It yields the theoretical side of the sum rule.
The hadronic (phenomenological) side of the sum rule is obtained by
saturating the correlator by the lowest-mass baryon state(s) via a
dispersion relation.
Reconciliation of the two sides of the sum rule with respect to
the Borel parameter within a continuous interval as large as
possible determines the variation of permissible values for a
particular moment.
We do not derive sum rules here.
For more details we refer to the original
works \cite{CZ84a,KS87,Ste89,COZ89a} and the review
article \cite{CZ84b}.
We merely use them as theoretical constraints imposed on the moments
of the nucleon {\DA}.
However, these constraints cannot fix the shape of the nucleon {\DA}
uniquely and one has to impose additional {\it global} constraints as
opposed to the {\MSR} which constitute {\it local}
constraints.\footnote{One can always add some oscillating function
which vanishes at the points fixed by the local constraints but which
contributes outside.}

The moments of the nucleon {\DA} $\Phi _{N}$ in terms of the expansion
coefficients $B_{n}$ are displayed in Table~\ref{tab:moments}.

%
\begin{table}
\caption{Analytical expressions for the moments of the nucleon
         distribution amplitude
         $\Phi _{N}^{(n_{1}n_{2}n_{3})}\equiv \Phi _{N}^{[k]}$
         in terms of the expansion coefficients $B_{n}$ up to order
         $M=n_{1}+n_{2}+n_{3}=3$.
\label{tab:moments}}
\begin{tabular}{r|l|l}
k&$\Phi_N^{[k]}$      & Moments       \\   \hline
0&$\Phi_N^{(000)}  $&${B_0} $\\
1&$\Phi_N^{(100)}  $&${{7\,{B_0} + {B_1} + {B_2}}\over {21}} $\\
2&$\Phi_N^{(010)}  $&${{7\,{B_0} - 2\,{B_2}}\over {21}}      $\\
3&$\Phi_N^{(001)}  $&${{7\,{B_0} - {B_1} + {B_2}}\over {21}} $\\
4&$\Phi_N^{(200)}  $&${{108\,{B_0} + 27\,{B_1} + 27\,{B_2} + 9\,{B_3} -
                   {B_4} - {B_5}}\over {756}}  $\\
5&$\Phi_N^{(020)}  $&${{18\,{B_0} - 9\,{B_2} + {B_3} + {B_5}}
\over {126}} $\\
6&$\Phi_N^{(002)}  $&${{108\,{B_0} - 27\,{B_1} + 27\,{B_2} + 9\,{B_3} +
{B_4} -  {B_5}}\over {756}}  $\\
7&$\Phi_N^{(110)}  $&${{72\,{B_0} + 9\,{B_1} - 9\,{B_2} - 3\,{B_3} +
{B_4} - 3\,{B_5}}\over {756}} $\\
8&$\Phi_N^{(101)}  $&${{36\,{B_0} + 9\,{B_2} - 3\,{B_3} +
2\,{B_5}}\over {378}} $\\
9&$\Phi_N^{(011)}  $&${{72\,{B_0} - 9\,{B_1} - 9\,{B_2} - 3\,{B_3} -
{B_4} - 3\,{B_5}}\over {756}} $\\
10&$\Phi_N^{(300)}  $&${{87120\,{B_0} + 29040\,{B_1} + 29040\,{B_2} +
17424\,{B_3} - 1936\,{B_4} - 1936\,{B_5}}\over {1219680}} $\\
  &                 &${{- 33\,{B_6} -
     33\,{\sqrt{97}}\,{B_6} - 33\,{B_7} + 33\,{\sqrt{97}}\,{B_7} +
     146\,{B_8} + 2\,{\sqrt{4801}}\,{B_8} + 146\,{B_9} -
     2\,{\sqrt{4801}}\,{B_9}}\over {1219680}} $\\
11&$\Phi_N^{(030)}  $&${{165\,{B_0} - 110\,{B_2} + 22\,{B_3} +
22\,{B_5} + {B_6} + {B_7}}\over {2310}} $\\
12&$\Phi_N^{(003)}  $&${{87120\,{B_0} - 29040\,{B_1} + 29040\,{B_2} +
17424\,{B_3} + 1936\,{B_4} - 1936\,{B_5}}\over {1219680}} $\\
  &                 &$ {{- 33\,{B_6} -
     33\,{\sqrt{97}}\,{B_6} - 33\,{B_7} + 33\,{\sqrt{97}}\,{B_7} -
     146\,{B_8} - 2\,{\sqrt{4801}}\,{B_8} - 146\,{B_9} +
     2\,{\sqrt{4801}}\,{B_9}}\over {1219680}} $\\
13&$\Phi_N^{(210)}  $&${{5940\,{B_0} + 1320\,{B_1} + 88\,{B_4} -
440\,{B_5} + 15\,{B_6} + 3\,{\sqrt{97}}\,{B_6} + 15\,{B_7} -
     3\,{\sqrt{97}}\,{B_7} - 12\,{B_8} - 12\,{B_9}}\over {166320}} $\\
14&$\Phi_N^{(201)}  $&${{130680\,{B_0} + 14520\,{B_1} + 43560\,{B_2} -
8712\,{B_3} - 968\,{B_4} + 10648\,{B_5}}\over {3659040}} $\\
  &                  &$ {{- 231\,{B_6} +
     33\,{\sqrt{97}}\,{B_6} - 231\,{B_7} - 33\,{\sqrt{97}}\,{B_7} -
     174\,{B_8} - 6\,{\sqrt{4801}}\,{B_8} - 174\,{B_9} +
     6\,{\sqrt{4801}}\,{B_9}}\over {3659040}} $\\
15&$\Phi_N^{(120)}  $&${{495\,{B_0} + 55\,{B_1} - 165\,{B_2} -
11\,{B_3} +      11\,{B_4} - 11\,{B_5} - 3\,{B_6} - 3\,{B_7} + {B_8} +
     {B_9}}\over {13860}}  $\\
16&$\Phi_N^{(102)}  $&${{130680\,{B_0} - 14520\,{B_1} + 43560\,{B_2} -
8712\,{B_3} + 968\,{B_4} + 10648\,{B_5}}\over {3659040}} $\\
  &                 &$  {{- 231\,{B_6} +
     33\,{\sqrt{97}}\,{B_6} - 231\,{B_7} - 33\,{\sqrt{97}}\,{B_7} +
     174\,{B_8} + 6\,{\sqrt{4801}}\,{B_8} + 174\,{B_9} -
     6\,{\sqrt{4801}}\,{B_9}}\over {3659040}} $\\
17&$\Phi_N^{(021)}  $&${{495\,{B_0} - 55\,{B_1} - 165\,{B_2} -
11\,{B_3} - 11\,{B_4} - 11\,{B_5} - 3\,{B_6} - 3\,{B_7} - {B_8} -
     {B_9}}\over {13860}} $\\
18&$\Phi_N^{(012)}  $&${{5940\,{B_0} - 1320\,{B_1} - 88\,{B_4} -
440\,{B_5} + 15\,{B_6} + 3\,{\sqrt{97}}\,{B_6} + 15\,{B_7} -
     3\,{\sqrt{97}}\,{B_7} + 12\,{B_8} + 12\,{B_9}}\over {166320}}$
\end{tabular}
\end{table}
%

Because of longitudinal momentum conservation,
$x_{1} + x_{2} + x_{3} =1$,
not all the moments at a given order $M=n_{1}+n_{2}+n_{3}$ are
linearly independent.
This implies \cite{KS87}
\begin{equation}
   \Phi _{N}^{(n_{1},n_{2},n_{3})}
=
   \Phi _{N}^{(n_{1}+1,n_{2},n_{3})}
 + \Phi _{N}^{(n_{1},n_{2}+1,n_{3})}
 + \Phi _{N}^{(n_{1},n_{2},n_{3}+1)} \; .
\label{eq:momconKS}
\end{equation}
For instance, at order $M=3$ there are $20$ moments out of which only
$10$ are strict.
Which combinations are actually taken, depends on the choice of the
polynomial basis in which the eigenfunctions are finally expressed.
We use throughout this report the powers of the monomial $x_{1}x_{3}$,
i.e., the basis $|k\, l \rangle$.
In terms of this basis, the moments of $\Phi _{N}$ read
\begin{equation}
  \Phi _N^{(n_{1}n_{2}n_{3})}
=
  \int_{0}^{1}\, dx_{1}\, \int_{0}^{1-x_1}\, dx_{3}\,
  \left[
        \sum_{i=0}^{n_2} \sum_{j=0}^{i}
        (-1)^{i}\,{n_{2} \choose i}\,{i \choose j}
        x_{1}^{n_1+i-j}x_{3}^{n_3+j}
  \right]
  \Phi _N(x_{1},x_{3}) \; .
\label{eq:momentsfinal}
\end{equation}
It was first shown in \cite{ELBA93,BANSKA93,KYF93} and then outlined
in \cite{Ste94} that it is possible to derive a {\it closed-form}
relation between the expansion coefficients $B_{n}$ and the strict
moments of $\Phi _{N}$, defined by
\begin{equation}
  \Phi _{N}^{(i0j)}
=
  \int_{0}^{1}[dx]\,x_{1}^{i}\,x_{2}^{0}\,x_{3}^{j}\,
  \Phi _{N}(x_{k}, \mu ^{2}) \; ,
\label{eq:strictmom}
\end{equation}
to arrive at
\begin{equation}
  \frac{B_ {n}(Q^{2})}{\sqrt{N_{n}}} =
  \frac{\sqrt{N_{n}}}{120}\,
                 \Biggl[
                        \frac {\ln (Q^{2}/\Lambda _{QCD}^{2})}
                        {\ln (\mu ^{2}/\Lambda _{QCD}^{2})}
                 \Biggr]^{-\gamma _{n}}
  \sum_{i,j=0}^{\infty}a_{ij}^{n}\
  \Phi _{N}^{(i0j)}(\mu ^{2}) \; ,
\label{eq:magic}
\end{equation}
where the normalization constants $N_{n}$, the matrix coefficients
$a_{ij}^{n}$, and the anomalous dimensions $\gamma _{n}$ (up to $M=4$)
are those tabulated in Table~\ref{tab:eigen}.
This expression enables the {\it analytical} calculation of the
coefficients to any desired order of polynomial expansion.
The utility of Eq.~(\ref{eq:magic}) is twofold: (1) As repeatedly
stated, the moments of hadron {\DA}s are not accurately determined.
Thus, without such an {\it explicit} relation between expansion
coefficients and strict moments, one has to perform a simultaneous
and self-consistent fit to the {\MSR} constraints, a procedure
which obviously becomes increasingly tedious as the moment-order
grows.
(2) As already outlined in the previous chapter, the Hilbert-Schmidt
orthogonalization procedure of polynomials with more than one
variable is not unique, and one is well-advised to look for
higher-order eigenfunctions which are as simple as possible,
absorbing numerical coefficients into the normalization constants.
This non-uniqueness entails that one can compare expansion
coefficients $B_{n}$, obtained in different approaches, only if they
are {\it normalized}.
Without the knowledge of the normalization constants $N_{n}$, the
values of $B_{n}$ are of no significance or practical usefulness.
However, having derived Eq.~(\ref{eq:magic}), the knowledge of the
specific normalization used in different approaches becomes
superfluous.
The coefficients $B_{n}$ can be self-consistently computed on the
basis of the strict moments alone, which are universal quantities,
modulo an overall normalization through the value of
$\Phi _{N}^{(000)}$.
In the following,
\begin{equation}
  \int_{0}^{1}[dx]\,
  \Phi _{N}\left(x_{i},\mu ^{2}\right)
=
  1
\label{eq:ndanorm}
\end{equation}
is used.
Hence, evaluating the strict moments of the nucleon {\DA} up to a
given order, one can determine the corresponding expansion coefficients
of the same order and vice versa.
All these advantages will become successively apparent through the
applications of the formalism in subsequent chapters.
The link between $B_{n}$ and the strict moments of $\Phi _{N}$ is
exemplified below (see also \cite{CORFU92,BS93}):
\begin{eqnarray}
  B_{1}(\mu ^{2})
& = &
  \frac{1260}{120}
\left[
      \Phi _{N}^{(100)} - \Phi _{N}^{(001)}
\right]\Big|_{\mu ^{2}}
\nonumber \\
  B_{2}(\mu ^{2})
& = &
  \frac{420}{120}
\left[
        2 \Phi _{N}^{(000)} -3 \Phi _{N}^{(100)} -3 \Phi _{N}^{(001)}
\right]\Big|_{\mu ^{2}}
\nonumber \\
  B_{3}(\mu ^{2})
& = &
  \frac{756}{120}
\left[
        2 \Phi _{N}^{(000)} -7 \Phi _{N}^{(100)} -7 \Phi _{N}^{(001)}
      + 8 \Phi _{N}^{(200)} +4 \Phi _{N}^{(101)} +8 \Phi _{N}^{(002)}
\right]\Big|_{\mu ^{2}}
\nonumber \\
  B_{4}(\mu ^{2})
& = &
  \frac{34020}{120}
\left[
          \Phi _{N}^{(100)} - \Phi _{N}^{(001)}
       -\frac{4}{3} \Phi _{N}^{(200)} + \frac{4}{3} \Phi _{N}^{(002)}
\right]\bigg|_{\mu ^{2}}
\nonumber \\
  B_{5}(\mu ^{2})
& = &
  \frac{1944}{120}
\left[
        2 \Phi _{N}^{(000)} -7 \Phi _{N}^{(100)} -7 \Phi _{N}^{(001)}
       +\frac{14}{3} \Phi _{N}^{(200)} +14 \Phi _{N}^{(101)}
       +\frac{14}{3} \Phi _{N}^{(002)}
\right]\bigg|_{\mu ^{2}}.
\label{eq:B1B2B3B4B5}
\end{eqnarray}
Note that $B_{0}$ is fixed to unity by the normalization of
$\Phi _{N}$ (cf. Eq.~(\ref{eq:ndanorm})).
Furthermore, recall once again that the notations of \cite{Ste89} are
used.
The next-to-leading order expansion coefficients ($M=3$) are:
\begin{eqnarray}
  B_{6}(\mu ^{2})
& = &
  \frac{4620}{120} \;
  \frac{485 + 11 {\sqrt{97}}}{97}
\biggl[
         \Phi _{N}^{(000)}
       - 6 \left( \Phi _{N}^{(100)} + \Phi _{N}^{(001)} \right)
       + \frac{41 + {\sqrt{97}}}{4}
         \left( \Phi _{N}^{(200)} + \Phi _{N}^{(002)} \right)
\nonumber \\
&&
       + 3 \frac{31 - {\sqrt{97}}}{4} \Phi _{N}^{(101)}
       - 5 \frac{17 + {\sqrt{97}}}{16}
         \left( \Phi _{N}^{(300)} +  \Phi _{N}^{(003)} \right)
\nonumber \\
&&
       - 5 \frac{31 - {\sqrt{97}}}{8}
         \left( \Phi _{N}^{(201)} +  \Phi _{N}^{(102)} \right)
\biggr]\bigg|_{\mu ^{2}}
\nonumber \\
  B_{7}(\mu ^{2})
& = &
  \frac{4620}{120} \;
  \frac{485 - 11 {\sqrt{97}}}{97}
\biggl[
           \Phi _{N}^{(000)}
       - 6 \left( \Phi _{N}^{(100)} + \Phi _{N}^{(001)} \right)
       + \frac{41 - {\sqrt{97}}}{4}
         \left( \Phi _{N}^{(200)} + \Phi _{N}^{(002)} \right)
\nonumber \\
&&
       + 3 \frac{31 + {\sqrt{97}}}{4} \Phi _{N}^{(101)}
       - 5 \frac{17 - {\sqrt{97}}}{16}
         \left( \Phi _{N}^{(300)} + \Phi _{N}^{(003)} \right)
\nonumber \\
&&
       - 5 \frac{31 + {\sqrt{97}}}{8}
         \left( \Phi _{N}^{(201)} + \Phi _{N}^{(102)} \right)
\biggr]\bigg|_{\mu ^{2}}
\nonumber \\
  B_{8}(\mu ^{2})
& = &
  \frac{27720}{120} \;
  \frac{33607 - 247 {\sqrt{4801}}}{4801}
\biggl[
           \Phi _{N}^{(100)} - \Phi _{N}^{(001)}
       - 3 \left( \Phi _{N}^{(200)} - \Phi _{N}^{(002)} \right)
\nonumber \\
&&
       +   \frac{601 + {\sqrt{4801}}}{264}
           \left( \Phi _{N}^{(300)} - \Phi _{N}^{(003)} \right)
\nonumber \\
&&
       +   \frac{59 - {\sqrt{4801}}}{44}
           \left( \Phi _{N}^{(201)} - \Phi _{N}^{(102)} \right)
\biggr]\bigg|_{\mu ^{2}}
\nonumber \\
  B_{9}(\mu ^{2})
& = &
  \frac{27720}{120} \;
  \frac{33607 + 247 {\sqrt{4801}}}{4801}
\biggl[
           \Phi _{N}^{(100)} - \Phi _{N}^{(001)}
       - 3 \left( \Phi _{N}^{(200)} - \Phi _{N}^{(002)} \right)
\nonumber \\
&&
       +   \frac{601 - {\sqrt{4801}}}{264}
           \left( \Phi _{N}^{(300)} -  \Phi _{N}^{(003)} \right)
\nonumber \\
&&
       +   \frac{59 + {\sqrt{4801}}}{44}
           \left( \Phi _{N}^{(201)} - \Phi _{N}^{(102)} \right)
\biggr]\bigg|_{\mu ^{2}} \; .
\label{eq:B6B7B8B9}
\end{eqnarray}
The values of the moments of $\Phi _{N}$ are extracted from the
correlator in Eq.~(\ref{eq:correlator}) for
$n_{1} + n_{2} + n_{3} \leq 3$ and $m=1$
at some self-consistently determined normalization point
$\mu = \mu _{\rm F}$ of order 1~GeV (see, e.g., \cite{CZ84a,Ste89}) at
which a short-distance {\OPE} can be safely performed and quark-hadron
duality is supposed to be valid.
Table~\ref{tab:sr} shows the range of values obtained by
COZ \cite{COZ89a} up to order $M=3$ together with those calculated
by KS \cite{KS87} for the first- and second-order moments.

%
\begin{table}
\caption[tab:COZ]
        {Numerical values of the moments
         $M=n_{1}+n_{2}+n_{3}\leq 3$ of the heterotic
         nucleon distribution amplitude in comparison with those
         of previous models versus the QCD sum-rule
         constraints derived by Chernyak, Ogloblin, and
         Zhitnisky~\cite{COZ89a}, and
         King and Sachrajda~\cite{KS87}.
\label{tab:sr}}
\begin{tabular}{cccccccc}
         ${(n_{1}n_{2}n_{3})}$ & COZ-SR  &  KS-SR  &
         $\Phi_{N/het}^{(n_{1}n_{2}n_{3})}$        &
         $\Phi_{N/COZ}^{(n_{1}n_{2}n_{3})}$        &
         $\Phi_{N/CZ}^{(n_{1}n_{2}n_{3})}$         &
         $\Phi_{N/GS}^{(n_{1}n_{2}n_{3})}$         &
         $\Phi_{N/KS}^{(n_{1}n_{2}n_{3})}$ \\
\hline
         (000)   & 1                &  1          & 1        &  1    & 1 &1     & 1   \\
         (100)   & 0.54---0.62      & 0.46---0.59 & 0.5721   &0.5790 &0.630   &0.6269 &0.550\\
         (010)   & 0.18---0.20      & 0.18---0.21 & 0.1837   &0.1920 &0.150   &0.1371 &0.210\\
         (001)   & 0.20---0.25      & 0.22---0.26 & 0.2442   &0.2290 &0.220   &0.2359 &0.240\\
         (200)   & 0.32---0.42      & 0.27---0.37 & 0.3380   &0.3690 &0.397   &0.2879 &0.350\\
         (020)   & 0.065---0.088    & 0.08---0.09 & 0.0661   &0.0680 &0.0233  &0.0321 &0.090\\
         (002)   & 0.09---0.12      & 0.10---0.12 & 0.1696   &0.0890 &0.080   &0.0079 &0.120\\
         (110)   & 0.08---0.10      & 0.08---0.10 & 0.1386   &0.0970 &0.110   &0.1080 &0.100\\
         (101)   & 0.09---0.11      & 0.09---0.11 & 0.0955   &0.1130 &0.123   &0.2309 &0.100\\
         (011)   & --0.03---0.03    & unreliable  & --0.0210 &0.0270 &0.017   &-0.0029 &0.020\\
         (300)   & 0.21---0.25      &             & 0.2101   &0.2445 &0.257   &0.1281 &0.2333\\
         (030)   & 0.028---0.04     &             & 0.0392   &0.0381 &0.0013  &0.0169 &0.0573\\
         (003)   & 0.048---0.056    &             & 0.1392   &0.0485 &0.0413  &-0.0515 &0.0813\\
         (210)   & 0.041---0.049    &             & 0.0789   &0.0587 &0.068   &0.0463 &0.0593\\
         (201)   & 0.044---0.055    &             & 0.0490   &0.0658 &0.0713  &0.1135 &0.0573\\
         (120)   & 0.027---0.037    &             & 0.0504   &0.0243 &0.0253  &0.0278 &0.030\\
         (102)   & 0.037---0.043    &             & 0.0372   &0.0331 &0.0353  &0.0836 &0.0320\\
         (021)   & --0.004---0.007  &             & --0.0235 &0.0056 &-0.003  &-0.0127 &0.0027\\
         (012)   & --0.005---0.008  &             & --0.0068 &0.0073 &0.003   &-0.0241 &0.0067\\
\end{tabular}
\end{table}
%

This exposition completes our discussion of the expansion coefficients
and the brief review of the conceptual essentials underlying their
non-perturbative determination.
Before moving on in the next chapter to the actual models for the
nucleon {\DA}, we briefly discuss now the momentum evolution of the
normalized expansion coefficients $B_{n}$ (recall
Eq.~\ref{eq:magic}) against the order of expansion in terms of
eigenfunctions.
The result is illustrated in Fig.~\ref{fig:comomorder}.
One observes that the ratio $\frac{B_{n}(Q^{2})}{B_{n}(\mu ^{2})}$
decreases both with increasing polynomial order and increasing
momentum transfer so that a truncation at low orders seems, in
principle, justifiable.
However, in order that the truncation of the infinite eigenfunctions
expansion, given by Eq.~(\ref{eq:Phieigen}), is computationally
useful, we have to ensure dominance of the lowest-order
contributions.
Given this premise, the guiding principle is to minimize the
influence of higher-order terms which we have discarded.
This procedure is in some sense analogous to the optimization of
the renormalization-scheme dependence of physical quantities
computed in a perturbative scheme ({\it principle of
minimum sensitivity} \cite{Ste81}).
Without enough understanding of the underlying non-perturbative
dynamics, it remains a challenge to develop a method of computing
hadron distribution amplitudes as a {\it whole}.
For the time being, we must content ourselves with an {\it effective}
description in which the {\it local} low-order moment constraints are
supplemented by additional {\it global} constraints to impose
restrictions on the shape of the nucleon {\DA} as a whole.
This may be achieved my means of a {\it hierarchical} (with respect to
moment order) $\chi ^{2}$-criterion  (see next section) or by demanding,
for instance, smoothness of the nucleon {\DA} \cite{Sch89}.
In this way, one can enforce the dominance of the lowest-order
contributions and minimize the influence of the disregarded higher-order
terms.
Since the extraction of reliable estimates of higher-order moments
from QCD sum rules is severely limited \cite{CZ84b} (the accuracy
of the moment values one can extract decreases as the moment order
increases), there is actually no other pragmatic alternative to this
type of approach.
So, if one is satisfied with a given accuracy (of observables
calculated with these nucleon {\DA}s) relative to existing data,
reconstructing an analytic representation of the nucleon {\DA} in
terms of higher and higher eigenfunctions is unnecessary --
maybe even irrelevant.

%
\begin{figure}
\begin{picture}(0,210)
  \put(67,-250){\psboxscaled{600}{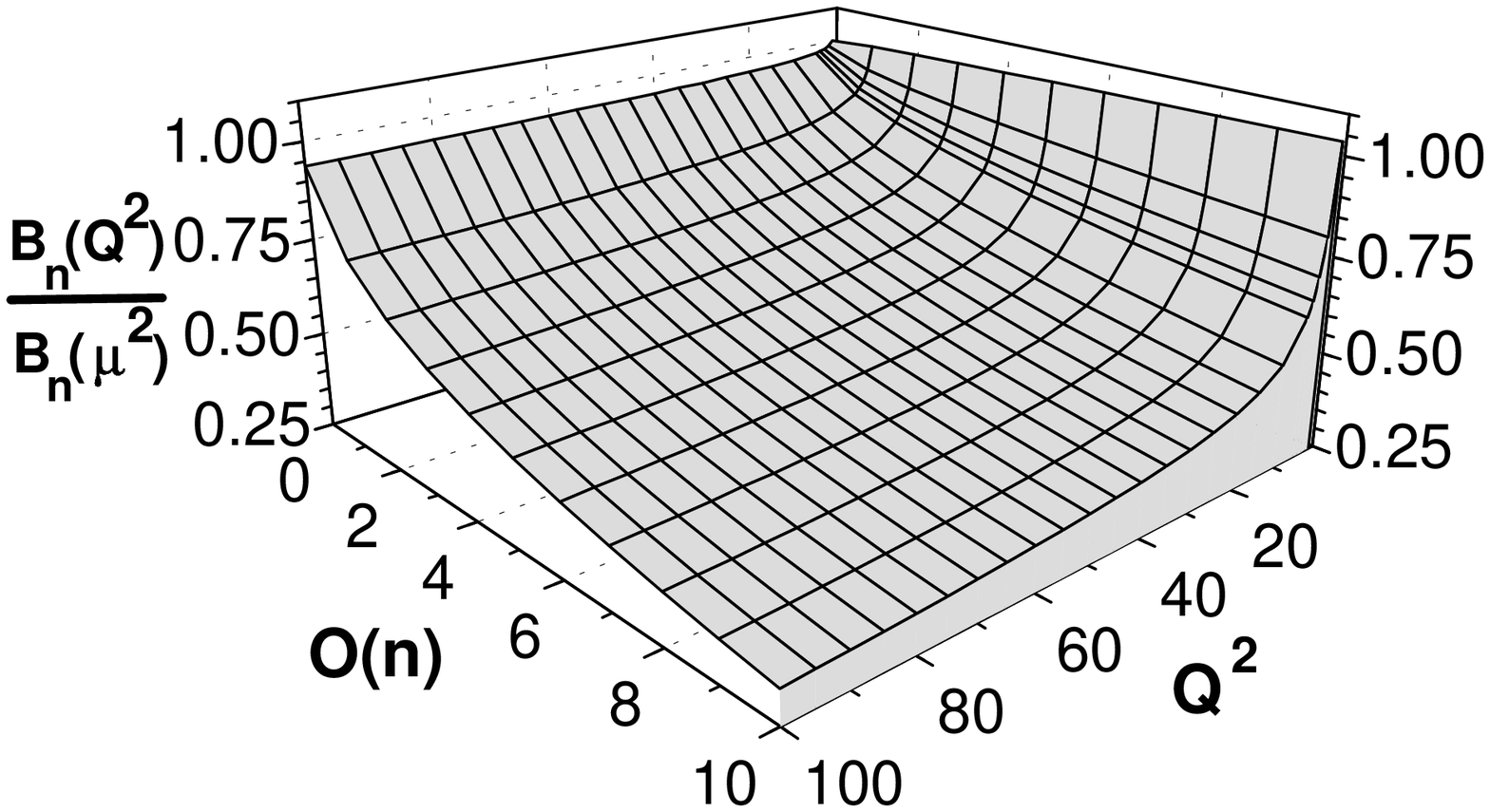}}
\end{picture}
\caption[fig:Bchange]
        {\tenrm
         Evolution behavior of the normalized expansion coefficients
         $B_{n}$ versus the tradeoff with the order of expansion in
         eigenfunctions.
\label{fig:comomorder}}
\end{figure}
%

\subsection{NUCLEON DISTRIBUTION AMPLITUDES}
\label{subsec:nucda}
It is evident from the discussion in the previous chapter that the
method of QCD sum rules enables the determination of the basic
characteristics of the nucleon {\DA}, namely: (1) its value at the
origin, $f_{\rm N}$, and (2) a model representation over longitudinal
momenta via a finite number of moments.
However, the reliability of this procedure depends crucially on the
particular way of using the QCD sum-rule constraints.
The basic assumptions in truncating the infinite series of
eigenfunctions should be:

\begin{itemize}
\item A truncated representation at relatively low orders
      provides a reasonable approximation of the true nucleon {\DA},
      given that the normalized expansion coefficients decrease
      with the (moment) order, albeit slowly.
\item The intrinsic inaccuracy of the sum rules for higher-order
      moments is taken into account.
\item The evolution equation is always satisfied.
\end{itemize}

The first premise says nothing about the convergence of the
eigenfunctions series.
We know that this series converges for very large values of $Q^{2}$
to the asymptotic solution $\phi _{\rm as}$, but only logarithmically.
Nevertheless, it seems reasonable to believe that the characteristic
properties of the nucleon {\DA} belong to the {\it entire} series and
that they do not first show up in higher orders.
On the contrary, the polynomial structure of the eigenfunctions
(comprising not only symmetric but also antisymmetric Appell
polynomials) will inevitably introduce oscillations (``wiggles'') at
every step of truncation because of the normalization condition
imposed on $\Phi _{N}$ (cf. Eq.~(\ref{eq:ndanorm})).
These wiggles are unphysical and should be washed out by destructive
interference at subsequent steps of truncation.
Hence, it is evident that the trial nucleon {\DA} at every step of
truncation should be chosen to have a shape as smooth as the
sum-rule constraints on its moments allow.
This procedure has close resemblance to the Tamm-Dancoff
method \cite{TD45} of truncating time-ordered products in quantum
field theory in seeking for a ground-state
solution.\footnote{I wish to thank Prof. D. Shirkov for discussions on
this point.}
In this method, as in our approach, the parameters of the low order
solutions are {\it effective}, meaning that they incorporate by
construction crucial high-order effects.

Following this strategy, we introduce a $\chi ^{2}$-criterion which
not only parameterizes deviations from the sum rules {\it laterally},
i.e., within some fixed moment order, but -- in addition -- which
weighs the quality of sum rules {\it vertically}, i.e., according to
their order \cite{CORFU92,SB93nuc,BS93,BS94,Ste94}.
This can be realized by defining
\begin{equation}
  \chi ^{2}_{k}
=
  \left( \chi ^{2}_{k,(a)} + \chi ^{2}_{k,(b)} \right) \,
  \left[
  1\, - \,\Theta \left( m_{k}-M_{k}^{\rm min} \right)
  \Theta \left( M_{k}^{\rm max}-m_{k} \right)
  \right]
\label{eq:chicrit1}
\end{equation}
with
\begin{equation}
  \chi ^{2}_{k,(a)}
=
  {\rm min} \left(
             \left\vert M_{k}^{{\rm min}}-m_{k} \right\vert ,
             \left\vert m_{k}-M_{k}^{{\rm max}} \right\vert
       \right) \, N_{k}^{-1}
\label{eq:chicrit2}
\end{equation}
($m_{k}$ ($k=1,2, \ldots , 18$) denoting collectively the moments),
where
$
 N_{k}= \left\vert M_{k}^{{\rm min}} \right\vert
$
or
$
 \left\vert M_{k}^{{\rm max}} \right\vert
$,
whether $m_{k}$ lies on the left- or on the right-hand side of the
corresponding sum-rule interval
($\chi ^{2}_{{\rm tot}}=\sum_{k}^{}\chi ^{2}_{k}$)
and
\begin{equation}
\chi ^{2}_{k,(b)}=
  \left\{\begin{array}{lll}
         100,   & 1\leq k\leq 3\\
         10,    & 4\leq k\leq 9\\
         1,     & 10\leq k\leq 18 \; .
\end{array}
\right.
\label{eq:chicrit3}
\end{equation}
Let us consider a concrete example in order to make this criterion
and its use more transparent.
Suppose a trial {\DA} corresponds to
$\chi ^{2}_{{\rm tot}}=132.85$.
This means that it breaks one first-order sum rule, three second-order
sum rules, and two third-order sum rules, whereas the total breaking is
$85\%$.

This ``hierarchical'' treatment of the sum rules (1) accounts for the
higher stability of the lower-level moments relative to the higher
ones \cite{Ste89}, and (2) does not overestimate the significance of
the unverified constraints \cite{COZ89a} for the third-order moments.
Consequently, our parameterization in Eq.~(\ref{eq:chicrit3}) favors
solutions that satisfy best the lowest-order moments and gives
(arbitrary) penalty points to those solutions which violate
them.
This relegates third-order terms to contribute only marginally,
i.e., to merely {\it refine} (if at all) the shape of the nucleon {\DA}
determined in the first step solely on the basis of the second-order
moment constraints.
It goes without saying that this procedure intrinsically suppresses
solutions with unphysical oscillations.
A physical analog of truncating the representation of the nucleon
{\DA} this way is perhaps provided by a holographic image which is not
destroyed when cut into smaller pieces (corresponding here to a lower
order of truncation) but becomes rather less sharp \cite{CORFU92}.
In contrast, a series truncation according to a simple (i.e.,
lateral) $\chi ^{2}$-criterion corresponds to a conventional image
which is really destroyed when cut into smaller pieces.

%
\begin{table}
\caption{Theoretical parameters to classify the nucleon distribution
         amplitudes discussed in the text. The samples shown refer to
         the COZ sum rules.
\label{tab:Bs}}
\begin{tabular}{lrrrrrrrrc}
 Model       & $B_{1}$\ \ &  $B_{2}$\ \ &   $B_{3}$\ \  &    $B_{4}$\ \ &     $B_{5}$\ \ &
 $\vartheta[deg]$\  &  R\ \ & $\chi^{2}$\ & Symbol \cr
\hline
 $Het     $ & 3.4437 &  1.5710 &   4.5937  &   29.3125 &    -0.1250  &    -1.89  &    .104 &  33.48  &{\Large $\bullet$} \cr
 $Het^\prime $& 4.3025&  1.5920 &   1.9675  &  -19.6580 &     3.3531  &    24.44  &    .448 &  30.63 &{\Large $\bullet$} \cr
 $COZ^{{\rm opt}}$ & 3.5268 &  1.4000 &   2.8736  &   -4.5227 &  0.8002  &     9.13  &    .465 &   4.49 &$\blacksquare$    \cr
 $COZ^{{\rm up}}$ & 3.2185 &  1.4562 &   2.8300  &  -17.3400 &  0.4700  &   5.83  &    .4881& 21.29 &$+$ \cr
 $COZ      $ & 3.6750 &  1.4840 &   2.8980  &   -6.6150 &     1.0260  &    10.16  &    .474 &  24.64 &$\Box$   \cr
 $CZ       $ & 4.3050 &  1.9250 &   2.2470  &   -3.4650 &     0.0180  &    13.40  &    .487 & 250.07 &$\blacklozenge$ \cr
 $KS^{{\rm low}}$ & 3.5818 &  1.4702 &   4.8831  &   31.9906 &  0.4313  &    -0.93  &    .0675&  36.27  &$\circ$ \cr
 $KS/COZ^{{\rm opt}}$  & 3.4242 &  1.3644 &   3.0844  &   -3.2656 &  1.2750  &     9.47  &    .453 &   5.66 &$\circ$    \cr
 $KS^{{\rm up}}$ & 3.5935 &  1.4184 &   2.7864  &  -13.3802 &   2.0594   &  13.82  &    .482 & 40.38 &$\circ$ \cr
 $KS       $ & 3.2550 &  1.2950 &   3.9690  &    0.9450 &     1.0260  &     2.47  &    .412 & 116.35 &$\Diamond$ \cr
 $GS^{{\rm opt}} $ & 3.9501 &  1.5273 &  -4.8174  &    3.4435 &  8.7534  &    80.87  &    .095 &  54.95 &$\blacktriangle$ \cr
 $GS^{{\rm min}} $ & 3.9258 &  1.4598 &  -4.6816  &    1.1898 &  8.0123  &    80.19  &    .035 &  54.11 &$\blacktriangledown$ \cr
 $GS       $ & 4.1045 &  2.0605 &  -4.7173  &    5.0202 &     9.3014  &    78.87  &    .097 & 270.82 &$\bigtriangleup$ \cr
\hline
Samples      &        &         &           &           &             &             &          &  &   \cr
\hline
 $   0     $ & 3.3125 &  1.4644 &   3.1438  &   -1.0000 &     0.8750  &     7.67  &    .441 &  4.63 &$+$ \cr
 $   1     $ & 3.2651 &  1.4032 &   3.5466  &    2.8685 &     1.7954  &     8.94  &    .405 &  5.11 &$+$ \cr
 $   2     $ & 3.4026 &  1.4917 &   3.0629  &    7.3430 &     0.6719  &     8.75  &    .385 & 16.07 &$+$ \cr
 $   3     $ & 3.7225 &  1.5030 &   3.6592  &   10.7265 &     1.5154  &     9.29  &    .355 & 17.78 &$+$ \cr
 $   4     $ & 3.8407 &  1.4968 &   3.2142  &   14.4093 &     0.8757  &    10.49  &    .325 & 19.41 &$+$ \cr
 $   5     $ & 3.6544 &  1.4000 &   3.0993  &   15.5614 &    -0.1329  &     6.35  &    .305 & 18.15 &$+$ \cr
 $   6     $ & 3.8607 &  1.4000 &   3.2375  &   19.8571 &    -0.1635  &     6.32  &    .255 & 20.57 &$+$ \cr
 $   7     $ & 3.9783 &  1.4000 &   3.2706  &   22.4194 &    -0.4805  &     5.29  &    .225 & 21.69 &$+$ \cr
 $   8     $ & 4.1547 &  1.4000 &   3.3756  &   26.1305 &    -0.5855  &     5.02  &    .175 & 23.52 &$+$ \cr
 $   9     $ & 3.4044 &  1.5387 &   4.3094  &   25.5625 &     0.0625  &      .01  &    .153 & 30.80 &$+$ \cr
\end{tabular}
\end{table}

All told, let us present the results.
They are gathered in Tables~\ref{tab:Bs} and \ref{tab:Bsks} in
correspondence with Figs.~\ref{fig:orbit} and \ref{fig:chi}.
We can see the utility of this procedure of fitting the sum rules as
follows: the parameter space of the decomposition coefficients
(``Appell'' coefficients) that project on to the nucleon eigenfunctions,
derived in the previous chapter, is systematically scanned seeking for
evolving morphologies of geometric features such as the peak pattern of
the amplitudes $V$, $A$, and $T$.
Quantification of a feature or region involves its isolation,
extraction, and stability.
Next, one tracks features corresponding to particular solutions to
determine how they change their form and how they compare with
features in other regions.
By testing the effects of various combinations of symmetric versus
antisymmetric components of the {\DA}s, the whole parameter space can
be systematically explored with respect to local minima of $\chi^{2}$
and a complete picture of their distribution pattern can be pieced
together.
Using for the first- and second-order moments either the COZ or the
KS sum-rule constraints in conjunction with those of COZ for the
third-order moments, a simple {\it scaling relation} between the
ratio $R\equiv |G_{\rm M}^{\rm n}|/G_{\rm M}^{\rm p}$ and the expansion
coefficient $B_{4}$ emerges as one progresses through the generated
solutions.
That smooth scaling behavior is inherent in the particular set of sum
rules used in the fit and does not rely on additional assumptions.

%
\begin{table}
\caption{Theoretical parameters to classify the nucleon distribution
         amplitudes discussed in the text. The samples shown refer
         to a combined use of the COZ and KS sum rules.
\label{tab:Bsks}}
\begin{tabular}{lrrrrrrrrc}
 Model       & $B_{1}$\ \ &  $B_{2}$\ \ &   $B_{3}$\ \  &    $B_{4}$\ \ &     $B_{5}$\ \ &
 $\vartheta[deg]$\  &  R\ \ & $\chi^{2}$\ & Symbol \cr
\hline
Samples
             &        &         &           &           &             &           &         &       &    \cr
\hline
 $   0     $ & 3.7520 &  1.3845 &   2.7000  &  -10.5000 &     1.8500  &    14.24  &    .480 & 28.77 &$\circ$ \cr
 $   1     $ & 3.7065 &  1.3298 &   2.9000  &  - 8.0500 &     1.1500  &    10.15  &    .475 & 18.28 &$\circ$ \cr
 $   2     $ & 3.4075 &  1.4191 &   3.3813  &  - 7.6500 &     1.4750  &     7.82  &    .465 &  9.02 &$\circ$ \cr
 $   3     $ & 3.6695 &  1.3186 &   2.8375  &  - 5.8875 &     1.1125  &    10.62  &    .470 & 16.90 &$\circ$ \cr
 $   4     $ & 3.4120 &  1.3906 &   3.2375  &  - 5.5875 &     1.3625  &     8.51  &    .460 &  6.76 &$\circ$ \cr
 $   5     $ & 3.4242 &  1.3644 &   3.0844  &  - 3.2656 &     1.2750  &     9.47  &    .453 &  5.66 &$\circ$ \cr
 $   6     $ & 3.3500 &  1.3710 &   3.1192  &  - 0.9556 &     1.2995  &     9.48  &    .440 &  5.75 &$\circ$ \cr
 $   7     $ & 3.3501 &  1.4045 &   3.3004  &    0.2513 &     1.4203  &     8.95  &    .430 &  5.86 &$\circ$ \cr
 $   8     $ & 3.3536 &  1.4327 &   3.4787  &    4.8241 &     1.5609  &     8.97  &    .395 & 15.11 &$\circ$ \cr
 $   9     $ & 3.3500 &  1.3303 &   3.1262  &    8.8918 &     1.3173  &    10.75  &    .360 & 15.19 &$\circ$ \cr
 $  10     $ & 2.9067 &  1.3664 &   4.0326  &   12.0701 &     1.0180  &     2.46  &    .300 & 26.37 &$\circ$ \cr
 $  11     $ & 2.9300 &  1.3899 &   4.4263  &   16.4589 &     0.8764  &   -0.019  &    .250 & 28.91 &$\circ$ \cr
 $  12     $ & 3.1760 &  1.4491 &   4.6009  &   22.9637 &     0.5213  &   -1.016  &    .180 & 31.06 &$\circ$ \cr
 $  13     $ & 3.2912 &  1.4545 &   4.6802  &   25.5161 &     0.4903  &   -1.014  &    .150 & 32.96 &$\circ$ \cr
 $  14     $ & 3.4017 &  1.4595 &   4.7558  &   27.9607 &     0.4594  &   -1.014  &    .120 & 34.20 &$\circ$ \cr
 $  15     $ & 3.5078 &  1.4638 &   4.8284  &   30.3070 &     0.4310  &   -1.012  &    .090 & 35.39 &$\circ$ \cr
\end{tabular}
\end{table}

Fig.~\ref{fig:chi} illustrates the $\chi^{2}$ quality of the COZ
samples.
Depending on the degree of matching with the corresponding sum rule,
a solution found this way appears as a local minimum of the
$\chi^{2}$ criterion.
We have plotted in the ($B_{4},R$) plane interpolating solutions to
the COZ sum rules ($+$ labels) and such to a combined set of KS/COZ
sum rules ($\circ$ labels).
The latter set is not quite consistent because the typical sum-rule
parameters, such as Borel intervals, energy thresholds, etc., are
treated differently in the two approaches.
Nevertheless, it is instructive to examine what changes the inclusion
of the KS sum rules may cause.
As it turns out, there is no significant difference between the two
treatments (see Fig.~\ref{fig:orbit}), and this insensitivity
justifies the whole approach and promotes the orbit structure to a
key element of the analysis.

%
\begin{figure}
\begin{picture}(0,160)
  \put(61,10){\psboxscaled{1600}{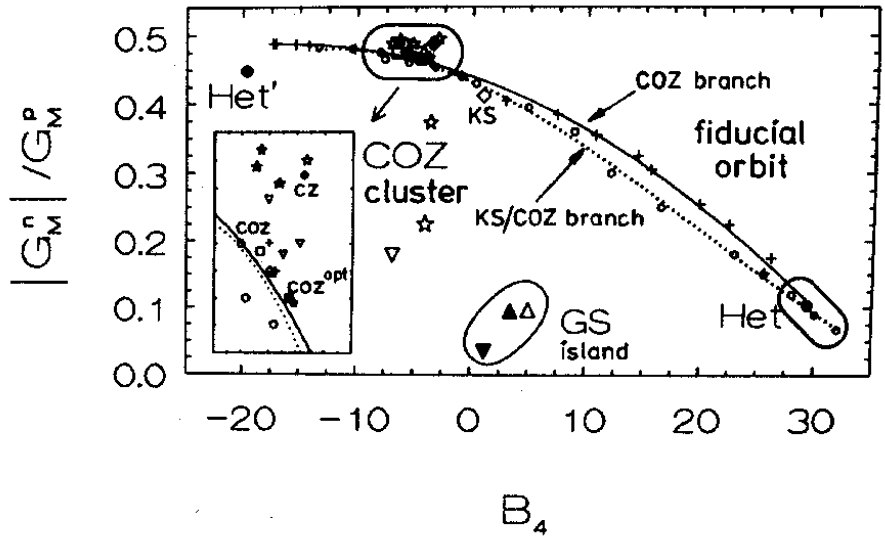}}
\end{picture}
\caption[fig:pattern]
        {\tenrm
         Pattern of nucleon distribution amplitudes matching the COZ
         sum-rule constraints (crosses) and such to a combined set of
         KS/COZ sum rules (open circles). A striking scaling relation
         between the ratio $R$ of the magnetic nucleon form factors
         and the coefficient $B_{4}$ that projects on to the
         corresponding eigenfunction is revealed.
         The various models of nucleon distribution amplitudes
         discussed in the text are compiled in Table~\ref{tab:Bs}.
         Interpolating samples of distribution amplitudes with respect
         to the COZ sum rules are denoted by crosses
         (see Table~\ref{tab:Bs}); those referring to the combined set
         of the KS/COZ sum rules, by circles (see Table~\ref{tab:Bsks}).
         The inset at the lower left expands the vertical scale between
         $0.455$ and $0.495$ (corresponding to the $B_{4}$ interval
         $[-10,0]$) to exhibit the close agreement between the fiducial
         orbit and a variety of proposed amplitudes with third-order
         Appell polynomials, {\it not} included in the fit.
         Significantly, those model amplitudes which appear as isolated
         points are exactly the ones which possess unphysical features,
         namely, either in the form of spurious oscillations, or because
         they violate the {\RG} equation, leading this way to a wrong
         QCD evolution behavior of form factors (see \cite{BS93}).
         The curves are fits to the local minima of the COZ sum rules
         (solid line) and the KS/COZ sum rules (dotted line).
         They constitute a {\it fiducial orbit} with respect to
         $\chi^{2}$ beginning in the heterotic region (small $R$ and
         large positive $B_{4}$) and terminating past the COZ cluster
         (large $R$ and large negative $B_{4}$).
\label{fig:orbit}}
\end{figure}
%

The solutions to the sum rules agree within less than $1\%$ with the
empirical fits :
\begin{equation}
 R = 0.437338 - 0.006016 B_{4} - 0.000176 B_{4}^{2}
\label{eq:cozfit}
\end{equation}
(only COZ constraints) or
\begin{equation}
 R = 0.431303 - 0.00752 B_{4} - 0.000241 B_{4}^{2} + 3.851221\times
10^{-6} B_{4}^{3}
\label{eq:cozksfit}
\end{equation}
(combined KS and COZ constraints), represented by the solid and dotted
curves, respectively.
For simplicity, we use the term ``orbit" to refer to both
characteristic curves.

Furthermore, we find that the magnitude of the ratio $R$ is restricted
at both ends of the orbit.
For the first discussed case, the upper bound is $0.4881$ and for the
second one, $0.482$.
The corresponding lower bounds are $0.104$ and $0.0675$.
In the course of this analysis \cite{BS93,BS94}, optimized versions
-- with respect to the sum rules -- of all previous
models \cite{CZ84a,GS86,KS87,COZ89a} have been determined.
These amplitudes, denoted by the superscript ``opt'', are
shown in Fig.~\ref{fig:Namplitudes}.
[The assignments of models to symbols are given in
Table~\ref{tab:Bs}.]

%
\begin{figure}
\begin{picture}(0,350)
  \put(-3,10){\psboxscaled{1100}{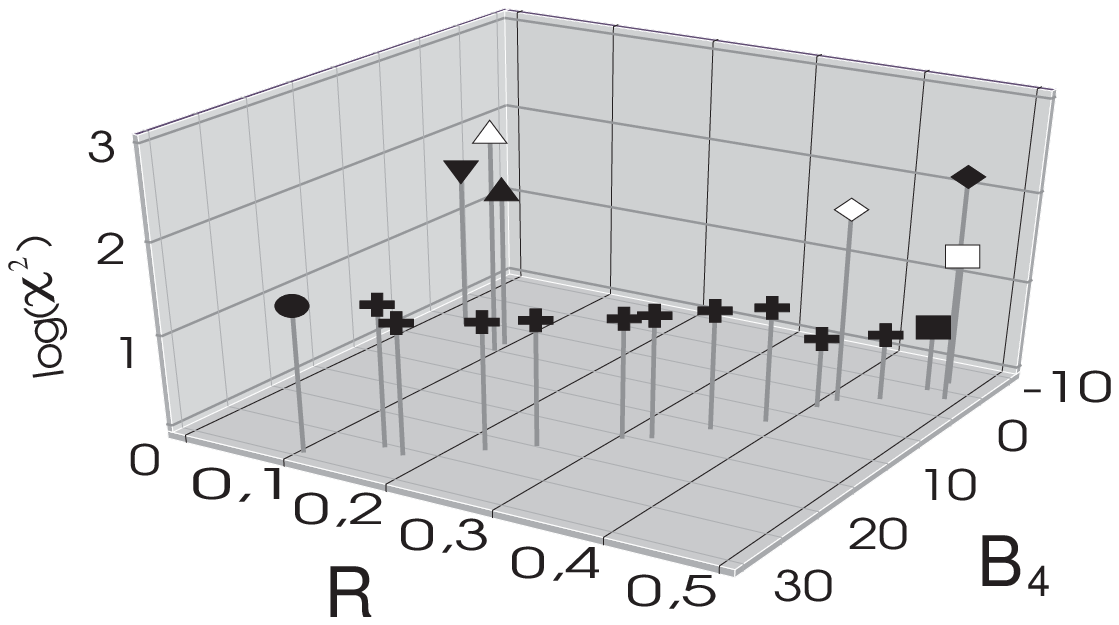}}
\end{picture}
\vspace{-4.0cm}
\caption[fig:Chi2def]
        {\tenrm
         Two-dimensional view on solutions to the QCD sum rules
         as successive local minima in the plane spanned by
         $R\equiv \left\vert G_{\rm M}^{\rm n} \right\vert /
         G_{\rm M}^{\rm p}$
         and $B_{4}$ with respect to the ``hierarchical''
         $\chi ^{2}$-criterion discussed in the text.
         Referring to the previous figure, only the COZ branch of
         the ``fiducial orbit'' is shown.
\label{fig:chi}}
\end{figure}
%

As mentioned previously, the lower part of the orbit is associated
with the heterotic solution \cite{SB93nuc} which gives the smallest
possible ratio still compatible with the sum-rule constraints.
This solution, although degenerated with respect to $R$, is {\it
distinct} from the GS one.
As we shall see in the next chapters, the predictions extracted from
the heterotic model are dramatically different compared to those
following from the GS model.
The upper region of the orbit controls COZ-type amplitudes and
contains a cluster of solutions densely populating the orbit in the
$R$-interval $0.455 \div 0.495$ (see the inset in
Fig.~\ref{fig:orbit}).
This cluster contains the amplitude denoted
$COZ^{{\rm opt}}$ (see Table~\ref{tab:Bs}), which is associated with
the absolute minimum of $\chi^{2}$ and plays the role of an attractor
for all other solutions with similar features.
Strictly speaking, also the CZ model and the KS one, although in the
vicinity of the orbit, are actually isolated points because they
correspond to much larger $\chi ^{2}$ values (cf. Fig.~\ref{fig:chi})
in correspondence with Table~\ref{tab:Bs}) than the proper solutions on
the orbit.
The heterotic amplitude and the original COZ amplitude correspond to
similar $\chi ^{2}$ values.
It is remarkable that the heterotic solution matches the KS \cite{KS87}
sum-rule constraints better than the original COZ amplitude.
This is worth noting because the KS results have been independently
verified in \cite{CP88}.

Analogously, the solution with the smallest $\chi^{2}$ value on the
KS/COZ branch of the orbit is the amplitude denoted
$KS/COZ^{{\rm opt}}$ (see Table~\ref{tab:Bs}).
GS-type amplitudes constitute an isolated region in the ($B_{4}, R$)
plane and are separated from the characteristic orbit by a large
$\chi^{2}$ barrier.
There is yet another type of solutions (we termed $H\!et^{\prime}$)
past the upper end of the orbit -- first discussed in \cite{BS93}.
$H\!et^{\prime}$ belongs to a secondary-branch of the main orbit which
can be reached by coercing the amplitudes $V$ and $T$ to have a
particular symmetry pattern of maxima and minima, corresponding to
the reversed combination as compared to the heterotic solution.
Remarkably, each of the discussed categories of solutions has unique
geometric characteristics specific for its class.
One has to exercise a certain amount of care to be sure that these
are indeed the only regions in the parameter space contributing to
the sum rules at the desired level of accuracy.

%
\begin{figure}
%
\centering
\epsfig{figure=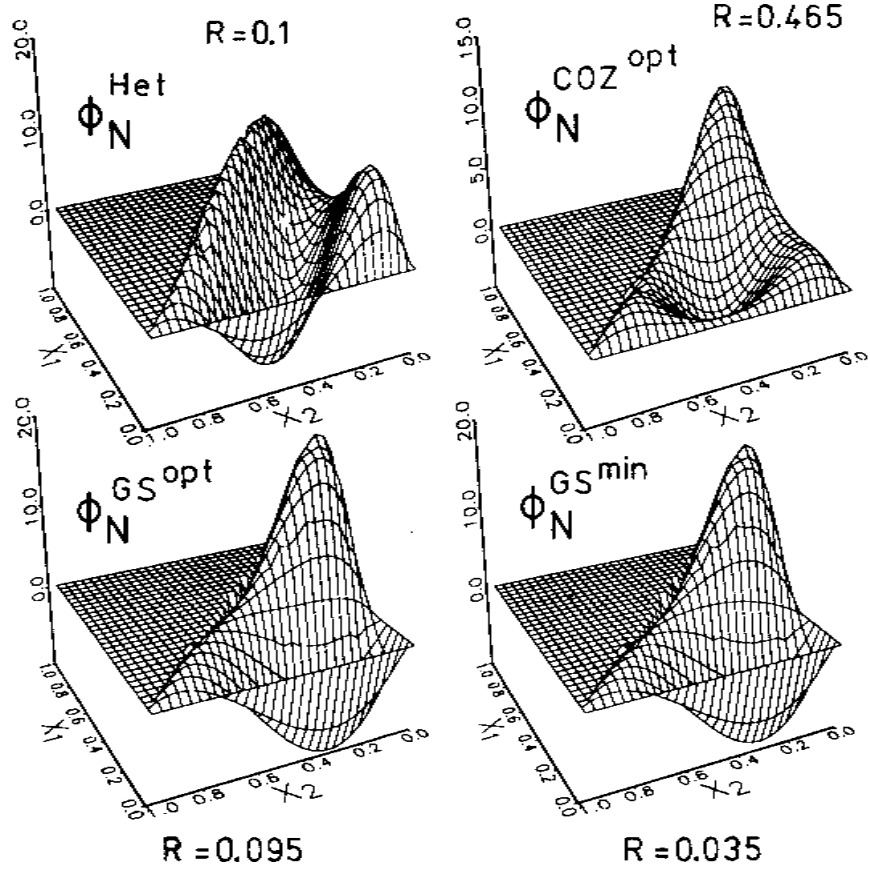,width=12cm,angle=-90,silent=}
\vspace{0.4cm}
\caption[fig:Nuclampl]
        {\tenrm
         Profiles of the optimized versions of model distribution
         amplitudes, commonly discussed in the literature, shown
         together with the heterotic one.
\label{fig:Namplitudes}}
\end{figure}
%

Having completed the classification of distribution amplitudes
rendering the degree of Appell polynomials $M\leq 2$, let us now turn
to models with functional representations which attempt to include
third-order eigenfunctions, in connection with additional
cutoff-parameters \cite{Sch89,HEG92}.
The inset in Fig.~\ref{fig:orbit} shows how such models \cite{Sch89}
(marked by stars) and \cite{HEG92} (marked by light upside-down
triangles) group around the optimum amplitudes $COZ^{{\rm opt}}$ and
$KS/COZ^{{\rm opt}}$, thus establishing the scaling relation between
$R$ and $B_{4}$ in a much more general context.
This result suggests that the inclusion of higher-order
eigenfunctions (Appell polynomials) in the nucleon distribution
amplitude is indeed a marginal effect, as argued above.
If so, the antisymmetric content of the nucleon distribution
amplitude is sufficiently accounted for by the eigenfunction
$
 \tilde \Phi _{4}(x_{i})
$
since higher antisymmetric components are offset by this term.
On the other hand, those model amplitudes proposed in
\cite{Sch89,HEG92} which appear as isolated points scattered
toward the GS ``island'' are unacceptable on physical grounds,
either because they exhibit unrealistic large oscillations in the
longitudinal momentum fractions \cite{Sch89} or because they entail
a wrong evolution behavior for the nucleon form factors \cite{HEG92}.
In addition, as was pointed out in \cite{CORFU92}, models which use
cutoffs, like those in \cite{HEG92}, have difficulties in preserving
the validity of the evolution equation.

The general picture emerging from the presented analysis is a pattern
of nucleon {\DA}s which develops into several regions of the ($B_{4},R$)
plane.
The main sector is characterized by an orbit which shows a striking
scaling behavior between the ratio $R$ (a measurable quantity) and the
(theoretical) expansion parameter $B_{4}$.
For a variety of amplitudes this result is unique, irrespective
of their functional representation.
Isolated samples in this plane are relegated to spurious solutions
owing to their unphysical features.
The profiles of the distribution amplitudes across the orbit
change in an orderly sequence of gradations with some mixture of COZ
and GS characteristics until the COZ amplitude is transmuted into the
heterotic solution.
The variation with shape of the nucleon {\DA} with $R$ is shown
graphically in Fig.~\ref{fig:metam}.

%
\begin{figure}
\begin{picture}(0,350)
  \put(30,10){\psboxscaled{900}{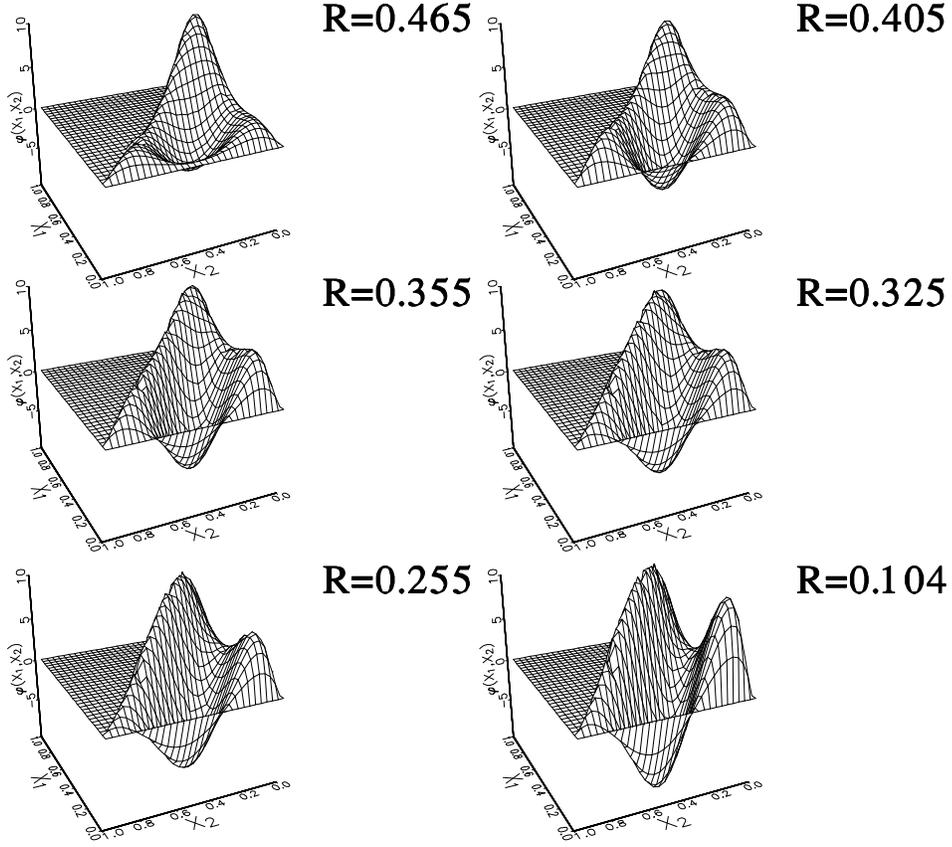}}
\end{picture}
\caption[fig:metamorphosis]
        {\tenrm
         A two-dimensional example showing how the profile of the
         nucleon distribution amplitude $\Phi _{N}$ changes along
         the $\chi ^{2}$ orbit. The amplitudes are labeled by the
         corresponding $R$ value.
\label{fig:metam}}
\end{figure}
%

Despite differing functional representations of the nucleon {\DA}s (use
of higher-order eigenfunctions with or without cutoffs), the results
plotted in Fig.~\ref{fig:orbit} show that all nucleon {\DA}s complying
with the sum-rule constraints actually collapse to the fiducial orbit,
thereby providing convincing evidence that the scaling relation
between the ratio $R$ and the expansion parameter $B_{4}$ reflects a
genuine non-perturbative feature of the true nucleon {\DA}.
To underline the hybridity character of the heterotic {\DA}, we show in
Fig.~\ref{fig:VAT} the invariant functions $V$, $A$, and $T$ associated
with it in comparison with their counterparts for the optimized versions
of the COZ and the GS models.
To leading order $M=2$, the algebraic expressions for these functions
in terms of the expansion coefficients $B_{n}$ are \cite{Ste89}:
\begin{eqnarray}
  V(x_{i})
& = &
  \phi _{\rm as}(x_{i})
  \biggl[
         \left(
               B_{0} + B_{2} - 5B_{3} - 5B_{5}
         \right)
      +  \frac{1}{2}
         \left(
               B_{1} - 3B_{2} + 11B_{3} + B_{4} + 21B_{5}
         \right) \left( x_{1} + x_{2} \right)
\nonumber \\
& &
      -  \left( B_{1} + B_{4}
         \right) x_{3}
      -  \left( 4B_{3} + 14B_{5}
         \right) x_{1}x_{2}
      +  \frac{1}{6}
         \left(
               12B_{3} - 4B_{4} - 28B_{5}
         \right) \left( x_{1}^{2} + x_{2}^{2} \right)
\nonumber \\
& &
      +  \frac{1}{3}
         \left(
               24B_{3} + 4B_{4} + 14B_{5}
         \right) x_{3}^{2}
  \biggr] \; ,
\label{eq:V}
\end{eqnarray}
\begin{eqnarray}
  A(x_{i})
& = &
  \phi _{\rm as}(x_{i})
  \biggl[
         \frac{1}{2}
         \left(
               - B_{1} - 3B_{2} + 3B_{3} - B_{4} - 7B_{5}
         \right) \left( x_{1} - x_{2} \right)
\nonumber \\
& &
      +  \frac{1}{6}
         \left(
               - 12B_{3} + 4B_{4} + 28B_{5}
         \right) \left( x_{1}^{2} - x_{2}^{2} \right)
  \biggr] \; ,
\label{eq:A}
\end{eqnarray}
\begin{eqnarray}
  T(x_{i})
& = &
  \phi _{\rm as}(x_{i})
  \biggl[
         \left(
               B_{0} + B_{2} - 5B_{3} - 5B_{5}
         \right)
      +  \left(
               -3B_{2} + 7B_{3} + 7B_{5}
         \right) x_{3}
\nonumber \\
& &
      +
         \left(
               4B_{3} + 14B_{5}
         \right) x_{1} x_{2}
      +
         \left(
                8B_{3} + \frac{14}{3}B_{5}
         \right) \left(x_{1}^{2} + x_{2}^{2} \right)
  \biggr] \; ,
\label{eq:T}
\end{eqnarray}
where the appropriate coefficients for every model have to be
inserted.\footnote{Note that at the central point $x_{i}=1/3$,
\begin{displaymath}
  V\left(x_{i}\right)
=
  T\left(x_{i}\right)
=
  \Phi _{N}\left(x_{i}\right)
=
  \frac{120}{729}\left(27 - 12B_{3} - 2B_{5}\right)
\end{displaymath}
to leading order $M=2$.}
While $V_{{\rm Het}}$ has a symmetry pattern similar to that of
$V_{{\rm GS}}$ (one main maximum in the central region),
$T_{{\rm Het}}$ is characterized by two maxima, much the same as
$T_{{\rm COZ}}$.
The inverse heterotic combination is realized by the ``mirror''
solution $\Phi _{N}^{{\rm Het}^{\prime}}$ (see Table~\ref{tab:Bs}).

%
\begin{figure}
\begin{picture}(0,420)
  \put(35,-60){\psboxscaled{700}{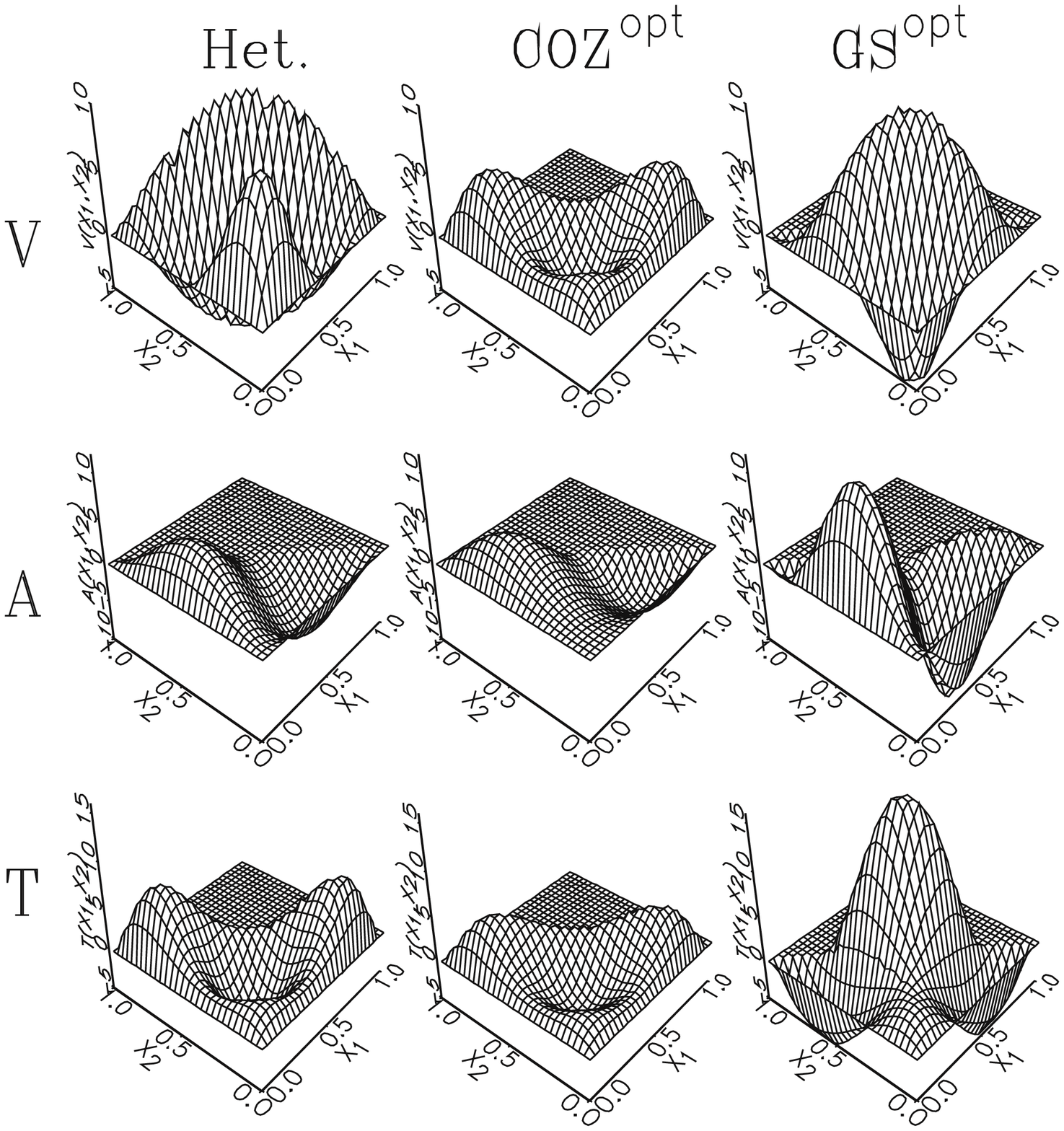}}
\end{picture}
\caption[fig:VATbild]
        {\tenrm
         Invariant distribution amplitudes $V$, $A$, and $T$
         associated with the optimized versions of the COZ and GS
         models in comparison with the heterotic model.
\label{fig:VAT}}
\end{figure}
%

These remarks can be put on more mathematical grounds by considering
a classification scheme of the various nucleon {\DA}s based on the
observation that the optimized versions of the COZ-type and GS-type
amplitudes, we derived, are almost orthogonal to each other with
respect to the weight $w(x_{i})=\phi _{\rm as}(x_{i})/120$.
Their normalized inner product
$\langle COZ^{{\rm opt}}\vert GS^{{\rm opt}} \rangle$
yields $0.1607$, which corresponds to an angle of $80.8^{\circ}$.
Thus these functions form a quasi-orthogonal basis that can be used
to continuously parameterize the nucleon {\DA}s in terms of a
``hybridity'' angle $\vartheta$, defined by
\begin{equation}
  \vartheta
=
  \arctan
  \left(
        \frac{\langle GS^{{\rm opt}} \vert i \rangle}
             {\langle COZ^{{\rm opt}} \vert i \rangle}
  \right) \; ,
\label{eq:hybrideangle}
\end{equation}
where the index $i$ denotes collectively any of the nucleon {\DA}s
listed in Tables~\ref{tab:Bs} and \ref{tab:Bsks}.
Here the bracket denotes the normalized inner product
\begin{equation}
  \langle i | j \rangle
\equiv
  \frac{\left( i,j \right)}{\parallel i\parallel \;
                            \parallel j\parallel } \; ,
\label{eq:norminnprod}
\end{equation}
where
\begin{equation}
  \left( i,j\right)
=
  \int_{0}^{1}[dx] \, w(x_{i})\, \Phi _{i}(x_{k}) \Phi _{j}(x_{k})
\label{eq:innprod}
\end{equation}
and
\begin{equation}
  \parallel i \parallel
=
  \sqrt{\left( i,i\right)} \; .
\label{eq:norm}
\end{equation}
The hybridity angle parameterizes the mingling of geometrical
characteristics attributed to COZ-like and GS-like amplitudes and
provides a quantitative measure for their presence in any solution
conforming with the sum-rule constraints (see Fig.~\ref{fig:class}).
The superimposed dashed line in Fig.~\ref{fig:class}(c) is a fit
given by
$
 \vartheta /[deg] = 8.5693 + 0.0160B_{4} + 0.0073B_{4}^{2}-
 0.00067B_{4}^{3}
$.
The dashed line in Fig.~\ref{fig:class}(a) represents the empirical
fit
$
 R = 0.436415 - 0.005374 B_{4} - 0.000197 B_{4}^{2}
$.
An improvement of this fit was presented above in connection with
the orbit structure of the nucleon {\DA}s (cf. Fig.~\ref{fig:orbit}).
The mixing of different geometrical characteristics is particularly
relevant for the heterotic solution, which generically amalgamates
features of both types of amplitudes (cf. Fig.~\ref{fig:VAT}).
In this role, the heterotic model has the special virtue of
simultaneously fitting the twin hopes for making reliable predictions
with respect to the experimental data while being in agreement with
the sum-rule constraints.
The other models can be tuned to fit some aspects of the data, but
never all aspects simultaneously.

To conclude this chapter, we remark that fixing the value of the ratio
$R$ by experiment, one could use the presented classification scheme
to identify the corresponding nucleon {\DA} determined by theory.

%
\begin{figure}
\begin{picture}(0,400)
  \put(100,10){\psboxscaled{1500}{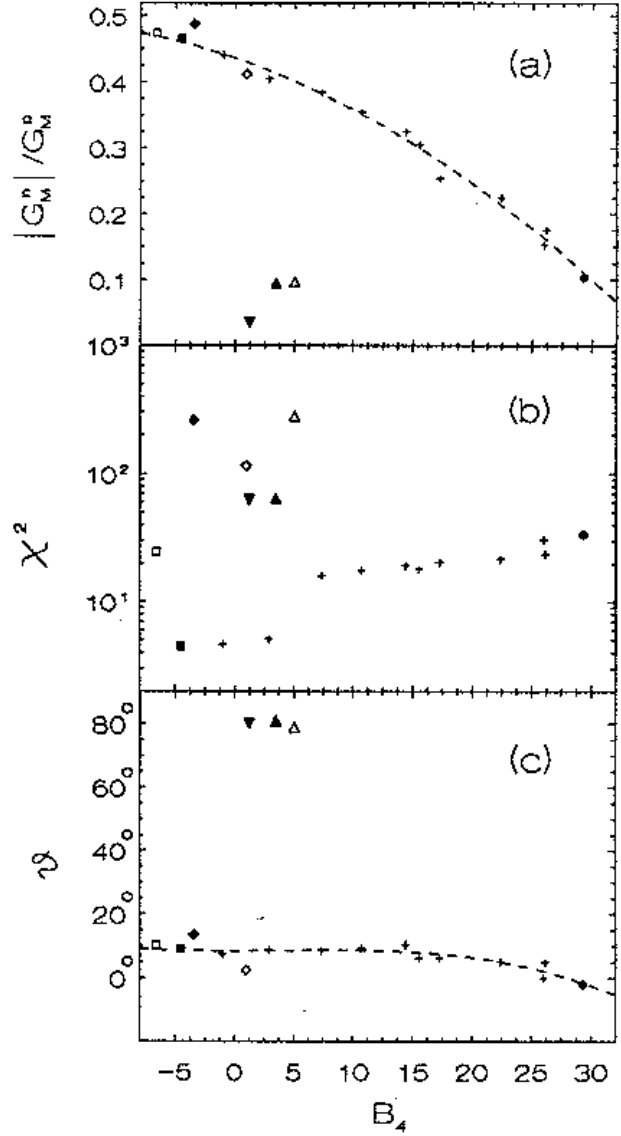}}
\end{picture}
\vspace{-0.5cm}
\caption[fig:classify]
        {\tenrm
         Classification scheme of nucleon distribution amplitudes
         complying with the constraints set by QCD sum rules
         (Table~\ref{tab:sr}).
         (a) The ratio
         $R=\vert G_{\rm M}^{\rm n}\vert /G_{\rm M}^{\rm p}$
         as a function of the expansion coefficient $B_{4}$.
         The positions of the $\chi ^{2}$ minima are indicated
         (see Fig.~\ref{fig:chi}).
         (b) Distribution of local minima of $\chi ^{2}$ (on a
         logarithmic scale) plotted vs $B_{4}$.
         (c) Pattern of nucleon distribution amplitudes parameterized by
         the hybridity angle $\vartheta$, defined in
         Eq.~(\ref{eq:hybrideangle}), vs $B_{4}$.
         The dashed curves are the fits described in the text.
\label{fig:class}}
\end{figure}
%

\subsection{DELTA DISTRIBUTION AMPLITUDES}
\label{subsec:deltada}
The heterotic concept developed for the nucleon {\DA} can be extended
to derive an optimal {\DA} for the $\Delta^{+}(1232)$ resonance as
well \cite{SB93del}.
The quark structure of this particle and the $N-\Delta$ transition
form factor were investigated within the method of QCD sum rules
by Farrar et al. ``FZOZ'' \cite{Far88} (denoted by the
acronyms of the authors), and independently by Carlson and Poor
(CP) \cite{CP88}.
Both groups derived sum-rule constraints on the moments of the
$\Delta$ {\DA} and used them to construct model {\DA}s in terms of
low-order eigenfunctions ($M=2$).
Note that only symmetric eigenfunctions under permutations
$x_{1} \leftrightarrow x_{3}$ can contribute to the $\Delta$ {\DA},
so that $B_{1}=B_{4}=0$.
Thus the general form of the {\DA} for $M=2$ is
\begin{eqnarray}
  \Phi _{\Delta}(x_{i})
& = &
  \phi _{\rm as}(x_{i})
  \biggl[
           \left(
                 8B_{3}^{\Delta} + \frac{14}{3}B_{5}^{\Delta}
           \right) \left(x_{1}^{2} + x_{3}^{2}\right)
         + \left(
                 -3B_{2}^{\Delta} + 7B_{3}^{\Delta} + 7B_{5}^{\Delta}
           \right) x_{2}
\nonumber \\
& &
         + \left(
                 4B_{3}^{\Delta} + 14B_{5}^{\Delta}
           \right) x_{1} x_{3}
         + \left(
                   B_{0}^{\Delta} + B_{2}^{\Delta}
                 - 5B_{3}^{\Delta} - 5B_{5}^{\Delta}
           \right)
  \biggr] \; .
\label{eq:DeltaDA}
\end{eqnarray}
Note also that
$
 T_{\Delta}(x_{1}, x_{2}, x_{3})
=
 \left[
       \Phi _{\Delta}(132) + \Phi _{\Delta}(231)
 \right]
$
and that in addition
$
 T_{\Delta}(x_{1}, x_{2}, x_{3})
=
 \Phi _{\Delta}(231)
$
because
$\Phi _{\Delta}(132)=\Phi _{\Delta}(231)$.
The shapes of these two model {\DA}s look quite different, and also the
predictions one can extract from them are different (see for details in
subsequent chapters).
While the CP amplitude has almost asymptotic profile, the FZOZ one
exhibits two maxima.
The corresponding sum rule estimates are compiled in
Table~\ref{tab:srdelta}}.
The same table contains also the moments of the proposed model {\DA}s.

%
\def\za{\phantom{1}}
\def\zb{\phantom{12}}
\def\zf{\phantom{(0.321)}}
\def\zm{\phantom{--}}
\begin{table}
\caption[tab:FZOZ]
        {Numerical values of the moments $n_{1}+n_{2}+n_{3}\le 3$ of
         model distribution amplitudes for the $\Delta ^{+}$-isobar
         in comparison with the sum-rule constraints of
         Carlson and Poor (CP) \cite{CP88} and those of
         Farrar, Zhang, Ogloblin, and Zhitnitsky (FZOZ) \cite{Far88}.
         The numbers in parentheses are those published by FZOZ.
\label{tab:srdelta}}
\begin{tabular}{cccccc}
         Moments & Sum rules &\multicolumn{4}{c}{Models} \\
         ${(n_{1}n_{2}n_{3})}$ & $T_{\Delta}(FZOZ)$ & CP & FZOZ &
         heterotic & FZOZ${}^{opt}$ \\
\hline
         (000)   & 1                & 1       &    1              & 1         & 1 \\
         (100)   & 0.31---0.35      & 0.350   & 0.325 (0.32)\za   & 0.321     & 0.325 \\
         (001)   & 0.35---0.40      & 0.300   & 0.350 (0.36)\za   & 0.357     & 0.350  \\
         (200)   & 0.14---0.16      & 0.160   & 0.150   \zf       & 0.140     & 0.156 \\
         (002)   & 0.15---0.18      & 0.123   & 0.160   \zf       & 0.151     & 0.154 \\
         (110)   & 0.07---0.1\za    & 0.101   & 0.080   (0.07)\za & 0.078     & 0.071  \\
         (101)   & 0.09---0.13      & 0.089   & 0.095 (0.1)\zb    & 0.103     & 0.098  \\
         (300)   & 0.06---0.09      & 0.085   & 0.083 (0.085)     & 0.073     & 0.090  \\
         (003)   & 0.06---0.10      & 0.060   & 0.085 (0.081)     & 0.071     & 0.078  \\
         (210)   & 0.025---0.04\za  & 0.039   & 0.030 (0.025)     & 0.027     & 0.026 \\
         (201)   & 0.04---0.06      & 0.035   & 0.037 (0.04)\za   & 0.040     & 0.040 \\
         (102)   & 0.035---0.06\za  & 0.031   & 0.037 (0.039)     & 0.040     & 0.038 \\
\hline
        & $V_{\Delta}(CP)$ &  &  & \\
\hline
         (001)   & 0.33---0.37      & 0.350   & 0.325          & 0.321       & 0.325 \\
         (002)   & 0.14---0.18      & 0.160   & 0.150          & 0.140       & 0.156 \\
         (101)   & 0.072---0.12\za  & 0.095   & 0.088          & 0.091       & 0.085 \\
\end{tabular}
\end{table}

\begin{figure}
\centering
\epsfig{figure=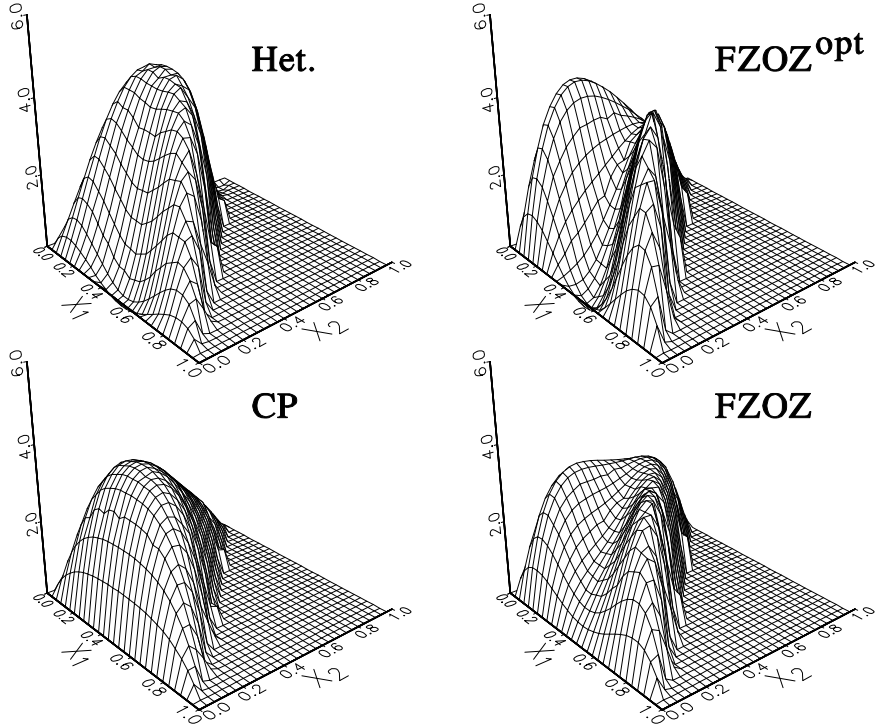,width=12cm,angle=-90,silent=}
\vspace{0.4cm}
\caption[deltas]
        {\tenrm
         Profiles of model distribution amplitudes, defined in
         Table~\ref{tab:DeltaBs}, for the $\Delta ^{+}$-isobar,
         derived from QCD sum rules.
\label{fig:amplitudesD}}
\end{figure}

Following again similar ideas of ``heteroticity'', as in the nucleon
case, one can attempt to combine the CP sum rules with those of FZOZ
in order to obtain an amplitude that is capable to satisfy both sets
simultaneously.
The incentive is to obtain an amplitude that is optimized with
respect to the sum-rule constraints as well as to improve its
phenomenological capabilities.
Both hopes can be satisfied by an amplitude termed again
``Het'', derived in \cite{SB93del}, and shown in
Fig.~\ref{fig:amplitudesD}.
%
%
%
%
This amplitude satisfies all FZOZ moment constraints while providing
the best possible compliance with those of CP (only one of their sum
rules is violated).
Concerning its shape, one sees that it is, nevertheless, close to the
CP amplitude, though its maximum is somehow shifted towards the center
of phase space.

There is yet another specific $\Delta$ amplitude \cite{ELBA93},
denoted ``FZOZ${}^{{\rm opt}}$, which was derived by demanding that
the nucleon-Delta transition form factor, which involves in the
nucleon channel the COZ${}^{{\rm opt}}$ {\DA}, is at least
positive.\footnote{All COZ-like {\DA}s yield N-$\Delta$ transition form
factors compatible with zero.}
Unfortunately, it turns out that this amplitude cannot reproduce any
data (see below), though it is perfectly acceptable from the point of
view of the sum rules.
Thus we reiterate that optimum agreement with the moment sum rules
-- provided by COZ${}^{{\rm opt}}$ -- does not necessarily entail
best phenomenological behavior.
The expansion coefficients for all considered {\DA}s are listed in
Table~\ref{tab:DeltaBs}.

%
\def\za{\phantom{1}}
\begin{table}
\caption[tab:symmetric]
        {Expansion coefficients for $\Delta$-isobar distribution
         amplitudes derived from QCD sum rules within a truncated
         expansion of eigenfunctions. Note that $B_{0}^{\Delta}=1$
         by normalization and that only eigenfunctions symmetric
         under $x_{1}\leftrightarrow x_{3}$ contribute, i.e.,
         $B_{1}^{\Delta}=B_{4}^{\Delta}\equiv 0$. As in the nucleon
         case, the notation of \cite{Ste89} is adopted.
\label{tab:DeltaBs}}
\begin{tabular}{ccccc}
            & \multicolumn{4}{c}{Models} \\
 Expansion coefficients & CP & FZOZ & heterotic & FZOZ${}^{opt}$\\
\hline
      $B_{2}^{\Delta}$  & 0.350\za & -0.175    & -0.2499    & -0.175 \\
      $B_{3}^{\Delta}$  & 0.4095   & \za 1.071 & \za 0.3297 & 1.4117 \\
      $B_{5}^{\Delta}$  & 0.1755   & -0.486    & -1.6205    &-1.620  \\
\end{tabular}
\end{table}

\section{ELECTROWEAK FORM FACTORS}
\label{sec:elff}
The technical apparatus discussed in the previous chapters can now
be used to calculate several electroweak form factors of the nucleon.

The standard differential inclusive cross section for electron-nucleon
scattering in terms of directly measured quantities is
\begin{equation}
  \frac{d^{2}\sigma}{d\Omega dE^{\prime}}
=
  \frac{\alpha_{e}^{2}}{4E^{2} \sin ^{4}\frac{\theta}{2}}
  \left[
         2W_{1}(Q^{2}, \nu )\sin ^{2} \frac{\theta}{2}
       + W_{2}(Q^{2}, \nu )\cos ^{2} \frac{\theta}{2}
  \right] \; ,
\label{eq:diffcrsec}
\end{equation}
where
$E^{\prime} = E - Q^{2}/2M_{N}^{2}$
and
$W_{1}$ and $W_{2}$ are Lorentz-invariant structure functions of the
target nucleon expressed in terms of $Q^{2}$,
$\nu=Q^{2}/2M_{N}=(E - E^{\prime}$) is the energy loss of the
incident lepton, and
$
 q^{2}
=
 (P^{\prime} - P)^{2}
=
 -4EE^{\prime} \sin ^{2} \frac{\theta}{2}
\leq
 0
$
($Q^{2}=-q^{2}>0)$ is the four-momentum transfer.
If the nucleon (with mass $M_{N}$) were a point-like particle like
the electron, say, we would have $P^{2}=P^{\prime 2}=M_{N}^{2}$ and
$P \cdot q =M_{N} \nu$ so that $\nu =Q^{2}/2M_{N}$; furthermore,
$W_{1}(Q^{2})$ and $W_{2}(Q^{2})$ would reduce respectively to
$Q^{2}/4M_{N}$ and 1.
Then Eq.~(\ref{eq:diffcrsec}) would simply be:
\begin{equation}
  \frac{d^{2}\sigma}{d\Omega dE^{\prime}}
=
  \frac{4\alpha_{e}^{2}E^{\prime 2}}{Q^{4}}
  \left[
          \cos ^{2} \frac{\theta}{2}
        + \frac{Q^{2}}{2M_{N}^{2}} \sin ^{2} \frac{\theta}{2}
  \right] \delta \left(\nu - \frac{Q^{2}}{2M_{N}} \right) \; .
\label{eq:pointnuc}
\end{equation}
Integrating over $dE^{\prime}$ one gets the familiar formula
\begin{equation}
  \frac{d\sigma}{d\Omega}
=
  \left(\frac{d\sigma}{d\Omega}\right)_{{\rm Mott}}
  \frac{E^{\prime}}{E}
  \left(
        1 + \frac{Q^{2}}{2M_{N}} \tan ^{2}\frac{\theta}{2}
  \right) \; ,
\label{eq:dsdo}
\end{equation}
where the Mott differential cross section for a singly-charged
point-like target, assumed to have an infinite mass, is
$
 \alpha_{e}^{2} \cos ^{2}
 \frac{\theta}{2} /4E^{2}
 \sin ^{4} \frac{\theta}{2}
$.
However, because of the internal (quark) structure of the nucleon,
this differential cross section does not describe the elastic
scattering of an electron off the nucleon.
Hence, one has to assume that the incident electron is scattered
elastically by a nucleon with a ``diffuse'' structure which can be
revealed by increasing the momentum transfer.
Then we can write Eq.~(\ref{eq:pointnuc}) in the form
\begin{eqnarray}
  \frac{d^{2}\sigma}{d\Omega dE^{\prime}}
& = &
 \frac{4\alpha_{e}^{2} E^{\prime 2}}{Q^{4}}
 \delta \left(\nu - \frac{Q^{2}}{2M_{N}} \right)
\nonumber \\
& & \times
  \left\{
         \left[
         \frac{G_{\rm E}^{2}(Q^{2})
         + \left(Q^{2}/4M_{N}^{2}\right)G_{\rm M}^{2}(Q^{2})}{1 +
         Q^{2}/4M_{N}^{2}}
         \right] \cos ^{2} \frac{\theta}{2}
         + \frac{G_{\rm M}^{2}(Q^{2})}{2M_{N}^{2}}
           \sin ^{2} \frac{\theta}{2}
  \right\} \; ,
\label{eq:diffcrosnucff}
\end{eqnarray}
where
$G_{\rm E}^{2}(Q^{2})$ and $G_{\rm M}^{2}(Q^{2})$ are the electric and
magnetic (``Sachs'') form factors of the nucleon, respectively,
parameterizing in some sense our ignorance about its non-perturbative
(binding) structure.
A phenomenologically successful parameterization is provided by
the empirical ``dipole'' form factor, {\it viz}:
\begin{equation}
  \frac{G_{\rm E}^{2}(Q^{2})}{G_{\rm E}^{2}(0)}
=
  \frac{G_{\rm M}^{2}(Q^{2})}{G_{\rm M}^{2}(0)}
=
  \left(
        1 + \frac{Q^{2}}{0.71 {\rm GeV}^{2}}
  \right)^{-2} \; .
\label{eq:dipole}
\end{equation}
Obviously, the finite-size (``diffuse'') nucleon causes the elastic
scattering cross section to decrease rapidly as $Q^{2}$ grows -- in
contrast to the behavior of a point-like nucleon described by
Eq.~(\ref{eq:pointnuc}).
Hence, as long as there is a single hadron in the final state,
there is an extremely rapid drop-off of the cross section for
$Q^{2} >> 0.7$~GeV${}^{2}$.
Clearly, the challenge is to calculate the nucleon's form factors
within QCD, employing for the intact nucleon wave functions of the
sort we discussed in the previous sections.

\subsection{NUCLEON FORM FACTORS}
\label{subsec:nucff}
From Lorentz covariance, charge conservation, and invariance under
space reflections there exist two electromagnetic form factors for
a spin-$1/2$ particle.
Then, in the Sachs parameterization, we have
\begin{equation}
  \langle P^{\prime} \vert J_{\mu}^{{\rm em}}(0) \vert P \rangle
=
  \frac{\tau}{2M_{N}}
  \bar{u}^{\prime}(P^{\prime})
  \left[
        G_{\rm E}(q^{2}) K_{\mu} + \frac{i}{2M_{N}} G_{\rm M}(q^{2})
        R_{\mu}
  \right] u(P) \; ,
\label{eq:Sachsff}
\end{equation}
where
$\tau = \left( 1 + \frac{q^{2}}{4M_{N}^{2}}\right)^{-1}$,
$K_{\mu} = \left( P_{\mu} + P_{\mu}^{\prime}\right)$,
and
$
 R_{\mu} = \frac{-i}{2}
   \left(\gamma _{\mu} \not\! P \not\! q
 - \not\! q \not\! P \gamma _{\mu}\right)
$
with normalizations as
$G_{\rm E}(0)=e_{N} \;\; (e_{p}=1, e_{n}=-1)$; $G_{\rm M}(0)=\mu$
being the total magnetic moment in units $e/2M_{N}$:
$\mu _{p}=2.7928$, $\mu _{n}=-1.9131$, and with the threshold condition
$G_{\rm E}(4M_{N}^{2})=G_{\rm M}(4M_{N}^{2})$.
The connection to the Dirac parameterization is established by the
on-shell relation
\begin{equation}
  \bar{u}^{\prime}(P^{\prime}) i R_{\mu} u(P)
=
  \bar{u}^{\prime}(P^{\prime})
  \left(
        2 i M_{N} \sigma _{\mu\nu} q^{\nu} + q^{2} \gamma _{\mu}
  \right) u(P)
\label{eq:onshellrel}
\end{equation}
with
\begin{eqnarray}
  F_{1}
& = &
  \tau \left( G_{\rm E} + \frac{q^{2}}{4M_{N}^{2}} G_{\rm M}\right)
\quad\quad {\rm Dirac \;\;\; form \;\; factor}
\cr
F_{2}
& = &
  \tau \left( G_{\rm M} - G_{\rm E}\right)
\quad\quad\quad\quad\;\;\; {\rm Pauli \;\; form \;\; factor}
\cr
 G_{\rm E}
& = &
  F_{1} - \frac{q^{2}}{4M_{N}^{2}} F_{2}
\quad\quad\quad\quad\;\;\;\; {\rm electric \;\; form \;\; factor}
\cr
G_{\rm M}
& = &
  F_{1} + F_{2}
\quad\quad\quad\quad\quad\quad\quad \;\,
{\rm magnetic \;\; form \;\; factor}
\label{eq:DiracSachscon}
\end{eqnarray}
so that the on-shell nucleon current in the Dirac parameterization
is given by
\begin{equation}
  \langle P^{\prime} \vert J_{\mu}^{{\rm em}}(0) \vert P \rangle
=
  \bar{u}^{\prime}(P^{\prime})
\left[
        F_{1}(q^{2}) \gamma _{\mu}
      + \frac{i}{2M_{N}} F_{2}(q^{2}) \sigma _{\mu\nu} q^{\nu}
\right] u(P) \; .
\label{eq:Diracff}
\end{equation}

Recalling Eq.~(\ref{eq:G(Q^2)}), the nucleon helicity-conserving form
factor can be written in convolution form as
\begin{equation}
  G_{\rm M}(Q^{2})
=
  \int_{0}^{1}[dx]\int_{0}^{1}[dy]\,
  \Phi ^{*}(y_{i},\tilde{Q}^{2}_{y})\,
  T_{\rm H}\left(x_{i}, y_{i}, Q^{2}, \alpha _{\rm s}(\mu ^{2})\right)\,
  \Phi (x_{i}, \tilde{Q}^{2}_{x}) \; .
\label{eq:magnucleonff}
\end{equation}
In the evaluation of this expression one may encounter singularities
owing to the fact that the gluon virtualities should enter the
arguments of the strong coupling constant in $T_{\rm H}$, i.e.,
$\alpha _{\rm s}(\mu ^{2}=Q^{2}x_{i}y_{j})$.
These virtualities depend on the longitudinal momentum fractions, and
as a result, $\alpha _{\rm s}$ becomes singular in the endpoint region.
Several options have been suggested to avert this singularity:

\begin{itemize}
\item The simplest choice is to assume that one can use two
      {\it collective} scales, one for the hard and another for
      the soft gluon propagator, and then take the geometric average
      of both couplings in the form factor.
      In this way the dependence of $\alpha _{\rm s}$ on the fractional
      momenta is avoided and $\alpha _{\rm s}$ can be taken outside the
      integral, amounting to a rescaled $Q^{2}$-dependent pre-factor
      (``peak approximation'').
      It seems reasonable to fix the gluon scales by the gluon
      virtualities from that Feynman graphs which contribute most
      for each particular nucleon {\DA} by substituting the longitudinal
      momenta at the position of the main maximum of the corresponding
      nucleon {\DA}.
      Then one has
      $
       \bar{\alpha}_{s}(Q^{2})
       =
       [\alpha_{s}(Q^{2}v_{{\rm hard}})
       \alpha_{s}(Q^{2}v_{{\rm soft}})]^{1/2},
      $
      where typically
      $v_{{\rm hard}}=\bar{x}_{i}\bar{y}_{i}$ and
      $v_{{\rm soft}}=x_{j}y_{j}$.
      This technique was used in several works, like
      \cite{CZ84a,GS86,GS87,Ste89,SB93nuc}, and the results are
      quite reasonable.
\item A refined version of this approximation was proposed by
      Stefanis in \cite{Ste89}.
      Recalling that there are in total 8 topologically distinct
      diagrams which give non-zero contributions to the form factor,
      one can use the virtualities of the hard and soft gluon
      propagators in each individual diagram to calculate the
      corresponding $\alpha _{\rm s}$ values.
      This procedure improves the predictions, though this
      improvement depends on the specific nucleon {\DA} used; for
      example, the effect when using the CZ-{\DA} is
      negligible \cite{Ste89}.
\item In both cases considered above, one can only change the
      absolute value of the form factor but not its slope.
      This can only be improved if $\alpha _{\rm s}$ can be retained
      inside the integral.
      To this end, one can either cutoff $\alpha _{\rm s}$ explicitly
      -- typically between 0.5 to 07 -- or saturate its momentum
      dependence by introducing, say, an effective gluon mass.
      Such an analysis was performed in \cite{JSL87}, having
      recourse to a modified expression for $\alpha _{\rm s}$, proposed
      by Cornwall \cite{Cor82} (see also \cite{PP80}):
      \begin{equation}
        \alpha _{\rm s}(Q^{2})
      =
        \frac{4\pi}{\beta _{0}
        \ln \left[\left(Q^{2}
        + 4 m_{\rm g}^{2}\right)/\Lambda _{{\rm QCD}}^{2}\right]} \; .
      \label{eq:modalphas}
      \end{equation}
      Here $m_{\rm g}$ stands for the effective gluon mass with values
      in the range ($500\pm200$)~MeV.
      Technically, $m_{\rm g}$ is an IR-cutoff serving implicitly to
      regularize that gluon propagator which becomes soft (compared
      to the large external momentum) in the endpoint region.
      It is evident that such a nonzero gluon mass will ``freeze''
      the value of the running coupling constant for
      $Q^{2}\leq 4 m_{\rm g}^{2}$.
      Since a dynamically generated (gluon) mass is not a constant
      but vanishes at large momentum, the asymptotic behavior is not
      affected.
      Strictly speaking, one should use to saturate $\alpha _{\rm s}$
      a scale-dependent gluon mass which is governed by the {\RG}
      equation with an appropriate anomalous dimension \cite{Cor82}.
      Such a saturation procedure was adopted in \cite{SB93del}
      and will be presented below.
      Physically, the IR regularization of $\alpha _{\rm s}$ connects
      to our previous discussion concerning the existence of a
      fundamental IR scale in the non-perturbative regime.
      This scale my be thought of as being the average transverse
      momentum of vacuum partons and is of order 300~MeV to 600~MeV,
      but cannot be reliably computed at present.
      Within the framework of quark/gluon condensates, this scale
      separates hard from soft momentum flows of gluon (or quark)
      propagators which go into the corresponding condensates.
      It is also worth noting that according to Cornwall, $m_{\rm g}$
      and $\Lambda _{{\rm QCD}}$ are interrelated by the consistency
      relation
      $m_{\rm g}/ \Lambda _{{\rm QCD}} \approx 1.5 - 2.0$.
      It was warned in \cite{Ste89} that the treatment in \cite{JSL87}
      violates this relation, handling $m_{\rm g}$ as an additional fit
      parameter for improving the slope of the form factors.
\item Another, theoretically more appealing, possibility is provided
      by including in the calculation of the form factor gluon
      radiative corrections in the form of Sudakov-type damping
      exponentials \cite{BS89}.
      This procedure is technically more complicated though
      theoretically more sound than, more or less plausible, cutoff
      or saturation prescriptions.
      The ``Sudakov option'' will be discussed in detail in the
      second main part of this report.
\end{itemize}
In the remainder of this chapter we shall present results obtained
within the peak approximation.
In this case, the nucleon form factor can be cast in the
form \cite{CZ84a}
\begin{equation}
  Q^{4}G_{\rm M}^{\rm N}(Q^{2})
=
  \frac{1}{54} \left[4\pi \bar{\alpha}_{s}(Q^{2})\right]^{2}
  \vert f_{\rm N} \vert^{2}
  \int_{0}^{1} [dx] \int_{0}^{1} [dy]
  \left[
           2 \sum_{i=1}^{7} e_{i}\, T_{i}(x_{j},y_{j})
        +  \sum_{i=8}^{14} e_{i}\, T_{i}(x_{j},y_{j})
  \right] \; ,
\label{eq:magffpeak}
\end{equation}
where the amplitudes
$
 T_{i}(x_{j},y_{j})
=
 \Phi _{i}(x_{j})T_{\rm H}(x_{j},y_{j})\Phi _{i}(y_{j})
$
represent convolutions of $T_{\rm H}$ with the appropriate {\DA}s
evaluated term by term for each contributing diagram (marked by the
index ``i'').
Inputting
\begin{eqnarray}
  \Phi _{N}(x_{i})
= & \phantom{} &\!\!\!\!\!\!
  \phi _{\rm as}(x_{i})
    \bigl[
            (B_{0} + B_{2} - 5B_{3} - 5B_{5})
          + (B_{1} + B_{4}) (x_{1} - x_{3})
          + (-3B_{2} + 7B_{3} + 7B_{5}) x_{2}
\nonumber\\
& + &       (4B_{3} + 14B_{5}) x_{1}x_{3}
          + (8B_{3} - \frac{4}{3}B_{4} + \frac{14}{3}B_{5}) x_{1}^{2}
          + (8B_{3} + \frac{4}{3}B_{4} + \frac{14}{3}B_{5}) x_{3}^{2}
    \bigr]
\label{eq:PhiBs}
\end{eqnarray}
and integrating over variables $x_{i}$ and $y_{i}$, the proton and
neutron form factors acquire the following form
\begin{equation}
  Q^{4}G_{\rm M}^{\rm p}(Q^{2})
=
  \frac{1}{54} \left[4\pi \bar{\alpha}_{s}(Q^{2})\right]^{2}
  \vert f_{\rm N} \vert^{2}
  I^{p} \; ,
\label{eq:magpffint}
\end{equation}
\begin{equation}
  Q^{4}G_{\rm M}^{\rm n}(Q^{2})
=
  \frac{1}{54} \left[4\pi \bar{\alpha}_{s}(Q^{2})\right]^{2}
  \vert f_{\rm N} \vert^{2}
  I^{n} \; ,
\label{eq:magnffint}
\end{equation}
where $I^{p}$ and $I^{n}$ are functions of the coefficients $B_{n}$.
Analytic expressions for $I^{p}$ and $I^{n}$ up to order $M=2$
($n=0,1,2, \ldots , 5$)
were published in \cite{Ste89}; such up to order $M=3$
($n=0,1,2, \ldots, 9$)
are given in Appendix \ref{ap:formfactors}.

\begin{figure}
\begin{picture}(0,350)
  \put(90,10){\psboxscaled{1300}{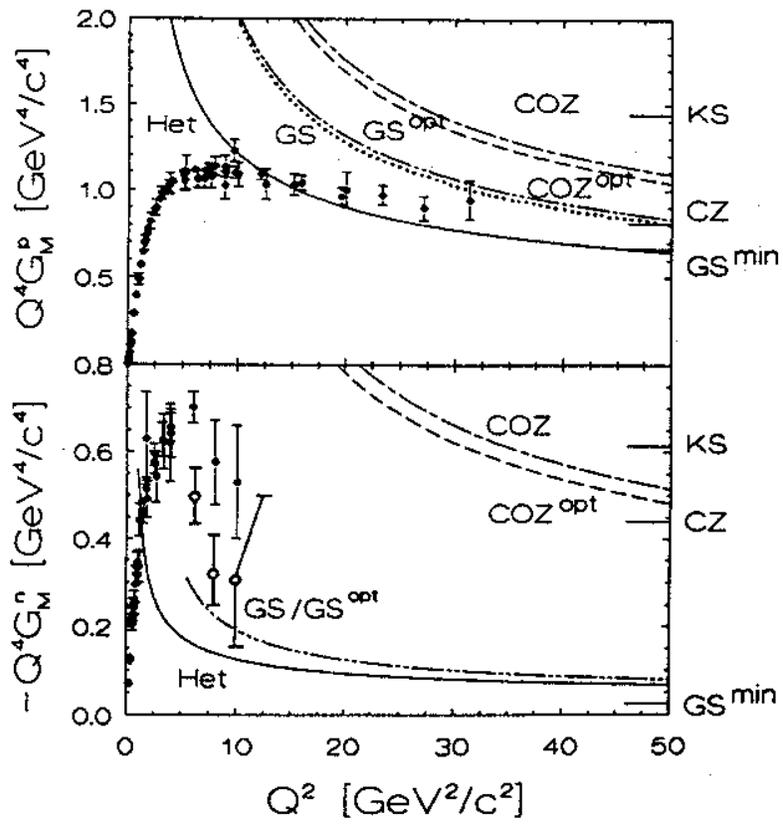}}
\end{picture}
\caption[fig:]
        {\tenrm
         Comparison with available data of the magnetic form factor of
         the proton and the neutron for the heterotic and other model
         distribution amplitudes, labeled by the acronyms of the
         corresponding authors.
\label{fig:GMff}}
\end{figure}

The results for the main models in use are shown in
Fig.~\ref{fig:GMff} in comparison with existing data \cite{SLAC86}
(for the proton) and \cite{Pla90} (for the neutron).
Note that the $Q^{2}$-evolution of the coefficients $B_{n}$ has been
neglected and that the average value
$
 \bar{\alpha}_{s}(Q^{2})
=
 [\alpha _{\rm s}(Q^{2} \bar{x}_{i}\bar{y}_{i})
  \alpha _{\rm s}(Q^{2} x_{j}y_{j})]^{1/2}
$
has been used to account for the different virtualities of the
involved gluon propagators.
For instance, for the heterotic amplitude we use
$
 \bar{\alpha}_{s}(Q^{2})
=
 [\alpha _{\rm s}(Q^{2}\times 0.427)
 \alpha _{\rm s}(Q^{2}\times 0.178)]^{1/2}
$.
Here and below the values
$\Lambda _{{\rm QCD}} = 180$~MeV \cite{GS87lambda} and
$|f_{\rm N}| = 5.0\times 10^{-3}$~GeV${}^{2}$ \cite{KS87,COZ89a}
are taken.

Since there are no direct data on $G_{\rm M}^{\rm n}$ beyond, say,
$5$~GeV${}^{2}$, it is virtually impossible to extract experimental
values of the neutron form factors in a model-independent way.
Therefore, we show in Fig.~\ref{fig:GMff} the theoretical predictions
along with $G_{\rm M}^{\rm n}$-data, extracted under the proviso of two
extreme, albeit reasonable assumptions:
(1) assuming that the electric neutron form factor is at least of the
    order of the magnetic one (open circles), and
(2) assuming $G_{\rm E}^{\rm n}=0$ (black dots).

%
%

The main conclusion to be inferred from these results is twofold:
(1) The heterotic model yields the best relative agreement with the
experimental data, though the prediction for the neutron form factor
is too small to account for the observed form factor.
(2) The COZ${}^{{\rm opt}}$ amplitude -- designed to yield optimum
agreement with the COZ sum rules -- fails in both cases to describe
the experimental data.

Another place to test these results is in the data for the elastic
cross sections
$\sigma_{p}$ and $\sigma_{n}$.
For small scattering angles, where the terms
$\propto tan^{2}(\theta/2)$
can be neglected, there are two main possibilities for the ratio
$
 \sigma_{n}/ \sigma_{p}.
$
If the Dirac form factor $F_{1}^{\rm n}$ is zero, or small compared
to the Pauli form factor $F_{2}^{n}$ \cite{KK77}, then
$\sigma_{n}$ should be due only to the higher-order term
$F_{2}^{n}$.
At large $Q^{2}$ the ratio would become
\begin{equation}
 \frac {\sigma_{n}} {\sigma_{p}} \Rightarrow
 \Bigl( \frac{C_{2}^{n}} {C_{1}^{p}} \Bigr)^{2}
 \frac{1}{4M_{N}^{2} Q^{2}}
\label{eq:crosectrat1}
\end{equation}
(modulo logarithmic corrections due to anomalous dimensions)
and would decrease with increasing $Q^{2}$ due to the extra power of
$1/Q^{2}$ of the Pauli form factor.
Alternatively, if $F_{1}^{\rm n}$ is comparable to $F_{2}^{\rm n}$, then
$\sigma_{n}$ would eventually be due to $F_{1}^{\rm n}$ at large
$Q^{2}$.
Then the ratio
$
 \sigma_{n}/ \sigma_{p}
$
would be given by some constant determined by the nucleon wave
functions
\begin{equation}
 \frac {\sigma_{n}} {\sigma_{p}} \Rightarrow
 \Bigl( \frac{C_{1}^{n}} {C_{1}^{p}}  \Bigr)^{2} \; .
\label{eq:crosectrat}
\end{equation}
In the above expressions, the wave-function characteristics are
parameterized by the (dimensionful) coefficients $C_{i}$, which are
functions of the expansion coefficients $B_{n}$ and the ``proton decay
constant''
$
 \vert f_{\rm N}\vert = (5.0\pm 0.3)\times 10^{-3}~{\rm GeV}^{2}
$ \cite{COZ89a,KS87}.

\begin{figure}
\begin{picture}(0,220)
  \put(50,10){\psboxscaled{1000}{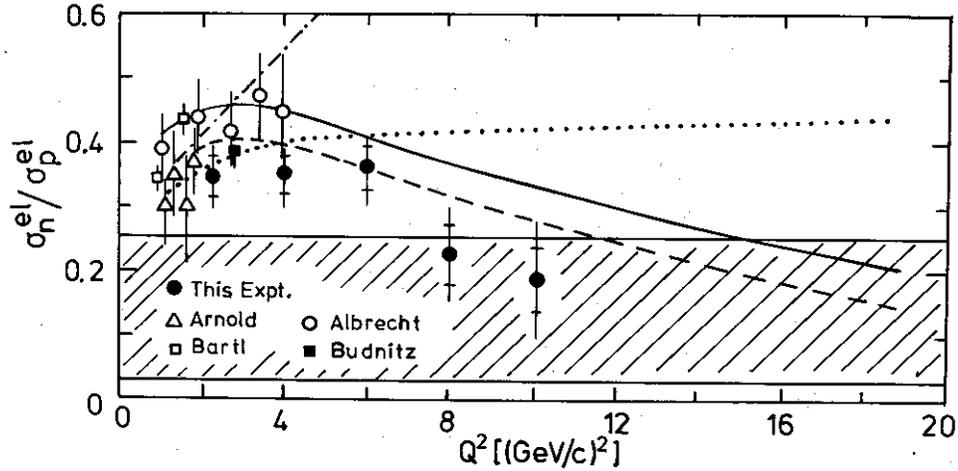}}
\end{picture}
\caption[fig:sigmadata]
        {\tenrm
         Bounds on the elastic cross sections
         $\sigma _{n}/\sigma _{p}$ obtained from QCD sum rules
         (shaded area).
         The data are taken from \cite{Pla90}.
\label{fig:cross}}
\end{figure}

From this analysis it follows that in the $Q^{2}$ domain where the
helicity-flipping parts of the form factors can be ignored,
$
 \sigma_{n}/ \sigma_{p}
$
should be within the range $0.238$ and $0.01$.
If the results of the combined KS/COZ sum rules are taken seriously,
then this range is somewhat shifted: $0.232 \div 0.005$.
Comparing with available data \cite{Pla90}, we see that the measured
value of $\sigma_{n}/\sigma_{p}$ enters the estimated range already at
$
 Q^{2}\approx 8~{\rm GeV}^{2}/c^{2}
$
(see Fig.~\ref{fig:cross}).
Here two concluding remarks are in order:
(1) The present accuracy of QCD sum rules (modulo technicalities) is
sufficient to provide bounds on
$
 \sigma_{n}/\sigma_{p}
$
within the observed region.
(2) The available data in the range
$
 Q^{2}\approx (8 \div 10)~{\rm GeV}^{2}/c^{2}
$
are well below the calculated upper bound and still decreasing.
This indicates that {\DA}s which give
$
 \vert G_{\rm M}^{\rm n}\vert / G_{\rm M}^{\rm p}
 \approx 0.5
$
may be in contradiction to experiment because they yield a Dirac form
factor $F_{1}^{\rm n}$ which starts to overestimate the data already at
$
 Q^{2}\approx 8~{\rm GeV}^{2}/c^{2}.
$
In particular, there is no room for contributions due to the Pauli
form factor.
On the contrary, models which give a small value of
$
 \vert G_{\rm M}^{\rm n}\vert / G_{\rm M}^{\rm p}
$
can explain the data only under the assumption that in this $Q^{2}$
region the Pauli contribution is still dominant.
For instance, the heterotic model requires at
$
 Q^{2}=10~{\rm GeV}^{2}/c^{2}
$
a Pauli contribution to the neutron form factor approximately three
times as large as that from the Dirac form factor.
The principal result from this discussion is that in the intermediate
$Q^{2}$ domain, $\sigma _{n}/\sigma _{p}$ should be within the range
0.238 and 0.001.

%
%

We close this section by presenting results for the axial-vector and
isoscalar axial-vector form factors of the nucleon.
These form factors were analyzed with perturbative QCD by Carlson and
Poor in \cite{CP86} and \cite{CP87}.
The axial-vector currents in question are:
\begin{equation}
  A_{\mu}
=
  \bar{d}\gamma _{\mu}\gamma _{5}u \, ,
\quad\quad
  A_{\mu}^{(3)}
=
  \frac{1}{2}\left(
                     {\bar u}\gamma _{\mu}\gamma _{5}u
                   - \bar{d}\gamma _{\mu}\gamma _{5}d
             \right) \; ,
\label{eq:axcurvec}
\end{equation}
\begin{equation}
  A_{\mu}^{(S)}
=
  \frac{1}{2}\left(
                     {\bar u}\gamma _{\mu}\gamma _{5}u
                   +
                     \bar{d}\gamma _{\mu}\gamma _{5}d
             \right) \; .
\label{eq:axcursca}
\end{equation}
The first two currents are related by an isospin rotation.
Then the axial-vector form factor $g_{A}$ and the pseudoscalar form
factor $g_{P}$ are defined by the matrix element with nucleon states
as follows
\begin{equation}
  \langle n(P^{\prime},\lambda ^{\prime}) \vert A^{\mu}
  p(P)(P,\lambda ) \rangle
=
  {\bar u}_{\lambda ^{\prime}}(P^{\prime})
  \left[
          g_{A}(Q^{2})\gamma ^{\mu}\gamma ^{5}
        + g_{P}(Q^{2})q^{\mu}\gamma ^{5}
  \right] u_{\lambda}(P) \; ,
\label{eq:axmael}
\end{equation}
where $q=P^{\prime} - P$, and $Q^{2}=-q^{2}$.
The axial-vector form factor can be isolated by taking the $A^{+}$
component and working in a frame with $q^{+}=0$.
Then
\begin{equation}
  \langle n \vert A^{+} \vert p \rangle
=
  2P^{+} g_{A} \; ,
\label{eq:axialvecexp}
\end{equation}
and
\begin{eqnarray}
  \langle p \vert A^{+(3)} \vert p \rangle
& = &
  2P^{+} G_{A}^{(3)} \; ,
\nonumber\\
  \langle p \vert A^{+(S)} \vert p \rangle
& = &
  2P^{+} G_{A}^{(S)} \; .
\label{eq:isovec.isosc}
\end{eqnarray}
The first two form factors are related by isospin invariance to give
\begin{equation}
  G_{A}^{(3)}
=
  \frac{1}{2} g_{A}
\label{eq:isospin}
\end{equation}
while the last one is the non-leading isoscalar form factor.
Both form factors can be cast in convolution form to read
\begin{equation}
  g_{A}(Q^{2})
=
  \int_{0}^{1} [dx]\int_{0}^{1} [dy]
  \Phi (y_{i}, Q^{2}) T_{H5}(x_{i},y_{i},Q^{2}) \Phi (x_{i},Q^{2}) \; ,
\label{eq:axiff}
\end{equation}
\begin{equation}
  G_{A}^{(S)}(Q^{2})
=
  \int_{0}^{1} [dx]\int_{0}^{1} [dy]
  \Phi (y_{i}, Q^{2})
  T_{H5}^{(S)}(x_{i},y_{i},Q^{2})
  \Phi (x_{i},Q^{2}) \; .
\label{eq:isoscaxiff}
\end{equation}
Assuming, as previously, constant arguments of the strong coupling
constants and expanding the nucleon {\DA}s in terms of their
eigenfunctions within the basis of Appell polynomials, the integrals
over the fractional momenta can be carried out to arrive at the
following expressions
\begin{equation}
  Q^{4}g_{A}(Q^{2})
=
  \left(\frac{4\pi}{27}\right)^{2}
  \left[\alpha _{\rm s}(Q^{2})\right]^{2} I_{A} \; ,
\label{eq:axfffin}
\end{equation}
\begin{equation}
  Q^{4}G_{A}^{(S)}(Q^{2})
=
  \left(\frac{4\pi}{27}\right)^{2}
  \left[\alpha _{\rm s}(Q^{2})\right]^{2} I_{S} \; ,
\label{eq:isoscaxfffin}
\end{equation}
where $I_{A}$ and $I_{S}$ are functions of the non-perturbative
expansion coefficients $B_{n}$.
Carlson and Poor truncate the eigenfunctions decomposition at
leading order $M=2$.
A more general expression for $g_{A}$ which includes the
next-to-leading order eigenfunctions ($M=3$) is given in
Appendix~\ref{ap:formfactors}.

Let us now present some results for the heterotic model.
The calculation for the nucleon axial form factor $g_{A}(Q^{2})$
according to \cite{CP86} yields at $Q^{2}\approx 10\,{\rm GeV}^{2}$,
$Q^{4}g_{A}(Q^{2})=0.90\,{\rm GeV}^{4}$ for
$\Lambda_{{\rm QCD}}=100$~MeV,
and $Q^{4}g_{A}(Q^{2})=1.44\, {\rm GeV}^{4}$ for
$\Lambda_{{\rm QCD}} = 180\, {\rm MeV}$.
These results compare fairly well with the value
$Q^{4}g_{A}(Q^{2})\approx 1.5\, {\rm GeV}^{4}$ extrapolated from the
data \cite{Bak81}.
Also the ratio ${g_{A}(Q^{2})}/{G_{\rm M}^{\rm p}(Q^{2})}
\approx 1.19$, in the region where the calculations can still be
trusted, is consistent with the (extrapolated) experimental value
${g_{A}(Q^{2})}/{G_{\rm M}^{\rm p}(Q^{2})}\approx 1.35$.
As for the isoscalar nucleon form factor \cite{CP87}, we find at
$Q^{2}\approx 10\, {\rm GeV}^{2}$,
$Q^{4}G_{A}^{(S)}(Q^{2})=0.83\, {\rm GeV}^{4}$ for
$\Lambda_{{\rm QCD}}=100\, {\rm MeV}$ and
$Q^{4}G_{A}^{(S)}(Q^{2})=1.34\, {\rm GeV}^{4}$
for $\Lambda_{{\rm QCD}}=180\, {\rm MeV}$.
Assuming isospin invariance, we combine these results with those for
$g_{A}$ to obtain ${G_{A}^{(S)}}/{G_{A}^{(3)}}\approx 1.85$, where
$G_{A}^{(3)}$ is the isovector axial-vector nucleon form factor.
If a dipole form
\begin{equation}
  G_{A}^{(S)}(Q^{2})
=
 \frac{G_{A}^{(S)}(0)}{\left( 1 + Q^{2}/M_{AS}^{2}\right)^{2}} \; ,
\label{eq:dipoleax}
\end{equation}
with $G_{A}^{(S)}(0)=0.38$ from $SU(6)$, is used to describe the
$Q^{2}$ dependence of $G_{A}^{(S)}$ \cite{CP87}, then, in the
high-$Q^{2}$ region, the heterotic model yields
$M_{AS}=(1.15-1.22)$~GeV for
$\Lambda_{{\rm QCD}}=100$~MeV and
$M_{AS}=(1.27-1.37)$~GeV for $\Lambda_{{\rm QCD}}=180$~MeV.
These values comply with the experimental value
$M_{A}=(1.032 \pm 0.036)$~GeV.
More systematic results are compiled in Table~\ref{tab:axialres}.

%
\vspace{-0.8 true cm}
\begin{table}
\caption[axialff]
        {Values of the axial-vector form factor $g_{A}$ and the
        isoscalar axial-vector form factor $G_{A}^{(S)}$ at
        $Q^{2} = 10\,{\rm GeV}^2/c^2$ for
        $\Lambda _{{\rm QCD}}=180$~MeV.
        The estimated empirical dipole masses $M_{A}$, $M_{AS}$ are
        also shown for some selected model {\DA}s for the nucleon.
\label{tab:axialres}}
\begin{tabular}{ccccccc}
Model           & $Q^4\,g_A$ & $M_A$ &  $Q^4\,G_A^{(S)}$ & $M_{AS}$ &
                        $G_A^{(S)}/G_A^{(3)}$ & $g_A/G_M^p$ \cr \hline
 Het       & 1.4374        & 1.0347  & 1.3379  & 1.3700  & 1.862  & 1.195 \cr
 $COZ^{{\rm opt}}$ & 4.1185  & 1.3462& 1.6104  & 1.4348  & 0.782  & 1.531 \cr
 $GS^{{\rm opt}}$    & 2.2419  & 1.2236  & 2.4376  & 1.5915  & 2.175  & 1.111   \cr
 CZ      & 3.5349  & 1.2957  & 1.2152  & 1.3373  & 0.688  & 1.529         \cr
 COZ     & 4.2747  & 1.3588  & 1.6404  & 1.4414  & 0.768  & 1.536         \cr
 KS      & 6.1508  & 1.4882  & 2.8808  & 1.6593  & 0.937  & 1.489         \cr
 GS      & 2.2130  & 1.1526  & 2.2436  & 1.5588  & 2.028  & 1.119
\end{tabular}
\end{table}

\subsection{NUCLEON-DELTA TRANSITION FORM FACTOR}
\label{subsec:nucdelff}
The electromagnetic $N-\Delta$ transition is another exclusive
process which offers the possibility to test the quality of the model
{\DA}s for the nucleon and the $\Delta$ resonance, obtained in the
previous chapters.
This process involves only interactions mediated by vector bosons
so that, neglecting quark masses, quark helicity is conserved.
Hence, analogously to the elastic electron-nucleon scattering, one
can define a helicity-conserving form factor
$G_{\rm M}^{{\rm N}\to\Delta}$
which according to perturbative QCD should fall off like $1/Q^{2}$ when
$Q^{2}$ becomes large.
On the experimental side, there are no exclusive data for $Q^{2}$
above 3~GeV${}^{2}$, though there are inclusive results, obtained
in SLAC experiment E133 spanning the $Q^{2}$ range
$2.5 - 10.0$~GeV${}^{2}$ \cite{Sto91}.

\begin{figure}
\centering
\epsfig{figure=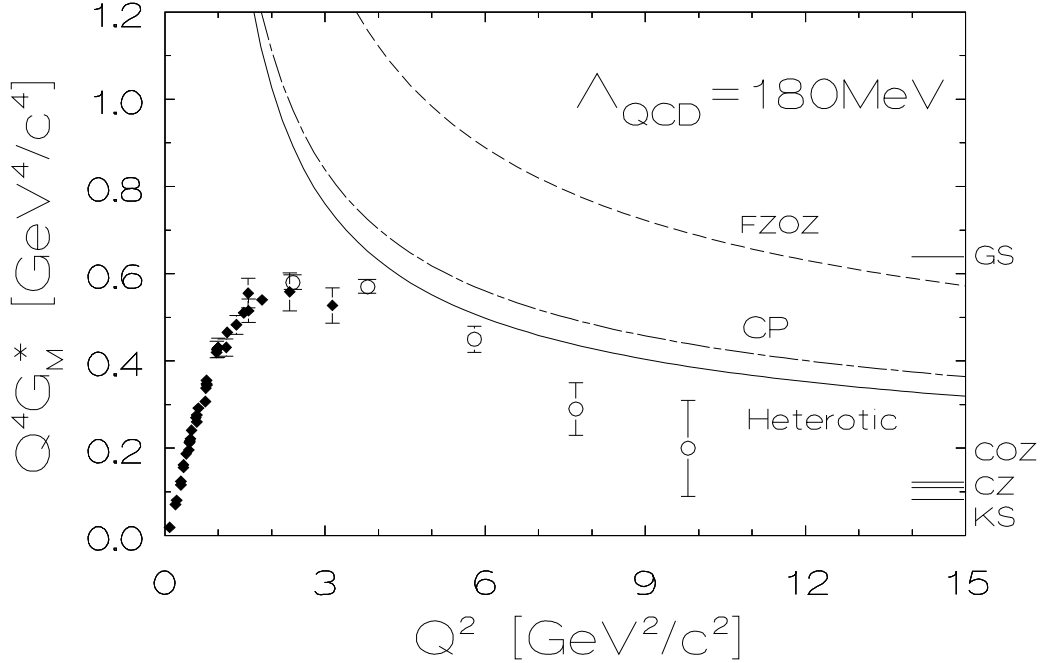,width=14cm,angle=90,silent=}
\vspace{0.4cm}
\caption[deltas]
        {\tenrm
         Transition form factor $\gamma p\Delta ^{+}$, calculated with
         the heterotic distribution amplitude for the nucleon
         \cite{SB93nuc,SB93del} and various $\Delta ^{+}$ distribution
         amplitudes in comparison with existing experimental data
         (see \cite{DATAdel}). The dashed-dotted line shows a
         calculation \cite{SB93del} which takes into account a dynamical
         gluon mass to saturate $\alpha _{\rm s}$ at low $Q^{2}$.
         The dashed line illustrates the effect of evolution of the
         coefficients $B_{n}^{\Delta}$. Predictions from previous
         nucleon distribution amplitudes are indicated on the right
         margin. The open circles denote the data from Stoler's
         \cite{Sto91} analysis. Note that for all curves the value
         $|f_{\Delta}|=11.5\times 10^{-3}$~GeV${}^{2}$ from \cite{CP88}
         has been used.
\label{fig:GM*het}}
\end{figure}

The form factor $G_{\rm M}^{{\rm N}\to\Delta}$ was calculated within
perturbative QCD by Carlson \cite{Car86}.
Starting from the convolution form
\begin{equation}
  G_{\rm M}^{{\rm N}\to\Delta}(Q^{2})
=
  \int_{0}^{1}[dx]\int_{0}^{1}[dy]\,
  \Phi _{\Delta}(y_{i},\tilde{Q}_{y})\,
  T_{\rm H}(x_{i}, y_{i}, Q)\,
  \Phi _{N}(x_{i}, \tilde{Q}_{x}) \; ,
\label{eq:transitionff}
\end{equation}
the result after integrating over longitudinal momenta can be expressed
in the following form
\begin{equation}
  Q^{4} G_{\rm M}^{*}(Q^{2})
=
  \left[
        \frac{4\pi \alpha _{\rm s}(Q^{2})}{27}\right]^{2}
        \frac{\sqrt 2}{3}
        I^{(\Delta)} \; ,
\label{eq:deltaff}
\end{equation}
where it is assumed that the argument of $\alpha _{\rm s}$ depends on
$Q^{2}$ only, and $I^{(\Delta)}$ is a function of the expansion
coefficients $B_{n}^{\Delta}$ which project the $\Delta$ {\DA} on to
the eigenfunctions of the interaction kernel:
\begin{equation}
  I^{\Delta}
\simeq
  \sum_{}^{} E_{ij} B_{i}^{(\Delta )} B_{j}^{(N)} \; .
\label{eq:Idelta}
\end{equation}
The hard part of the process derives from $T_{\rm H}$, encountered in
the calculation of the nucleon form factor, so that $\Phi _{\Delta}$
can be expanded over the same eigenfunctions basis, i.e., the Appell
polynomials.
The difference here is that, as stated in
Subsection~\ref{subsec:deltada}, only symmetric terms under
$x_{1} \leftrightarrow x_{3}$, contribute.
The coefficients $E_{ij}$ in Eq.~(\ref{eq:Idelta}) are calculable with
perturbative QCD.
Their values for $i,j=0,1, \ldots ,5$ are tabulated in \cite{Car86}.
The non-perturbative expansion coefficients $B_{n}^{(N)}$ can be taken
from Table~\ref{tab:Bs}, those for $B_{n}^{(\Delta )}$ from
Table~\ref{tab:DeltaBs}.
Several options are possible, depending on the favored choice among
the various {\DA}s.
In Fig.~\ref{fig:GM*het} we show predictions (solid lines) for
$\gamma p\Delta ^{+}$, calculated with the heterotic {\DA} for the
nucleon and various $\Delta$ {\DA}s, in comparison with existing
experimental data (see \cite{Sto91,DATAdel}).
The short horizontal lines on the right margin are predictions for
the absolute value of
$G_{\rm M}^{*}$ at $Q^{2}\,=\, 15\, {\rm GeV}^{2}$ from other nucleon
{\DA}s, listed in Table~\ref{tab:Bs}.
In all considered cases the CP value
$\vert f_{\Delta}\vert = 11.5\times 10^{-3}\ {\rm GeV}^{2}$
has been used, which is within the spread of the FZOZ estimate.
We emphasize that the sign of $G_{\rm M}^{*}$ predicted by
CZ \cite{CZ84a}, COZ \cite{COZ89a}, and GS \cite{GS86} comes out
negative for all $\Delta$ {\DA}s discussed here, accept for the
amplitude FZOZ${}^{{\rm opt}}$ which yields for COZ${}^{{\rm opt}}$
a small positive form factor by construction.
In contrast, convolution of FZOZ${}^{{\rm opt}}$ with the heterotic
nucleon {\DA} yields a form factor which overshoots the data by orders
of magnitude.
Only the heterotic nucleon {\DA} and the KS \cite{KS87} one yield a
positive sign for $G_{\rm M}^{*}$ and only the heterotic $\Delta$ {\DA}
comes close to the data.

%
%

%
\begin{figure}
\begin{picture}(0,350)
  \put(50,10){\psboxscaled{1100}{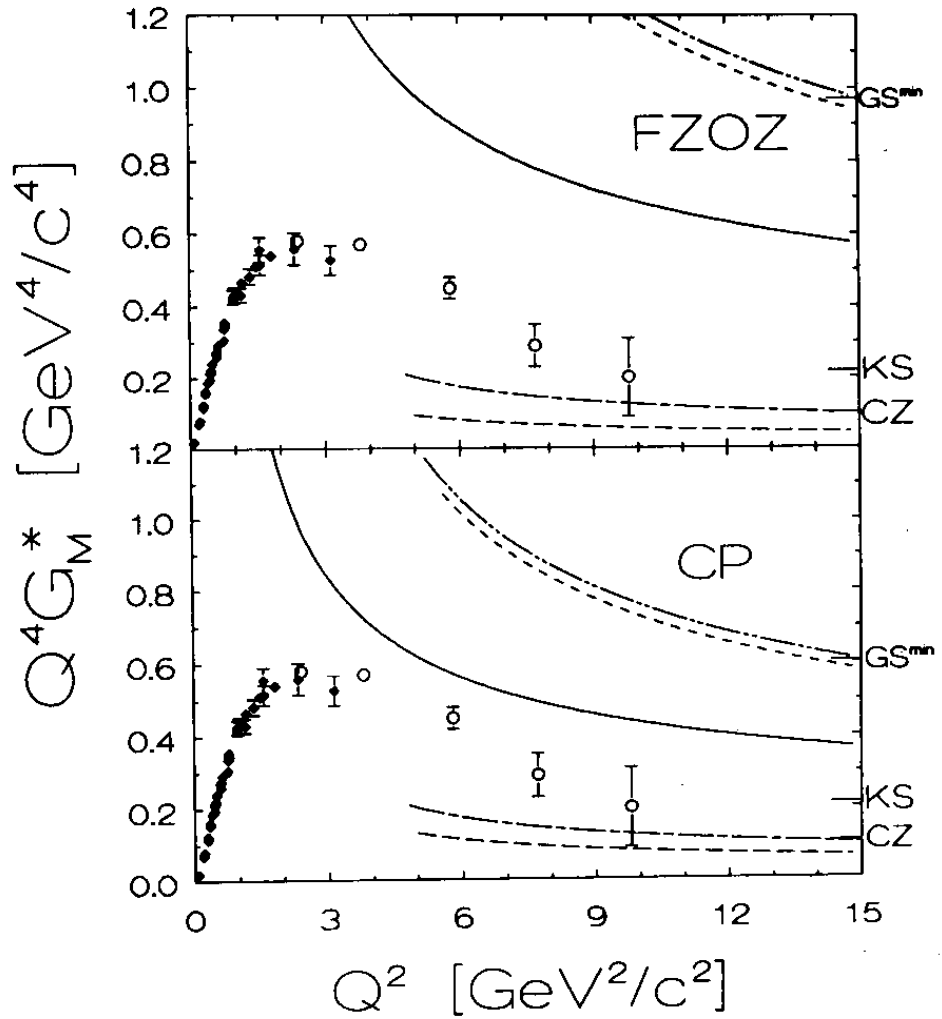}}
\end{picture}
\vspace{-0.5cm}
\caption[fig:transdata]
        {\tenrm
         Transition form factor $\gamma p\Delta ^{+}$, calculated with
         the FZOZ (top figure) and the CP distribution amplitude (lower
         figure) for the $\Delta$ resonance, and various nucleon
         distribution amplitudes in comparison with existing
         experimental data (see \cite{DATAdel}). The dashed-dotted line
         shows the calculation with the COZ distribution amplitude; the
         dashed line that one with the COZ${}^{{\rm opt}}$ {\DA}.
         Predictions from previous nucleon distribution amplitudes are
         again indicated on the right margin. The open circles denote
         the data from Stoler's \cite{Sto91} analysis. Note that for all
         curves the value $|f_{\Delta}|=11.5\times 10^{-3}$~GeV${}^{2}$
         from \cite{CP88} has been used.
\label{fig:GM*all}}
\end{figure}
%
This agreement can be significantly improved.
In order to account for (unknown) confinement effects at low $Q^{2}$,
we saturate $\alpha _{\rm s}$ in the one-loop approximation by
introducing an effective gluon mass \cite{Cor82}:
$
 \alpha _{\rm s}(Q^{2}) \mapsto
 \alpha _{\rm s}(Q^{2} + 4\, m_{\rm g}(Q^{2}))
$.
In contrast to other approaches of this type \cite{JSL87,HEG92}, we
use a dynamical, i.e., scale-dependent gluon mass derived by
Cornwall \cite{Cor82}:
\begin{equation}
  m_{\rm g}^{2}(Q^{2})
=
  m_{\rm g}^{2}
  \left[
  \ln \left(\frac{Q^{2} + 4 m_{\rm g}^{2}}{{\rm \Lambda}_{QCD}^{2}}
      \right)
             {\Bigg /} \ln
             \left(\frac{4 m_{\rm g}^{2}}{{\rm \Lambda}_{QCD}^{2}}
             \right)
           \right]^{-12/11}\; .
\label{eq:gluonmass}
\end{equation}
Due to the positivity of the anomalous dimension of the mass
operator, this gluon mass vanishes asymptotically.
This soft behavior at short distances leaves the validity of form-factor
evolution at large momentum transfer virtually unaffected (as it should,
if the dynamical mass generation is to be consistent with the {\RG}).
In the fit represented by the dashed-dotted line
(Fig.~\ref{fig:GM*het}) we use $m_{\rm g}=380$~MeV, which matches
Cornwall's consistency relation, mentioned previously.
Referring to the same figure, we see that including the perturbative
$Q^{2}$ evolution of the coefficients $B_{n}^{\Delta}$, it is
sufficient to provide a good fit to the data above
$Q^{2}\approx 3~{\rm GeV}^{2}/c^{2}$ (dashed line) (see, the recent
data analysis in \cite{Stu93,Stu96}).
At lower $Q^{2}$ values, additional non-perturbative parameters have
to be introduced, e.g., parton transverse momenta, effective parton
masses, higher-twist contributions, quark clustering, etc. to account
for the limitations of the leading-order formalism.
In a recent analysis \cite{BR95} Belyaev and Radyushkin calculated
on the basis of local quark-hadron duality the purely non-perturbative
soft contribution to the $\gamma ^{*}p \to \Delta$ form factor and
showed that their result can account for the observed behavior of
the form factor at $Q^{2}$ values as low as 3~GeV${}^{2}$.
They also argued that this contribution remains leading up to the
highest laboratory $Q^{2}$ values without any need for the
contribution due to hard-gluon exchange (embedded in $T_{\rm H}$).
It is clear that experimental efforts to measure $G_{\rm M}^{*}$ beyond
$10$~GeV${}^{2}/c^{2}$ would be extremely helpful to discriminate
among these options.

To complete our discussion, we show in Fig.~\ref{fig:GM*all} the
form-factor predictions for $G_{\rm M}^{*}$ obtained from other nucleon
{\DA}s in convolution with the model $\Delta$ {\DA}s of CP (top figure)
and FZOZ (lower figure).
One may interpret these figures as indicating that none of the
shown amplitudes neither for the nucleon nor for the $\Delta$ can
account for the observed behavior of the form factor.
COZ-type {\DA}s for the nucleon exhibit strong cancellation of symmetric
vs antisymmetric contributions \cite{CGS87}, while GS-type {\DA}s yield
unrealistically large negative contributions.

\section{CHARMONIUM DECAYS}
\label{sec:charm}
The last phenomenological application we consider here within the
{\SCS} here deals with exclusive decays of charmonium states to
$p\bar p$ and $\Delta\bar\Delta$.
Such decays are sensitive to the nucleon ($\Delta$) distribution
amplitude and hence may serve to discriminate among proposed models.
Calculations of charmonium decays have been performed by many
authors \cite{COZ89b,BL81,And84,DTB85} within the QCD convolution
framework.
We follow \cite{COZ89b}.
We consider first the ${}^{3}P_{J}$ states with $J=1,2$.
The branching ratio for the decay of the
$\chi_{c1}$ state ($J^{PC}=1^{++}$) into $p\bar p$ is given
by\footnote{I wish to thank Victor Chernyak for bringing to my
attention that the factor $(\pi \alpha _{\rm s})^{3}$ was missing in
\cite{COZ89b,SB93nuc,Ste94}.}
\begin{equation}
  {\rm BR}
          \left(
                 \frac{{}^{3}P_{1}\to p\bar p}{{}^{3}P_{1}\to {\rm all}}
          \right)
\approx
   \frac{0.75}{\ln ({\bar M}/\Delta )}
   \left( \pi \alpha _{\rm s} \right)^{3}
   \frac{16\pi ^{2}}{729}
   \Bigg |
          \frac{f_{\rm N}}{\bar{M}^{2}}
   \Bigg |^{4}\,M_{1}^{2} \; ,
\label{eq:BR3P1}
\end{equation}
where $\bar M\approx 2 m_{c} \approx 3$~GeV and $\Delta = 0.4$~GeV
(the last value from \cite{BGK76}).
The non-perturbative content of Eq.~(\ref{eq:BR3P1}) is due to $f_{\rm
N}$ and the decay amplitude for the process
$
 {}^{3}P_{1}\to p\bar p,
$
denoted $M_{1}$, which involves $\Phi _{N}$.
Inserting expression (\ref{eq:PhiBs}), in connection with
Table~\ref{tab:Bs}, the decay amplitude $M_{1}$ for each model is
computed with an elaborate integration routine which properly takes
account of contributions near singularities \cite{Ber94}.
The results for different model distribution amplitudes are compiled
in Table~\ref{tab:charmN}.

%
\def\za{\phantom{1}}
\def\zb{\phantom{12}}
\begin{table}
\caption[tab:FNAL]
        {Exclusive charmonium decays into $p\bar p$ for a variety of
         nucleon distribution amplitudes. The data are taken from
         \cite{Arm92}. The numbers in parentheses are those given
         in \cite{COZ89b}. The decay amplitudes termed $M_{0}$ were
         independently verified by Bolz \cite{Bol96}.
\label{tab:charmN}}
\begin{tabular}{l|ccc}
Models & $M({}^3P_{1}\to p\bar{p})=M_1$ & $M({}^3P_{2}\to p\bar{p})=M_2$ & $M(^3S_{1}\to p\bar p)=M_0$ \\
\hline
{\rm heterotic}    &   99849.6 &   515491.2  &  13726.8  \\
CZ           &   28310.4 \,  (0.63$\times 10^{5}$) & 246052.8\,
                (2.87$\times 10^{5}$) & 7545.6\, (0.72$\times 10^{4}$)\\
COZ          &   53625.6 \,  (0.88$\times 10^{5}$) & 298123.2\,
                (3.4$\times  10^{5}$) & 8758.8\, (0.79$\times 10^{4}$)\\
COZ$^{{\rm opt}}$  &   55137.6 &   289728.0  &   8499.6  \\
GS           &   26366.4 &   232632.0  & 928.8\, (0.7$\times 10^{3}$)\\
GS$^{{\rm opt}}$   &   19915.2 &   230659.2  &    986.4  \\
GS$^{{\rm min}}$   &   17193.6 &   210729.6  &    964.8  \\
KS           &   94723.2\,   (1.35$\times 10^{5}$) & 416937.6\,
                (4.84$\times 10^{5}$) & 11484.0\, (1.15$\times 10^{4}$)\\
{rm asympt.}      &   20086.6\,   (0.2$\times 10^{5}$)  & 43099.2\,
                (0.43$\times 10^{5}$) & 1517.4\, (0.14$\times 10^{4}$)\\
\hline
 Observables &
BR$\Bigl(\frac{{}^3P_{1}\to p\bar{p}}{{}^3P_{1}\to all}\Bigr)$ in \% &
BR$\Bigl(\frac{{}^3P_{2}\to p\bar{p}}{{}^3P_{2}\to all}\Bigr)$ in \% & $ \Gamma ({}^3S_{1}\to p\bar{p}) $ [eV] \\
\hline
{\rm heterotic}    & 0.22 $\!\times 10^{-2}$ & 0.89 $\!\times 10^{-2}$ & 138.37\\
CZ           & 0.017 $\!\times 10^{-2}$ & 0.20 $\!\times 10^{-2}$ &\za 41.81\\
COZ          & 0.06 $\!\times 10^{-2}$ & 0.30 $\!\times 10^{-2}$ &\za 56.34\\
COZ$^{{\rm opt}}$  & 0.066 $\!\times 10^{-2}$ & 0.28 $\!\times 10^{-2}$  &\za 53.07\\
GS           & 0.014 $\!\times 10^{-2}$ & 0.18 $\!\times 10^{-2}$ &\zb 0.63\\
GS$^{{\rm opt}}$   & 0.086 $\!\times 10^{-3}$ & 0.18 $\!\times 10^{-2}$ &\zb 0.71\\
GS$^{{\rm min}}$   & 0.066 $\!\times 10^{-3}$ & 0.15 $\!\times 10^{-2}$ &\zb 0.68\\
KS           & 0.198 $\!\times 10^{-2}$ & 0.59 $\!\times 10^{-2}$ &\za96.84\\
{\rm asympt.}      & 0.086 $\!\times 10^{-3}$ & 0.01 $\!\times 10^{-2}$ &\zb 1.69\\
\hline
E760         & ($0.78\pm 0.10\pm 0.11$)$\times 10^{-2}$ &
               ($0.91\pm 0.08\pm 0.14$)$\times 10^{-2}$ &
               $180\pm 16\pm 26$ \\
\end{tabular}
\end{table}

The analogous expression to (\ref{eq:BR3P1}) for the $\chi _{c2}$
state ($J^{PC}=2^{++}$) has the form
\begin{equation}
  {\rm BR}
          \left(
                 \frac{{}^3P_{2}\to p\bar p}{{}^3P_{2}\to {\rm all}}
          \right)
\approx
  0.85 (\pi\alpha _{\rm s})^{4}
  \frac{16}{729}
  \Bigg |\frac{f_{\rm N}}{{\bar M}^{2}}
  \Bigg |^{4} M_{2}^{2}\;,
\label{eq:BR3P2}
\end{equation}
which is Eq.~(20) of \cite{COZ89b} with an obvious minor correction.
The results for the branching ratio of this process, shown in
Table~\ref{tab:charmN}, have been calculated with
$\alpha _{\rm s}(m_{c})=0.210\pm 0.028$
(see third paper of \cite{BGK76}).

Similar considerations apply also to the charmonium decay into
$p\bar p$
of the level ${}^{3}S_{1}$ with $J^{PC}=1^{--}$.
The partial width of $J/\psi$ (i.e., $\chi_{c0}$) into $p\bar p$ is
defined by
\begin{equation}
  \Gamma({}^3S_{1}\to p\bar p)
=
  (\pi\alpha _{\rm s})^{6}
  \frac{1280}{243\pi}
  \frac{|f_{\psi}|^{2}}{\bar M}
\Bigg
     |\frac{f_{\rm N}}{{\bar M}^{2}}
\Bigg|^{4}M_{0}^{2}\;,
\label{eq:Gamma3S1}
\end{equation}
where $f_{\psi}$ determines the value of the ${}^3S_{1}$-state wave
function at the origin, {\it viz},
\begin{equation}
  \langle  0 | \bar{c}(0) \gamma _{\mu} c(0) | {}^{3}S_{1} \rangle
=
  \psi _{\mu} f_{\psi} M_{\psi} \; ,
\label{eq:origin}
\end{equation}
where $\psi _{\mu}$ is the $J/\psi$ polarization vector.
The value of $f_{\psi}$ can be extracted from the leptonic width
$\Gamma ({}^3S_{1}\to e^{+}e^{-}) = (5.36 \pm 0.29)$~keV \cite{PDG92}
via the Van Royen-Weisskopf formula:
\begin{equation}
  \Gamma \left({}^{3}S_{1} \to e^{+}e^{-} \right)
=
  \frac{64}{9}
  \pi \alpha _{\rm s}^{2}\vert \psi_{J/\psi}(0)\vert ^{2}
  \frac{1}{\bar{M}^{2}} \; .
\label{eq:vanRoyWeis}
\end{equation}
The result is
$|f_{\psi}|=409$~MeV with $m_{J/\psi}=3096.93$~MeV.
(Note that $\bar{M}\approx M_{J/\psi}$.)
Then, using the same values of parameters as before, we obtain the
results shown in Table~\ref{tab:charmN}.
One sees from this table that the agreement between the predictions
of the heterotic model and the recent high-precision data of the
E760 experiment at Fermilab \cite{Arm92} is quite good.
The only exception is the branching ratio for the process
${}^{3}P_{1}\to p\bar{p}$ which comes out too small, though the
heterotic amplitude yields the largest value, compared to all other
models.
The experimental result can be reproduced using
$\alpha _{\rm s}\simeq 0.32$.

In a completely analogous way, one can calculate the corresponding
exclusive decays of (helicity-conserving) charmonium states in
$\Delta \bar \Delta$.
The branching ratios follow by replacing $f_{\rm N}$ by
$f_{\Delta}/\sqrt{3}$.
The results of this calculation are summarized in
Table~\ref{tab:charmdel}.
As regards the branching ratio of the ${}^3S_{1}$ state, the heterotic
{\DA} yields
$ BR({{{}^3S_{1}\to \Delta\bar\Delta} / {{}^3S_{1}\to all}})
  = 0.30 \times 10^{-2} \% $, where
$\Gamma_{tot} = 85.5^{+6.1}_{-5.8}$~keV is used.
Note that in all considered decays
$\alpha _{\rm s}(m_{c}) = 0.210 \pm 0.028$
as before.

%
\begin{table}
\caption{Exclusive charmonium decays in $\Delta\bar{\Delta}$ for the
         models discussed in the text.
\label{tab:charmdel}}
\begin{tabular}{cccc}
 Amplitudes   &  CP & FZOZ &  {\rm Heterotic} \\
\tableline
         $M^{\Delta}(^3P_{1})$  & 11418.23   &   21585.99  &   11651.26  \\
         $M^{\Delta}(^3P_{2})$  & 30924.07   &   48233.23  &   26277.40 \\
         $M^{\Delta}(^3S_{1})$  &\za 1480.67 &\za 1882.31  &\za 1134.94 \\
\tableline
 Observables & & & \\
\tableline
$BR\Bigl({{^3P_{1}\to \Delta\bar{\Delta}}\over {^3P_{1}\to all}}\Bigr)$
& $0.089\times 10^{-3}\%\ \ \ $ & $0.3196\times 10^{-3}\%\ \ \ $ &
$0.093\times 10^{-3}\%\ \ \ $
\vspace{.1cm} \\
$BR\Bigl({{^3P_{2}\to \Delta\bar{\Delta}}\over {^3P_{2}\to all}}\Bigr)$
& $0.100\times 10^{-3}\%\ \ \ $ & $0.106\times 10^{-3}\%\ \ \ $ & $0.072\times 10^{-3}\%\ \ \ $
\vspace{.1cm} \\
$ \Gamma(^3S_{1}\to \Delta\bar{\Delta}) $
& $0.439\times 10^{-2}\, {\rm keV}$& $0.709\times 10^{-2}\, {\rm keV}$& $0.258\times 10^{-2}\, {\rm keV}$ \\
\end{tabular}
\end{table}

\section{MODIFIED CONVOLUTION SCHEME}
\label{sec:mod}
\begin{figure}
\begin{picture}(0,40)
  \put(35,-80){\psboxscaled{700}{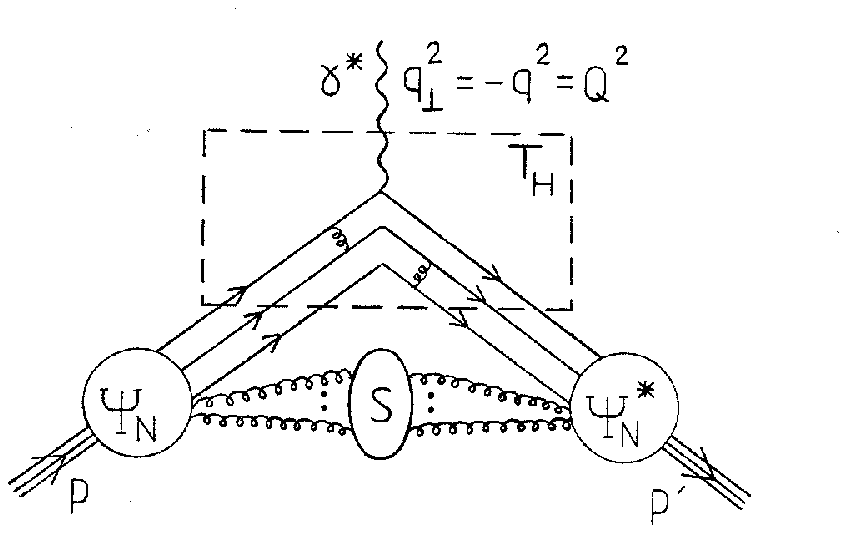}}
\end{picture}
\begin{picture}(0,40)
  \put(248,-68){\psboxscaled{700}{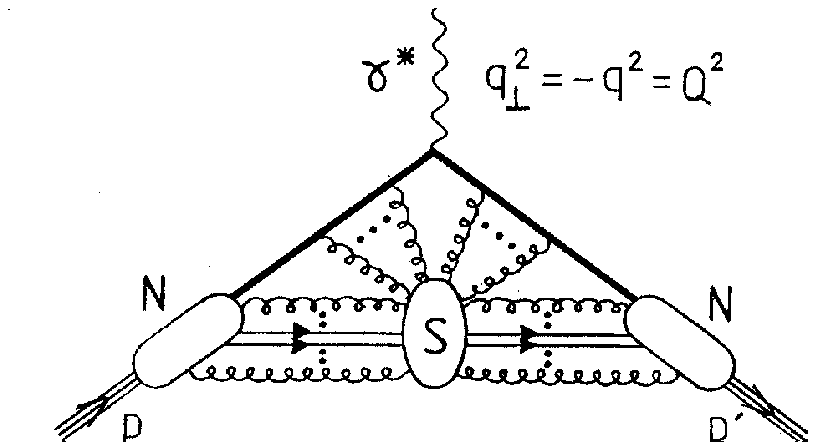}}
\end{picture}
\vspace{3.0cm}
\caption[fig:mech]
        {\tenrm
         Mechanisms for momentum transfer during elastic scattering.
         The lhs shows hard-gluon exchange within pQCD.
         The blob $S$ containing soft gluon lines (and an analogous
         one with soft quark-anti-quark lines not shown here) spoils
         factorization but is power-suppressed, i.e., non-leading.
         The (rhs) shows the Feynman mechanism using, for purposes
         of illustration, quark and gluon lines. The leading quark
         is denoted by a heavy line, while all other lines represent
         wee quarks and soft gluons.
\label{fig:momtranmechs}}
\end{figure}

In casting the hadron form factors in convolution form
(cf. Eq.~(\ref{eq:magnucleonff})), we tacitly assumed that the
$k_{\perp}$-dependence of the quark and gluon propagators in $T_{\rm H}$
can be ignored.
This is tantamount to factorizing the $k_{\perp}$-dependence into the
distribution amplitudes which are the wave functions integrated over
$k_{\perp}$ up to the factorization scale.
Thus, in the limit $Q^{2}\to\infty$, the only gluon radiative
corrections remaining uncanceled are those giving rise to
wave-function renormalization.
Indeed, recalling Eqs.~(\ref{eq:piexp}), (\ref{eq:solnuc}), one sees
that asymptotically the most likely configurations are those in which
the valence quarks share longitudinal momentum in a uniform way, i.e.,
$x_{i}=1/2$ for the pion and $x_{i}=1/3$ for the nucleon.
When confinement sets in, a quark is not able to venture too far
from the anti-quark (in the pion) or the other two valence quarks
(in the nucleon), and this poses constraints on the off-shellness of
the involved propagators (see lhs of Fig.~\ref{fig:momtranmechs}).
However, in the end-point region, the parton transverse momenta in
$T_{\rm H}$ cannot be {\it a priori} ignored, since, say, for the pion,
$
 \left({\bf k}^{}_{\perp i} + {\bf k}_{\perp j}^{\prime}\right)^{2}
\gg
 x^{}_{i}x_{j}^{\prime}Q^{2}
$.
As a result, the transverse distance between the quark and the
anti-quark becomes large compared to $1/Q$ and the corresponding
gluon line is no more part of the hard-scattering process but
should be counted to its soft part.
This is in contrast to the perturbative QCD treatment in which the
struck quark connects to the other valence quarks via highly
off-shell gluon propagators, meaning that the transverse inter-quark
distances are rather small, {\it viz.} of order $1/Q$ and that
{\it all} partons share comparable fractions of longitudinal
momentum.
In other words, the mechanism of hard-gluon exchange becomes
inapplicable and should be replaced by that of Feynman \cite{Fey72}.
According to the latter mechanism, almost all of the hadron's momentum
is carried off by a single parton, the others being ``wee''.
This picture is consistent with a configuration in which only the
struck quark is within an impact distance $1/Q$ of the electron while
all other partons have rather random positions in the transverse
direction, building a soft ``cloud'' with transverse size
$\gg 1/Q$ \cite{STARA96}.
Once the elastic scattering has happened, rearrangements are
necessary to change quarks and gluons into an unscathed hadron.
This conversion procedure (visualized on the rhs of
Fig.~\ref{fig:momtranmechs}) is controlled by the overlap of the
initial and final state wave functions and cannot be computed
within pQCD.

%
%

\subsection{MODIFIED FACTORIZATION}
\label{subsec:modfact}
\begin{figure}
\begin{picture}(0,340)
  \put(120,10){\psboxscaled{1500}{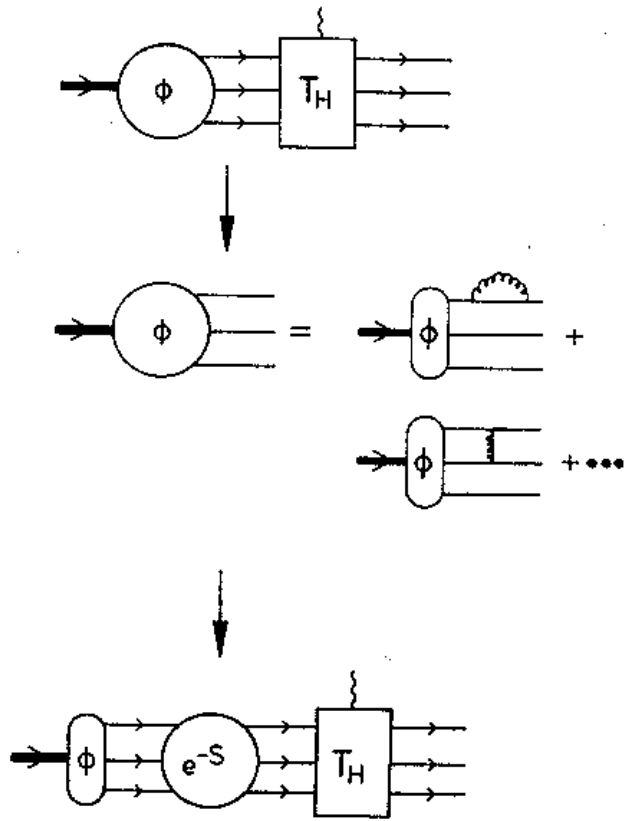}}
\end{picture}
\caption[fig:modfact]
        {\tenrm
         Modified factorization of soft-gluon contributions
         tantamount to a finite renormalization of the nucleon
         wave function.
\label{fig:modPhi}}
\end{figure}

The physical basis of the {\MCS} is to dissect the process in such a
way as that for transverse distances large compared to $1/Q$
(the latter being the playground of the hard-scattering mechanism) but
still small relative to the true confinement regime -- characterized by
$1/\Lambda _{{\rm QCD}}$ -- the hadron wave function is modified to
exhibit the effect of Sudakov enhancements {\it explicitly} up to the
transverse scale retained in $T_{\rm H}$ (see Fig.~\ref{fig:modPhi}).
Going over to the transverse configuration space, the modified wave
function in the axial gauge $A^{+}=0$ reads
\begin{equation}
  \hat{\Psi} ^{(H)}_{({\rm mod})}
  (x_{i}, 1/\tilde{b}_{i}, Q^{2}, \mu _{{\rm ren}}^{2})
=
  {\rm e}^{-S}\hat{\Psi} ^{(H)}(x_{i}, Q^{2}, 1/\tilde{b}_{i})
\label{eq:modPsi} \; .
\end{equation}
The Sudakov exponential factor re-sums contributions from two-particle
reducible diagrams (giving rise to double logarithms), whereas
two-particle irreducible diagrams (giving rise to single logarithms)
are absorbed into the hard scattering amplitude
$T_{\rm H}$ \cite{BS89}.
Hence, ${\rm e}^{-S}$ can be conceived of as being a {\it finite} IR
renormalization factor to the hadron's wave function \cite{Ste95},
encoding the exponentiation of the probability for no-emission of
soft gluons.
This factor renormalizes the wave function in addition to the
conventional renormalization factor $Z_{2}^{\rm q}$ due to UV
divergences, encountered before.
While $Z_{2}^{\rm q}$ contains single logarithms in leading order,
the Sudakov renormalization factor is dominated by double logarithms.
The leading double logarithms derive from those momentum regions
where soft gluons (all four-momentum components small) and collinear
gluons to the external quark lines overlap.
These contributions are numerically dominated by the term
\begin{equation}
 \exp
     \left\{
            - \frac{2C_{\rm F}}{\beta _{0}}
              \ln \frac{\xi _{i}Q}{\sqrt{2}\Lambda _{{\rm QCD}}}
              \ln\frac{\ln \left(\xi _{i}Q/\sqrt{2}\Lambda _{{\rm QCD}}
                           \right)
                      }
              {\ln\left(1/\tilde{b}_{i}\Lambda _{{\rm QCD}}\right)}
     \right\} \; ,
\label{eq:doublelog}
\end{equation}
where $\xi _{i}$ is one of the fractions
$x_{i}, \overline{x}_{i}\equiv 1-x_{i}$
(for incoming valence quarks) or
$x_{i}', \overline{x}^{\prime}_{i}\equiv 1-x_{i}'$
(for outgoing valence quarks), and $\beta _{0}$ is the first-order
term of the Gell-Mann and Low function encountered before.
The single logarithm in Eq.~({eq:doublelog}) stems from the running of
the coupling constant, and the double logarithm contains the
exponentiated higher-order corrections -- required by the {\RG} --
rendered finite by the {\it inherent} IR-cutoff $1/\tilde{b}_{i}$.
This issue marks the crucial difference between the {\MCS} and previous
approaches dealing with {\it isolated} quarks where such IR-cutoff
parameters had to be introduced as {\it external} regulators.
Here IR regularization is provided {\it in situ} without additional
assumptions.
For small transverse distances (or equivalently,
$1/\tilde{b}_{i}\gg \xi _{i}Q$),
gluonic radiative corrections are treated as being part of $T_{\rm H}$
and are excluded from the Sudakov form factor.
Consequently, for
$\xi _{i}\le\sqrt{2}/\tilde{b}_{i}Q$ the Sudakov functions
$s(\xi _{i},\tilde{b}_{i},Q)$ are set equal to zero.
On the other hand, as $\tilde{b}_{i}$ increases, $e^{-S}$ decreases,
reaching zero at $\tilde{b}_{i}\Lambda _{{\rm QCD}}=1$.

%
%

The full expression for the Sudakov exponent was calculated by Botts
and Sterman \cite{BS89}.
It is given by\footnote{A corrected form of this expression was
derived by Bolz \cite{Bol95}; see also \cite{DJK95}.}
\begin{eqnarray}
  S_{j}
= & \phantom{} &
\!\!\!\!\!\!\!
  \sum_{l=1}^{3}
  \left[
          s(x_{l},\tilde{b}_{l},Q)
        + \int_{1/\tilde{b}_{l}}^{t_{j1}}
                    \frac{d\bar{\mu}}{\bar{\mu}}
                    \gamma _{q}(g(\bar{\mu} ^{2}))
  \right]
\nonumber \\
& + &
  \sum_{l=1}^{3}
  \left[
          s(x_{l}^{\prime},\tilde{b}_{l},Q)
        + \int_{1/\tilde{b}_{l}}^{t_{j2}}
                    \frac{d\bar{\mu}}{\bar{\mu}}
                    \gamma _{q}(g(\bar{\mu} ^{2}))
  \right] \; ,
\label{eq:S}
\end{eqnarray}
wherein the Sudakov functions $s(\xi _{l},\tilde{b}_{l},Q)$ are given
by
\begin{eqnarray}
  s(\xi _{l},\tilde{b}_{l},Q)
= & \phantom{} &
\!\!\!\!\!\!\!\!
   \frac{A^{(1)}}{2\beta _{1}}\;
   \hat{q}_{l}
   \ln\!\left(
                 \frac{\hat{q}_{l}}{\hat{b}_{l}}
        \right)
 +
   \frac{A^{(2)}}{4\beta _{1}^{2}}\,
           \left(
                 \frac{\hat{q}_{l}}{\hat{b}_{l}} - 1
           \right)
 - \frac{A^{(1)}}{2\beta _{1}}\;
            (\hat{q}_{l}-\hat{b}_{l})
\nonumber \\
& - &
   \frac{A^{(1)}\beta _{2}}{16\beta _{1}^{3}}\,
   \hat{q}_{l}
           \left[
                   \frac{\ln (2\hat{b}_{l}) + 1}{\hat{b}_{l}}
                 -
                   \frac{\ ln (2\hat{q}_{l}) + 1}{\hat{q}_{l}}
           \right]
\nonumber \\
& - &
  \left[
          \frac{A^{(2)}}{4\beta _{1}^{2}}
        - \frac{A^{(1)}}{4\beta _{1}}
                   \ln \Bigl({\rm e}^{2\gamma -1}/2\Bigr)
  \right]
  \ln \left(
                \frac{\hat{q}_{l}}{\hat{b}_{l}}
      \right)
\nonumber \\
& - &
  \frac{A^{(1)}\beta _{2}}{32\beta _{1}^{3}}
           \left[
                   \ln ^{2}(2\hat{q}_{l})
                 - \ln ^{2}(2\hat{b}_{l})
           \right] \; .
\label{eq:s}
\end{eqnarray}
Here
$\xi _{l}=x_{l}, \overline{x}_{l}$ or $x_{l}^{\prime},
\overline{x}^{\prime}_{l}$
($l=1,2,3$)
and the variables $\hat{q}$ and $\hat{b}$ are defined as follows
\begin{equation}
  \hat{q}_{l}
=
  \ln [\xi _{l}Q/(\sqrt{2}\Lambda _{{\rm QCD}})] \; ,
\end{equation}
\begin{equation}
  \hat{b}_{l}
=
  \ln [1/\tilde{b}_{l}\Lambda _{{\rm QCD}}].
\end{equation}
The coefficients $A^{(i)}$ and $\beta _{i}$ are
\[
  A^{(1)} =  \frac{4}{3}, \;\;\;
  A^{(2)} =   \frac{67}{9} - \frac{1}{3}\pi ^{2} - \frac{10}{27}n_{\rm f}
            + \frac{8}{3}\beta _{1}\ln \!
              \left(
                    \frac{1}{2}{\rm e}^{\gamma}
              \right),
\]
\begin{equation}
  \beta _{1}=\frac{33-2n_{\rm f}}{12}=\frac{1}{4}\beta _{0},\;\;\;
  \beta _{2}=\frac{153-19n_{\rm f}}{24},
\end{equation}
where $n_{\rm f}$ is the number of quark flavors and
$\gamma =0.5772\ldots$ is the Euler-Mascheroni constant.

The Sudakov function, $s(\xi _{l},\tilde{b}_{l},Q)$, in Eq.~(\ref{eq:s})
takes into account leading and next-to-leading order gluonic radiative
corrections of the form shown in Fig.~\ref{fig:modPhi} and accounts
for {\RG} evolution from the IR-scale $1/\tilde{b_{i}}$ to the
renormalization scale $\mu _{ren}$ via the quark anomalous dimension.
The quantities $\tilde{b}_{l}$ ($l=1,2,3$) are IR cutoff parameters,
naturally related to, but not uniquely determined by the mutual
separations of the three quarks \cite{CS81}.
A physical perspective on the choice of the IR cutoff is provided
by the following analogy to ordinary QED.
One expects that because of the color neutrality of a hadron, its
quark distribution cannot be resolved by gluons with a wavelength
much larger than a characteristic quark separation scale.
Hence, long-wavelength gluons probe the color singlet proton and
radiation is damped.
On the other hand, radiative corrections with wavelengths between the
IR cutoff and an upper limit (related to the physical momentum Q) yield
to suppression.
It is understood that still softer gluonic corrections are already
taken care of in the hadron wave function, whereas harder gluons are
considered as being part of $T_{\rm H}$.

In the pion case, there is only one transverse scale, notably, the
quark--anti-quark separation $b$, so that suppression is automatically
accomplished.
Indeed, when it happens that one Sudakov function $s(\xi ,b,Q)=0$
(meaning that the corresponding exponential is set equal to unity) the
other (negative) Sudakov function in the exponent, $s(1-\xi ,b,Q)$,
diverges, thus providing sufficient suppression.
This IR-protecting behavior due to the Sudakov form factor, makes it
possible to choose a renormalization scale which depends on the initial
and final longitudinal momenta.
It was shown in Ref. \cite{DR81}, within the collinear approximation,
that such a subtraction point minimizes the contributions of the
next-to-leading order corrections to the pion form factor, ensuring
dominance of the leading-order perturbative contribution.
In the {\SCS}, such a choice leads to singularities in the running
coupling constant $\alpha _{\rm s}$, unless it is saturated by
additional IR regulators; e.g., a dynamical gluon mass.
Here the modification of $\alpha _{\rm s}$ becomes superfluous because
the Sudakov effect inhibits soft-gluon emission, thus effectively
{\it screening} the $\alpha _{\rm s}$-singularities.
Physically, the Sudakov suppression enhances the dominance of quark
configurations with a small color dipole moment and this enhancement
becomes more pronounced as $Q^{2}$ increases.
Concerning the nucleon, the situation is more complicated because
several transverse scales are involved and therefore a careful IR
regularization is required.
Now the Sudakov function comprises six terms
($\xi _{l}=x_{l}$ or $\xi _{l}=x_{l}'$ for $l=1,2,3$),
each depending on its own transverse scale and with corresponding
QCD-evolution contributions, driven by the anomalous dimensions
associated with quark self energy.
Recall that the IR cutoff to be used is the factorization scale,
below which {\OPE} becomes a poor approximation.
It is also understood that genuine non-perturbative momenta below the
IR cutoff are implicitly accounted for in the nucleon wave function.

Different choices of the IR cutoff have been used in the literature:
Thus, Li \cite{Li93} chooses $\tilde{b}_{l}=b_{l}$ (this choice is
termed hereafter the ``L'' prescription), whereas Hyer \cite{Hye93} in
his analysis of the proton-antiproton annihilation into two photons and
of the time-like proton form factor as well as Sotiropoulos and Sterman
\cite{SS94} take
$\tilde{b}_{1}=b_{2}$, $\tilde{b}_{2}=b_{1}$, $\tilde{b}_{3}=b_{3}$
(this choice is denoted the ``H-SS'' prescription).
Still another possibility, proposed in \cite{BJKBS95pro}, is to use
for reasons that will be explained below as IR cutoff the maximum of
the three inter-quark separations, i.e., to set
\begin{equation}
  \tilde{b}\equiv {\rm max}\{b_{1},b_{2},b_{3}\}
=
  \tilde{b}_{1}=\tilde{b}_{2}=\tilde{b}_{3} \; .
\label{eq:MAX}
\end{equation}

This choice, designated by ``MAX'', is analogous to that in the meson
case, wherein the quark-anti-quark distance naturally provides a secure
IR cutoff.
The specific features of each particular cutoff choice will be
discussed in detail later.

The integrals in Eq.~(\ref{eq:S}) arise from the application of the
{\RG} equation.
The evolution from one scale value to another is governed by the
anomalous dimensions of the involved operators.
The integrals combine the effects of the application of the {\RG}
equation on the wave functions and on the hard scattering amplitude.
The range of validity of Eq.~(\ref{eq:s}) for the Sudakov functions is
limited to not too small $\tilde{b_{l}}$ values.
Whenever $1/\tilde{b}_{l}$ is large relative to the hard (gluon)
scale $\xi _{l}Q$, the gluonic corrections are to be considered as
higher-order corrections to $T_{\rm H}$ and hence are not included in
the Sudakov factor but are absorbed into $T_{\rm H}$ (hierarchy of
scales).
For that reason, Li \cite{Li93} sets any Sudakov function
$s(\xi _{l},\tilde{b}_{l},Q)$
equal to zero whenever $\xi _{l}\leq \sqrt{2}/(Q\tilde{b}_{l})$.
Moreover, Li holds the Sudakov factor ${\rm e}^{-S_{j}}$ equal to
unity whenever it exceeds this value, which is the case in the small
$\tilde{b}_{l}$-region.
Actually, the full expression Eq.~(\ref{eq:S}) shows in this region a
small enhancement resulting from the interplay of the next-to-leading
logarithmic contributions to the Sudakov exponents and the integrals
over the anomalous dimensions.

\begin{figure}
\centering
\epsfig{figure=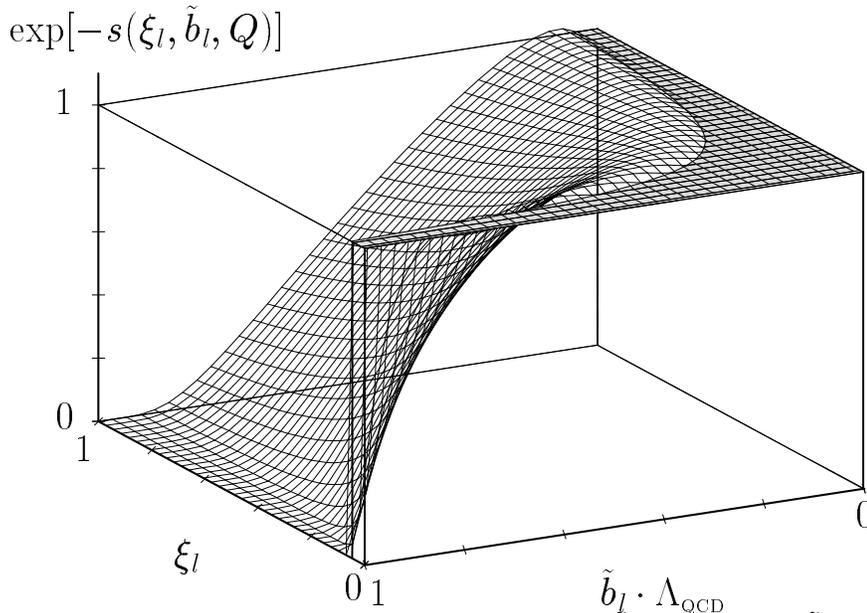,width=12cm,silent=}
\hspace{1.2cm}
\vspace{-0.4cm}
\caption[fig:Sud]
        {\tenrm
         The exponential of the Sudakov function
         $s(\xi _{l},\tilde{b}_{l},Q)$ vs $\xi _{l}$ and
         $\tilde{b}_{l}\Lambda _{{\rm QCD}}$ for
         $Q=30 \Lambda _{{\rm QCD}}$. In the hatched area the Sudakov
         function is set equal to zero according to Li's requirement
         \cite{Li93}.
\label{fig:sudakov}}
\end{figure}

The IR cutoffs $1/\tilde{b}_{l}$ in the Sudakov exponents mark the
interface between the purely non-perturbative soft momenta, which are
implicitly accounted for in the proton wave function, and the
contributions from soft gluons, incorporated in a perturbative way in
the Sudakov factors.
Obviously, the IR cutoff serves at the same time as the gliding
factorization scale $\mu _{\rm F}$ to be used in the evolution of the
wave function.
For that reason, Li \cite{Li93} as well as Sotiropoulos and
Sterman \cite{SS94} take
$\mu _{\rm F} = {\rm min}\{1/\tilde{b}_{l}\}$.
The ``MAX'' prescription (\ref{eq:MAX}), adopted in \cite{BJKBS95pro},
naturally complies with the choice of the evolution scale proposed
in \cite{Li93,SS94}.
In Fig.~\ref{fig:sudakov} we display the exponential of the Sudakov
function
$\exp[-s(\xi _{l},\tilde{b}_{l},Q)]$ for $Q=30\: \Lambda _{{\rm QCD}}$
by imposing Li's requirement \cite{Li93}:
$s(\xi _{l},\tilde{b}_{l},Q)=0$ whenever
$\xi _{l}\leq \sqrt{2}/Q\tilde{b}_{l}$.

%
%

\subsection{MODIFIED NUCLEON FORM FACTOR}
\label{subsec:modnucleonff}
The crucial element to incorporate radiative corrections in the
calculation of form factors within the {\MCS} is to retain the explicit
$k_{\perp}$-dependence in the convolution of the wave functions with
the hard scattering amplitude.
This new convolution formula can formally be derived by using the
methods described in detail by Botts and Sterman \cite{BS89}.
According to Li \cite{Li93}, the convolution formula for the proton
form factor can be written in the form
\begin{equation}
  G_{\rm M}(Q^{2})
=
  \frac{16}{3}
  \int_{0}^{1}[dx][dx']
  \int_{}^{}[d{}^{2}{\bf k}_{\perp}][d{}^{2}{\bf k}_{\perp}^{\prime}]
  \sum_{j=1}^{2}
  T_{{\rm H}_{j}}(x,x',{\bf k}_{\perp},{\bf k}'_{\perp},Q,\mu )
  Y_{j}(x,x',{\bf k}_{\perp},{\bf k}'_{\perp},\mu _{\rm F}) \; .
\label{eq:G_M}
\end{equation}
One recognizes that this formula appears as an intermediate step in
the derivation of the standard hard-scattering formula for the proton
form factor \cite{LB80}.
Note, however, that the notation adopted in the present exposition is
slightly different from that used by Li.
Making use of the symmetry properties of the proton wave function
under permutation, discussed previously, the contributions from the
diagrams involved in the calculation of the proton form factor at
Born level can be arranged into two reduced hard-scattering amplitudes
of the form
\begin{equation}
  T_{{\rm H}_{1}}
=
  \frac{2}{3} \,C_{\rm F} \,
  \frac{(4\pi \alpha _{\rm s}(\mu ))^{2}}
       {
        \left[
                (1-x_{1})(1-x_{1}')Q^{2}
              + ({\bf k}_{\perp 1} - {\bf k}_{\perp 1}')^{2}
        \right]
        \left[
                x_{2}x_{2}'Q^{2}
              + ({\bf k}_{\perp 2} - {\bf k}_{\perp 2}')^{2}
        \right]
       } \; ,
\label{eq:T_H_1}
\end{equation}
\begin{equation}
  T_{{\rm H}_{2}}
=
  \frac{2}{3} \,C_{\rm F} \,
  \frac{(4\pi \alpha _{\rm s}(\mu ))^{2}}
       {
        \left[
                x_{1}x_{1}'Q^{2}
              + ({\bf k}_{\perp 1} - {\bf k}_{\perp 1}')^{2}
        \right]
        \left[
                x_{2}x_{2}'Q^{2}
              + ({\bf k}_{\perp 2} - {\bf k}_{\perp 2}')
        \right]
       } \; .
\label{eq:T_H_2}
\end{equation}
In the hard scattering amplitudes only the $k_{\perp}$-dependence of
the gluon propagators is included, whereas that of the quark
propagators has been neglected.
There is, in principle, no problem to include in the calculation all
$\vec k_\perp$-dependence of $T_{\rm H}$.
In the case of the pion form factor this has been explicitly
demonstrated by Li \cite{Li93}.
He found an additional suppression of the final result of about
$10\%$.
However, in the proton case, such terms would lead to an 11-dimensional
integration which cannot be carried out to sufficient accuracy with
present day computers.\footnote{Explorative studies by using, for
instance, the $k_{\perp}$-dependence of only a certain subset of
quark propagators in addition to those of the gluon propagators (leading
to a 9-dimensional integration) show that the inclusion of such terms
yields an additional suppression of the perturbative result.
Since we are mainly interested in estimating rather the maximum
perturbative contribution, we dispense with such terms.}

The functions $Y_{j}$ in Eq.~(\ref{eq:G_M}) are short-hand notations
for linear combinations of products of the initial- and final-state
wave functions
$
 \Psi _{ijk} \; ,\Psi ^{\star\prime} _{i^{\prime}j^{\prime}k^{\prime}}
$,
weighted by $x_{i}$-dependent factors arising from the fermion
propagators, namely:
\begin{eqnarray}
  \hat{Y}_{1}
=
  \frac{1}{\overline{x}_{1}\overline{x}^{\prime}_{1}}
\Bigl\{  & \phantom{} &
\!\!\!\!\!\!\!
             4\hat{\Psi}^{\star\prime}_{123}\hat{\Psi}_{123}
           + 4\hat{\Psi}^{\star\prime}_{132}\hat{\Psi}_{132}
           + \hat{\Psi}^{\star\prime}_{231}\hat{\Psi}_{231}
           + \hat{\Psi}^{\star\prime}_{321}\hat{\Psi}_{321}
\nonumber \\
\!\!&  +  &\!\! 2\hat{\Psi}^{\star\prime}_{231}\hat{\Psi}_{132}
           + 2\hat{\Psi}^{\star\prime}_{132}\hat{\Psi}_{231}
           + 2\hat{\Psi}^{\star\prime}_{321}\hat{\Psi}_{123}
           + 2\hat{\Psi}^{\star\prime}_{123}\hat{\Psi}_{321}
  \Bigr\}\, ,
\label{eq:Y_1}
\end{eqnarray}
\begin{eqnarray}
  \hat{Y}_{2}
= & \phantom{} &
\!\!\!\!\!\!\!
  \frac{1}{2\overline{x}_{2}\overline{x}^{\prime}_{1}}
  \left\{
            3\hat{\Psi}^{\star\prime}_{132}\hat{\Psi}_{132}
         -  \hat{\Psi}^{\star\prime}_{231}\hat{\Psi}_{231}
         -  \hat{\Psi}^{\star\prime}_{231}\hat{\Psi}_{132}
         -  \hat{\Psi}^{\star\prime}_{132}\hat{\Psi}_{231}
  \right\}
\nonumber \\
& - & \frac{1}{\overline{x}_{3}\overline{x}^{\prime}_{1}}
  \left\{
           4\hat{\Psi}^{\star\prime}_{321}\hat{\Psi}_{321}
         +  \hat{\Psi}^{\star\prime}_{123}\hat{\Psi}_{123}
         + 2\hat{\Psi}^{\star\prime}_{321}\hat{\Psi}_{123}
         + 2\hat{\Psi}^{\star\prime}_{123}\hat{\Psi}_{321}
  \right\}
\label{eq:Y_2}
\end{eqnarray}
for the proton \cite{BJKBS95pro} and by
\begin{eqnarray}
  \hat{Y}_{1}^{n}
=
  \frac{1}{\overline{x}_{1}\overline{x}^{\prime}_{1}}
\Bigl\{  & \phantom{} &
\!\!\!\!\!\!\!
          -  2\hat{\Psi} ^{\star\prime}_{123}\hat{\Psi} _{123}
          -  2\hat{\Psi} ^{\star\prime}_{132}\hat{\Psi} _{132}
          +  \hat{\Psi} ^{\star\prime}_{231}\hat{\Psi} _{231}
          +  \hat{\Psi} ^{\star\prime}_{321}\hat{\Psi} _{321}
\nonumber \\
\!\!& - & \!\! \hat{\Psi} ^{\star\prime}_{231}\hat{\Psi} _{132}
          -  \hat{\Psi} ^{\star\prime}_{132}\hat{\Psi} _{231}
          -  \hat{\Psi} ^{\star\prime}_{321}\hat{\Psi} _{123}
          -  \hat{\Psi} ^{\star\prime}_{123}\hat{\Psi} _{321}
  \Bigr\}\, ,
\label{eq:Y_1^n}
\end{eqnarray}
\begin{eqnarray}
  \hat{Y}_{2}^{n}
= & \phantom{} &
\!\!\!\!\!\!\!
  \frac{1}{\overline{x}_{2}\overline{x}^{\prime}_{1}}
  \left\{
           \hat{\Psi} ^{\star\prime}_{231}\hat{\Psi} _{231}
         + \hat{\Psi} ^{\star\prime}_{231}\hat{\Psi} _{132}
         + \hat{\Psi} ^{\star\prime}_{132}\hat{\Psi} _{231}
  \right\}
\nonumber \\
& + & \frac{1}{\overline{x}_{3}\overline{x}^{\prime}_{1}}
  \left\{
            2\hat{\Psi} ^{\star\prime}_{321}\hat{\Psi} _{321}
         -  \hat{\Psi} ^{\star\prime}_{123}\hat{\Psi} _{123}
         +  \hat{\Psi} ^{\star\prime}_{321}\hat{\Psi} _{123}
         +  \hat{\Psi} ^{\star\prime}_{123}\hat{\Psi} _{321}
  \right\}
\label{eq:Y_2^n}
\end{eqnarray}
for the neutron \cite{BJKBS95neu}.
The subscripts on $\Psi$ refer to the order of momentum arguments,
for example
$
 \Psi _{123}(x,{\bf k}_{\perp})
=
 \Psi (x_{1},{\bf k}_{\perp 1};
 x_{2},{\bf k}_{\perp 2};
 x_{3},{\bf k}_{\perp 3})
$.
Note that, in general, the wave function depends on the factorization
scale $\mu _{\rm F}$.
We make the following convenient ansatz for the wave function:
\begin{equation}
  \Psi _{123}(x,{\bf k}_{\perp})
=
  \frac{1}{8\sqrt{N_{\rm c}!}}\,
  f_{\rm N}(\mu _{\rm F})
  \Phi _{N}(x,\mu _{\rm F}=1/\tilde{b}_{l})\,
  \Omega (x,{\bf k}_{\perp}) \; .
\label{eq:ansatz}
\end{equation}

The $k_{\perp}$-dependence of the wave function is contained in the
function $\Omega$ which is normalized according to
\begin{equation}
  \int_{}^{}[d{}^{2}{\bf k}_{\perp}]
  \Omega _{123}(x,{\bf k}_{\perp})
=
  1 \; .
\label{eq:omeganorm}
\end{equation}
Due to Eqs.~(\ref{eq:ndanorm}) and (\ref{eq:omeganorm}), $f_{\rm N}$ is
the value of the wave function at the origin of the configuration
space.
Recall that in contrast to the pion decay constant $f_{\pi}$, which
has zero anomalous dimension, $f_{\rm N}$ exhibits evolution behavior
driven by the leading anomalous dimension $\gamma _{0}$ according to
\begin{equation}
  f_{\rm N}(\mu _{\rm F})
=
  f_{\rm N}(\mu _{0})
  \left(
  \frac{\alpha _{\rm s}
       (\mu _{\rm F}^{2})}{\alpha _{\rm s}(\mu _{0}^{2})}
  \right)^{2/3\beta_{0}} \; .
\label{eq:f_N}
\end{equation}
At the starting point of evolution it has the value obtained from
QCD sum rules \cite{COZ89a,KS87}:
$
  f_{\rm N}(\mu _{0})
=
  (5.0\pm 0.3)\times 10^{-3}~{\rm GeV}{}^{2}
$.

In Eq.~(\ref{eq:ansatz}) $\Psi $ represents the soft part of the
proton wave function that results by removing the perturbative
part and absorbing it into the hard-scattering amplitude $T_{\rm H}
$.
The perturbative tail of the full wave function behaves as
$1/k_{\perp}^{4}$ for large $k_{\perp}$, whereas the soft part
vanishes as $1/k_{\perp}^{6}$ or faster.
The non-perturbative or intrinsic $k_{\perp}$-dependence of the soft
wave function, being related to confinement, is parameterized in our
analysis as a simple Gaussian distribution according to
\begin{equation}
  \Omega _{123}(x,{\bf k}_{\perp})
=
  (16\pi ^{2})^{2}
  \frac{a^{4}}{x_{1}x_{2}x_{3}}
  \exp
     \left [
            -a^{2} \sum_{i=1}^{3}{\bf k}_{\perp i}^{2}/x_{i}
     \right ] \; .
\label{eq:BLHM-Omega}
\end{equation}

This parameterization of the intrinsic $k_{\perp}$-dependence of the
wave function, which is due to Brodsky, Huang, and Lepage \cite{BHL83},
seems to be more favorable than the standard form of factorizing $x$-
and $k_{\perp}$-dependences.
At least for the case of the pion wave function, this has recently
been effected by Zhitnitsky \cite{Zhi94} on the basis of QCD sum
rules.
He found that a factorizing wave function is in conflict with some
general theoretical constraints with which any reasonable wave
function should comply.
Zhitnitsky's QCD sum-rule analysis of the pion wave function seems to
indicate that the $k_{\perp}$-distribution may also show a
double-hump structure (like the distribution over longitudinal
momenta in the CZ pion {\DA}), which means that small and large values
of $k_{\perp}$ are favored relative to intermediate values.
It is likely that the proton wave function may exhibit a similar
behavior, though this kind of analysis has yet to be done.

In Eq.~(\ref{eq:BLHM-Omega}) the parameter $a$ controls the root
mean square transverse momentum (r.m.s.),
$\langle k_{\perp}^{2}\rangle ^{1/2}$, and the r.m.s. transverse
radius of the proton valence Fock state.
From the known charge radius of the proton, we expect the r.m.s.
transverse momentum to be larger than about $250$~MeV.
The actual value of $\langle k_{\perp}^{2}\rangle ^{1/2}$ may be much
larger than $250$~MeV, even as large as 600~MeV.
Indeed, Sotiropoulos and Sterman \cite{SS94} show that application
of the {\MCS} to proton-proton elastic scattering leads to an
approximate $t^{-8}$-behavior of the differential cross section at
moderate $|t|$.
The behavior $d\sigma /dt \sim t^{-10}$, predicted by dimensional
counting, appears only at very large $|t|$.
At precisely which value of $|t|$ the transition from the $t^{-8}$ to
the $t^{-10}$ behavior occurs, depends on the transverse size of the
valence Fock state of the proton.
Since the ISR \cite{ISR79} and the FNAL \cite{FNAL77} data are rather
compatible with a $t^{-8}$-behavior of the differential cross
section, Sotiropoulos and Sterman conclude that the transverse size of
the proton is small, perhaps $\leq 0.3$ fm.
Correspondingly, the r.m.s. transverse momentum is then larger than
600~MeV.
It is worth noting that such a large value is supported by the findings
of the EMC group \cite{EMC80} in a study of the transverse momentum
distribution in semi-inclusive deep inelastic $\mu p$ scattering.
A phenomenologically successful approach to the {\SCS}, in which
baryons are viewed as bound states of a quark and an effective
diquark particle, also uses a value of this size for
$\langle k_{\perp}^{2}\rangle ^{1/2}$ \cite{Kro88,KPSS93,JKSS93}.
There is a second constraint on the wave function, namely the
probability for finding three valence quarks in the proton:
\begin{equation}
  P_{3{\rm q}}
=
  \frac{|f_{\rm N}|^{2}}{3}
  (\pi a)^{4}
  \int_{0}^{1}[dx]
  \frac{2\left(\Phi _{123}(x)\right)^{2} + \Phi _{132}(x)
  \Phi _{231}(x)}{x_{1}x_{2}x_{3}}
\leq 1 \; .
\label{eq:P3q}
\end{equation}
Similarly to Sotiropoulos and Sterman \cite{SS94}, we write the
valence quark component of the proton state with positive helicity
in the form
\begin{eqnarray}
  \vert P,+\!\rangle \;
=
  \frac{1}{\sqrt{N_{\rm c}!}}
  \int_{0}^{1}
  [dx]
  \int_{}^{}
  [d^{2}{\bf k}_{\perp}]
  \Bigl\{  & \phantom{} &
\!\!\!\!\!\!
            \Psi _{123}\,{\cal M}_{+-+}^{a_{1}a_{2}a_{3}} +
            \Psi _{213}\,{\cal M}_{-++}^{a_{1}a_{2}a_{3}}
\nonumber \\
& - &
            \Bigl(\Psi _{132}\, + \,
            \Psi _{231}\Bigr){\cal M}_{++-}^{a_{1}a_{2}a_{3}}
  \Bigr\}
  \epsilon _{a_{1}a_{2}a_{3}} \; ,
\label{eq:|P,+>}
\end{eqnarray}
where we assume the proton to be moving rapidly in the
$3$-direction.
Hence, the ratio of transverse to longitudinal momenta of
the quarks is small.
The 3-quark state with helicities
$
\lambda _{1}, \lambda _{2}, \lambda _{3}
$
and colors
$
a_{1}, a_{2}, a_{3}
$
is given by
\begin{equation}
  {\cal M}_{\lambda _{1}\lambda _{2}\lambda _{3}}^{a_{1}a_{2}a_{3}}
=
  \frac{1}{\sqrt{x_{1}x_{2}x_{3}}}
  \vert u_{a_{1}};x_{1},{\bf k}_{\perp 1},\lambda _{1}\rangle
  \vert u_{a_{2}};x_{2},{\bf k}_{\perp 2},\lambda _{2}\rangle
  \vert d_{a_{3}};x_{3},{\bf k}_{\perp 3},\lambda _{3}\rangle \; .
\end{equation}
Since the orbital angular momentum is assumed to be zero, the proton
helicity is the sum of the quark helicities.
The quark states are then normalized as follows
\begin{equation}
  \langle q_{a_{i}'};x_{i}',{\bf k}_{\perp i}',\lambda _{i}'
  \vert q_{a^{}_{i}};x^{}_{i},{\bf k}^{}_{\perp i},
  \lambda ^{}_{i}\rangle \;
=
  2x_{i}(2\pi )^{3}
  \delta _{a'_{i}a^{}_{i}}
  \delta _{\lambda _{i}'\lambda ^{} _{i}}
  \delta (x_{i}'-x^{}_{i})
  \delta ({\bf k}_{\perp i}'-{\bf k}^{}_{\perp i}) \; .
\end{equation}

In the numerical analysis of \cite{BJKBS95pro,BJKBS95neu}, to be
presented below, we make use of two different values of the r.m.s.
transverse momentum: (1) One option is to use that value which is
obtained by requiring $P_{3{\rm q}}=1$ for a given wave function.
[This corresponds to the minimum value of the r.m.s. transverse
momentum.]
(2) As another option for the r.m.s. transverse momentum, we
consider the rather large value of 600~MeV.
In the latter case, the probability for the valence quark Fock state
is much smaller than unity and depends on the structure of the
particular wave function.

In order to include the Sudakov corrections, it is advantageous to
re-express Eq.~(\ref{eq:G_M}) in terms of the variables ${\bv}_{i}$,
which are canonically conjugate to ${\bf k}_{\perp i}$ and span the
transverse configuration space.
Then
\begin{equation}
  G_{\rm M}(Q^{2})
=
  \frac{16}{3}
  \int_{0}^{1}[dx][dx']
  \int_{}^{}\frac{d{}^{2}{\bv}_{1}}{(4\pi )^{2}}
            \frac{d{}^{2}{\bv}_{2}}{(4\pi )^{2}}
  \sum_{j}^{}\, \hat{T}_{j}(x,x',{\bv},Q,\mu )
  \hat{Y}_{j}(x,x',{\bv},\mu _{\rm F}) \,
  {\rm e}^{-S_{j}} \; ,
\label{eq:G_M(b)}
\end{equation}
where the Fourier transform of a function
$f({\bf k}_{\perp})=f({\bf k}_{\perp 1},{\bf k}_{\perp 2})$
is defined by
\begin{equation}
  \hat{f}({\bv})
=
  \frac{1}{(2\pi )^{4}}
  \int_{}^{}d{}^{2}{\bf k}_{\perp 1}d{}^{2}{\bf k}_{\perp 2}
  {\rm exp}\{-i{\bv}_{1}\!\!\cdot {\bf k}_{\perp 1}
 - i{\bv}_{2}\!\cdot {\bf k}_{\perp 2}\}
  f({\kv}) \; .
\label{eq:Fourier}
\end{equation}

Since the hard scattering amplitudes depend only on the differences
of initial- and final-state transverse momenta, there are only two
independent Fourier-conjugate vectors
${\bv}_{1}\;(={\bv}_{1}')$
and
${\bv}_{2}\;(={\bv}_{2}')$.
They are, respectively, the transverse separation vectors between
quarks 1 and 3 and between quarks 2 and 3.
Accordingly, the transverse separation of quark 1 and quark 2 is
given by
\begin{equation}
  {\bv}_{3}
=
  {\bv}_{2} - {\bv}_{1} \; .
\label{eq:b123}
\end{equation}
[Note that Sotiropoulos and Sterman \cite{SS94} define the transverse
separations in a cyclic way which results in the interchange
${\bv}_{1} \longleftrightarrow -{\bv}_{2}$, as compared to the
definition used here.]

The fact that there are only two independent transverse separation
vectors is a consequence of the approximation made in the treatment
of the hard scattering amplitudes (given by Eqs.~(\ref{eq:T_H_1}) and
(\ref{eq:T_H_2})) which disregards the $k_{\perp}$-dependence of the
quark propagators.
This approximation is justified by the enormous technical
simplification it entails, given that the thereby introduced errors
are very small.
Then by virtue of rotational invariance of the system with respect
to the longitudinal axis, the form factor (\ref{eq:G_M(b)}) can be
expressed in terms of a seven-dimensional integral instead of an
eleven-dimensional one.
Physically, the relations
$
 {\bv}_{1}={\bv}_{1}'
$,
$
 {\bv}_{2}={\bv}_{2}'
$
mean that the physical probe (i.e., the photon) mediates only such
transitions from the initial- to the final-proton state, which have
the same transverse configurations of the quarks.

The Fourier-transformed hard scattering amplitudes appearing in
Eq.~(\ref{eq:G_M(b)}) read
\begin{equation}
  \hat{T}_{1}
=
  \frac{8}{3}\,C_{\rm F}\,
  \alpha _{\rm s}(t_{11}) \alpha _{\rm s}(t_{12})
  K_{0}
       \left(
             \sqrt{(1-x_{1})(1-x_{1}')}Qb_{1}
       \right)
  K_{0}\left(
             \sqrt{x_{2}x_{2}'}Qb_{2}
       \right) \; ,
\label{eq:FourierT_1}
\end{equation}
\begin{equation}
  \hat{T}_{2}
=
  \frac{8}{3}\,C_{\rm F}\,
  \alpha _{\rm s}(t_{21}) \alpha _{\rm s}(t_{22})
  K_{0}
       \left(
             \sqrt{x_{1}x_{1}'}Qb_{1}
       \right)
  K_{0}
       \left(
             \sqrt{x_{2}x_{2}'}Qb_{2}
       \right) \; ,
\label{eq:FourierT_2}
\end{equation}
where $K_{0}$ is the modified Bessel function of order 0, and
$b_{l}$ denotes the length of the corresponding vector.
We have chosen the renormalization scale in such a way that each
hard gluon carries its own individual momentum scale $t_{ji}$ which
in turn appears as the argument of the corresponding $\alpha _{\rm s}$.
The $t_{ji}$ are defined as the maximum scale of either the
longitudinal momentum or the inverse transverse separation,
associated with each of the gluons:
\begin{eqnarray}
& t_{11} &
=
  {\rm max} \left[
                  \sqrt{(1-x_{1})(1-x_{1}^{\prime})}Q, 1/b_{1}
            \right] \; ,
\nonumber \\
& t_{21} &
=
  {\rm max} \left[
                  \sqrt{x_{1}x_{1}^{\prime}}Q, 1/b_{1}
            \right] \; ,
\nonumber \\
& t_{12} &
=
  t_{22}
=
  {\rm max} \left[
                  \sqrt{x_{2}x_{2}^{\prime}}Q, 1/b_{2}
            \right] \; ,
\label{eq:t_ij}
\end{eqnarray}
One may think of other choices.
However, they are not expected to lead to significantly different
predictions for the form factor \cite{Li93}.

The quantities $\hat{Y}_{j}$ contain the same combinations
of initial- and final-state wave functions as those in
Eq.~(\ref{eq:Y_1}) and Eq.~(\ref{eq:Y_2}), the only difference being
that now the products
$
 \Psi ^{\star\prime}_{i^{\prime}j^{\prime}k^{\prime}}
 \Psi _{ijk}
$
are replaced by corresponding products of Fourier-transformed wave
functions:
$
 \hat{\Psi}^{\star\prime}_{i^{\prime}j^{\prime}k^{\prime}}
           (x',{\bv},\mu _{\rm F})
 \hat{\Psi}_{ijk}(x,{\bv},\mu _{\rm F})
$.
Using Eq.~(\ref{eq:ansatz}) and Eq.~(\ref{eq:BLHM-Omega}), the
Fourier transform of the wave function reads
\begin{equation}
  \hat{\Psi}_{123}(x,{\bv},\mu _{\rm F})
=
  \frac{1}{8\sqrt{N_{\rm c}!}}
  f_{\rm N}(\mu _{\rm F})
  \Phi _{123}(x,\mu _{\rm F})
  \hat{\Omega}_{123}(x,{\bv}) \; ,
\label{eq:FourierPsi}
\end{equation}
where the Fourier-transform of the $k_{\perp}$-dependent part is
given by
\begin{equation}
  \hat{\Omega}_{123}(x,{\bv})
=
  (4\pi )^{2}
  {\rm exp}
           \left\{
                  - \frac{1}{4a^{2}}
                  \Bigl[
                          x_{1}x_{3}b_{1}^{2} + x_{2}x_{3}b_{2}^{2}
                        + x_{1}x_{2}b_{3}^{2}
                  \Bigr]
           \right\} \; .
\label{eq:FourierOmega}
\end{equation}

The exponentials ${\rm e}^{-S_{j}}$ in Eq.~(\ref{eq:G_M(b)}) are the
Sudakov factors, discussed previously, which encapsulate the effects
of gluonic radiative corrections.
This makes it apparent that Eq.~(\ref{eq:G_M(b)}) is not simply
the Fourier transform of Eq.~(\ref{eq:G_M}) but an expression
comprising additional physical input.
Thus, with hindsight, Eq.~(\ref{eq:G_M(b)}) may be termed the
``modified convolution formula'' in which hard-gluon re-scattering
can still be isolated (factored out).

\subsubsection{Screening of $\alpha _{\rm s}$ singularities}
\label{susubsec:screen}
Using as argument in $\alpha _{\rm s}$ a renormalization scale, which
is independent of the longitudinal momentum fractions, may induce
large contributions from higher orders in the endpoint region.
Indeed, for the pion form factor this has been explicitly shown, at
least at next-to-leading order \cite{DR81}.
Surely such large higher-order contributions would render the
leading-order calculation useless.
For this reason, one may be tempted to use as a more appropriate
choice the renormalization scale
$\sqrt{x_{2}x^{\prime}_{2}}Q$,
for such a scale would eliminate the large logarithms arising from
the higher-order contributions.
Unfortunately, this is achieved at the expense that $\alpha _{\rm s}$
becomes singular in the endpoint regions and the convolution form
of the form factors diverges.
To render the form factors finite, additional external parameters,
like an effective gluon mass \cite{Cor82} or a cutoff prescription
have to be introduced, as we discussed before.
One of the crucial advantages of the {\MCS}, is that there is no need
for external regulators because the Sudakov factor may suppress the
singularities of the ``bare'' (one-loop) $\alpha _{\rm s}$ inherently.
Indeed, in the pion case, it was shown \cite{LS92} that the transverse
quark-anti-quark separation is tantamount to an IR regulator which
suffices to suppress all singularities arising from the soft region.

Concerning the proton form factor, we shall effect in the following
that the screening of the $\alpha _{\rm s}$-singularities by the
Sudakov factor depends sensitively on the choice of the IR
regularization prescription in transverse configuration space.

Ultimately, the suppression of the $\alpha _{\rm s}$-singularities
relies on the fact that whenever one of the $\alpha _{\rm s}$ couplings
tends to infinity (owing to the limit $t_{ji}\to \Lambda _{{\rm QCD}}$),
the Sudakov factor ${\rm e}^{-S_{j}}$ rapidly decreases to zero.
As it can be observed from Fig.~\ref{fig:sudakov}, this is not the
case in the region determined by
$\xi _{l}\le \sqrt{2}\Lambda _{{\rm QCD}}/Q$
and simultaneously
$\tilde{b}_{l}\Lambda _{{\rm QCD}}\to 1$,
where
$\exp[-s(\xi _{l},\tilde{b}_{l},Q)]$
is fixed to unity.
In the pion case this does not matter, since the other
$\exp[-s(1-\xi ,\tilde{b},Q)] \to 0$
faster than any power of
$\ln[1/(\tilde{b}\Lambda _{{\rm QCD}})]$
and, consequently, the Sudakov factor drops to zero ensuring IR
protection.
The treatment of the proton form factor is technically more subtle.
In that case,
${\rm e}^{-S_{j}}$
does not necessarily vanish fast enough to enforce the suppression of
the $\alpha _{\rm s}$-singularities.
This can be illustrated by the following string-like configurations:
$x_{1}< \sqrt{2}\Lambda _{{\rm QCD}}/Q$
and
${b}_{1}\Lambda _{{\rm QCD}}\to 1$
and
${b}_{2}\Lambda_{{\rm QCD}} \simeq {b}_{3}\Lambda_{{\rm QCD}}
\simeq 1/2$.
Using the ``L'' prescription, it is obvious that
$s(x_{1},\tilde{b}_{1},Q) = 0$
and the other two Sudakov functions are finite.
Therefore, there is no suppression of the
$\alpha _{\rm s}$-singularities owing to the limit $\tilde{b}_{1}
\Lambda _{{\rm QCD}} \rightarrow 1$.
Sufficient suppression of the $\alpha _{\rm s}$-singularities is
provided only if all three $\tilde{b}_{l}$ are coerced to be equal.
Allowing for the three $\tilde{b}_{l}$ to be different, the Sudakov
factor provides suppression only through the contributions of
the anomalous dimensions which are only logarithmic, and hence cannot
provide sufficient suppression.
Similar observations hold for the ``H-SS'' prescription.
Since the ``L'' and ``H-SS'' prescriptions allow for different
$\tilde{b}_{l}$ values in the Sudakov functions, the integrand in
(\ref{eq:G_M(b)}) has singularities behaving as
\begin{equation}
  \sim \ln^{-\kappa} \left(
                 \frac{1}{\tilde{b}_{l}\Lambda _{{\rm QCD}}}
                     \right)
\label{eq:sing}
\end{equation}
for $\tilde{b}_{l}\Lambda _{{\rm QCD}} \simeq 1$ and $x_{l}$ hold
fixed.
The maximum degree of divergence is given by
\begin{equation}
  \kappa
=
  \frac{1}{\beta _{0}}
                      \left(
                            \frac{4}{3} + 2\tilde{\gamma}_{{\rm max}}
                             - 2
                      \right)
                              + 1 \; ,
\label{eq:kappa}
\end{equation}
where the first term $4/3$ comes from the evolution of $f_{\rm N}$,
Eq.~(\ref{eq:f_N}), and the constant $\tilde{\gamma}_{{\rm max}}$
is related to the anomalous dimension driving the evolution behavior
of the proton {\DA}
(see Eq.~(\ref{eq:solnuc}) and Table~\ref{tab:sudBs}).
Here $\tilde{\gamma}_{{\rm max}}$ is the maximum value of the
$\{\tilde{\gamma}_{n}\}$ within a given polynomial order of the
eigenfunctions expansion of the {\DA}.
We reiterate that the $\tilde{\gamma}_{n}$ are positive fractional
numbers increasing with $n$.
Thus, the singular behavior of the integrand becomes worse as the
expansion in terms of eigenfunctions extends to higher and higher
orders.
The term $-2$ in Eq.~(\ref{eq:kappa}) stems from the integrations over
the anomalous dimensions in the Sudakov factor ${\rm e}^{-S_{j}}$ (see
Eq.~(\ref{eq:S})).
Finally, the term 1 originates from that
$\alpha _{\rm s}(t_{jk})$ which becomes singular
in Eq.~(\ref{eq:G_M(b)}) (cf. Eq.~(\ref{eq:t_ij})).
Which one of the $\alpha _{\rm s}$ couplings becomes actually singular
depends on the prescription imposed on the IR cutoff parameters
$\tilde{b}_{l}$.
The integral in Eq.~(\ref{eq:G_M(b)}) does not exist if
$\tilde{\gamma}_{{\rm max}}\geq \frac{1}{3}$.
As Table~\ref{tab:sudBs} reveals, this happens already for proton
{\DA}s which include (Appell) polynomials of order 1, i.e., for all
{\DA}s except for the asymptotic one.
Thus application of the ``L'' and ``H-SS'' prescriptions to choose the
IR cutoff parameters $\tilde{b}_{l}$ in the proton form factor entails
the {\MCS} to be invalid.
In view of these results, Li's numerical analysis of the proton form
factor \cite{Li93} is seriously flawed.
Nevertheless, his general approach is a decisive step towards a deeper
understanding of the electromagnetic baryon form factors.

%
\begin{table}
\caption[DAsgamma]
        {Expansion coefficients for selected nucleon {\DA}s, taken from
         \cite{COZ89a} and \cite{BS93,BS94}, and used in the {\MCS}. The
         notation of \cite{Ste89} is adopted.
         The $\{\tilde{\gamma}_{n}\}$ are related to those given in
         Table~\ref{tab:Bs} as follows:
         $
          \tilde{\gamma}_{n}^{{\rm Table~I}}
         =
          \frac{1}{\beta _{0}}(\frac{2}{3} + \tilde{\gamma}_{n})
         $, where $\frac{2}{3}=\gamma _{0}$ in this table.
         The associated r.m.s transverse momentum and the oscillator
         parameter for each model wave function, normalized via
         $P_{3{\rm q}}=1$, are shown.
\label{tab:sudBs}}
\vspace{0.5cm}
\begin{tabular}{lrrrrrr}
$\;n$ & $\tilde \gamma_n$
  & $B_n$(COZ$^{\mbox{\scriptsize up}}$) & $B_n$(COZ)
  & $B_n$(COZ$^{\mbox{\scriptsize opt}}$)
  & $B_n$({\rm Het}) & $B_n$({\rm low}) $\;$ \cr
\tableline
$\;1$ & 20/9  &  3.2185  &  3.6750  &  3.5268 &  3.4437
      &  4.1547 $\;$\cr
$\;2$ & 24/9  &  1.4562  &  1.4840  &  1.4000 &  1.5710
      &  1.4000 $\;$\cr
$\;3$ & 32/9  &  2.8300  &  2.8980  &  2.8736 &  4.5937
      &  3.3756 $\;$\cr
$\;4$ & 40/9  & -17.3400 & -6.6150  & -4.5227 & 29.3125
      & 26.1305 $\;$\cr
$\;5$ & 42/9  &  0.4700  &  1.0260  &  0.8002 & -0.1250
      & -0.5855 $\;$\cr
\tableline \tableline
\multicolumn{2}{l}{$\langle k_{\perp}^2\rangle ^{1/2}$
     [{\rm MeV}]} & 271 & 271 & 267 & 317 & 299 $\;$\cr
\multicolumn{2}{l}{$a$ [{\rm GeV}$^{-1}$]} & 0.9893 & 0.9939 & 1.0069 &
     0.8537 & 0.9217 $\;$\cr
\end{tabular}
\end{table}

A simple recipe to bypass the singular behavior of the integrand is
to ignore completely the evolution of the {\DA} or to ``freeze'' it at
any (arbitrary) value larger than $\Lambda _{{\rm QCD}}$.
Hyer \cite{Hye93} suggests, for example, to take for the factorization
scale
$\mu _{\rm F}={\rm max}\left\{1/b_{l}\right\}$.
In this case, $\tilde{\gamma}_{{\rm max}}$ appears in
Eq.~(\ref{eq:kappa}) only if all three $\tilde{b}_{l}$ tend to
$1/\Lambda _{{\rm QCD}}$ at once.
But then at least one of the
$\exp[-s(\xi _{l},\tilde{b}_{l},Q)]$
drops to $0$ faster than any power of
$\ln\left(1/\tilde{b}_{l}\Lambda _{{\rm QCD}}\right)$.
Apparently, Hyer's choice of the factorization scale avoids
singularities of the form of Eq.~(\ref{eq:sing}), but seems physically
implausible.
Since he gives numerical results for the proton form factor in
the time-like region only, his results cannot be compared directly
with those presented here.

%
\begin{figure}
\centering
\epsfig{figure=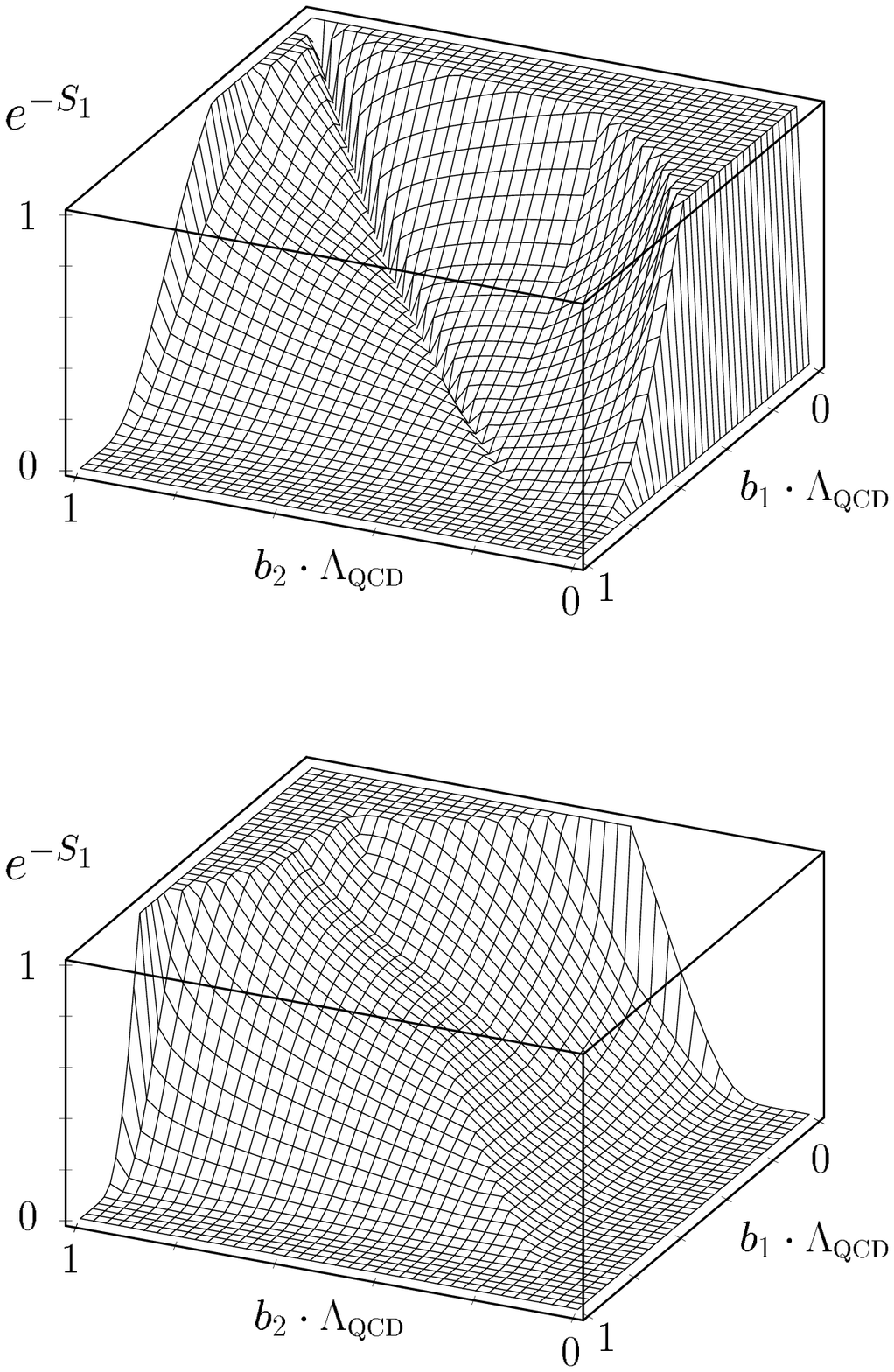,width=12cm,silent=}
\vspace{-0.4cm}
\caption[fig:Sudakovs]
        {\tenrm
         The Sudakov factor ${\rm e}^{-S_{1}}$ vs
         $b_{1}\Lambda _{{\rm QCD}}$ and
         $b_{2}\Lambda _{{\rm QCD}}$
         evaluated at $Q=30\Lambda _{{\rm QCD}}$, and
         $x_{1}=x_{1}^{\prime}=0.9$,
         $x_{2}=x_{3}=x_{2}^{\prime}=x_{3}^{\prime}=0.05$
         assuming a linear quark configuration
         (i.e., ${\bv}_{1}$ and ${\bv}_{2}$ are parallel to each other).
         The upper and lower figures correspond to the ``L''
         and ``MAX'' prescriptions, respectively.
\label{fig:bcplots}}
\end{figure}
%

Another option, and actually the one favored here, is to use a common
IR cutoff not only for the evolution of the wave function but also
in the Sudakov exponent \cite{BJKBS95pro,BJKBS95neu,Ste95}.
Indeed, for a common cutoff $\tilde{b}$, the Sudakov factors always
compensate the $\alpha _{\rm s}$-singularities, i.e., if, for a given
$l$, it happens that one Sudakov function is in the dangerous region
of phase space,
$\xi _{l}<\sqrt{2}\Lambda _{{\rm QCD}}/Q$,
$\tilde{b}\Lambda _{{\rm QCD}} \to 1$,
at least one of the other two Sudakov functions lies in the region
$\xi _{l^{\prime}}>\sqrt{2}\Lambda _{{\rm QCD}}/Q$,
$\tilde{b}\Lambda _{{\rm QCD}} \to 1$ ($l^{\prime}\neq l$) and
therefore provides sufficient suppression, as outlined above.
In particular, we favor $\tilde{b}={\rm max}\{b_{l}\}$ as the optimum
choice (``MAX'' prescription), since it does not only lead to a
regular integral but also to a non-singular {\it integrand} of the form
factor as well.
The Sudakov factor ${\rm e}^{-S_{1}}$ subject to the ``L'' and
``MAX'' prescriptions is plotted for a specific quark configuration
in Fig.~\ref{fig:bcplots}.
This figure makes it apparent that the Sudakov factor in connection
with the ``MAX'' prescription is unencumbered by singularities in the
soft regions.
As a consequence of the regularizing power of the ``MAX'' prescription,
the perturbative contribution to the proton form factor
Eq.~(\ref{eq:G_M(b)}) saturates, i.e., the results become insensitive
to the inclusion of the contributions of the soft regions.
A saturation as strong as possible should be regarded as a prerequisite
for the self-consistency of the perturbative contribution.

\begin{figure}
\begin{picture}(0,325)
  \put(90,10){\psboxscaled{1000}{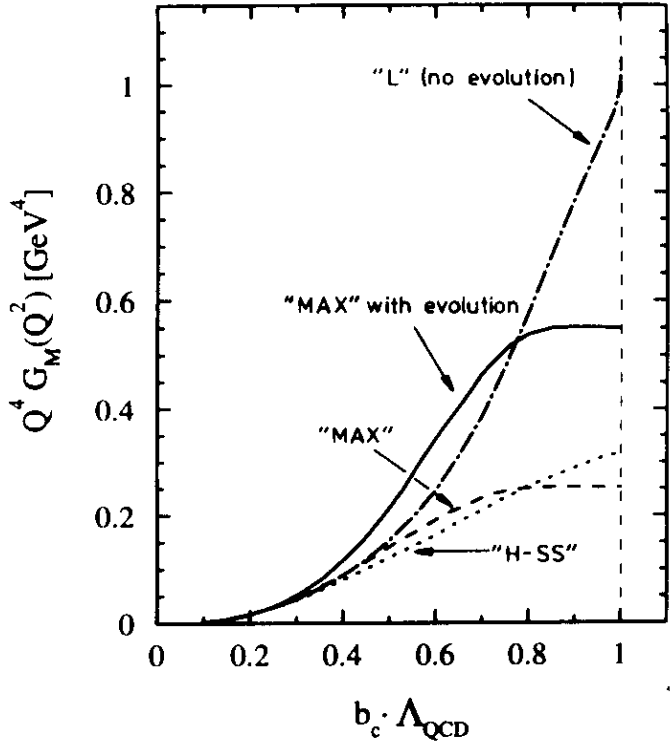}}
\end{picture}
\caption[fig:saturation]
        {\tenrm
         The proton magnetic form factor as a function of
         $b_{c}\Lambda _{{\rm QCD}}$. The curves shown are
         obtained at $Q=30\Lambda _{{\rm QCD}}$ for the COZ {\DA}.
         The solid line corresponds to the ``MAX'' prescription
         including evolution. The dotted (dashed, dashed-dotted) line
         represents results using the ``H-SS'' (``MAX'', ``L'')
         prescription, by ignoring the intrinsic $k_{\perp}$-dependence,
         and also evolution.
\label{fig:G_M(b_c)}}
\end{figure}

To demonstrate the amount of saturation, we calculate the proton form
factor through Eq.~(\ref{eq:G_M(b)}), employing a cutoff procedure to
the $b_{l}$-integrations at a maximum value $b_{c}$.
In Fig.~\ref{fig:G_M(b_c)} the dependence of $G_{\rm M}^{\rm p}$ on
$b_{c}$ for the three choices, labeled ``L'', ``H-SS'', and ``MAX''
is shown, using, for reasons of an easier comparison with previous
works, the COZ {\DA} and ignoring evolution.\footnote{Evolution has
been dispensed with to avoid a concomitant singularity in
$Q^{4}G_{\rm M}^{\rm p}$ as $b_{c}\Lambda _{{\rm QCD}}\to 1$
when imposing the ``L'' and ``H-SS'' prescriptions.}

%
%

As one sees from this figure, the ``MAX'' prescription empowers
saturation, i.e., the soft region
$b_{c}\Lambda _{{\rm QCD}}>0.7$
does not contribute to the form factor substantially.
In fact, already $50\,\%$ of the result are obtained from the regions
with $b_{c}\Lambda _{{\rm QCD}}<0.48$, while $\alpha _{\rm s}$
increases to a value of $0.95$ at
$b_{c}\Lambda _{{\rm QCD}} \approx 0.48$.
This indicates that a sizeable fraction of the contributions to the
form factor is indeed accumulated in the perturbative region.

Unfortunately, this saturation is achieved at the expense of
a rather strong damping of the perturbative contribution to the
proton (and neutron) form factors.
Using the other two prescriptions (``L'' and ``H-SS'') and
ignoring evolution, larger results for $G_{\rm M}^{\rm p}$, and
$G_{\rm M}^{\rm n}$ can be obtained, but there is no indication for
saturation:
The additional contributions to the form factor gained this way are
accumulated exclusively in the soft regions, i.e., for values of
$b_{c}\Lambda _{{\rm QCD}}$ near $1$.
Hence, this would be a rather deceptive improvement since the
criticism of Isgur and Llewellyn-Smith \cite{ILS84} and of
Radyushkin \cite{Rad84,Rad90} would again apply.

These findings are in evident contradiction to Li's results
(figure 5 in \cite{Li93}) for which an acceptable saturation was
claimed.
On the other hand, the saturation behavior of the proton form factor
calculated by Hyer \cite{Hye93} in the time-like region is confirmed.
A saturating behavior of the form factor should be regarded as a
stringent test for the self-consistent applicability of pQCD.
Therefore, calculations, which accumulate large contributions from soft
regions (large $b_{c}$), can hardly be considered as theoretically
legitimate, even if they seem to be phenomenologically successful.

The role of the evolution effect subject to the ``MAX'' prescription
is also effected in Fig.~\ref{fig:G_M(b_c)}.
It is evident from this figure that the effect of evolution is large,
though finite, owing to the strong suppression provided by the
Sudakov factor (recall that the factorization scale is
$\mu _{\rm F}=1/\tilde{b}$).
The significant feature of the evolution effect is that it tends to
neutralize the influence of the IR cutoff.
Thus one may trade larger values of the proton (neutron) form factor
for a slightly worse saturation.

One may criticize the use of the one-loop formula for $\alpha _{\rm s}$
in the soft region.
The ``true'' $\alpha _{\rm s}$ coupling may likely differ from that in
the soft region.
However, the saturation behavior due to the ``MAX'' prescription
(see Fig.~\ref{fig:G_M(b_c)}) reveals that there is practically no
contribution to the form factor from soft regions
(say for $\tilde b\Lambda_{QCD}\geq 0.8$).
The Sudakov factor together with the intrinsic transverse-momentum
dependence of the wave function suppresses this region completely.
Therefore, it is irrelevant what description of $\alpha _{\rm s}$ one is
using in that region, and hence more involved $\alpha _{\rm s}$
parameterizations can be avoided.

We now comment on the connection between the {\MCS}, employed here,
and approaches (e.g., \cite{JSL87}) which make use of an effective
gluon mass \cite{Cor82} to regularize the running coupling constant
and gluon propagators at small values of $Q^{2}$.
The purpose of the following discussion is to show the utility of the
$k_{\perp}$ (or transverse configuration space) representation in
compensating the divergent contributions of the soft region of phase
space.
According to Eq.~(\ref{eq:G_M}) or equivalently Eq.~(\ref{eq:G_M(b)}),
all values of ${\bv}$ have been integrated out, in particular those
corresponding to large distances, i.e., such of order
$1/\Lambda _{{\rm QCD}}$, which are not governed by perturbative QCD.
As discussed above, once
$Q^{2} \sim \Lambda _{{\rm QCD}}^{2}$,
the external probe (the virtual photon) no longer ``sees'' the
substructure of the nucleon, it rather sees the ``diffuse'' nucleon
as a whole.
This means that the strength of the dynamical coupling constant
ceases to increase and ``freezes'' at some scale
$Q^{2}\leq M^{2}=4m_{\rm g}^{2}$ (recall Eq.~(\ref{eq:modalphas})),
below which the long-range forces become saturated \cite{Ste89}.
The scale $M^{2}$ can be fixed by the requirement that it should lead
to the experimentally observed form factors.
Then the leading-order coupling constant is replaced by the modified
expression \cite{Cor82,PP80} given by Eq.~(\ref{eq:modalphas}),
where the arguments of $\alpha _{\rm s}$ are
$
 Q_{\xi}^{2}
=
 Q^{2}\xi _{l}\xi _{l}^{\prime}
$
or
$
 Q^{2}\xi _{l}\bar{\xi}_{l}^{\prime}
$
with $\bar{\xi}_{l}=(1-\xi _{l})$.
In this case, the IR-protected $\alpha _{\rm s}$ remains small enough
that perturbation theory still applies.
Furthermore, when the longitudinal momentum fractions
$\xi _{l}$, $\bar{\xi}_{l}$ tend to zero, the gluon virtuality
remains finite and equal to $M^{2}$, and well above the Landau ghost
pole.
On the other hand, when $Q^{2}$ becomes large enough, the form factor
becomes insensitive to the large distances or small momenta,
characterized by $M$.
Thus, Eq.~(\ref{eq:modalphas}) simulates the effect of the Sudakov
factor in suppressing contributions from large values of $|{\bv}|$.
However, in contrast to the Sudakov damping factor, which has
universal character, the IR-cutoff parameter $M$ has to be adjusted
to the external momentum.
This becomes evident from Table~\ref{tab:simul}, where the results
for the Sudakov-suppressed magnetic proton form factor are simulated
in the range of momentum transfers 4 to 30~GeV${}^{2}$, using again
as reference model the COZ {\DA}.
Evolution is taken into account via the factor
$
 \left(
       \frac{\alpha _{\rm s}\big(Q_{\xi}^{2} + M^{2} \big)}
            {\alpha _{\rm s}\big(\mu _{0}^{2} + M^{2} \big) }
 \right)^{\gamma _{n}/\beta _{0}},
$
where $\mu _{0}=1$ GeV${}$.
Indeed, while the form factor begins to scale at
$Q^{2} \approx$ 14 GeV${}^{2}$,
the values of the {\it ad hoc} parameter $M$ still increase with
$Q^{2}$, in order to provide sufficient IR-protection.
Such a $Q^{2}$ behavior of $M$ can hardly be associated with the
{\RG} controlled evolution of a dynamical gluon mass with a definite
anomalous dimension \cite{Cor82}.

%
\begin{table}
\caption[simmass]
                {Simulation of Sudakov suppression and intrinsic
                 transverse momentum by an IR-cutoff parameter $M$ in
                 the running coupling constant $\alpha _{\rm s}$ at
                 the one-loop order.
                 The values of $Q^{4}G_{\rm M}^{\rm p}$ are calculated
with the
                 COZ {\DA} and the r.m.s. transverse momentum
                 $\langle k_{\perp}^2 \rangle^{1/2} = 271$~MeV.
\label{tab:simul}}
\vspace{0.5cm}
\begin{tabular}{cccc}
 $Q^2$ [GeV$^2$] & $Q^4\,G_M^p$ [GeV$^4$] & $M$ [MeV] &
 $\alpha {\rm _s}(g^2)$ \cr \tableline

    4.0 &  0.3593 &  330.9 & 1.147  \cr
    6.0 &  0.3823 &  343.8 & 1.079  \cr
    8.0 &  0.3962 &  353.4 & 1.035  \cr
   10.0 &  0.4042 &  361.6 & 1.001  \cr
   12.0 &  0.4094 &  368.5 & 0.974  \cr
   14.0 &  0.4134 &  374.6 & 0.953  \cr
   16.0 &  0.4164 &  380.0 & 0.934  \cr
   18.0 &  0.4186 &  384.9 & 0.919  \cr
   20.0 &  0.4195 &  389.6 & 0.904  \cr
   22.0 &  0.4197 &  394.1 & 0.891  \cr
   24.0 &  0.4193 &  398.4 & 0.879  \cr
   26.0 &  0.4185 &  402.5 & 0.868  \cr
   28.0 &  0.4176 &  406.4 & 0.857  \cr
   30.0 &  0.4165 &  410.2 & 0.848  \cr
%
\end{tabular}
\end{table}

This exercise demonstrates that IR regularization by such techniques
is conceptually inferior to the inclusion of the explicit dependence
of $k_{\perp}$ within the {\MCS}.
Perhaps the most significant feature of the latter approach is that it
remains valid in any regime of momentum transfer without recourse to
{\it ad hoc} assumptions.

\subsubsection{Numerical form-factor analysis}
\label{subsubsec:numan}
\begin{figure}
\begin{picture}(0,360)
  \put(55,10){\psboxscaled{700}{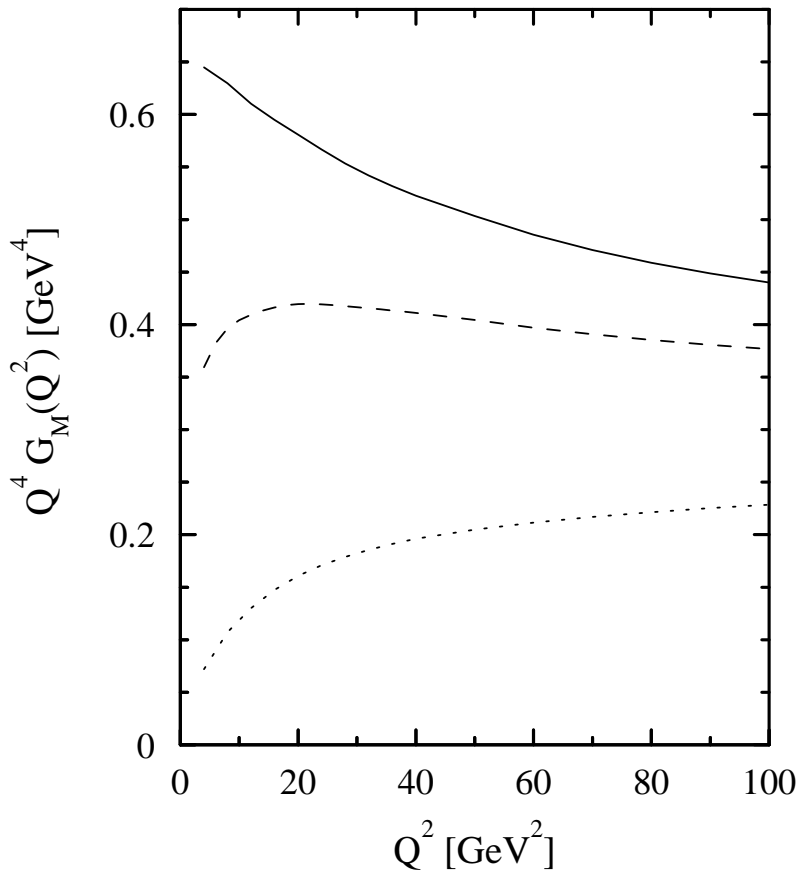}}
\end{picture}
\caption[fig:intrinsic]
        {\tenrm
         The influence of the intrinsic transverse momentum on the
         proton magnetic form factor within the {\MCS}. The curves
         shown are obtained for the COZ {\DA} by imposing the ``MAX''
         prescription, and including evolution. The solid line
         represents the result without $k_{\perp}$-dependence, whereas
         the dashed and dotted lines are obtained with
         $\langle k^{2}_{\perp}\rangle ^{1/2}=271$~MeV and $600$~MeV,
respectively.
\label{fig:G_M(Q^2)}}
\end{figure}

In this section we present numerical results for the proton and the
neutron form factors obtained in \cite{BJKBS95pro,BJKBS95neu,Ste95}.
In these calculations the ``MAX'' prescription is employed, with
evolution included, using $\Lambda _{{\rm QCD}}=$180~MeV and
$\mu _{0}=1$~GeV.
Before proceeding with the presentation of the results, let us first
investigate the effect of including the intrinsic transverse
momentum in form-factor calculations.
The $k_{\perp}$-dependence of the nucleon wave function effectively
introduces a confinement scale in the formalism, the importance of
which may be appreciated by looking at Fig.~\ref{fig:G_M(Q^2)}.

%
%

\begin{figure}
\begin{picture}(0,360)
  \put(55,10){\psboxscaled{700}{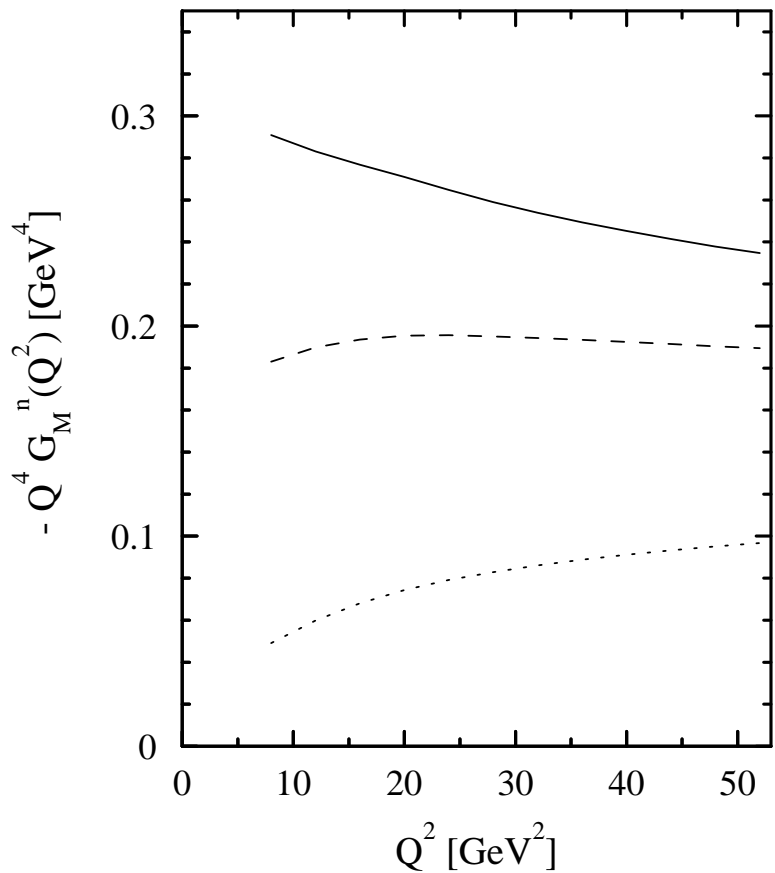}}
\end{picture}
\caption[fig:intrinsicneu]
        {\tenrm
         The influence of the intrinsic transverse momentum on the
         neutron magnetic form factor within the {\MCS}.
         The curves shown are obtained for the COZ {\DA} by imposing
         the ``MAX'' prescription, and including evolution.
         The solid line represents the result without
         $k_{\perp}$-dependence, whereas the dashed and dotted lines
         are obtained with
         $\langle k^{2}_{\perp}\rangle ^{1/2}=271$~MeV
         and $600$~MeV, respectively.
\label{fig:G_M^n(Q^2)}}
\end{figure}

This figure shows results, obtained for the COZ {\DA} without and with
$k_{\perp}$-dependence, using two different values of
$\langle k^{2}_{\perp}\rangle ^{1/2}$.
To describe the intrinsic $k_{\perp}$-dependence, one can use
Eq.~(\ref{eq:BLHM-Omega}) or, after Fourier-transforming to the
transverse configuration space, Eq.~(\ref{eq:FourierOmega}).
Notice that in Li's approach the Gaussian in
Eq.~(\ref{eq:FourierOmega}) has been replaced by unity.
The oscillator parameter $a$ is determined in such a way that either
the normalization of the wave function $P_{3{\rm q}}$ is unity
(resulting into $\langle k^{2}_{\perp}\rangle ^{1/2}=271$~MeV for the
COZ {\DA}), or by inputting the value of the r.m.s. transverse momentum.
In the second case, a value of $600$~MeV is used, which implies
$P_{3{\rm q}}=0.042$.
As can be seen from this and the analogous figure for the neutron
(Fig.~\ref{fig:G_M^n(Q^2)}), the predictions for the form factor are
quite different for the three considered cases.

%
%

In fact, the intrinsic $k_{\perp}$-dependence of the nucleon wave
function leads to further suppression of the perturbative contribution,
which becomes substantial if the r.m.s. transverse momentum is large.
On the other hand, this suppression is accompanied by an increasing
amount of saturation because the Gaussian distribution,
Eq.~(\ref{eq:FourierOmega}), also provides suppression; predominantly
of contributions from the soft regions, {\it viz.}, the large
$b$-regions.
In contrast to the Sudakov factor, however, this suppression is
$Q$-independent (there is no evolution).
The interplay of the two effects, Sudakov suppression and intrinsic
transverse momentum, leads to a different $Q$-behavior of the form
factor, depending on the value of the r.m.s transverse momentum, as
can be seen from Figs.~\ref{fig:G_M(Q^2)},~\ref{fig:G_M^n(Q^2)}.
The $Q$-dependence beyond $10$~GeV${}^{2}$ is rather weak, being
approximately compatible with dimensional counting (modulo {\RG}
generated logarithmic corrections).
For very large values of $Q^{2}$, say, beyond $1000$~GeV${}^{2}$, the
three curves have already approached each other within $10\%$ accuracy.
This happens when the Sudakov factor dominates the Gaussian
$k_{\perp}$-distribution and selects those configurations with
small inter-quark separations.
In this region, which one may consider as the pure perturbative
region, the results for the form factor are independent of the
confinement scale introduced by the r.m.s. transverse momentum.

The penalty of the additional suppression of the perturbative
contribution caused by Eq.~(\ref{eq:FourierOmega}) is
mitigated by the advantage of promoting the perturbative contribution
to become self-consistent by the {\it in situ} IR protection due to the
Sudakov form factor.
This is clearly indicated in the enhanced amount of saturation with
increasing r.m.s. transverse momentum.
Adapting the criterion of self-consistency, originally suggested by
Li and Sterman \cite{LS92} for the pion case, namely that $50\%$ of
the results should accumulate at moderate values of the coupling
constant, say, $\alpha _{\rm s}^{2}\leq 0.5$, we find for the nucleon
form factors self-consistency for $Q^{2}\approx 7$~GeV${}^{2}$ (still
referring to the COZ {\DA}).

\begin{figure}
\begin{picture}(0,330)
  \put(55,10){\psboxscaled{1000}{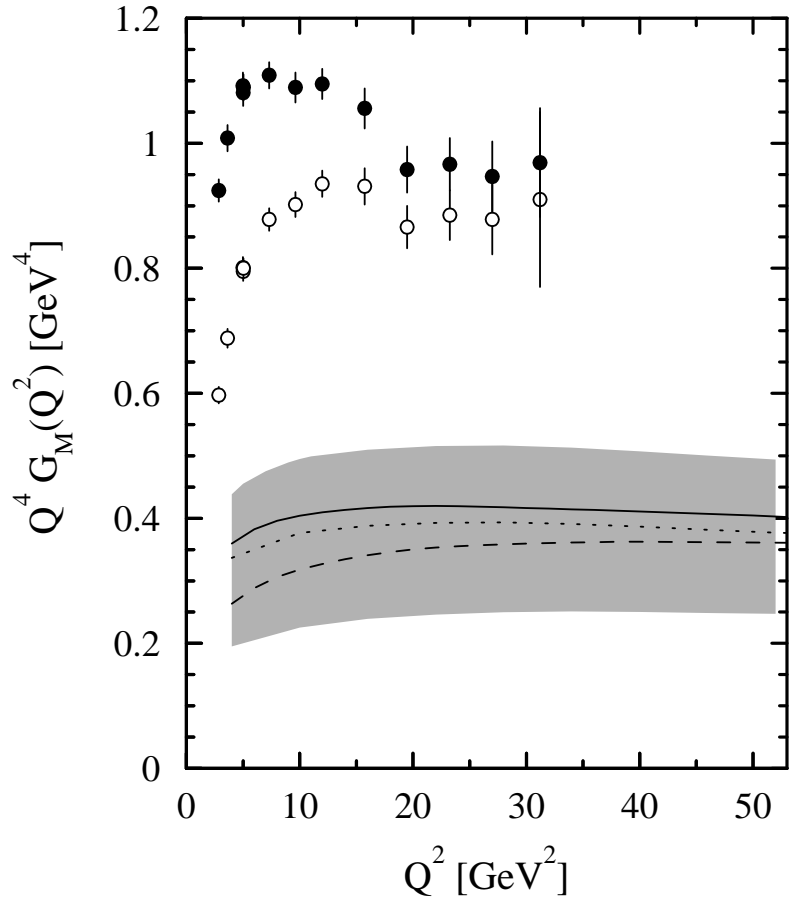}}
\end{picture}
\caption[fig:proband]
        {\tenrm
         The proton magnetic form factor vs $Q^{2}$. Data are taken
         from \cite{Roc82,SLAC86}. The $G_{\rm M}^{\rm p}$ data are
         represented by black dots, whereas those for $F_{1}^{\rm p}$
         are indicated by open circles. The theoretical results are
         obtained using the ``MAX'' prescription including evolution and
         normalizing the wave functions to unity. The shadowed strip
         indicates the range of predictions derived from the set of
         {\DA}s determined in \cite{BS93,BS94} in the context of QCD sum
         rules (see text). The solid (dashed, dotted) line corresponds
         to the COZ (heterotic, optimized COZ) {\DA}.
\label{fig:strip}}
\end{figure}

\begin{figure}
\begin{picture}(0,360)
  \put(55,10){\psboxscaled{700}{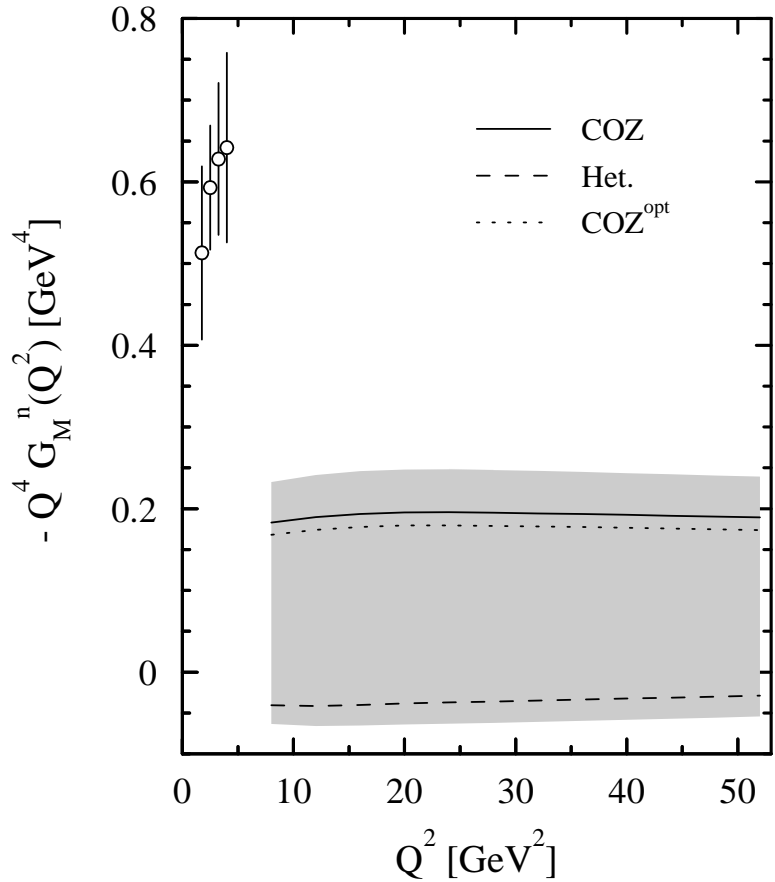}}
\end{picture}
\caption[fig:nutsud]
         {\tenrm
          The neutron magnetic form factor vs $Q^{2}$. The theoretical
          results are obtained using the ``MAX'' prescription including
          evolution and normalizing the wave functions to unity. The
          shadowed strip indicates the range of predictions derived from
          the set of {\DA}s determined in \cite{BS93} in the context of
          QCD sum rules (see text). The solid (dashed, dotted) line
          corresponds to the COZ (heterotic, optimized COZ) {\DA}. The
          data are taken from \cite{Roc82}.
\label{fig:stripn}}
\end{figure}

The final results are shown in Fig.~\ref{fig:strip} (proton) and in
Fig.~\ref{fig:stripn} (neutron) for different nucleon {\DA}s.

%
%

%
%

\begin{figure}
\begin{picture}(0,280)
  \put(55,10){\psboxscaled{700}{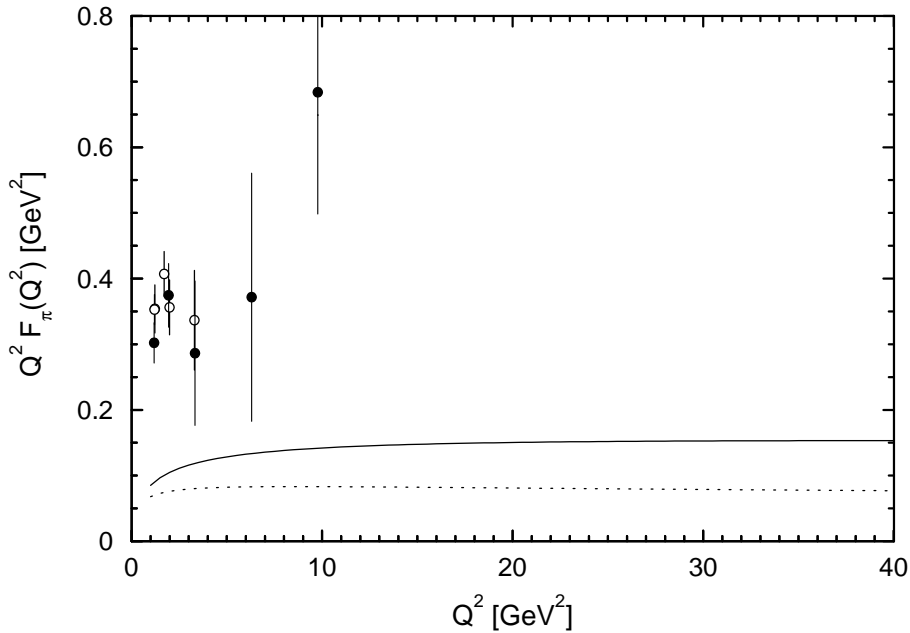}}
\end{picture}
\caption[fig:Bebec]
         {\tenrm
          The pion form factor in the space-like region including Sudakov
          effects and those due to the intrinsic $k_{\perp}$-dependence
          of the pion wave function \cite{JK93}, evaluated for
          $\Lambda_{{\rm QCD}}=200$~MeV. The solid line shows the result
          for the Chernyak-Zhitnitsky amplitude
          $\Phi _{CZ}=30x_{1}x_{2}(x_{1}-x_{2})^{2}$,
          and the dotted line that for the asymptotic solution
          to the evolution equation, $\Phi _{\rm as}=6x_{1}x_{2}$.
          The data are taken from \cite{Beb76}.
\label{fig:pionsud}}
\end{figure}

Though we have not considered the pion form factor in the {\MCS},
we find it instructive to include it in Fig.~\ref{fig:pionsud} which
shows the (corrected) calculation by Jakob and Kroll \cite{JK93}.

%
%

A set of 45 nucleon {\DA}s \cite{BS93,BS94} have been investigated,
{\DA}s which respect the QCD sum-rule constraints \cite{COZ89a,KS87}
with comparable $\chi ^{2}$ values.
The results for the various {\DA}s -- or more precisely wave functions,
since their intrinsic transverse-momentum dependence is taken into
account -- obtained under the ``MAX'' prescription with evolution
included, form the shaded area shown in Figs.~\ref{fig:strip},
\ref{fig:stripn}.
Note that all wave functions are normalized to unity and that the
corresponding r.m.s. transverse momenta vary between $267$~MeV and
$317$~MeV (see Table~\ref{tab:sudBs}).

The theoretical form-factor predictions span a ``band'' congruent
to the ``orbit'' of solutions found previously in \cite{BS93,BS94} and
discussed already in detail in this report.
The upper bound of the ``band'' corresponds to the {\DA}
COZ${}^{{\rm up}}$, which yields the maximum value of the
form-factor ratio
$|G_{\rm M}^{\rm n}|/G_{\rm M}^{\rm p}=0.4881$ in the {\SCS}.
The lower limit of the ``band'' is obtained using the {\DA} ``low''
(sample 8 in \cite{BS94}) (cf. Table~\ref{tab:Bs}) with
$|G_{\rm M}^{\rm n}|/G_{\rm M}^{\rm p}=0.175$.
Explicitly shown are the results for the COZ {\DA}, its optimized
version COZ${}^{{\rm opt}}$, and the heterotic {\DA} \cite{SB93nuc}.
We note that the differences among these curves practically
disappear already at about $Q^{2}=80$~GeV${}^{2}$, despite the fact
that these amplitudes have distinct geometrical characteristics.
This behavior effects once more that momentum evolution at larger
$Q^{2}$ values is enough to wash out shape peculiarities due to the
truncation of the eigenfunctions series.

Because the true valence Fock state probability is likely much
smaller, or invariably the r.m.s. transverse momentum larger than a
value of order $300$~MeV, the ``band'' describes rather {\it maximal}
expectations for the (leading-order) perturbative contributions to
the form factor; at least for proton wave functions of the type we
considered.
Comparison with the experimental data reveals that the theoretical
predictions amount, at best, to approximately $50\%$ of the measured
values.
This is the benchmark against which we have to discern novelties and
aberrations.

Concerning the proton form factor, we note that, since we are
calculating only the helicity-conserving part of the current
matrix element, it is not obvious whether we should compare the
theoretical predictions with the data for the Sachs form factor
$G_{\rm M}^{\rm p}$ or with those for the Dirac form factor
$F_{1}^{\rm p}$.
Therefore, we exhibit in Fig.~\ref{fig:strip} both sets of
data \cite{Roc82} for comparison.
However, since the two sets of data differ by only $10\%$, our
conclusions concerning the smallness of the theoretical results
remain unaffected.

The results for the neutron magnetic form factor are shown in
Fig.~\ref{fig:stripn}.
Unfortunately, there is no form-factor data available in the region
where the perturbative contribution is self-consistent.
Yet the trend of the low $Q^2$ data \cite{Roc82} seems to indicate
that the size of the perturbative contribution is rather small.
Measurements of the neutron magnetic form factor beyond 5~GeV${}^{2}$
are extremely important in order to check the validity of the
theoretical predictions.

One place to test these results beyond 5 GeV$^2$ is in the data
\cite{Pla90} for the differential cross sections for elastic
electron-proton and electron-neutron scattering,
$\sigma _{p}$ and $\sigma _{n}$, respectively.
For small scattering angles, where the terms
$\propto\tan^{2}(\theta/2)$ can be neglected, and for large $Q^{2}$,
the ratio
$
 \sigma _{n}/\sigma _{p}
$
becomes approximately proportional to the square of the ratio of the
neutron to the proton magnetic form factor, provided the electric
form factors are negligible.
For the electric form factor of the proton this assumption has
recently been verified experimentally \cite{Pla90}.
Its neutron counterpart is experimentally still unknown beyond
5~GeV$^{2}$, but many phenomenological models currently in use
predict a small $G_{\rm E}^{\rm n}$.
We note that the low $Q^2$ data on $G_{\rm E}^{\rm n}$ \cite{Roc82} are
compatible with zero.

\begin{figure}
\begin{picture}(0,360)
  \put(55,10){\psboxscaled{700}{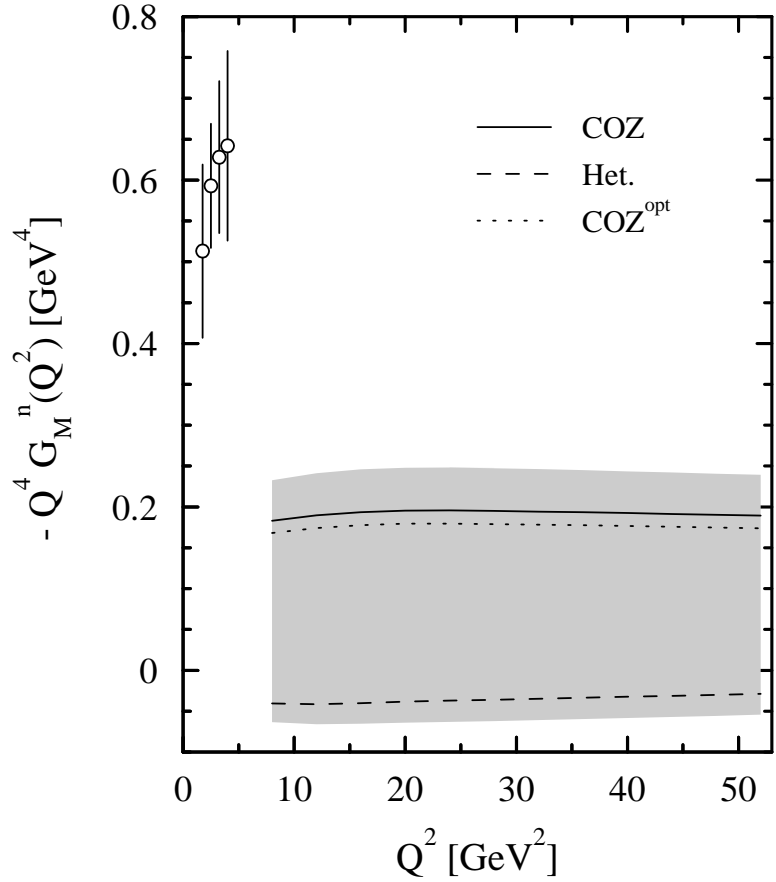}}
\end{picture}
\caption[fig:sigma]
         {\tenrm
          The ratio $\sigma _{n} / \sigma _{p}$ of the differential
          elastic electron-neutron to electron-proton cross section vs
          $Q^{2}$ at scattering angles of $10^{\circ}$. The shaded area
          and the model {\DA}s for the nucleon correspond to those shown
          in Figs.~\ref{fig:strip} and \ref{fig:stripn}. The data are
          taken from \cite{Pla90}.
\label{fig:stripsigma}}
\end{figure}

Combining the calculations for the proton with those for the neutron,
we can extract theoretical predictions for
$
 \sigma _{n}/\sigma _{p}
$,
using the same set of nucleon model {\DA}s as before.
The results are shown in Fig.~\ref{fig:stripsigma} (shaded area) in
comparison with available data \cite{Pla90}.
From this figure we see that the measured values of
$\sigma _{n}/\sigma _{p}$ enter the estimated range
already at $Q^{2}\approx 8$~GeV${}^{2}$.
[The corresponding values of the ratio
$-G_{\rm M}^{\rm n}/G_{\rm M}^{\rm p}$,
allowed by this analysis, range between -0.2 and 0.5.]
It is important to emphasize that the ratio of the magnetic form
factors is the first observable for which the {\MCS} yields predictions
which, albeit in a slightly model-dependent way, indicate overlap with
the existing data \cite{Pla90}.
This tentative agreement occurs at data points corresponding to the
largest momentum transfers measured, where, incidentally, the
presented theoretical calculations become self-consistent.

%
%

The various model (nucleon) wave functions considered lead to
self-consistency of the perturbative contribution, meaning that
$50\%$ of the results are accumulated in regions where
$\alpha _{\rm s}^{2}\leq 0.5$, i.e., in the range of $Q^{2}$ between
$6$ and $10$~GeV${}^{2}$.

Closing this chapter we note that using other values of the r.m.s.
transverse momentum but the same set of {\DA}s, similar ``bands'' are
obtained with about the same relative widths.
Even for zero r.m.s. transverse momentum
(see Fig.~\ref{fig:G_M(Q^2)}), the ``band'' does not overlap
with the data.

Our present knowledge of the proton wave function resides only on QCD
sum-rule calculations and is thus rather limited -- contrary to the
pion case.
Therefore, we present two sets of results, such with a minimal r.m.s.
transverse momentum (corresponding to $P_{3{\rm q}}=1$), and such with
600~MeV (corresponding to $P_{3{\rm q}}\simeq 0.04$) in order to
demonstrate the role of this parameter.
The probability of the ``true'' valence Fock state wave function,
say, at $Q^{2}=10~{\rm GeV}^{2}$, is certainly not 1 but smaller,
probably close to the value of the other set.
However, according to Botts and Sterman \cite{BS89} (see also
\cite{Ste95}), one may view the Sudakov form factor as part of the wave
function, describing the perturbative part of the transverse momentum
distribution.
Accepting this idea and interpreting the case $P_{3{\rm q}}=1$ as the
soft wave function in the region $Q^{2}\simeq 1~{\rm GeV}^{2}$, the
product of wave function and Sudakov form factor at about
$Q^{2}=10~{\rm GeV}^{2}$ leads to $P_{3{\rm q}}=0.21$ and
$\langle {k_\perp}^2\rangle^{1/2}\simeq 500~{\rm MeV}$.
Thus, in order to avoid double counting, it is perhaps reasonable
to start with the intrinsic $k_{\perp}$-dependence of about
$\langle {k_{\perp}}^{2}\rangle^{1/2}\simeq 300~{\rm MeV}$ and
$P_{3{\rm q}}=1$.

We finally note that the value of the proton wave function at the
origin, $f_{\rm N}$, is also burdened by large errors, which in turn
induce further uncertainties in the form factor predictions.
In present lattice theory calculations \cite{RSS87}, the value
$f_{\rm N}^{\rm lat}=(0.29\pm 0.6)\times 10^{-3}$~GeV${}^{2}$ was found.
Literal use of such a value would lead to form factor results even
smaller than the predictions shown in Figs.~\ref{fig:strip},
\ref{fig:stripn}.

\subsubsection{Higher-order nucleon distribution amplitudes}
\label{subsubsec:higher}
\begin{figure}
\centering
\epsfig{figure=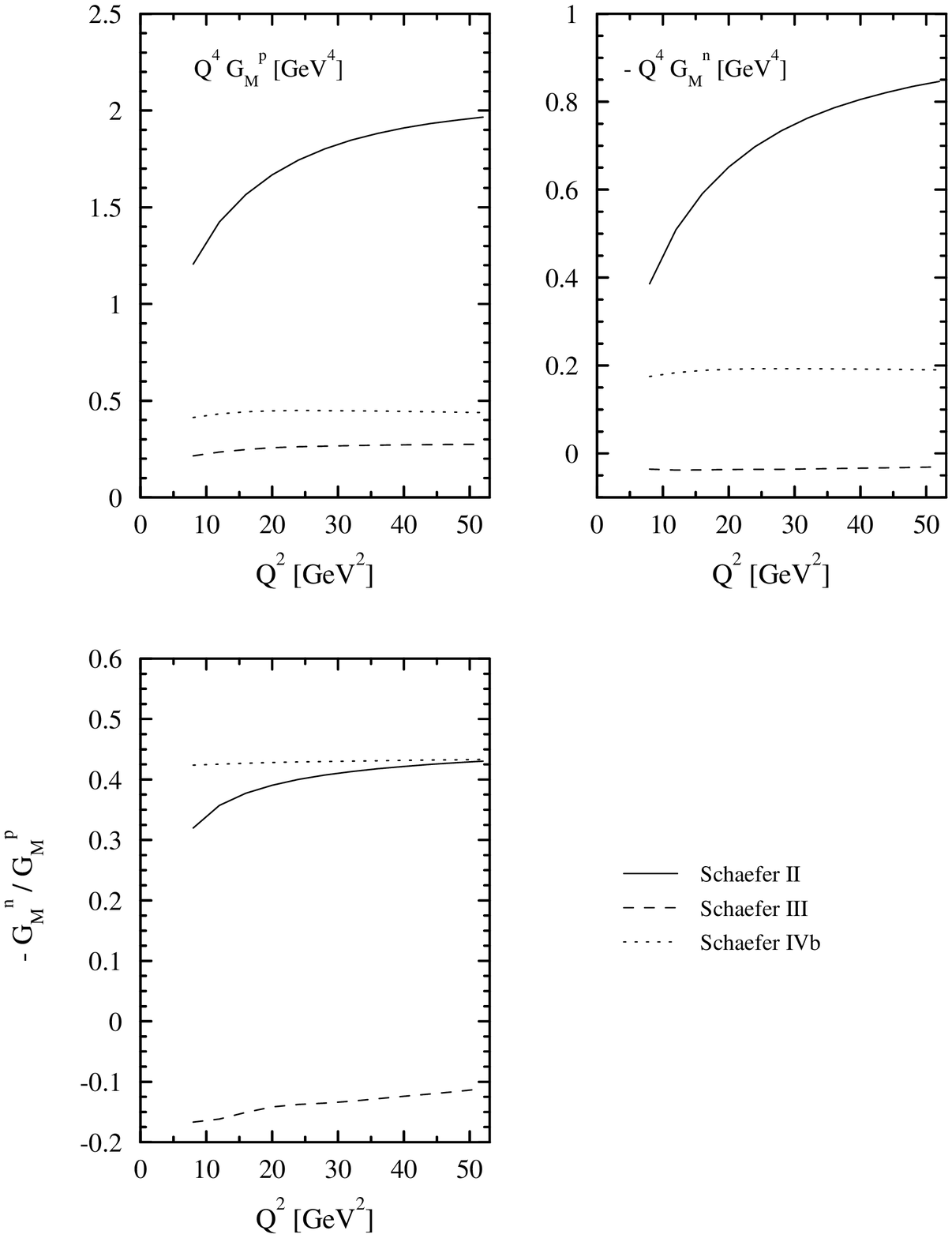,width=13.5cm,height=9.3cm,rwidth=13.5cm,rheight=9.5cm,clip=}
\vspace{0.2cm}
\caption[Sch]
         {\tenrm
          Plots for the proton and neutron magnetic form factor
          within the {\MCS} calculated with nucleon distribution
          amplitudes, determined by Sch\"afer \cite{Sch89}, which
          include third-order eigenfunctions (Appell polynomials).
          As in the previous figures, the ``MAX'' prescription and the
          value $\Lambda _{{\rm QCD}}=180$~MeV were used.
\label{fig:orderthree}}
\end{figure}

To effect that the depletion of the perturbative contribution is
not the consequence of truncating the nucleon distribution
amplitude at the level of second-order eigenfunctions,
we show in Fig.~\ref{fig:orderthree} predictions for the proton
and the neutron magnetic form factors calculated with nucleon {\DA}s,
determined by Sch\"afer \cite{Sch89}, which incorporate Appell
polynomials of order three.\footnote{The expansion coefficients
$B_{n}$ for each of these amplitudes can be readily obtained from the
corresponding moments via Eqs.~(\ref{eq:B1B2B3B4B5}),
(\ref{eq:B6B7B8B9}).}
The solid lines show the results for the amplitude Sch~II (in
Sch\"afer's notation) which deliberately incorporates third-order terms.
It is clearly obvious that both form factors $G_{\rm M}^{\rm p}$ and
$G_{\rm M}^{\rm n}$ overshoot the data and have, in addition, a wrong
$Q^{2}$ evolution.
Besides, their saturation behavior deteriorates, for they still
increase at $Q^{2}$ as large as $50$~GeV${}^{2}$ and insensitivity
to the soft region $b_{l}\Lambda _{QCD}\to 1$ sets in at
$Q^{2}\approx 17$~GeV${}^{2}$, which is a much larger scale compared to
that found for the second-order amplitudes:
$Q^{2}\le 10$~GeV${}^{2}$.
As we have discussed in detail in previous publications
\cite{Ste89,Ste94,SB93nuc,CORFU92,BS93}, and also in previous sections
of this report, the moment sum rules are not stringent enough to
exclude this sort of amplitudes.
Additional criteria have to be imposed in order to filter out
physically meaningful solutions, i.e., those which ensure dominance
of the lowest leading-order contributions.

%
%

\begin{figure}
%
\centering
\epsfig{figure=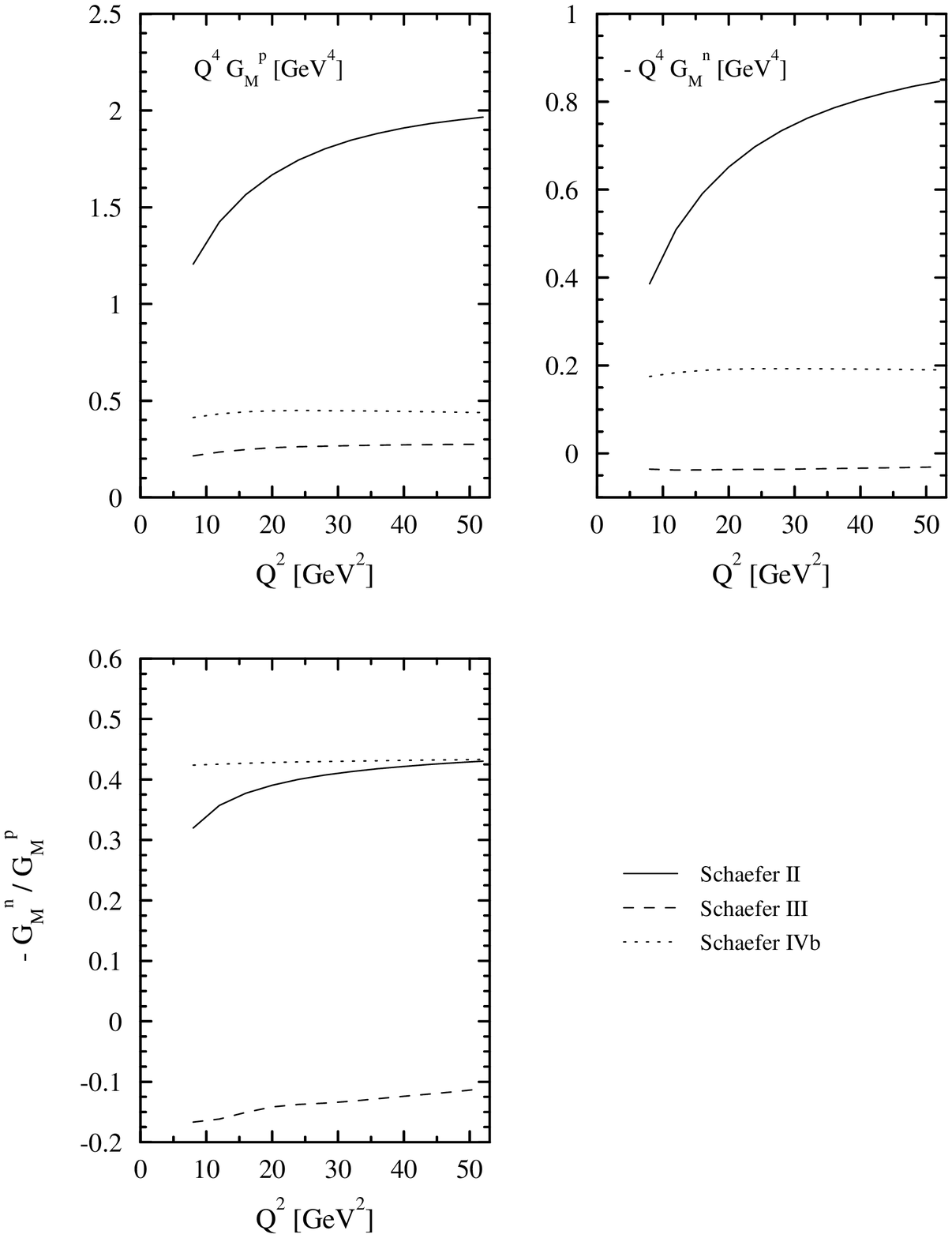,width=8.0cm,height=9.5cm,rwidth=8.0cm,rheight=9.5cm,clip=}
\hspace{2cm}
\vspace{0.2cm}
\caption[ratioSch]
         {\tenrm
          Ratio of the nucleon magnetic form factors computed with
          the Sch\"afer amplitudes vs the momentum transfer $Q^{2}$.
\label{fig:ratioSch}}
\end{figure}

Along similar lines of thought, Sch\"afer \cite{Sch89} has
determined a collection of amplitudes which, although including
third-order Appell polynomials, effectively resemble those
of polynomial order two.
The dotted lines in Fig.~\ref{fig:orderthree} exemplify the
form-factor predictions obtained with such amplitudes.
To be specific, the underlying amplitude (denoted Sch~IVb) was
determined under the proviso of a ``smoothness'' criterion with the
purpose of excluding spurious oscillations (``wiggles'') caused by
the third-order (Appell) polynomials.
This criterion is quite restrictive and, as one sees, the obtained
results have the same overall quality as those computed with
amplitudes truncated after taking into account bilinear combinations
of longitudinal momentum fractions.
This indicates that the truncation at the second polynomial
order is justifiable, since the shape of the corresponding
amplitudes is a characteristic property of the {\it entire} series
and that the errors (and amount of cutoff dependence involved)
are of sub-leading importance if properly incorporated.
For completeness, Fig.~\ref{fig:ratioSch} shows also form factors,
computed with an amplitude (termed Sch~III and represented by dashed
lines) which exhibits unphysical oscillations.
The form-factor ratio $-G_{\rm M}^{\rm n}/G_{\rm M}^{\rm p}$, as a
function of the momentum transfer $Q^{2}$, for the three amplitudes
Sch~II, Sch~III, and Sch~IVb is shown in Fig.~\ref{fig:ratioSch}.

%
%

\subsubsection{Global pattern of nucleon distribution amplitudes}
\label{subsubsec:modorbit}
\begin{figure}
\begin{picture}(0,300)
  \put(75,10){\psboxscaled{1500}{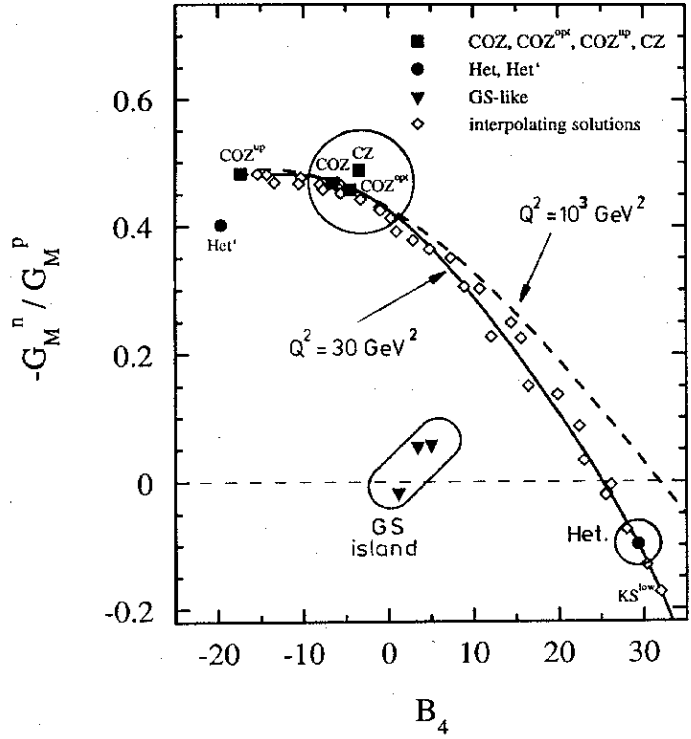}}
\end{picture}
\caption[fig:globpatt]
         {\tenrm
          The relation between the ratio
          $R\equiv |G_{\rm M}^{\rm n}|/G_{\rm M}^{\rm p}$
          of the magnetic nucleon form factors and the expansion
          coefficient $B_{4}$ of the Appell polynomial decomposition of
          the nucleon {\DA}, including the effect of the Sudakov form
          factor and intrinsic transverse-momentum dependence.
          The results are obtained at $Q^2=30$ GeV${}^2$, employing the
          ``MAX'' prescription. The superimposed solid line is an
          empirical polynomial fit similar to the original one given in
          \cite{BS93} within the {\SCS}. The dashed line serves to
          illustrate the dependence on the momentum scale
          ($Q^{2}=10^{3}$ GeV${}^{2}$).
\label{fig:sudorbit}}
\end{figure}

It is remarkable that the collective pattern of solutions to the QCD
sum rules \cite{COZ89a,KS87}, found within the {\SCS} \cite{BS93,BS94},
pertains to the inclusion of transverse-momentum contributions
comprising the Sudakov factor and those due to the intrinsic
transverse momentum \cite{JK93} (see Fig.~\ref{fig:sudorbit}).

%
%

Indeed, the solutions (nucleon {\DA}s) arrange themselves again across
an ``orbit'' in the ($B_{4},-G_{\rm M}^{\rm n}/G_{\rm M}^{\rm p}$) plane
which is, however, somewhat shifted as compared to the original one
(Fig.~\ref{fig:orbit}).
Now the ``orbit'' is slightly $Q^{2}$-dependent, though its
coherent structure is not destroyed but changed in a global way.
The solid and dashed lines correlate to the calculations within the
{\MCS} presented above at two different scales, {\it viz.}, at
$Q^{2}=30$~GeV${}^{2}$ and $Q^{2}=10^{3}$~GeV${}^{2}$, respectively.
The orbit at $Q^{2}=30$~GeV${}^{2}$ (solid line) can be characterized
by the empirical relation
$
 -G_{\rm M}^{\rm n}/G_{\rm M}^{\rm p}
=
  0.426 - 9.91 \times 10^{-3} B_{4} - 4.27 \times 10^{-4} B_{4}^{2}
  + 4.59 \times 10^{-6} B_{4}^{3} ,
$
which complies with that found within the {\SCS}.
The dashed line in Fig.~\ref{fig:sudorbit} represents a similar fit
for
$Q^{2}=10^{3}$~GeV${}^{2}$.
We observe that as $Q^{2}$ increases the modified orbit approaches
again the original one.
This observation is double-edged: it means not only that the orbit
structure is stable and can be used in both convolution schemes
in order to exclude unphysical solutions, for example, Sch~III, it
also supports the view that in the axial gauge, $e^{-S}$ operates
like a finite wave-function renormalization factor \cite{Ste95}, as
outlined above.
This conception may be used reversely to model hadron distribution
amplitudes with an improved, i.e., ``softened'' endpoint behavior
relative to those determined by QCD sum rules alone.

\section{CONCLUSIONS AND OUTLOOK}
\label{sec:concl}
We have reviewed a large body of exclusive processes which relates to
testing the validity of perturbative QCD, in particular factorization
theorems, and the universality of non-perturbative hadron distribution
amplitudes modeled on the basis of QCD sum rules.

Within the standard convolution scheme, our ``heterotic''
distribution amplitude for the nucleon \cite{SB93nuc} -- which
amalgamates the pivotal features of COZ-type \cite{COZ89a} with
GS-type \cite{GS86,GS87,Ste89} distribution amplitudes --
successfully correlates a wide variety of unrelated observables to
the data without tuning free parameters from case to case and
without using additional phenomenological constraints \cite{Ste94}.
This agreement is non-trivial; no other theoretical model matches the
available experimental data nearly as well.

By analyzing the QCD sum-rule constraints on the moments of nucleon
distribution amplitudes \cite{COZ89a,KS87} with the aid of a
``hierarchical'' $\chi ^{2}$ criterion \cite{SB93nuc} -- that takes
into account the higher accuracy \cite{CZ84b,Ste89} of lower-order
sum rules -- we were able to systematically explore the
characteristic features of distribution amplitudes complying with
the sum-rule constraints \cite{BS93}.
Our extensive analysis does reveal definite evidence for a global
pattern of such distribution amplitudes across an orbit in the plane
spanned by the ratio
$R\equiv |G_{\rm M}^{\rm n}|/G_{\rm M}^{\rm p}$ of the magnetic form
factors of the nucleon and the expansion coefficient $B_{4}$, the
latter projecting on to an antisymmetric eigenfunction of order two of
the nucleon evolution equation.
This orbit is finite, ranging from distribution amplitudes with a
ratio value $R$ around 0.1 (heterotic region) up to COZ-type solutions
with $R\lesssim 0.49$, where the orbit terminates (fixed point).
Proposed nucleon distribution amplitudes which fall outside the
orbit turn out to be unphysical, irrespectively of their functional
representation in terms of eigenfunctions.
Hence, we are inclined to believe that the orbit captures a genuine
feature of the true nucleon distribution amplitude, i.e., of the
whole eigenfunctions series, resulting in turn into an insensitivity
to particular schemes of truncation.
We were able to formulate the concept of ``heteroticity'' in
mathematical terms by introducing a ``hybridity'' angle to quantify
the mingling of COZ- and GS-type distribution amplitudes, thus
completing the classification scheme of nucleon distribution
amplitudes.

Factorizing the exclusive-scattering amplitude for the nucleon form
factors according to the modified convolution scheme -- which
includes gluon radiative corrections in the form of Sudakov-like
damping exponentials -- the orbit retains its structure and changes
only slightly as a whole.
For momentum-transfer values on the order of 10$^{3}$~GeV${}^{2}$ it
rotates back to the one obtained within the standard convolution
scheme \cite{Ste95,BJKBS95neu}.
On the phenomenological side, limiting the value of the form-factor
ratio by experiment, one could use the orbit to extract the
theoretically appropriate nucleon distribution amplitude.

Similar ideas of ``heteroticity'' could be applied to QCD sum-rule
analyses \cite{Far88,CP88} which constrain the moments of the $\Delta
^{+}$ distribution amplitude.
A theoretical model was derived \cite{SB93del}, termed again
``heterotic'', that provides the best agreement with existing data
\cite{Stu93,Stu96} on the (electromagnetic) transition form factor
$G_{\rm M}^{*}$.

Overall, the current status of the QCD sum rules program to exclusive
high-momentum processes produced a rich body of theoretical ideas and
experimental tests that could be utilized to scrutinize new
developments, like the use of nonlocal condensates.

But there are problems, both technical and conceptual.
The inversion of few moments cannot fix hadron distribution
amplitudes uniquely \cite{GS87,Ste89,Rad90}.
On the other hand, the calculation of higher-order moment
constraints is burdened by intrinsic theoretical uncertainties
\cite{CZ84b,Rad90}; e.g., power corrections grow with the moment order
thus overwhelming the perturbative contribution.
Clearly, from the theoretical point of view, a QCD sum-rule method
to calculate hadron distribution amplitudes as a {\it whole} would be
desirable, but is still to be developed.
First steps in this direction \cite{Rad90,BM95} are promising but
have to be further pursued to include baryons.

On the perturbative side, we were able to calculate a total of 55
next-to-leading eigenfunctions of the nucleon evolution equation,
tabulating results up to polynomial order $M=4$ (Tables~\ref{tab:Bs},
\ref{tab:Bsks}).
We investigated the corresponding anomalous dimensions of
leading-twist three and estimated their large-order behavior in
successive steps from $M=9$, $M=20$, $M=38$, up to $M=150$.
This behavior can be best reproduced by a logarithmic fit, thus
confirming previous expectations based on exponentiation of
one-loop results.

Repeating the form-factor calculations for the nucleon within the
modified convolution scheme \cite{LS92,Li93}, which incorporates
the effect of (soft) gluon radiative corrections by means of
Sudakov-type form factors \cite{BS89}, we showed how the
screening of $\alpha _{\rm s}$ singularities in the endpoint region can
be enforced by employing an infrared regularization prescription,
based on the assumption that quarks at large transverse separations
act coherently and thus cannot radiate soft gluons (``MAX''
prescription).
This prescription renders the calculations finite and reinstates the
validity of the perturbative treatment providing sufficient infrared
protection, since, even with evolution, the integrand in
Eq.~(\ref{eq:G_M(b)}) remains
finite \cite{BJKBS95pro,BJKBS95neu,Ste95}.
A significant feature of this treatment is that the nucleon form
factors saturate, i.e., become insensitive to the contributions
from large transverse separations.
The other choices of infrared-cutoff prescriptions (``L'',
``H-SS'')~\cite{Li93,Hye93,SS94} do not lead to saturation.

However, a heavy price is paid.
A reliable saturation and infrared protection of the form factors
are achieved at the expense of a strong reduction of the perturbative
contribution.
The damping of the nucleon form factors becomes enhanced if the
intrinsic transverse momentum dependence of the nucleon wave function
(see Fig.~\ref{fig:G_M(b_c)} and Fig.~\ref{fig:G_M(Q^2)}) is taken
into account.
This has been done by assuming a non-factorizing $x$ and
$k_{\perp}$-dependence of the wave function of the
Brodsky-Lepage-Huang type \cite{BHL83}, and fixing the value
of $\langle k^{2}_{\perp}\rangle ^{1/2}$ either via the valence quark
probability $P_{3{\rm q}}$ or by inputting the value $600$~MeV by
hand \cite{SS94}.
A remarkable finding is that the form factors calculated this way
show only a mild dependence on the particular model distribution
amplitude used.

The perturbative contribution to the form factor becomes
self-consistent in all cases for momentum transfers larger than $6$
to $10$~GeV${}^{2}$, albeit the actual value of the onset of
self-consistency depends on the particular wave function and the r.m.s.
transverse momentum chosen.
Self-consistency is defined such that $50\%$ of the result are
accumulated in regions where $\alpha _{\rm s}^{2}$ is smaller than
$0.5$.

Comparing these theoretical results with the data, it turns out that
they fall short by at least $50\%$.
This is true not only for the COZ distribution amplitude (which we have
exemplarily used as reference model) but actually for the whole
spectrum of amplitudes determined in \cite{BS93,BS94} that satisfy
existing sum-rule requirements.
Depending on the actual value of the r.m.s. transverse momentum, the
reduction of the perturbative contribution may be even stronger than
$50\%$.

This damping may perhaps be counteracted by taking into account
pairing effects in the hadron, effects which go beyond the simple
impulse approximation.
The latter assumes that the interaction of the external photon takes
place with just one parton at a time.
One may argue that non-perturbative vacuum fluctuations, e.g., due
to instantons \cite{Dor96}, give way to parton correlations and
provide additional finite renormalizing factors \cite{Ste95} greater
than 1 to bridge the gap to the experimental data at laboratory
momentum transfer.
Or perhaps higher-order corrections of perturbation theory giving
rise to a rather large K-factor of the order 2, as found for other
large-momentum transfer (inclusive) processes \cite{Ant83},
may improve the theoretical predictions.
Furthermore, the failing of the leading-order form-factor contribution
to reproduce the existing data, may be a signal that soft contributions
(non-logarithmic contributions) not accounted for so far in the
formalism should be included.
Such contributions include, e.g., improved and/or more complicated
wave functions comprising higher-twists, orbital angular momentum,
higher Fock components, quark-quark correlations (diquarks),
higher-dimensional quark/gluon condensates, quark masses, etc.
Also remainders of genuine soft contributions, like
vector-meson-dominance terms or the non-factorizing overlap of the
soft parts of the wave functions (Feynman contributions), may still
be large at accessible momentum transfers.
The rather large value of the Pauli form factor $F_{2}$ around
$10$~GeV${}^{2}$, as found experimentally (second reference
in\cite{Pla90}), indicates that sizeable higher-twist contributions
still exist in that region of momentum transfer.
One may suspect similar or even larger higher-twist contributions to
the helicity non-flip current matrix element controlling $F_{1}$ and
$G_{\rm M}$.

Once all this is accomplished, we shall probably be able to make
contact with the data in a more meaningful way.

A real theoretical breakthrough is, however, not a project that
simply reconciles form-factor calculations with experimental data.
The basic challenge is finding a way to describe an exclusive
process in regimes where the impulse approximation becomes invalid.
This happens when for some reason the interaction time becomes
comparable with that of hadron formation.
Then, speaking in terms of pictures, the dynamics is not a ``snapshot''
(the essence of the impulse approximation) but rather a ``blurred''
(i.e., ``smeared'') field configuration which extends to a large range
of scales linking the initial with the final hadron state (scale
intrusions).
The consequence is that factorization of the amplitude breaks down
and a short-distance part amenable to perturbation theory can no more
be isolated.
This is the case in the kinematic endpoint region when partons
can become very ``wee''.
The appearance of soft propagators in the hard part of the amplitude
signals infrared sensitivity and means that scale locality is lost.
These problems cannot be resolved insisting to use valence quarks.
These are not asymptotic fields and hence cannot be part of the true
asymptotic dynamics (Hamiltonian of the system).
One should use instead quark lines which are quark-gluon composites
and can inherently accommodate soft modes within an effective
approach.
The goal is to transcend perturbation theory by providing all-order
expressions in the coupling constant for dynamical quantities, like
Green and universal vertex functions.
Work in this direction is in progress \cite{Stenew,KKS95}.


\bigskip

\acknowledgements
I would like to thank Michael Bergmann, Jan Bolz, Rainer Jakob,
and Peter Kroll for valuable contributions at several stages of this
work.
It is a pleasure to thank Alexander Bakulev, Alexander Dorokhov,
Sergey Mikhailov, and Prof. Dmitrij V. Shirkov for useful discussions.
I am especially grateful to the members of The Bogoliubov Laboratory
of Theoretical Physics, JINR, Dubna for their warm hospitality during
several research stays, where and when parts of this manuscript were
prepared, and the Landau-Heisenberg-Foundation for travel grants.
Finally, I wish to express my gratitude to my wife Annemarie Stefanis
for her patience and encouragement during the course of this work, and
to thank Peter Druck for technical assistance with the figures.

\newpage   

\appendix
\section{EIGENFUNCTIONS OF THE NUCLEON EVOLUTION EQUATION}
\label{ap:nucevolkern}
The explicit form of the matrix $U_{ij,kl}$ which relates the nucleon
evolution kernel to the polynomial basis $|k \, l \rangle$ is
\begin{eqnarray}
  \frac{\hat{V}|k \, l>}{2 w(x_{i})}
& = &
  \frac{1}{2} \,
  \int_{0}^{1} [dy] \,
  \frac{V(x_{i},y_{i})}{w(x_{i})} \,
  |y_{1}^{k} y_{3}^{l} \rangle
  \cr
& = &
  x_{1}^{k} x_{3}^{l}
  \left(
               \frac{1}{(k+1)(k+2)} \,
        - \, 3 \sum_{j=2}^{k+1}\frac{1}{j} \,
        + \,   \frac{1}{(l+1)(l+2)}
        - \, 3 \sum_{j=2}^{l+1}\frac{1}{j} \right) +
\cr
&&
  \left[
        \sum_{i=1}^{k} \frac{k-i+2}{i(k+2)} x_{1}^{k-i}
        \sum_{j=0}^{i} \left( \begin{array}{c} i \cr j
\end{array}
  \right)
  (-1)^{j} \, x_{3}^{l+j} + \right.
  \cr
&& \,\ \left.
  \sum_{i=1}^{l} \frac{l-i+2}{i(l+2)} x_{3}^{l-i}
  \sum_{j=0}^{i}
  \left( \begin{array}{c} i \cr j \end{array} \right)
  (-1)^j \, x_1^{k+j}
  \right] -    \cr
&&
  \left[
        \sum_{i=1}^{l} x_{1}^{k+i}x_{3}^{l-i} \, \sum_{j=0}^{i}
  \left( \begin{array}{c} l \cr j \end{array} \right)
  \left( \begin{array}{c} l-j \cr l-i \end{array} \right) (-1)^j
  \sum_{m=2}^{k+j+1} \frac{1}{m}  \right. +
\cr
&& \,\ \left.
  \sum_{i=1}^{k} x_{3}^{l+i}x_1^{k-i} \, \sum_{j=0}{i}
  \left( \begin{array}{c} k \cr j \end{array} \right)
  \left( \begin{array}{c} k-j \cr k-i \end{array} \right) (-1)^j
  \sum_{m=2}^{l+j+1} \frac{1}{m}  \right]
\cr
& = &
  \frac{1}{2} \sum_{i,j}^{i+j\leq M} |i\, j \rangle U_{ij,kl} \;
\label{eq:ux1x3}
\end{eqnarray}
with
$w(x_{i})=x_{1}x_{2}x_{3}=x_{1}(1-x_{1}-x_{3})x_{3}$.
The operation of $\hat V$ on a polynomial $|k\, l\rangle $ of
degree $M=k+l$ projects on another polynomial of degree $i+j\leq M$.
This projection is expressed through the matrix $U$.
In the present work $U$ is represented in terms of Appell polynomials
which are defined on the triangle
$T=T(x_{1},x_{3})$ with $x_{1}>0$, $x_{3}>0$, $x_{1} + x_{3}<1$
as follows
\begin{eqnarray}
  {\cal F}_{mn} (\chi _{0},\chi _{1},\chi _{3};x_{1},x_{3}) \,
= \,
  w^{-1}(\chi _{0},\chi _{1},\chi _{3};x_{1},x_{3}) \,
  \frac{\Gamma (\chi _{1} + m) \,
  \Gamma (\chi _{3} + n)}{\Gamma (\chi _{1}) \, \Gamma (\chi _{3})}
\cr
  \frac{\partial^{m+n}}{\partial x_{1}^{m} \partial x_{3}^{n}} \,
  \left[
        x_{1}^{m+\chi _{1} - 1}\, x_{3}^{n+\chi _{3} - 1} \,
        (1-x_{1}-x_{3})^{m+n+\chi _{0} - \chi _{1} - \chi _{3}}
  \right] \; .\ \ \
\label{eq:Adgl}
\end{eqnarray}
Here $\Gamma$ denotes the $\Gamma$-function with
\begin{equation}
  \frac{\Gamma (\chi _{i} + j)}{\Gamma (\chi _{i})}
= \chi _{i}\,
  (\chi _{i}-1) \cdots (\chi _{i} + j - 1)
\label{eq:fracgammafun}
\end{equation}
and the weight function is defined by
\begin{equation}
  w(\chi _{0},\chi _{1},\chi _{3};x_1,x_3)\,
= \,
  x_{1}^{\chi _{1} - 1}\, x_{3}^{\chi _{3} - 1} \,
  (1 - x_{1} - x_{3})^{\chi _{0} - \chi _{1} - \chi _{3}} \; .
\label{eq:weight}
\end{equation}
The polynomials ${\cal F}_{mn}$ are of order $M=m+n$ and analytical,
provided the $\chi _{i}$ are real, i.e.,
\begin{equation}
  \Re{(\chi _{1})} > 0\, , \hspace{0.5cm}
  \Re{(\chi _{3})} > 0\, , \hspace{0.5cm}
  \Re{(\chi _{0})} > \Re{(\chi _{1}+\chi _{3})} -1 \; .
\label{eq:chireal}
\end{equation}
The scalar product of an arbitrary function $f(x_{1}, x_{3})$ on the
triangle $T$ with ${\cal F}_{mn}$ gives
\begin{eqnarray}
  \langle f|{\cal F}_{mn} \rangle \,
& := \,
  \frac{\Gamma (\chi _{1} + m) \,
  \Gamma (\chi _{3} + n)}{\Gamma (\chi _{1}) \,
  \Gamma (\chi _{3})}\, \int_{G}^{} \,dx_{1}\,dx_{3} \,
  f(x_{1},x_{3}) \,
  \frac{\partial ^{m+n}}{\partial x_{1}^{m} \partial x_{3}^{n}}
\cr
& \left[
        x_{1}^{m+\chi _{1} - 1}\, x_{3}^{n+\chi _{3} - 1} \,
        (1-x_{1}-x_{3})^{m+n+\chi _{0} - \chi _{1} - \chi _{3}}
  \right] \; ,
\label{eq:scalarappell}
\end{eqnarray}
where a ``bracket'' notation, resembling that in
Chapter~\ref{subsubsec:nucevoleq}, is used.
If the function $f(x_{1}, x_{3})$ is another Appell polynomial,
then the following two relations hold:
\begin{itemize}
\item (1)  \ \ $m+n \ne m^{\prime}+n^{\prime}$
\begin{equation}
  \langle {\cal F}_{m^{\prime}n^{\prime}}|{\cal F}_{mn}\rangle
\, = \, 0
\label{eq:scalarprodAppApp1}
\end{equation}
meaning that Appell polynomials of different degree are mutually
orthogonal.
\item  (2) \ \ $m+n = m^{\prime}+n^{\prime}$ \ \ \ \ \
\begin{eqnarray}
  \langle {\cal F}_{mn}|{\cal F}_{m^{\prime}n^{\prime}}\rangle
\, = \,
  (-1)^{m+n}\,\frac{\Gamma (\chi _{1} + m) \,
                    \Gamma (\chi_ {3} + n)}
                   {\Gamma (\chi _{1}) \,
                    \Gamma (\chi _{3})} \,
  \frac{\partial^{m+n}{\cal F}_{m^{\prime}n^{\prime}}}
       {\partial x_{1}^{m} \partial x_{3}^{n}}
\cr
  \int_{G}^{} \,dx_{1}\,dx_{3} \,
  \left[
        x_{1}^{m+\chi _{1} - 1}\, x_{3}^{n+\chi _{3} - 1} \,
        (1-x_{1}-x_{3})^{m+n+\chi _{0} - \chi _{1} - \chi_{3}}
  \right]
\cr
=\,
  \frac{\Gamma (\chi_ {1}) \,\Gamma (\chi _{3}) \,
        \Gamma (\chi_ {0} + m+n- \chi _{1} - \chi _{3} + 1)}
       {\Gamma (\chi_ {0} + 2m+2n+1) } \,
       (-1)^{m+n}\,
  \frac{\partial^{m+n}{\cal F}_{m^{\prime}n^{\prime}}}
       {\partial x_{1}^{m} \partial x_{3}^{n}}
\cr
\label{eq:scalarprodAppApp2}
\end{eqnarray}
meaning that Appell polynomials of the same degree are not
{\it a priori} orthogonal to each other.
\end{itemize}
\par

In the nucleon case, $M=m+n=0,1,2,3, \ldots$ equals the total
number of derivatives in the interpolating three-quark operators
between the nucleon and the vacuum.
In order to ensure that the asymptotic behavior of the nucleon {\DA}
predicted by perturbative QCD is described by the amplitude
$\Phi _{\rm as}(x_{i})=120x_{1}x_{2}x_{3}$, the weight function of the
Appell polynomials has to be $w(x_{i})=x_{1}x_{2}x_{3}$ with
$x_{2}=1-x_{1}-x_{3}$ because of momentum conservation.
Hence, $\chi _{1}=\chi_{3}=2$ and $\chi _{0}=0$, so that
the relevant Appell polynomials are
\begin{eqnarray}
  {\cal F}_{mn}^{(M)}(5,2,2;x_{1},x_{3})
& = &
  w(x_{i})^{-1}\, \frac{1}{(m+1)!(n+1)!} \,
  \frac{\partial ^{(M)}}{\partial x_{1}^{m}\partial x_{3}^{n}} \,
  \left[
        x_{1}^{m+1}\, x_{3}^{n+1}\, (1-x_{1}-x_{3})^{m+n+1}
  \right]
\nonumber \\
& = &
  \sum_{k,l}^{k+l\leq m+n} a_{kl}^{(mn)}\, |k\, l \rangle \; .
\label{eq:defAppelpoly}
\end{eqnarray}
The symmetrized Appell polynomials used in the text are defined by the
differential equation
\begin{eqnarray}
  \tilde{\cal F}_{mn}(x_{1},x_{3})
& = &
  x_{1}x_{3}(1-x_{1}-x_{3})
  \frac{\Gamma (2+m) \,\Gamma (2+n)}{2(\Gamma (2))^2} \,
\cr
& &
   \frac{\partial ^{m+n}}
        {\partial x_{1}^{m} \partial x_{3}^{n}
         \pm \partial x_{1}^{n} \partial x_{3}^{m}}
  \left[
        x_{1}^{m+1}\, x_{3}^{n1}\, (1-x_{1}-x_{3})^{m+n+1}
  \right] \; ,
\label{eq:symAppellpol}
\end{eqnarray}
where the plus sign refers to the case $m\geq n$ and the minus sign
to the case $m<n$.

\section{ELASTIC FORM FACTORS OF THE NUCLEON}
\label{ap:formfactors}
The functions $I^{p}(B_{n})$ and $I^{n}(B_{n})$, entering,
respectively, the calculation of the proton and neutron Dirac form
factors are:
\begin{eqnarray}
I^{p}&=& 1400\,{B_0}\,{B_1} + {{2000\,{{{B_1}}^2}}\over 9} +
  1800\,{B_0}\,{B_2} + {{2800\,{B_1}\,{B_2}}\over 3} +
  1200\,{{{B_2}}^2} +
  \cr &&
  6600\,{B_0}\,{B_3} +
  {{22000\,{B_1}\,{B_3}}\over 9} + 4800\,{B_2}\,{B_3} +
  {{18800\,{{{B_3}}^2}}\over 3} - {{1000\,{B_0}\,{B_4}}\over 3} -
  \cr &&
  {{2600\,{B_1}\,{B_4}}\over {27}} -
  {{2200\,{B_2}\,{B_4}}\over 9} - {{4600\,{B_3}\,{B_4}}\over 9} +
  {{2600\,{{{B_4}}^2}}\over {243}} - 1400\,{B_0}\,{B_5} -
  \cr &&
  {{3500\,{B_1}\,{B_5}}\over 9} - {{1100\,{B_2}\,{B_5}}\over 3} -
  {{5900\,{B_3}\,{B_5}}\over 3} +
  {{4100\,{B_4}\,{B_5}}\over {81}} +
  {{7700\,{{{B_5}}^2}}\over {27}} -
  \cr &&
  {{135\,{B_0}\,{B_6}}\over 8} -
  {{855\,{\sqrt{w_1}}\,{B_0}\,{B_6}}\over 8} -
  {{715\,{B_1}\,{B_6}}\over {24}} -
  {{265\,{\sqrt{w_1}}\,{B_1}\,{B_6}}\over 8} -
  \cr &&
  {{1025\,{B_2}\,{B_6}}\over 8} -
  {{625\,{\sqrt{w_1}}\,{B_2}\,{B_6}}\over 8} -
  {{3775\,{B_3}\,{B_6}}\over {24}} -
  {{4175\,{\sqrt{w_1}}\,{B_3}\,{B_6}}\over {24}} +
  \cr &&
  {{35\,{B_4}\,{B_6}}\over 3} +
  {{65\,{\sqrt{w_1}}\,{B_4}\,{B_6}}\over 9} -
  {{1025\,{B_5}\,{B_6}}\over {48}} +
  {{325\,{\sqrt{w_1}}\,{B_5}\,{B_6}}\over {16}} +
  \cr &&
  {{94901\,{{{B_6}}^2}}\over {768}} +
  {{2629\,{\sqrt{w_1}}\,{{{B_6}}^2}}\over {768}} -
  {{135\,{B_0}\,{B_7}}\over 8} +
  {{855\,{\sqrt{w_1}}\,{B_0}\,{B_7}}\over 8} -
  \cr &&
  {{715\,{B_1}\,{B_7}}\over {24}} +
  {{265\,{\sqrt{w_1}}\,{B_1}\,{B_7}}\over 8} -
  {{1025\,{B_2}\,{B_7}}\over 8} +
  {{625\,{\sqrt{w_1}}\,{B_2}\,{B_7}}\over 8} -
  \cr &&
  {{3775\,{B_3}\,{B_7}}\over {24}} +
  {{4175\,{\sqrt{w_1}}\,{B_3}\,{B_7}}\over {24}} +
  {{35\,{B_4}\,{B_7}}\over 3} -
  {{65\,{\sqrt{w_1}}\,{B_4}\,{B_7}}\over 9} -
  \cr &&
  {{1025\,{B_5}\,{B_7}}\over {48}} -
  {{325\,{\sqrt{w_1}}\,{B_5}\,{B_7}}\over {16}} -
  227\,{B_6}\,{B_7} + {{94901\,{{{B_7}}^2}}\over {768}} -
  \cr &&
  {{2629\,{\sqrt{w_1}}\,{{{B_7}}^2}}\over {768}} +
  {{8309\,{B_0}\,{B_8}}\over {44}} +
  {{109\,{\sqrt{w_2}}\,{B_0}\,{B_8}}\over {44}} +
  {{67525\,{B_1}\,{B_8}}\over {1188}} +
  \cr &&
  {{925\,{\sqrt{w_2}}\,{B_1}\,{B_8}}\over {1188}} +
  {{17197\,{B_2}\,{B_8}}\over {132}} +
  {{197\,{\sqrt{w_2}}\,{B_2}\,{B_8}}\over {132}} +
  {{120643\,{B_3}\,{B_8}}\over {396}} +
  \cr &&
  {{1643\,{\sqrt{w_2}}\,{B_3}\,{B_8}}\over {396}} -
  {{10939\,{B_4}\,{B_8}}\over {891}} -
  {{139\,{\sqrt{w_2}}\,{B_4}\,{B_8}}\over {891}} -
  {{94717\,{B_5}\,{B_8}}\over {2376}} -
  \cr &&
  {{1717\,{\sqrt{w_2}}\,{B_5}\,{B_8}}\over {2376}} -
  {{331907\,{B_6}\,{B_8}}\over {63360}} -
  {{265651\,{\sqrt{w_1}}\,{B_6}\,{B_8}}\over {63360}} -
  \cr &&
  {{2507\,{\sqrt{w_2}}\,{B_6}\,{B_8}}\over {63360}} -
  {{3451\,{\sqrt{w_1w_2}}\,{B_6}\,{B_8}}\over {63360}} -
  {{331907\,{B_7}\,{B_8}}\over {63360}} +
  \cr &&
  {{265651\,{\sqrt{w_1}}\,{B_7}\,{B_8}}\over {63360}} -
  {{2507\,{\sqrt{w_2}}\,{B_7}\,{B_8}}\over {63360}} +
  {{3451\,{\sqrt{w_1w_2}}\,{B_7}\,{B_8}}\over {63360}} +
  \cr &&
  {{2390087\,{{{B_8}}^2}}\over {348480}} +
  {{33287\,{\sqrt{w_2}}\,{{{B_8}}^2}}\over {348480}} +
  {{8309\,{B_0}\,{B_9}}\over {44}} -
  {{109\,{\sqrt{w_2}}\,{B_0}\,{B_9}}\over {44}} +
  \cr &&
  {{67525\,{B_1}\,{B_9}}\over {1188}} -
  {{925\,{\sqrt{w_2}}\,{B_1}\,{B_9}}\over {1188}} +
  {{17197\,{B_2}\,{B_9}}\over {132}} -
  {{197\,{\sqrt{w_2}}\,{B_2}\,{B_9}}\over {132}} +
  \cr &&
  {{120643\,{B_3}\,{B_9}}\over {396}} -
  {{1643\,{\sqrt{w_2}}\,{B_3}\,{B_9}}\over {396}} -
  {{10939\,{B_4}\,{B_9}}\over {891}} +
  {{139\,{\sqrt{w_2}}\,{B_4}\,{B_9}}\over {891}} -
  \cr &&
  {{94717\,{B_5}\,{B_9}}\over {2376}} +
  {{1717\,{\sqrt{w_2}}\,{B_5}\,{B_9}}\over {2376}} -
  {{331907\,{B_6}\,{B_9}}\over {63360}} +
  {{2507\,{\sqrt{w_2}}\,{B_6}\,{B_9}}\over {63360}} -
  \cr &&
  {{265651\,{\sqrt{w_1}}\,{B_6}\,{B_9}}\over {63360}} +
  {{3451\,{\sqrt{w_1w_2}}\,{B_6}\,{B_9}}\over {63360}} -
  {{331907\,{B_7}\,{B_9}}\over {63360}} +
  \cr &&
  {{265651\,{\sqrt{w_1}}\,{B_7}\,{B_9}}\over {63360}} +
  {{2507\,{\sqrt{w_2}}\,{B_7}\,{B_9}}\over {63360}} -
  {{3451\,{\sqrt{w_1w_2}}\,{B_7}\,{B_9}}\over {63360}} +
  \cr &&
  {{791\,{B_8}\,{B_9}}\over {1188}} +
  {{2390087\,{{{B_9}}^2}}\over {348480}} -
  {{33287\,{\sqrt{w_2}}\,{{{B_9}}^2}}\over {348480}} \; ,
\label{eq:IpM3}
\end{eqnarray}

\begin{eqnarray}
I_n&=& 1800\,{{{B_0}}^2} - 1400\,{B_0}\,{B_1} +
  {{2200\,{{{B_1}}^2}}\over 9} - 1800\,{B_0}\,{B_2} -
  {{2800\,{B_1}\,{B_2}}\over 3} -
  \cr &&
  200\,{{{B_2}}^2} -
  1000\,{B_0}\,{B_3} - {{22000\,{B_1}\,{B_3}}\over 9} -
  2000\,{B_2}\,{B_3} - {{17000\,{{{B_3}}^2}}\over 9} +
  \cr &&
  {{1000\,{B_0}\,{B_4}}\over 3} -
  {{2600\,{B_1}\,{B_4}}\over {27}} +
  {{2200\,{B_2}\,{B_4}}\over 9} + {{4600\,{B_3}\,{B_4}}\over 9} +
  {{2600\,{{{B_4}}^2}}\over {243}} +
  \cr &&
  {{2000\,{B_0}\,{B_5}}\over 3} +
  {{3500\,{B_1}\,{B_5}}\over 9} + {{500\,{B_2}\,{B_5}}\over 9} +
  {{6500\,{B_3}\,{B_5}}\over 9} -
  {{4100\,{B_4}\,{B_5}}\over {81}} -
  \cr &&
  {{7250\,{{{B_5}}^2}}\over {81}} +
  {{725\,{B_0}\,{B_6}}\over 8} +
  {{325\,{\sqrt{w_1}}\,{B_0}\,{B_6}}\over 8} +
  {{715\,{B_1}\,{B_6}}\over {24}} +
  {{265\,{\sqrt{w_1}}\,{B_1}\,{B_6}}\over 8} +
  \cr &&
  {{115\,{B_2}\,{B_6}}\over 8} +
  {{195\,{\sqrt{w_1}}\,{B_2}\,{B_6}}\over 8} +
  {{5815\,{B_3}\,{B_6}}\over {72}} +
  {{4295\,{\sqrt{w_1}}\,{B_3}\,{B_6}}\over {72}} -
  \cr &&
  {{35\,{B_4}\,{B_6}}\over 3} -
  {{65\,{\sqrt{w_1}}\,{B_4}\,{B_6}}\over 9} +
  {{1705\,{B_5}\,{B_6}}\over {144}} -
  {{935\,{\sqrt{w_1}}\,{B_5}\,{B_6}}\over {144}} -
  \cr &&
  {{91813\,{{{B_6}}^2}}\over {2304}} -
  {{2357\,{\sqrt{w_1}}\,{{{B_6}}^2}}\over {2304}} +
  {{725\,{B_0}\,{B_7}}\over 8} -
  {{325\,{\sqrt{w_1}}\,{B_0}\,{B_7}}\over 8} +
  \cr &&
  {{715\,{B_1}\,{B_7}}\over {24}} -
  {{265\,{\sqrt{w_1}}\,{B_1}\,{B_7}}\over 8} +
  {{115\,{B_2}\,{B_7}}\over 8} -
  {{195\,{\sqrt{w_1}}\,{B_2}\,{B_7}}\over 8} +
  \cr &&
  {{5815\,{B_3}\,{B_7}}\over {72}} -
  {{4295\,{\sqrt{w_1}}\,{B_3}\,{B_7}}\over {72}} -
  {{35\,{B_4}\,{B_7}}\over 3} +
  {{65\,{\sqrt{w_1}}\,{B_4}\,{B_7}}\over 9} +
  \cr &&
  {{1705\,{B_5}\,{B_7}}\over {144}} +
  {{935\,{\sqrt{w_1}}\,{B_5}\,{B_7}}\over {144}} +
  77\,{B_6}\,{B_7} - {{91813\,{{{B_7}}^2}}\over {2304}} +
  \cr &&
  {{2357\,{\sqrt{w_1}}\,{{{B_7}}^2}}\over {2304}} -
  {{8309\,{B_0}\,{B_8}}\over {44}} -
  {{109\,{\sqrt{w_2}}\,{B_0}\,{B_8}}\over {44}} +
  {{69973\,{B_1}\,{B_8}}\over {1188}} +
  \cr &&
  {{973\,{\sqrt{w_2}}\,{B_1}\,{B_8}}\over {1188}} -
  {{17197\,{B_2}\,{B_8}}\over {132}} -
  {{197\,{\sqrt{w_2}}\,{B_2}\,{B_8}}\over {132}} -
  {{120643\,{B_3}\,{B_8}}\over {396}} -
  \cr &&
  {{1643\,{\sqrt{w_2}}\,{B_3}\,{B_8}}\over {396}} -
  {{10939\,{B_4}\,{B_8}}\over {891}} -
  {{139\,{\sqrt{w_2}}\,{B_4}\,{B_8}}\over {891}} +
  {{94717\,{B_5}\,{B_8}}\over {2376}} +
  \cr &&
  {{1717\,{\sqrt{w_2}}\,{B_5}\,{B_8}}\over {2376}} +
  {{331907\,{B_6}\,{B_8}}\over {63360}} +
  {{265651\,{\sqrt{w_1}}\,{B_6}\,{B_8}}\over {63360}} +
  \cr &&
  {{3451\,{\sqrt{w_1w_2}}\,{B_6}\,{B_8}}\over {63360}} +
  {{331907\,{B_7}\,{B_8}}\over {63360}} -
  {{265651\,{\sqrt{w_1}}\,{B_7}\,{B_8}}\over {63360}} +
  \cr &&
  {{2507\,{\sqrt{w_2}}\,{B_7}\,{B_8}}\over {63360}} -
  {{3451\,{\sqrt{w_1w_2}}\,{B_7}\,{B_8}}\over {63360}} +
  {{4062433\,{{{B_8}}^2}}\over {580800}} +
  \cr &&
  {{169699\,{\sqrt{w_2}}\,{{{B_8}}^2}}\over {1742400}} -
  {{8309\,{B_0}\,{B_9}}\over {44}} +
  {{109\,{\sqrt{w_2}}\,{B_0}\,{B_9}}\over {44}} +
  {{2507\,{\sqrt{w_2}}\,{B_6}\,{B_8}}\over {63360}} +
  \cr &&
  {{69973\,{B_1}\,{B_9}}\over {1188}} -
  {{973\,{\sqrt{w_2}}\,{B_1}\,{B_9}}\over {1188}} -
  {{17197\,{B_2}\,{B_9}}\over {132}} +
  {{197\,{\sqrt{w_2}}\,{B_2}\,{B_9}}\over {132}} -
  \cr &&
  {{120643\,{B_3}\,{B_9}}\over {396}} +
  {{1643\,{\sqrt{w_2}}\,{B_3}\,{B_9}}\over {396}} -
  {{10939\,{B_4}\,{B_9}}\over {891}} +
  {{139\,{\sqrt{w_2}}\,{B_4}\,{B_9}}\over {891}} +
  \cr &&
  {{94717\,{B_5}\,{B_9}}\over {2376}} -
  {{1717\,{\sqrt{w_2}}\,{B_5}\,{B_9}}\over {2376}} +
  {{331907\,{B_6}\,{B_9}}\over {63360}} -
  {{2507\,{\sqrt{w_2}}\,{B_6}\,{B_9}}\over {63360}} +
  \cr &&
  {{265651\,{\sqrt{w_1}}\,{B_6}\,{B_9}}\over {63360}} -
  {{3451\,{\sqrt{w_1w_2}}\,{B_6}\,{B_9}}\over {63360}} +
  {{331907\,{B_7}\,{B_9}}\over {63360}} -
  \cr &&
  {{2507\,{\sqrt{w_2}}\,{B_7}\,{B_9}}\over {63360}} -
  {{265651\,{\sqrt{w_1}}\,{B_7}\,{B_9}}\over {63360}} +
  {{4062433\,{{{B_9}}^2}}\over {580800}} +
  \cr &&
  {{3451\,{\sqrt{w_1w_2}}\,{B_7}\,{B_9}}\over {63360}} +
  {{695\,{B_8}\,{B_9}}\over {1188}} -
  {{169699\,{\sqrt{w_2}}\,{{{B_9}}^2}}\over {1742400}} \; ,
\label{eq:InM3}
\end{eqnarray}
where $w_{1}=97$ and $w_{2}=4801$.

Particularly useful expressions are obtained by recasting these
formulae in terms of strict moments
$\Phi _{N}^{(n_{1}0\,n_{3})}\equiv \left(n_{1}0\,n_{3}\right)$
via Eq.~(\ref{eq:magic}).
The results up to order $M=2$ (in correspondence to the nucleon
distribution amplitudes on the orbit) are
\begin{eqnarray}
  I_{p}
& = &
  196\biggl[  4899 - 9851 \,(001) + 47231 \,(001)^{2} + 24954 \,(002)
            - 163098 \,(001)(002)
\nonumber \\
& & +
              138696 \,(002)^{2} - 29425 \,(100) + 95825 \,(001)(100)
            - 175500 \,(002)(100) + 67625 \,(100)^{2}
\nonumber \\
& & +
              20790 \,(101) + 30720 \,(001)(101) - 30780 \,(002)(101)
            - 22350 \,(100)(101)
\nonumber \\
& & +
              37800 \,(101)^{2} + 29310 \,(200) - 132120 \,(001)(200)
            + 231780 \,(002)(200)
\nonumber \\
& & -
              161550 \,(100)(200) + 101400 \,(200)^{2}
     \biggr] \; ,
\label{eq:Ipmom}
\end{eqnarray}
\begin{eqnarray}
  I_{n}
& = &
  490\biggl[- 485 - 982 \,(001) + 3799 \,(001)^{2} - 5028 \,(002)
            + 30300 \,(001)(002) - 26544 \,(002)^{2}
\nonumber \\
& & +
              4792 \,(100) - 16184 \,(001)(100) + 34176 \,(002)(100)
  -           12680 \,(100)^{2} - 1728 \,(101)
\nonumber \\
& & -
              6744 \,(001)(101) + 5976 \,(002)(101) + 1200 \,(100)(101)
            - 3420 \,(101)^{2} - 5400 \,(200)
\nonumber \\
& & +
              26880 \,(001)(200) - 44520 \,(002)(200)
            + 30720 \,(100)(200) + 2520 \,(101)(200)
\nonumber \\
& & -
              19020 \,(200)^{2}
     \biggr] \; .
\label{eq:Inmom}
\end{eqnarray}
Similar expressions can be obtained also for the other nucleon form
factors displayed below.

The axial form factor of the nucleon in terms of nucleon
eigenfunctions, including next-to-leading ($M=3$) contributions, is
\begin{eqnarray}
Q^4g_A(Q^2) &=& \left(\frac{16\pi\bar{\alpha}_S}{3}\right)^2 |f_N|^2
\left[{{75\,{{{ B_0}}^2}}\over 4} + {{175\,{ B_0}\,{ B_1}}\over 6} +
  {{25\,{{{ B_1}}^2}}\over {108}} + {{25\,{ B_0}\,{ B_2}}\over 2} +
\right.   \cr && \left.
  {{175\,{ B_1}\,{ B_2}}\over 9} + {{75\,{{{ B_2}}^2}}\over 4} +
  {{625\,{ B_0}\,{ B_3}}\over 6} +
  {{1375\,{ B_1}\,{ B_3}}\over {27}} +
  {{125\,{ B_2}\,{ B_3}}\over 2} +
\right.   \cr && \left.
  {{9625\,{{{ B_3}}^2}}\over {108}} -
  {{125\,{ B_0}\,{ B_4}}\over {18}} -
  {{275\,{ B_2}\,{ B_4}}\over {54}} -
  {{575\,{ B_3}\,{ B_4}}\over {54}} -
  {{625\,{ B_0}\,{ B_5}}\over {36}} -
\right.   \cr && \left.
  {{875\,{ B_1}\,{ B_5}}\over {108}} -
  {{625\,{ B_2}\,{ B_5}}\over {108}} -
  {{2875\,{ B_3}\,{ B_5}}\over {108}} +
  {{1025\,{ B_4}\,{ B_5}}\over {972}} +
  {{15625\,{{{ B_5}}^2}}\over {3888}} +
\right.   \cr && \left.
  {{125\,{ B_0}\,{ B_6}}\over {192}} -
  {{275\,{\sqrt{97}}\,{ B_0}\,{ B_6}}\over {192}} -
  {{715\,{ B_1}\,{ B_6}}\over {1152}} -
  {{265\,{\sqrt{97}}\,{ B_1}\,{ B_6}}\over {384}} -
\right.   \cr && \left.
  {{1195\,{ B_2}\,{ B_6}}\over {576}} -
  {{635\,{\sqrt{97}}\,{ B_2}\,{ B_6}}\over {576}} -
  {{3265\,{ B_3}\,{ B_6}}\over {1728}} -
  {{4145\,{\sqrt{97}}\,{ B_3}\,{ B_6}}\over {1728}} +
\right.   \cr && \left.
  {{35\,{ B_4}\,{ B_6}}\over {144}} +
  {{65\,{\sqrt{97}}\,{ B_4}\,{ B_6}}\over {432}} -
  {{95\,{ B_5}\,{ B_6}}\over {384}} +
  {{985\,{\sqrt{97}}\,{ B_5}\,{ B_6}}\over {3456}} +
\right.   \cr && \left.
  {{31891\,{{{ B_6}}^2}}\over {18432}} +
  {{899\,{\sqrt{97}}\,{{{ B_6}}^2}}\over {18432}} +
  {{125\,{ B_0}\,{ B_7}}\over {192}} +
  {{275\,{\sqrt{97}}\,{ B_0}\,{ B_7}}\over {192}} -
\right.   \cr && \left.
  {{715\,{ B_1}\,{ B_7}}\over {1152}} +
  {{265\,{\sqrt{97}}\,{ B_1}\,{ B_7}}\over {384}} -
  {{1195\,{ B_2}\,{ B_7}}\over {576}} +
  {{635\,{\sqrt{97}}\,{ B_2}\,{ B_7}}\over {576}} -
\right.   \cr && \left.
  {{3265\,{ B_3}\,{ B_7}}\over {1728}} +
  {{4145\,{\sqrt{97}}\,{ B_3}\,{ B_7}}\over {1728}} +
  {{35\,{ B_4}\,{ B_7}}\over {144}} -
  {{65\,{\sqrt{97}}\,{ B_4}\,{ B_7}}\over {432}} -
\right.   \cr && \left.
  {{95\,{ B_5}\,{ B_7}}\over {384}} -
  {{985\,{\sqrt{97}}\,{ B_5}\,{ B_7}}\over {3456}} -
  {{113\,{ B_6}\,{ B_7}}\over {36}} +
  {{31891\,{{{ B_7}}^2}}\over {18432}} -
\right.   \cr && \left.
  {{899\,{\sqrt{97}}\,{{{ B_7}}^2}}\over {18432}} +
  {{8309\,{ B_0}\,{ B_8}}\over {2112}} +
  {{109\,{\sqrt{4801}}\,{ B_0}\,{ B_8}}\over {2112}} +
  {{17\,{ B_1}\,{ B_8}}\over {792}} +
\right.   \cr && \left.
  {{{\sqrt{4801}}\,{ B_1}\,{ B_8}}\over {2376}} +
  {{17197\,{ B_2}\,{ B_8}}\over {6336}} +
  {{197\,{\sqrt{4801}}\,{ B_2}\,{ B_8}}\over {6336}} +
\right.   \cr && \left.
  {{120643\,{ B_3}\,{ B_8}}\over {19008}} +
  {{1643\,{\sqrt{4801}}\,{ B_3}\,{ B_8}}\over {19008}} -
  {{94717\,{ B_5}\,{ B_8}}\over {114048}} -
\right.   \cr && \left.
  {{1717\,{\sqrt{4801}}\,{ B_5}\,{ B_8}}\over {114048}} -
  {{331907\,{ B_6}\,{ B_8}}\over {3041280}} -
  {{265651\,{\sqrt{97}}\,{ B_6}\,{ B_8}}\over {3041280}} -
\right.   \cr && \left.
  {{2507\,{\sqrt{4801}}\,{ B_6}\,{ B_8}}\over {3041280}} -
  {{3451\,{\sqrt{w_1w_2}}\,{ B_6}\,{ B_8}}\over {3041280}} -
  {{331907\,{ B_7}\,{ B_8}}\over {3041280}} +
\right.   \cr && \left.
  {{265651\,{\sqrt{97}}\,{ B_7}\,{ B_8}}\over {3041280}} -
  {{2507\,{\sqrt{4801}}\,{ B_7}\,{ B_8}}\over {3041280}} +
  {{3451\,{\sqrt{w_1w_2}}\,{ B_7}\,{ B_8}}\over {3041280}} +
\right.   \cr && \left.
  {{3701\,{{{ B_8}}^2}}\over {2613600}} +
  {{17\,{\sqrt{4801}}\,{{{ B_8}}^2}}\over {871200}} +
  {{8309\,{ B_0}\,{ B_9}}\over {2112}} -
  {{109\,{\sqrt{4801}}\,{ B_0}\,{ B_9}}\over {2112}} +
\right.   \cr && \left.
  {{17\,{ B_1}\,{ B_9}}\over {792}} -
  {{{\sqrt{4801}}\,{ B_1}\,{ B_9}}\over {2376}} +
  {{17197\,{ B_2}\,{ B_9}}\over {6336}} -
  {{197\,{\sqrt{4801}}\,{ B_2}\,{ B_9}}\over {6336}} +
\right.   \cr && \left.
  {{120643\,{ B_3}\,{ B_9}}\over {19008}} -
  {{1643\,{\sqrt{4801}}\,{ B_3}\,{ B_9}}\over {19008}} -
  {{94717\,{ B_5}\,{ B_9}}\over {114048}} +
\right.   \cr && \left.
  {{1717\,{\sqrt{4801}}\,{ B_5}\,{ B_9}}\over {114048}} -
  {{331907\,{ B_6}\,{ B_9}}\over {3041280}} -
  {{265651\,{\sqrt{97}}\,{ B_6}\,{ B_9}}\over {3041280}} +
\right.   \cr && \left.
  {{2507\,{\sqrt{4801}}\,{ B_6}\,{ B_9}}\over {3041280}} +
  {{3451\,{\sqrt{w_1w_2}}\,{ B_6}\,{ B_9}}\over {3041280}} -
  {{331907\,{ B_7}\,{ B_9}}\over {3041280}} +
\right.   \cr && \left.
  {{265651\,{\sqrt{97}}\,{ B_7}\,{ B_9}}\over {3041280}} +
  {{2507\,{\sqrt{4801}}\,{ B_7}\,{ B_9}}\over {3041280}} -
  {{3451\,{\sqrt{w_1w_2}}\,{ B_7}\,{ B_9}}\over {3041280}} -
\right.   \cr && \left.
  {{{ B_8}\,{ B_9}}\over {1188}} +
  {{3701\,{{{ B_9}}^2}}\over {2613600}} -
  {{17\,{\sqrt{4801}}\,{{{ B_9}}^2}}\over {871200}} \right] \; .
\label{eq:gAM3}
\end{eqnarray}
This expression verifies the leading-order results obtained by
Carlson and Poor \cite{CP86}.

\newpage   

\newpage   

\end{document}